\documentclass[twocolumn]{aastex62}

\usepackage{natbib}
\usepackage[style=american]{csquotes}
\usepackage{epsfig}
\usepackage{apjfonts}
\usepackage{lmodern} 
\usepackage{graphicx}
\usepackage{tabularx}   
\usepackage{xcolor}   
\usepackage{mathrsfs}
\usepackage{amsmath,amssymb}
\usepackage{verbatim}
\usepackage{wrapfig}
\usepackage{enumitem}
\usepackage{longtable}
\usepackage{bm}
\usepackage{threeparttable}
\usepackage{booktabs}
\usepackage{multirow}
\usepackage{comment}
\usepackage{hyperref}

\usepackage[morefloats=256]{morefloats}
\definecolor{cerulean}{rgb}{0,0.482,0.655}

\definecolor{darkmagenta}{rgb}{0.55, 0.0, 0.55}
 
\definecolor{warmorange}{rgb}{1.0, 0.49, 0.0}

\newcommand{\mps}{m s$^{-1}$}

\def\mps{m/s}

\def\leq{\leqslant}
\def\geq{\geqslant}

\shorttitle{MTS Mid-Survey}
\shortauthors{Teske, Wang, \& Wolfgang et al.}

\begin{document}

\title{The Magellan-TESS Survey I: Survey Description and Mid-Survey Results \footnote{This paper includes data gathered with the 6.5 meter Magellan Telescopes located at Las Campanas Observatory, Chile.} \footnote{Some of the data presented herein were obtained at the W. M. Keck Observatory, which is operated as a scientific partnership among the California Institute of Technology, the University of California and the National Aeronautics and Space Administration. The Observatory was made possible by the generous financial support of the W. M. Keck Foundation.}}

\correspondingauthor{Johanna Teske}
\email{jteske@carnegiescience.edu}

\correspondingauthor{Sharon~Xuesong~Wang}
\email{sharonw@mail.tsinghua.edu.cn}


\author{Johanna Teske} 
\altaffiliation{Much of this work was completed while this author \\ was a NASA Hubble Fellow at the Observatories of \\ the Carnegie Institution for Science.}
\affiliation{Earth and Planets Laboratory, Carnegie Institution for Science, 5241 Broad Branch Road, NW, Washington, DC 20015, USA}
\affil{Observatories of the Carnegie Institution for Science, 813 Santa Barbara Street, Pasadena, CA 91101, USA}

\author{Sharon~Xuesong~Wang}
\affil{Department of Astronomy, Tsinghua University, Beijing 100084, People's Republic of China}
\affil{Observatories of the Carnegie Institution for Science, 813 Santa Barbara Street, Pasadena, CA 91101, USA}

\author{Angie Wolfgang}
\affiliation{Department of Astronomy \& Astrophysics, The Pennsylvania State University, 525 Davey Laboratory, University Park, PA 16802, USA}
\affiliation{Center for Exoplanets and Habitable Worlds, The Pennsylvania State University, 525 Davey Laboratory, University Park, PA 16802, USA}

\author[0000-0002-4503-9705]{Tianjun~Gan} 
\affil{Department of Astronomy, Tsinghua University, Beijing 100084, People's Republic of China}

\author{Mykhaylo Plotnykov} 
\affiliation{Department of Physics, University of Toronto, Toronto, ON M5S 3H4, Canada}

\nocollaboration{}
\author[0000-0002-5080-4117]{David J. Armstrong} 
\affiliation{Centre for Exoplanets and Habitability, University of Warwick, Gibbet Hill Road, Coventry, CV4 7AL, UK}
\affiliation{Department of Physics, University of Warwick, Gibbet Hill Road, Coventry CV4 7AL, UK}

\author{R. Paul Butler}
\affiliation{Earth and Planets Laboratory, Carnegie Institution for Science, 5241 Broad Branch Road, NW, Washington, DC 20015, USA}

\author{Bryson Cale}  
\affiliation{George Mason University, 4400 University Dr, Fairfax, VA 22030}

\author{Jeffrey D. Crane}
\affil{Observatories of the Carnegie Institution for Science, 813 Santa Barbara Street, Pasadena, CA 91101, USA}

\author{Ward Howard}
\affiliation{Department of Physics and Astronomy, University of North Carolina at Chapel Hill, Chapel Hill, NC 27599-3255, USA}

\author[0000-0002-4625-7333]{Eric L. N. Jensen} 
\affiliation{Dept.\ of Physics \& Astronomy, Swarthmore College, Swarthmore PA 19081, USA}

\author{Nicholas Law}
\affiliation{Department of Physics and Astronomy, University of North Carolina at Chapel Hill, Chapel Hill, NC 27599-3255, USA}

\author{Stephen A. Shectman}
\affil{Observatories of the Carnegie Institution for Science, 813 Santa Barbara Street, Pasadena, CA 91101, USA}

\author{Peter Plavchan} 
\affiliation{George Mason University, 4400 University Dr, Fairfax, VA 22030}

\author{Diana Valencia} 
\affiliation{Department of Physical \& Environmental Sciences, University of Toronto at Scarborough, Toronto, ON M1C 1A4, Canada}
\affiliation{Department of Astronomy \& Astrophysics, University of Toronto, Toronto, ON M5S 3H4, Canada}

\author[0000-0001-7246-5438]{Andrew~Vanderburg} 
\altaffiliation{NASA Sagan Fellow}
\affiliation{Department of Astronomy, University of Wisconsin-Madison, Madison, WI 53706, USA}

\nocollaboration{}

\author{George R. Ricker}
\affiliation{Kavli Institute for Astrophysics and Space Research, Massachusetts Institute of Technology, Cambridge, MA 02139, USA}

\author{Roland Vanderspek}
\affiliation{Kavli Institute for Astrophysics and Space Research, Massachusetts Institute of Technology, Cambridge, MA 02139, USA}

\author{David W. Latham}
\affiliation{Center for Astrophysics \textbar \ Harvard \& Smithsonian, 60 Garden Street, Cambridge, MA 02138, USA}

\author[0000-0002-6892-6948]{Sara Seager}
\affiliation{Department of Physics and Kavli Institute for Astrophysics and Space Research, Massachusetts Institute of Technology, Cambridge, MA 02139, USA}
\affiliation{Department of Earth, Atmospheric and Planetary Sciences, Massachusetts Institute of Technology, Cambridge, MA 02139, USA}
\affiliation{Department of Aeronautics and Astronautics, MIT, 77 Massachusetts Avenue, Cambridge, MA 02139, USA}

\author{Joshua N. Winn}
\affiliation{Department of Astrophysical Sciences, Princeton University, Princeton, NJ 08544, USA}

\author{Jon M. Jenkins}
\affiliation{NASA  Ames  Research  Center,  Moffett  Field,  CA  94035, USA }

\nocollaboration{}

\author[0000-0002-0601-6199]{Vardan Adibekyan} 
\affiliation{Instituto de Astrof\'isica e Ci\^{e}ncias do Espa\c co, Universidade do Porto, CAUP, Rua das Estrelas, 4150-762 Porto, Portugal}

\author[0000-0002-5971-9242]{David Barrado}	
\affiliation{Centro de Astrobiolog\'ia (CAB, CSIC-INTA), Depto. de Astrof\'isica, ESAC campus 28692 Villanueva de la Ca\~nada (Madrid), Spain}

\author[0000-0003-2434-3625]{Susana C. C. Barros}
\affiliation{Instituto de Astrof\'isica e Ci\^{e}ncias do Espa\c co, Universidade do Porto, CAUP, Rua das Estrelas, 4150-762 Porto, Portugal}
\affiliation{Departamento de Fisica e Astronomia, Faculdade de Ci\^{e}ncias, Universidade do Porto, Rua Campo Alegre, 4169-007 Porto, Portugal}

\author[0000-0001-6285-9847]{Zouhair Benkhaldoun} 
\affiliation{Oukaimeden Observatory, High Energy Physics and Astrophysics Laboratory, Cadi Ayyad University, Marrakech, Morocco }

\author[0000-0003-1098-2422]{David J. A. Brown}  
\affiliation{Centre for Exoplanets and Habitability, University of Warwick, Gibbet Hill Road, Coventry, CV4 7AL, UK}
\affiliation{Department of Physics, University of Warwick, Gibbet Hill Road, Coventry CV4 7AL, UK}

\author[0000-0001-7904-4441]{Edward M. Bryant} 
\affiliation{Department of Physics, University of Warwick, Gibbet Hill Road, Coventry CV4 7AL, UK}
\affiliation{Centre for Exoplanets and Habitability, University of Warwick, Gibbet Hill Road, Coventry, CV4 7AL, UK}

\author{Jennifer Burt} 
\affiliation{Jet Propulsion Laboratory, California Institute of Technology, 4800 Oak Grove Drive, Pasadena, CA 91109, USA}

\author[0000-0003-1963-9616]{Douglas A. Caldwell} 
\affiliation{Carl Sagan Center, SETI Institute, Moffett Field, CA, United States; NASA Ames Research Center, Moffett Field, CA, USA}

\author[0000-0002-9003-484X]{David Charbonneau} 
\affiliation{Center for Astrophysics \textbar \ Harvard \& Smithsonian, 60 Garden Street, Cambridge, MA 02138, USA}

\author[0000-0001-5383-9393]{Ryan Cloutier} 
\affiliation{Center for Astrophysics \textbar \ Harvard \& Smithsonian, 60 Garden Street, Cambridge, MA 02138, USA}
\altaffiliation{Banting Fellow}

\author[0000-0001-6588-9574]{Karen A.\ Collins} 
\affiliation{Center for Astrophysics \textbar \ Harvard \& Smithsonian, 60 Garden Street, Cambridge, MA 02138, USA}

\author[0000-0003-2781-3207]{Kevin I.\ Collins} 
\affiliation{George Mason University, 4400 University Drive, Fairfax, VA, 22030 USA}

\author{Knicole~D.~Colon} 
\affiliation{NASA Goddard Space Flight Center, Exoplanets and Stellar Astrophysics Laboratory (Code 667), Greenbelt, MD 20771, USA}

\author[0000-0003-2239-0567]{Dennis M.\ Conti} 
\affiliation{American Association of Variable Star Observers, 49 Bay State Road, Cambridge, MA 02138, USA}

\author[0000-0001-7918-0355]{Olivier D. S. Demangeon} 
\affiliation{Instituto de Astrof\'isica e Ci\^{e}ncias do Espa\c co, Universidade do Porto, CAUP, Rua das Estrelas, 4150-762 Porto, Portugal}
\affiliation{Departamento de Fisica e Astronomia, Faculdade de Ci\^{e}ncias, Universidade do Porto, Rua Campo Alegre, 4169-007 Porto, Portugal}

\author[0000-0003-3773-5142]{Jason D.\ Eastman} 
\affiliation{Center for Astrophysics \textbar \ Harvard \& Smithsonian, 60 Garden St, Cambridge, MA 02138, USA}

\author{Mohammed Elmufti} 
\affiliation{George Mason University, 4400 University Dr, Fairfax, VA 22030}

\author{Fabo Feng} 
\affiliation{Earth and Planets Laboratory, Carnegie Institution for Science, 5241 Broad Branch Road, NW, Washington, DC 20015, USA}

\author{Erin Flowers} 
\altaffiliation{NSF GRFP Fellow}
\affiliation{Department of Astrophysical Sciences, Princeton University, Princeton, NJ 08544, USA}

\author[0000-0002-5169-9427]{Natalia~M.~Guerrero} 
\affiliation{Department of Physics and Kavli Institute for Astrophysics and Space Research, Massachusetts Institute of Technology, Cambridge, MA 02139, USA}

\author[0000-0002-0417-1902]{Saeed Hojjatpanah} 
\affiliation{Instituto de Astrof\'isica e Ci\^{e}ncias do Espa\c co, Universidade do Porto, CAUP, Rua das Estrelas, 4150-762 Porto, Portugal}
\affiliation{Departamento de Fisica e Astronomia, Faculdade de Ci\^{e}ncias, Universidade do Porto, Rua Campo Alegre, 4169-007 Porto, Portugal}

\author{Jonathan M. Irwin} 
\affiliation{Center for Astrophysics \textbar \ Harvard \& Smithsonian, 60 Garden Street, Cambridge, MA 02138, USA}

\author{Giovanni Isopi} 
\affiliation{Campo Catino Astronomical Observatory, Regione Lazio, Guarcino (FR), 03010 Italy}

\author[0000-0003-3742-1987]{Jorge Lillo-Box} 
\affiliation{Centro de Astrobiolog\'ia (CAB, CSIC-INTA), Depto. de Astrof\'isica, ESAC campus 28692 Villanueva de la Ca\~nada (Madrid), Spain}

\author{Franco Mallia}  
\affiliation{Campo Catino Astronomical Observatory, Regione Lazio, Guarcino (FR), 03010 Italy}

\author[0000-0001-8879-7138]{Bob Massey} 
\affiliation{Villa '39 Observatory, Landers, CA 92285, USA}

\author{Mayuko Mori} 
\affiliation{The University of Tokyo, 7-3-1 Hongo, Bunkyō, Tokyo 113-8654, Japan}

\author[0000-0001-7106-4683]{Susan E. Mullally} 
\affiliation{Space Telescope Science Institute, 3700 San Martin Drive, Baltimore, MD 21218, USA}

\author[0000-0001-8511-2981]{Norio Narita} 
\affiliation{Komaba Institute for Science, The University of Tokyo, 3-8-1 Komaba, Meguro, Tokyo 153-8902, Japan}
\affiliation{JST, PRESTO, 3-8-1 Komaba, Meguro, Tokyo 153-8902, Japan}
\affiliation{Astrobiology Center, 2-21-1 Osawa, Mitaka, Tokyo 181-8588, Japan}
\affiliation{Instituto de Astrof\'{i}sica de Canarias (IAC), 38205 La Laguna, Tenerife, Spain}

\author[0000-0003-1510-8981]{Taku Nishiumi} 
\affiliation{Department of Astronomical Science, The Graduated University for Advanced Studies, SOKENDAI, 2-21-1, Osawa, Mitaka, Tokyo, 181-8588, Japan}
\affiliation{Astrobiology Center, 2-21-1 Osawa, Mitaka, Tokyo 181-8588, Japan}

\author[0000-0002-5899-7750]{Ares Osborn} 
\affiliation{Department of Physics, University of Warwick, Gibbet Hill Road, Coventry CV4 7AL, UK}
\affiliation{Centre for Exoplanets and Habitability, University of Warwick, Gibbet Hill Road, Coventry, CV4 7AL, UK}

\author[0000-0001-8120-7457]{Martin Paegert} 
\affiliation{Center for Astrophysics \textbar \ Harvard \& Smithsonian, 60 Garden Street, Cambridge, MA 02138, USA}

\author{Jerome Pitogo de Leon} 
\affiliation{The University of Tokyo, 7-3-1 Hongo, Bunkyō, Tokyo 113-8654, Japan}

\author[0000-0002-8964-8377]{Samuel N. Quinn} 
\affiliation{Center for Astrophysics \textbar \ Harvard \& Smithsonian, 60 Garden Street, Cambridge, MA 02138, USA}

\author{Michael Reefe} 
\affiliation{George Mason University, 4400 University Dr, Fairfax, VA 22030}

\author[0000-0001-8227-1020]{Richard P. Schwarz} 
\affiliation{Patashnick Voorheesville Observatory, Voorheesville, NY 12186, USA}

\author[0000-0002-1836-3120]{Avi Shporer} 
\affiliation{Department of Physics and Kavli Institute for Astrophysics and Space Research, Massachusetts Institute of Technology, Cambridge, MA 02139, USA}

\author{Abderahmane Soubkiou} 
\affiliation{Oukaimeden Observatory, High Energy Physics and Astrophysics Laboratory, Cadi Ayyad University, Marrakech, Morocco }

\author[0000-0001-9047-2965]{S\'ergio G. Sousa} 
\affiliation{Instituto de Astrof\'isica e Ci\^{e}ncias do Espa\c co, Universidade do Porto, CAUP, Rua das Estrelas, 4150-762 Porto, Portugal}

\author[0000-0003-2163-1437]{Chris Stockdale} 
\affiliation{Hazelwood Observatory, Australia}

\author[0000-0002-7823-1090]{Paul A. Str\o m} 
\affiliation{Department of Physics, University of Warwick, Gibbet Hill Road, Coventry CV4 7AL, UK}
\affiliation{Centre for Exoplanets and Habitability, University of Warwick, Gibbet Hill Road, Coventry, CV4 7AL, UK}

\author[0000-0001-5603-6895]{Thiam-Guan Tan} 
\affiliation{Perth Exoplanet Survey Telescope, Perth, Western Australia}

\author[0000-0002-7563-7618]{Jiaxin Tang} 
\affiliation{Department of Astronomy, Tsinghua University, Beijing 100084, People's Republic of China}

\author{Peter Tenenbaum} 
\affiliation{Carl Sagan Center, SETI Institute, Moffett Field, CA, United States; NASA Ames Research Center, Moffett Field, CA, USA}

\author[0000-0003-1452-2240]{Peter J.\ Wheatley} 
\affiliation{Department of Physics, University of Warwick, Gibbet Hill Road, Coventry CV4 7AL, UK}
\affiliation{Centre for Exoplanets and Habitability, University of Warwick, Gibbet Hill Road, Coventry, CV4 7AL, UK}

\author{Justin Wittrock} 
\affiliation{George Mason University, 4400 University Dr, Fairfax, VA 22030}

\author[0000-0003-4755-584X]{Daniel A. Yahalomi} 
\affiliation{Department of Astronomy, Columbia University, 550 W 120th Street, New York, NY 10027, USA}
\affiliation{Center for Astrophysics \textbar \ Harvard \& Smithsonian, 60 Garden Street, Cambridge, MA 02138, USA}

\author{Farzaneh Zohrabi} 
\affiliation{LSU Department of Physics \& Astronomy, 202 Nicholson Hall, Baton Rouge, LA 70803}

\begin{abstract}

 \textit{Kepler} revealed that roughly one-third of Sun-like stars host planets orbiting within 100 days and between the size of Earth and Neptune. How do these planets form, what are they made of, and do they represent a continuous population or multiple populations? To help address these questions, we began the Magellan-TESS Survey (MTS), which uses Magellan II/PFS to obtain radial velocity (RV) masses of 30 TESS-detected exoplanets and develops an analysis framework that connects observed planet distributions to underlying populations. In the past, small planet RV measurements have been challenging to obtain due to host star faintness and low RV semi-amplitudes, and challenging to interpret due to the potential biases in target selection and observation planning decisions. The MTS attempts to minimize these biases by focusing on bright TESS targets and employing a quantitative selection function and observing strategy. In this paper, we (1) describe our motivation and survey strategy, (2) present our first catalog of planet density constraints for 27 TESS Objects of Interest (TOIs; 22 in our population analysis sample, 12 that are members of the same systems), and (3) employ a hierarchical Bayesian model to produce preliminary constraints on the mass-radius (M-R) relation. We find that the biases causing previous M-R relations to predict fairly high masses at $1~R_\oplus$ have been reduced. This work can inform more detailed studies of individual systems and offer a framework that can be applied to future RV surveys with the goal of population inferences.

\end{abstract}

\section{Introduction \label{sec:intro}}

\subsection{Background on Small Planet Formation}

The \textit{Kepler} mission revolutionized our understanding of planet formation by indicating that small (1-3~R$_{\oplus}$), relatively short period ($P <100$ days) planets are much more frequent around solar-type stars in the Galaxy than larger, Saturn and Jupiter-sized planets \citep{howard2012,dressing&charbonneau2013, fressin2013, petigura2013, burke2015}. There is great interest in understanding how similar or different 1-3~R$_{\oplus}$ planets are from the solar system's smaller terrestrial and larger ice giant planets, and in particular the mechanism(s) driving 1-3~R$_{\oplus}$ planet formation. As discussed below, there is evidence that these small planets may naturally fall into two subgroups, super-Earths ($\sim$1-1.6~R$_{\oplus}$) and sub-Neptunes ($\sim$2-3~R$_{\oplus}$). A natural question is, do super-Earth and sub-Neptune planets represent a continuous population that has been influenced by post-formation processes but formed in a similar way, or did they form differently from the start?

The proposed formation scenarios for 1-3~R$_{\oplus}$ planets fall into two broad categories -- planet or planetesimal migration (e.g., \citealt{raymond&cossou2014,cossou2014,schlichting2014}) or in situ formation (e.g., \citealt{lee2014,lee&chiang2015,lee&chiang2016,hansen&murray2013,chiang&laughlin2013}), although issues remain with both of these scenarios \citep[e.g.,][]{lissauer2011,fabrycky2014,raymond2014,schlichting2014}. Perhaps small planet formation could occur through some combination of both, such that embryos migrate to increase the inner disk solid surface density, and resonant chains are then altered when the gas dissipates and the embryos collide to form the final planets \citep{raymond2008,hansen&murray2012,izidoro2017,inamdar&schlichting2016}. Whether small planets form in situ, via disk migration, or a combination of the two, additional processes may also influence the final range of small planet gas mass fractions, such as photoevaporation \citep{lopez2012,owen&wu2013,owen&wu2017,jin2014,chen&rogers2016}, tidal heating \citep{ginzburg&sari2017}, the timing of gas accretion \citep{Lee&Connors2020}, and giant impacts \citep{inamdar&schlichting2015,inamdar&schlichting2016,liu2015}.
From the theoretical perspective, it remains unclear whether the combination of different formation scenarios and post-formation processes is consistent with one or multiple small planet populations. 

Luckily, the number of precise small planet observations is increasing and leading to tests of different formation scenarios. Studies of the \textit{Kepler} sample revealed a clear bimodality in the population of short period ($<$100 day), small radius planets around solar-type stars \citep{Fulton2017,fulton&petigura2018}, with a decrease in occurrence rates (a gap) around $\sim$1.8 R$_{\oplus}$. 
The presence and location of the gap is consistent with a model of photoevaporation via X-ray and extreme ultraviolet radiation of planets' volatile envelopes after formation  \citep{owen&wu2013,lopezfortney2013,owen&wu2017,jin&mordasini2018,Lopez&Rice2018}, which naturally herds planets into two groups, one of bare cores and one of planets with H/He envelopes roughly the same size as the core (and a few \% mass). 
Most of the planets in the \cite{Fulton2017} (California Kepler Survey) sample do not have measured masses, so their core mass fractions are unknown. However, their planet radius distribution is well matched by a modeled planet population having uniformly rocky cores, similar to the Earth's bulk density \citep{owen&wu2017,Rogers&Owen2020}, and not by planets having a substantial mass fraction ($\gtrsim 75$\%) of ice/water or made purely of iron. 

However, photoevaporation may not be the only mechanism shaping small planets into super-Earths and sub-Neptunes. The observed \textit{Kepler} distribution, including the location of the radius gap, how it changes with planet orbital period, and trends with stellar properties, can also be matched with models of core-powered mass loss, in which the initial energy of the planet's formation can overcome the gravitational binding energy of the planet's atmosphere \citep{Ginzburg2018, Gupta&Schlichting2019,Gupta&Schlichting2020}. \cite{Gupta&Schlichting2019}'s results also suggest most super-Earths and sub-Neptunes have rocky interior compositions composed of mostly iron and rock, although the interiors can contain up to $\sim 20$\% water-ice and still match the observations reasonably well. Heterogeneity in the interior composition ``smears out'' the gap in the radius distribution \citep{lopez&fortney2013,owen&wu2017,jin&mordasini2018}, and \cite{teske2018b} found observational evidence of this ``smearing out'' by accounting for detected and undetected (in a probabilistic sense) stellar companions to \textit{Kepler} planet host stars. More variation in the interior composition of super-Earth and sub-Neptune planets, perhaps a non-zero fraction of their interiors being ice/water, could open up the possibility that some of them formed beyond the ice line and migrated inwards. Alternatively or in addition to this, there appears to be evidence of a population of super-Earth planets that were ``born rocky'' \citep{Neil&Rogers2020,Rogers&Owen2020}, and theoretical models suggest that after undergoing core-powered mass loss, super-Earths can retain small H/He atmospheres \citep{Misener&Schlichting2021}. Thus, from an observational perspective, the question of whether super-Earth and sub-Neptune planets are a continuous population or distinct populations (or even sub-populations) remains open. 

\subsection{Mass-Radius Relations}

Some population-level insights have already come from examining small planet mass-radius relations. For example,  \cite{Dressing2015} first pointed out that the five $R_p < 1.6$ R$_{\oplus}$ planets with well-measured masses at the time all fell very close to a composition curve representing a simplified Earth-like composition model, suggesting a single mass-radius relation may encompass many rocky exoplanets. Subsequent publications of $R_p < 1.6$ R$_{\oplus}$ transiting planet masses have started to show a larger scatter in mass at a given radius (e.g., \citealt{marcy2014,Bonomo2019}), although some also fall along an Earth-like composition curve (e.g., \citealt{Lopez-Morales2016,Dai2019}), perhaps supporting the work above that suggests little variation in interior composition. On the other hand, \cite{weiss&marcy2014} found in their seminal study of the masses and radii of 65 exoplanets smaller than 4 R$_{\oplus}$ that the weighted mean density increased for planets up to 1.5 R$_{\oplus}$, at which point it started decreasing with increasing radius, pointing to a natural break in the small planet population. They also found moderate scatter in the mass-radius relation between 1.5-4 R$_{\oplus}$ that was not explained fully by measurement error, which they suggested was due to a diversity in compositions at a given radius. \cite{wolfgang2016} presented the first probabilistic mass-radius relationships evaluated within a Bayesian framework, which quantifies astrophysical dispersion in the population and can incorporate upper limits, and also found evidence for significant (non-zero $\sigma$, standard deviation of Gaussian distribution of planet masses) intrinsic, astrophysical scatter. Building on \cite{wolfgang2016}, \cite{chenkipping2017} extended the idea of hierarchical Bayesian modeling to a larger range of planet radii by fitting a series of probablistic broken power laws. These authors found the first empirical break in the mass-radius relation to be at about 2.04 M$_{\oplus}$ or 1.23$^{+0.44}_{-0.22}$ R$_{\oplus}$, slightly lower than but still consistent with previous works. Most recently \cite{Neil&Rogers2020} combined hierarchical Bayesian modeling with a mixture model to examine the mass-radius-period distribution of \textit{Kepler} planets with mass measurements, finding that a three-component model with gas envelope, evaporated envelope, and intrinsically rocky planets was preferred over a model with only the first two components.  

However, a major hurdle that has been thus far unaccounted for in studies of small planet mass-radius relations is the observational biases introduced in mass measurements. There may be a ``publication bias'' wherein upper limits are not published, as often the decision to publish a planet's mass is based on the RV precision reaching a certain threshold, like $K/\sigma_{RV} > 6$. 
This leads to a measurable bias in the population's mass-radius relation \citep{montet2018,Burt2018}, such that very small planets (1-2~R$_{\oplus}$) are predicted to be more massive than would be physically expected for that population. Additionally, the decisions for following up transiting planets are not sufficiently described in the literature, but robust statistical analyses require this process to be reproducible to infer accurate population distributions from the observed sample. No published papers describing the mass-radius relation or the small-planet composition distribution take these selection effects into account, partly due to the difficulty in quantifying the time-varying selection function used to create the existing heterogeneous data set. Overall, the criteria of target selection, design of RV coverage/cadence, and publication of data vary significantly from target to target and from team to team, limiting the power of statistical or population studies of the sample of small planets.

\subsection{Motivating Questions for This Work \label{subsec:motivation}}
Motivated by these factors in small planet mass measurements, as well as the open questions of super-Earth and sub-Neptune planet formation, we embarked on a survey with the goal of building a statistically-robust sample of well-characterized small planets that can be used for population inference. Fortunately, the Transiting Exoplanet Survey Satellite
 (TESS, \citealt{ricker2015}) was predicted to find thousands of small planets around bright stars ($4 \le V \le 12$) across the entire sky \citep{sullivan2015,barclay2018}, amenable to RV mass measurement and planet density constraints. TESS has been making good on that prediction -- there have already been dozens of small TESS planet mass measurements published \citep[e.g.,][]{Huang2018,Dragomir2019,Winters2019,Luque2019,Cloutier2019,Espinoza2020,Astudillo-Defru2020,Shporer2020}. 
Our work, the Magellan-TESS Survey (MTS), aims to measure the masses of transiting planets between 1--3 R$_{\oplus}$ detected by TESS to help robustly address the population-level question: Do super-Earth and sub-Neptune planets have disparate or similar formation pathways? The MTS consists primarily of RV follow-up observations using the Planet Finder Spectrograph (PFS; \citealt{crane2006,crane2008,crane2010}) on the Magellan II Clay telescope at Las Campanas Observatory in Chile, in combination with results from other ground-based observations. 
The MTS will produce a probablistic mass-radius relation in similar fashion to \cite{wolfgang2016}, but with the important differences of (1) accounting for the sample selection and (2) being derived from RV observations specifically designed to mitigate bias. 

In addition, our survey will go beyond a mass-radius relation to study 
three specific questions about 
potentially distinguishing effects in the exoplanet population:

\begin{enumerate}
    \item How do planetary bulk densities depend on stellar insolation? The gap in the radius distribution appears to drift toward lower radii at lower incident fluxes in systems with solar-type stars, a trend consistent with both photoevaporation and core-powered mass loss \cite{vaneylen2018,Martinez2019}. However, mass measurements provide a more complete picture, as photoevaporation depends primarily on a planet’s surface gravity and the core-powered mass loss rate also includes an exponential term with planet mass. We want to know whether the radius bimodality results in a sharp transition in mass-radius-insolation flux space, or whether mass and radius values reflect a more continuous population over insolation flux. One way to quantify this is to assess how the empirical mass-radius (M-R) relation changes between highly vs. moderately irradiated planets. 
    \item How do planetary bulk densities depend on host star composition? While most of the spread in the small planet M-R relation may be explained by volatile envelope variation, it may also indicate diversity in interior composition, and by proxy, host star composition. Stellar abundances can be important in breaking the degeneracies in constraining planetary interior structure using mass and radius (\citealt{dorn2015}, although see also \citealt{Plotnykov&Valencia2020}), and stellar abundance estimates are natural products from the high resolution stellar spectra obtained for the RV observations by MTS. By comparing the densities of a larger sample of small planets to their detailed host star abundances, we can probe the role of ``birth composition'' in distinguishing super-Earths and sub-Neptunes (e.g., \citealt{brewer&fischer2018}).
    \item How do planetary bulk densities depend on system architecture? Follow-up observations on single transiting planet systems can help uncover the system architectures in terms of eccentricities and additional planets. Higher eccentricities would indicate a more dynamically rich history (e.g., \citealt{naoz2016}), while a high occurrence rate of additional, non-transiting planets would indicate high mutual inclinations, perhaps indicating a formation scenario in passive disks at later times (e.g., \citealt{bitsch2019}). Furthermore, it is also useful to map out the outer part of systems, especially the existence of gas giants, which recent studies suggest are common in systems hosting small inner planets \citep{zhu&wu2018,bryan2019}. How outer gas giant exoplanets facilitate/prohibit the channeling of disk materials inward to form inner small exoplanets \citep[e.g.,][]{bae2016} and thus how the giant planets affect the compositions of inner planets \citep[e.g.,][]{Schlecker2020} is not observationally constrained.
\end{enumerate}

The rest of the paper outline is as follows: In \S\ref{sec:survey} we detail how we designed a survey strategy, including target selection and observation cadence, to help address these questions as well as the issues related to robust population-level analyses described above. In \S\ref{sec:radii} and \S\ref{sec:mass} we describe how we uniformly derived the radii of our targets from the TESS light curves, how we derived the masses of our targets from PFS data, and how they compare to masses from the literature. Our hierarchical Bayesian modeling (HBM) framework is described and applied to our measured radii and masses in \S\ref{sec:hbm}; we also apply the framework to a heterogeneous data set and compare the resulting HBM hyperparameter posteriors. Our discussion in \S\ref{sec:discussion} includes a comparison to previous small planet mass-radius relations, implications of our results for the detailed planet compositions, limitations of our work, and future plans. We end with a summary of conclusions and take-away points in \S\ref{sec:summary}. In Appendix \ref{sec:individual_systems} we also describe background information and previous publications about the planets in our sample. In Appendix \ref{sec:appendix_tess_lc_detrends} and \ref{sec:appendix_tess_lc_fits} and we show our fits to the TESS light curves, and in Appendix \ref{sec:appendix_rv_fits} we show our fits to the radial velocities. In Appendix \ref{sec:ancillary_tables} we list ancillary parameters to the light curve and RV fits. In Appendix \ref{sec:eliminated_tois} we describe three TOIs (262, 271, \& 560) that were initially in our target list but later excluded for various reasons; this appendix also includes the PFS RVs of TOI-560.


\section{Description of Survey}\label{sec:survey}

The survey was designed to address our three science questions outlined above, minimize observational and selection biases where possible, and be documented at each step. In this section, we describe how we vetted TESS-discovered planet candidates and selected the targets for our survey (Section~\ref{subsec:target_selection}), as well as our procedure for assigning a RV observing cadence to each target for each observing run (Section~\ref{subsec:cadence_choice}). Importantly, both of these aspects of the survey are driven by quantitative criteria that were decided prior to the start of RV observations.

    \subsection{Target Selection}\label{subsec:target_selection}

The principles of our target selection are: (1) we follow a set procedure with quantified criteria at each step for cuts and rankings, and (2) we stick to our decisions and document our entire target selection process. 

To summarize our selection process, we start by drawing from the TOIs released by the TESS Science Office\footnote{\hyperlink{https://tess.mit.edu/toi-releases/}{https://tess.mit.edu/toi-releases/} }, which we assume are all \textit{bona fide} planet candidates\footnote{We made this assumption at the beginning of our selection process because at the time of MTS's first observations, few of the planets had been vetted with imaging and reconnaissance spectroscopy.  However, as more information emerged and some planets were identified as false positives, they were removed from both the parent sample and the MTS sample.  Due to conservative choices (see \S \ref{subsubsec:elimination}) made within the constraints of our predetermined cadence plan (see \S \ref{subsec:cadence_choice}), we have in practice lost very little observing time to eventual false positives.}. We update the TOI list prior to every PFS observing run; this is not so consequential at this point in the survey but was during the first year of the TESS mission as new TOIs were added. We evaluate these planet candidates with a merit function (\S \ref{subsubsec:ranking}), and then among the top ranked $\sim$60 planets we perform a more careful vetting to select targets that are validated and orbit host stars that are viable for RV follow-up observations. We describe our target selection in detail in the following subsections. Importantly, all of our targets are drawn from the TOI list announced by the TESS team, and are then subject to a large ground-based follow-up effort. The biases in these vetting and follow-up procedures are not well documented or easily quantifiable and are left to be addressed in future works; in principle the analyses presented here could be combined with such works. See \S\ref{subsec:limitations} for further discussion. 

For the MTS, we chose to pursue 30 TESS-discovered small planets for mostly practical reasons -- we wanted to finish our survey within approximately three years for maximum relevance to the community and the TESS mission, given a realistic assumption of available telescope time. At the time we began the MTS, adding 30 small planet masses would have roughly doubled the number of planets of this size with well-measured RV masses. We also wanted to cover planets across the entire sky, ideally a few in each TESS sector, to help with scheduling our observations in the PFS queue. The number 30 is the number of planets, not planetary systems; in reality, we will have fewer than 30 planetary systems in our final sample, as some stars have multiple highly-ranked planets. 

\begin{figure*}[htbp]
    \centering
    \includegraphics[width=\textwidth]{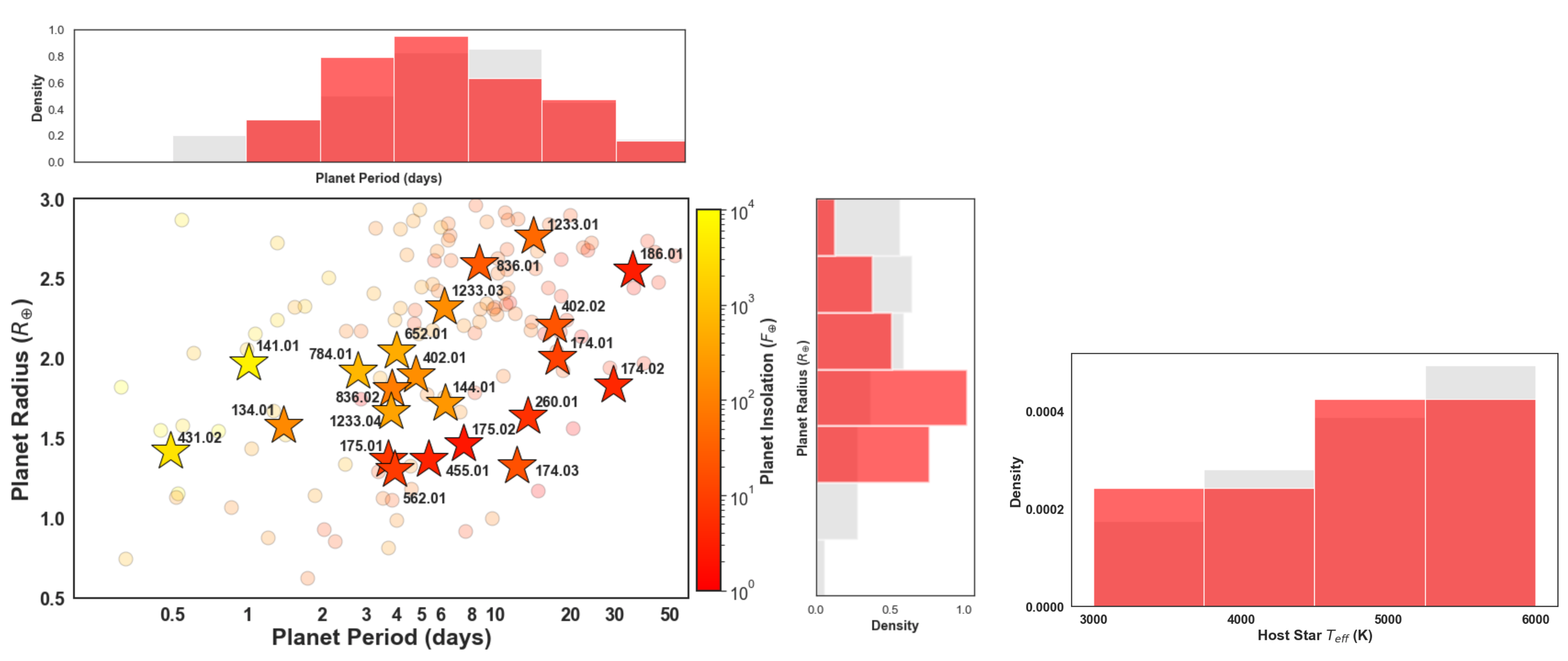}
    \caption{A summary of the properties of our preliminary sample, which consists of 22 TOIs ranked highly by our metric. The planet radii and periods shown here are derived from ExoFOP-TESS and were used in our ranking for target selection; the planet radii and periods in Table \ref{trasit_fit_posteriors} were derived from our own analysis and are included in our mass-radius modeling. The final survey sample will have 30 small planets. \textbf{Left:} Periods and radii of the planets included in our metric-selected sample (star symbols) along with the parent sample of TOIs (faint circles), colored according to insolation flux relative to the Earth. Normalized histograms show the current MTS sample in red and the parent sample in grey. \textbf{Right}: A normalized histogram of the current MTS sample (red) and the TOI parent sample (grey) host star effective temperatures.}
    \label{fig:sample_properties}
    \end{figure*}

On the other hand, there are planets that are \textit{not} among our top ranked 30 planets but are in the systems that we are following up; we will acquire mass constraints on these additional planets (some of which are currently listed in Tables \ref{trasit_fit_posteriors} and \ref{tab:pfs_K}) although we do not explicitly use them in our cadence decisions (see \S \ref{subsec:cadence_choice}) or in the mass-radius relation analysis (see \S \ref{sec:hbm}). In this paper, we present masses for 22 planets, or the first two-thirds of our sample. A summary of the properties of those 22 planets, as compared to the parent sample of TOIs (described in \S\ref{subsubsec:elimination}), is shown in Figure \ref{fig:sample_properties}. We note that this figure reflects the planet radii and periods\footnote{from ExoFOP-TESS as accessed on April 19, 2020, with radii derived from scaling the transit depths using TICv8 stellar radii} used in our target ranking (\S\ref{subsubsec:ranking}), while Table \ref{trasit_fit_posteriors} lists the radii and periods we derived in this work and that are included in our mass-radius modeling (\S\ref{sec:hbm}). The rest of the sample is not yet finalized due to incomplete vetting observations, but will come from the top ranked $\sim$60 planets. Once planets are ranked by our metric in the top 30 (after disqualifying targets as described below), they are effectively ``equal''; the ranking metric serves to make a cut between the 30 planets included in the survey, not to prioritize or weight planets within that subset. 

        \subsubsection{Target Ranking Metric}\label{subsubsec:ranking}

We rank the alerted TESS TOIs according to a merit function:
\begin{equation} \label{eqn:merit}
\textrm{merit} = 
\begin{cases}
F_{\rm insol}^{-1/3} \mspace{5mu} T_{\rm exp}^{-1} \mspace{4mu}  e^{-1.25(R_p-1.8)^2} & {\rm for\ R_p<3 R_{\oplus}} \\
0 & {\rm otherwise}
\end{cases}
\end{equation}
where \noindent $F_{\rm{insol}}$ is the stellar insolation flux received by the planet calculated using the TESS Input Catalog (TIC) v8 \citep{Stassun2017tic,Stassun2019tic} stellar parameters, $T_{\rm{exp}}$ is the total exposure time needed for RV characterization (described in the next paragraph), and $R_p$ is the planet radius taken from the TESS alert. This merit function was chosen to prioritize planets (1) receiving low stellar insolation flux, which may be preferentially avoided in RV surveys due to their longer periods; (2) with radii around $\sim$1.7 R$_{\oplus}$, most effective in addressing the origin of the radius gap; and (3) requiring less observing time to characterize. The specific shape and parametrization of this merit function was chosen by ranking the simulated TESS targets produced by \cite{barclay2018} and visually examining the distributions of $F_{\rm insol}$, $R_p$, and estimated $T_{\rm exp}$ of the top-ranked targets. We emphasize that this function has not been derived from any `first-principles' argument; it is purely a quantitative, reproducible way for us to select the targets that are best suited to address this survey's scientific goals.

The $T_{\rm{exp}}$ term is calculated as follows: We first estimate the single-exposure time $t_{exp}$ needed to achieve the photon-limited RV precision, $\sigma_{\rm{RV}}$, of 2 m~s$^{-1}$ based on previous performance of Magellan II/PFS. We set a minimum $t_{exp}$ of 10 minutes to help address the effects of stellar oscillation \citep[e.g.,][]{dumusque2011}. Second, we estimate the RV semi-amplitude $K$ of the planet using the \cite{Ning2018} nonparametric mass-radius relation to get the mass and convert that into $K$ using the TESS-measured planet period and the TICv8 stellar mass. Third, we estimate $N$, the total number of RV observations needed to meet the criterion $\sqrt{N} \times K$/$\sigma_{\rm{RV}} > 6$ (e.g. \citealt{howard2012,dumusque2017}). Note that while we use $K$/$\sigma_{\rm{RV}}$ to estimate $N$, we do not actually impose any $K$-significance criterion on the masses we report here nor the masses we use for the population modeling. We require $N \geq 20$ RV observations no matter what the predicted $K$ value based on the minimum number of observations needed in \cite{marcy2014} for \textit{Kepler} systems. The median and mean predicted $N$ values for our top 60 ranked targets (prior to the elimination process in \S\ref{subsubsec:elimination}) are 34 and 40 observations, respectively. Finally, $T_{\rm{exp}}$ for each target is calculated as $t_{exp} \times N \times 2$, where the factor of two is to account for extra time that may be needed to address the effects of stellar jitter (e.g., caused by stellar rotation).

This merit function is frozen for the course of our 3-year survey to preserve our ability to perform a population inference analysis (\S\ref{sec:hbm}). At the start of the survey, we applied the merit function to the TOIs released in each TESS sector as they became available and selected the top $\sim$2--3 planets that were suitable for RV observations. Most of the planets selected in this way from early TESS sectors have consistently stayed within our top 30 ranked list as we have added targets from subsequent sectors in TESS year 1 (for more detailed discussion of the few exceptions, see Appendix \ref{sec:eliminated_tois}). We do not include the community provided TOIs or TOIs released beyond TESS year 1 so that we do not have a constantly changing target list, and we limit the parent sample to only southern targets. Finally, in practice we not only use the merit calculated from the planet's reported ``best-fit" radius and flux to determine a planet's rank --- and therefore whether it constrains the mass-radius relation --- but the full measurement uncertainty distribution of these planet parameters; see \S \ref{subsec:HBM} for details.

        \subsubsection{Target Elimination Criteria}\label{subsubsec:elimination}

Once we have our top ranked sample of 60 targets, we proceed to eliminate those that show evidence of not being real planet candidates. This evidence comes from the TESS Follow-up Observation Group (TFOP), particularly sub-group 1 (SG1), which performs ground-based seeing-limited photometric follow-up to rule out false positive scenarios (e.g., a background eclipsing binary) and detect additional transits on the presumed host star. We typically wait to start any RV observations until the planet candidate has reached at least the ``cleared-planet-candidate'' (CPC; no NEBs found on stars close enough and bright enough to have caused the TESS detection) or ``verified-planet-candidate'' (VPC; transit event verified in aperture and no other contaminating stars bright enough to cause the TESS detection) SG1 follow-up status. For targets without CPC or VPC status, we keep them ``on hold'' for RV observations until further vetting has been conducted. 

As we aim to measure precise masses of small planets using RVs, we need to discard stars for which this would not be feasible. We eliminate any systems with host stars with rotation periods shorter than 15 days to avoid a large stellar RV jitter caused by strong surface magnetic activity such as star spots (as well as rotational broadening at a level that limits RV precision). Such a cut-off in rotation period means that our sample could be skewed towards relatively older stars in general according to gyrochronology \citep{Barnes2007,Mamajek2008}. This is unlikely to cause any significant bias in our results, since our sample will probably entirely consist of host stars well into their main sequence lifetime and relatively mature planetary systems; our primary goals do not include capturing young planetary systems or distinguishing the effects of stellar age in the mass-radius relation. Our estimate of stellar rotation period comes primarily from examining the light curves of our targets from TESS and Evryscope, as detailed in Howard et al. (submitted). When these light curves are inconclusive, we look for measurements of the star's $v$~sin$i$, often from TFOP SG2 (reconnaissance spectroscopy), and combine that with the radius (assuming sin~$i$=1) to estimate a rotation period.\footnote{We note while this estimate is not robust for deriving the stellar rotation period, it serves our purpose as a back-up method for vetting out the faster rotators. At least for a relatively fast rotating star with periods shorter than 15 days, the $v$~sin$i$ is usually large and easy to measure precisely, relatively speaking.}

We also remove from the sample planet candidates that orbit stars with radii $> 1.2 R_{\odot}$ or effective temperatures $T_{\rm{eff}} > 6000$ K, as these are likely to have a substantial amount of RV jitter due to stellar oscillation and granulation (e.g., \citealt{Luhn2020}). In addition, we eliminate systems whose host stars are spectroscopic binaries or visual binaries that are too close to resolve with slit spectroscopy (similarly for stars in a crowded field).

There are some TOIs that do not have reported planetary radii with the TESS year 1 data release, and therefore these targets do not receive a ranking at this step. However, we have examined them on an individual basis and decided to eliminate them all, because they are either confirmed false positives, or very likely false positives, or their host stars have radii $> 1.2 R_{\oplus}$. We also do not rank any targets without period estimates from TESS year 1 data (e.g., single-transit systems).  

There are three TOIs that were initially in our target list but later eliminated: TOI-262, TOI-271, and TOI-560. We describe why they were eliminated in Appendix \ref{sec:eliminated_tois}, and present the RV data of TOI-560 (the RVs of TOI-262 and TOI-271 will be published elsewhere).

A summary of prior knowledge of each the TOI systems currently in the MTS is given in Appendix \ref{sec:individual_systems}, which includes 22 TESS planets/planet candidates -- 134.01, 141.01, 144.01, 174.01 \& 174.02 \& 174.03, 175.01 \& 175.02, 186.01, 260.01, 402.01 \& 402.02, 431.02, 455.01, 562.01, 652.01, 784.01, 836.01 \& 836.02, 1233.01 \& 1233.03 \& 1233.04. In this work, we do not attempt to validate the TOIs that, as of this publication, remain planet candidates (as noted in the individual summaries in Appendix \ref{sec:individual_systems}); that is beyond the scope of this work. Based on our vetting and elimination criteria, we suggest there is a high likelihood that the to-be-validated TOIs included here are true planets. The status of these TOIs will be reassessed in our next survey paper, when we will also update our analysis to include a larger sample of planets/planet candidates. The TOIs are also listed in Table \ref{tab:stellar_params}, where we summarize key parameters of the host stars as listed in TICv8 \citep{Stassun2019tic}. 

\begin{deluxetable*}{cccccccc}
\centering 
\tabletypesize{\footnotesize}
\tablewidth{\columnwidth}
 \tablecaption{Host Star Key Parameters from TICv8 \label{tab:stellar_params}}
 \tablehead{ 
 \colhead{TOI} & \colhead{TIC} & \colhead{Alternate} & \colhead{$T_{\rm{eff}}$}  & \colhead{[M/H]} & \colhead{R$_*$}  & \colhead{M$_*$} & \colhead{L$_*$}
\\
\colhead{} & \colhead{} &\colhead{name(s)} & \colhead{(K)}  & \colhead{(dex)} & \colhead{(R$_{\odot}$)}  & \colhead{(M$_{\odot}$)} & \colhead{(L$_{\odot}$)}}
\startdata 
134 & 234994474 &  HIP 115211, GJ 4332, L 168-9 & 3746$\pm$157 & $...$ & 0.597$\pm$0.018 & 0.586$\pm$0.021 & 0.063$\pm$0.015 \\
141 & 403224672&  HD 213885, HIP 111553 & 5795$\pm$111 & $...$ & 1.132$\pm$0.049 & 1.040$\pm$0.129 & 1.302$\pm$0.034 \\
144 & 261136679 &  HD 39091, HIP 26394, $\pi$ Men& 5992$\pm$104 & 0.089$\pm$0.007 & 1.149$\pm$0.045 & 1.100$\pm$0.140 & 1.533$\pm$0.046 \\
174 & 425997655 &  HD 23472, HIP 17264 & 4813$\pm$112 & -0.190$\pm$0.080 & 0.691$\pm$0.042 & 0.780$\pm$0.089 & 0.231$\pm$0.007 \\
175 & 307210830 &  L 98-59 & 3429$\pm$157 & $...$ & 0.314$\pm$0.009 & 0.293$\pm$0.020 & 0.012$\pm$0.003 \\
186 & 279741379 & HD 21749, HIP 16069, GJ 143 & 4629$\pm$151 & 0.003$\pm$0.053 & 0.705$\pm$0.064 & 0.730$\pm$0.088 & 0.206$\pm$0.010 \\
260 & 37749396 &  HIP 1532 & 4049$\pm$121 & $...$ & 0.618$\pm$0.060 & 0.630$\pm$0.081 & 0.093$\pm$0.007 \\
402 & 120896927 &  HD 15337, HIP 11433 & 5175$\pm$107 & 0.060$\pm$0.021 & 0.842$\pm$0.042 & 0.880$\pm$0.111 & 0.458$\pm$0.010 \\
431 & 31374837 &  HIP 26013 & 4891$\pm$112 & 0.020$\pm$0.080 & 0.727$\pm$0.034 & 0.800$\pm$0.089 & 0.272$\pm$0.008 \\
455 & 98796344 &  HIP 14101, GJ 3193A, LTT 1445A & 3384$\pm$157 & $...$ & 0.276$\pm$0.008 & 0.249$\pm$0.020 & 0.009$\pm$0.002 \\
562 & 413248763 &  HIP 47103, GJ 357, L 678-39 & 3490$\pm$157 & $...$ & 0.360$\pm$0.011 & 0.346$\pm$0.020 & 0.017$\pm$0.004 \\
652 & 22221375 &  HD 86226, HIP 48739 & 5940$\pm$110 & 0.017$\pm$0.017 & 1.034$\pm$0.042 & 1.090$\pm$0.133 & 1.99$\pm$0.035 \\
784 & 460984940 &  HD 307842 & 5464$\pm$130 & $...$ & 0.896$\pm$0.050 & 0.960$\pm$0.123 & 0.645$\pm$0.016 \\
836 & 440887364 &  HIP 73427 & 4250$\pm$120 & $...$ & 0.691$\pm$0.061 & 0.660$\pm$0.077 & 0.143$\pm$0.009 \\
1233 & 260647166 &  HD 108236, HIP 60689 & 5724$\pm$111 & $...$ & 0.864$\pm$0.039 & 1.020$\pm$0.125 & 0.722$\pm$0.019 \\
\enddata
\end{deluxetable*}

\subsubsection{Why Quantify Target Selection?}\label{subsubsec:whyquantselect}

There are multiple reasons why a quantifiable selection process is advantageous.  First, it increases the transparency of potential biases in the sample and therefore which conclusions about the population are more likely to be robust to the selection process.  Take, for example, the MTS sample.  We emphasize that the sample itself is not unbiased, as our selection function 1) focuses on small planets around the radius gap and 2) draws from the human-vetted list of TOIs whose completeness and reliability has not been quantified.  Consequently, 1) our results become more uncertain as the radius increases toward $3~R_\oplus$ and 2) we cannot currently comment on the occurrence rate of planets in different parts of the joint mass-radius space (but see point \#5 below).  That said, MTS is \emph{less} biased than previous M-R datasets for the smallest planets due to the effect that this selection function has had on our data acquisition process (see point \#4 below), and so the population results for the smallest planets are more robust.  Furthermore, the fact that this bias is quantified means that it can be reliably corrected for in future work, which allows the population result to be less biased than the sample itself.

Second, along with clear identification of the utilized planet and stellar parameters, the selection function is reproducible and testable with other planet samples, thereby enabling more apples-to-apples comparisons across population studies.  
    
Third, a quantifiable selection process more clearly delineates which RV measurements were ``targets of opportunity'' and which were chosen for population studies. Since targets of opportunity tend to be novel and therefore extreme in some aspect of parameter space, distinguishing these from the population science sample causes the population results to be less subject to novelty bias -- the bias in which one-of-a-kind planets that do not occur in the population sample at the same rate as their actual frequency in nature contribute disproportionately to the result, depending on the population model and the location of the planet in parameter space.

Fourth, a quantifiable selection process incentivizes continued observations of systems where, because the planet is intrinsically low mass or in a multi-planet system, statistically significant RV variation may emerge more slowly as a function of the number of observations.  This incentive is particularly important for population analyses of observations that have significant human involvement over time, as evolving choices can change the existence and precision of measurements in ways that are difficult to correct for on the population level.  Much of the power and utility of a selection function for RV follow-up lies here, i.e. in forcing the acquisition of data which may not otherwise have been gathered.  Since the RV signatures of low-mass planets and multi-planet systems require more observations to distinguish from noise, an RV selection function reduces the significant observational and publication biases toward high-mass, single-planet systems.

Fifth, when combined with information about the completeness and reliability of the parent sample -- in this case, the TESS transit detections -- a selection function can be used to perform robust, quantitative occurrence rate results.\footnote{Occurrence rates, i.e. the joint mass-radius distribution $f(m,r) = f(m|r)f(r)$, requires knowledge of both the completeness and reliability of the TESS survey, which allows a ``true'' marginal radius distribution $f(r)$ to be computed, and the completeness of the follow-up mass measurement effort, which allows a ``true'' mass-radius relation $f(m|r)$ to be computed.  MTS provides a $f(m|r)$ with the mass-given-radius completeness quantitatively defined; if future work produces a reliably completeness-corrected $f(r)$ for TESS, then the full $f(m,r)$ occurrence rate may be computed.}  For the exoplanet mass-radius space, \citet{Neil&Rogers2020} combined the well-quantified \emph{Kepler} completeness information (\citealt {Christiansen2017, Burke2017&Catanzarite2017,Thompson2018}; see also \citealt{Christiansen2020}) with a less well-quantified RV selection function to produce a first look at the occurrence rate of small-planet compositions. The completeness and reliability of the TESS parent sample is currently less well-quantified, while we endeavor with MTS to provide a very well-defined RV selection function.  Therefore, we anticipate the final results of this survey to be complementary to prior work and provide insight into the relative importance of the different components in planetary composition occurrence rate analyses.

    \subsection{Cadence Choice for RV Observations}\label{subsec:cadence_choice}

The spirit of our cadence design is to avoid human biases in deciding which targets receive more observations relative to the others in any given run.  We strive to decouple the quantity and the quality of the RV observations for each system from prior knowledge of their radii and estimated masses in our decision tree when planning observations.

In practice, decisions of whether and how each target shall be observed are made for each observing run of Magellan II/PFS, as we do not have unlimited or continuous access to our RV instrument. The Magellan II telescope is classically scheduled, and PFS is a PI instrument, so it is mounted in blocks of time through the semester, typically between one and two weeks during bright and gray time. Similar to the California-Kepler Survey, the PFS team pools all awarded PFS nights from partner institutions together and manages a queue across the entire semester. The original goal of our survey -- masses of 30 small TESS planets in six semesters -- amounted to $\sim$12 nights per semester. Through the competitive Carnegie telescope allocation process, we have been awarded 10--12 nights per semester for MTS since 2018B.

We use a cadence scheme for our RV observations that would enable a ground-up simulation for our survey, and we document our decision making process for each observing run. The basic elements in our observing cadence design are: (1) Should we include this target in this observing run? (2) How many nights should we observe this target during this observing run? (3) If we do observe the target on a particular night, how many times should we visit this target, and how long should the exposure time be?

For elements (1) and (2), we have adopted a ``grade'' system with four levels of observing cadence: high, medium, low, and trailing cadence. Ideally, we would observe each target with the highest possible cadence (nightly visit or even higher) throughout our three year survey, because this would sample two of the most critical timescales for measuring masses very well -- the planetary orbital timescales (typically days), and the timescales of the stellar rotation (the most common source of stellar RV jitter among our targets, typically $\sim$tens of days). However, we have to run an efficient survey given the limited telescope time and the fact that we also need to cover a variety of RV time baselines to search for additional planets in the system. Therefore, we have decided to adopt a mixed cadence scheme, where each target receives some amount of high-cadence coverage, presumably enough to enable a robust mass measurement, in combination with a suite of medium, low, and trailing cadence to sample a variety of timescales for further characterization of the planetary systems (e.g., eccentricities and additional planets). 

\begin{figure}[ht]
    \centering
    \includegraphics[width=\columnwidth]{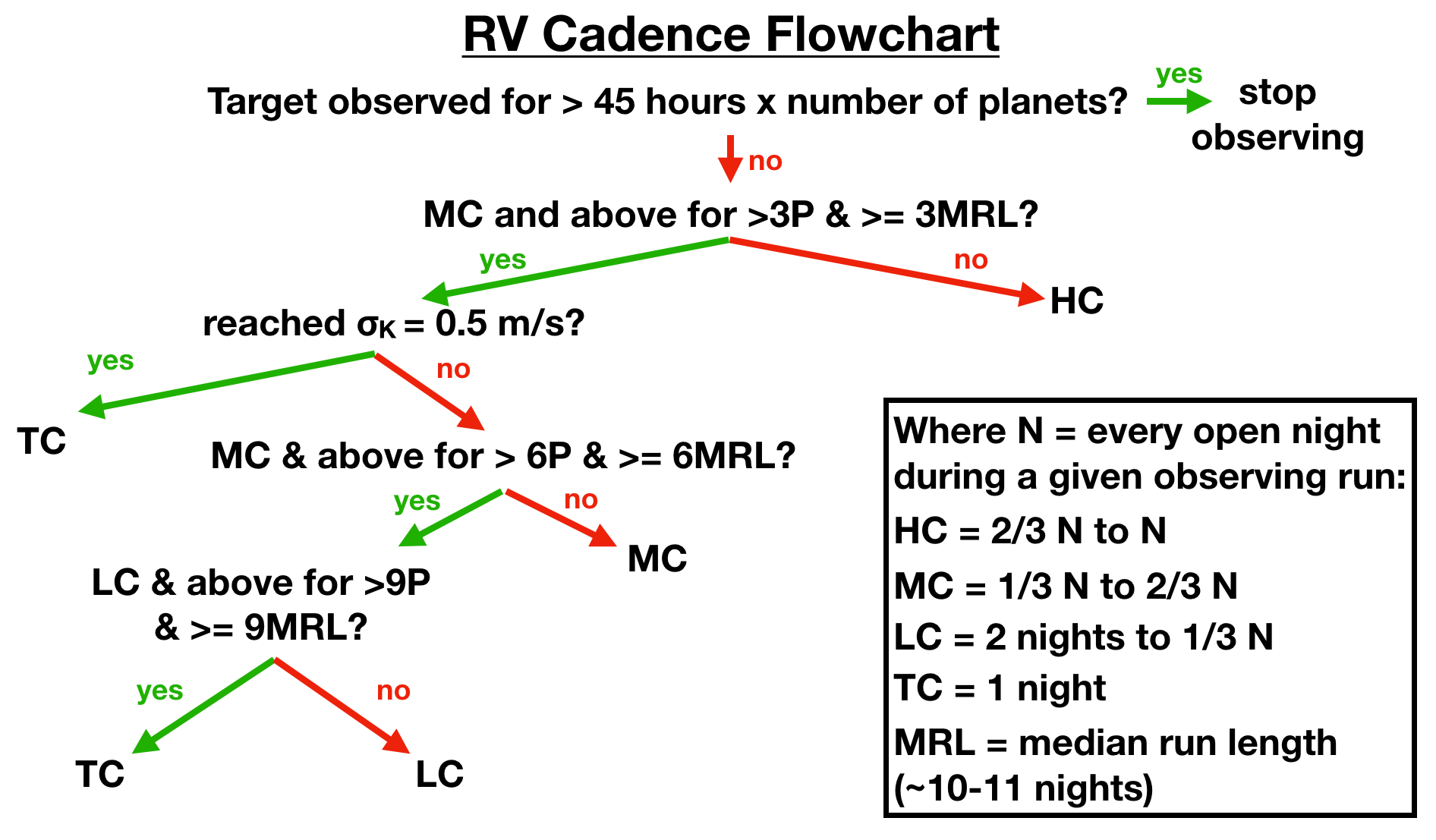}
    \caption{Graphical representation of our cadence choice scheme for conducting RV observations of our sample targets. P means the orbital period of the planet. The flowchart starts from the top, and continues down to each decision point, with a `yes' or 'no' decision. Abbreviations are explained in the right corner inset box.}
    \label{fig:cadence_flowchart}
    \end{figure}

The flowchart in Figure \ref{fig:cadence_flowchart} shows our decision tree for each target's cadence grade in each observing run. For an observing run with $N$ nights: high cadence means observing the target every night (on $N$ nights); medium cadence means observing the target on random nights with a total of $m$ nights, where $m$ is a random number between $\max{(N/2, N/P)}$ and $N$ (where $P$ is the planet period); low cadence means observing the target on $n$ random nights, where $n$ is a random number between 1 and $\max{(N/2, N/P)}$; and trailing cadence means only observing this target once during this observing run. Practically, as noted in the figure, this amounts to observations from $\sim$ 2/3 to 1 $N$, $\sim$ 1/3 to 2/3 $N$, and $\sim$ 2 nights to 1/3 $N$ for high, medium, and low cadence, respectively. The randomness of the assigned nights for the medium and low cadence is to avoid aliasing in our RV data, and the choice of $\max{(N/2, N/P)}$ is to balance between lowering the cadence and achieving a good coverage on the planet's orbital period. In systems with multiple planets, we adopt the highest cadence grade among all of the planets in a given system. 

As shown in Figure \ref{fig:cadence_flowchart}, the cadence priority for each target in an upcoming observing run is determined by how much coverage the target has received so far (e.g., we do not observe any star for more than 45 hours per planet to keep things reasonable), and how well the planet's $K$ value is constrained by data collected so far. The presence of additional larger or lower-ranked planets in the system have the potential to influence the mass determinations of the planets around which we design our cadence. However, given the architectures of the planets in our sample -- mostly short period and compact -- it is unlikely that the additional planets will be less well sampled in phase. In addition, while our cadence \textit{ranking} is related to the planetary period, the actual observing cadence (which determines how our RV data populates the phase space of our selected planets) is in fact randomized and decoupled from the planetary period. This is opposed to the commonly seen strategy of more dense sampling near the RV semi-amplitude maximum and minimum. Therefore, we do not anticipate strong bias in our mass measurements due to non-MTS planets or unseen planets in the system. We will further evaluate this statement at the end of our survey.

We decrease the cadence significantly when the absolute error on $K$ reaches 0.5~m~s$^{-1}$, following the recommendation by \cite{montet2018} to avoid using a $K$/$\sigma_K$ stopping criterion as is used in typical RV campaigns, since this can bias RV masses towards larger values. The criteria for downgrading a given target's cadence are listed in Figure  \ref{fig:cadence_flowchart} in conditional statements such as ``MC and above for > 6P''. In this example, if the total time span over which the target was observed at Medium Cadence (MC) from all previous observing runs exceeds six times the planet's period (P) and equals or exceeds six times the median run length, then the observations of this star should drop to a lower cadence. Note that this is different than a condition based on the total number of nights on which the star was observed, as it includes nights between observations.
The specific choices of ``3P'', ``6P'', or ``9P'' were more or less arbitrary choices, with the general goal of achieving good coverage of planetary orbits and stellar rotation periods (typically days to tens of days for our survey). 

In reality, when we had too many observable targets with high cadence for an upcoming observing run (which was more common at the beginning of our survey), we retained the high cadence status for the top ranked targets, and decreased the cadence grades for targets at the bottom of the ranking in a gradual fashion, until the planned observation would fit into the survey's allocated time for the given run. This meant that lower ranked targets initially received lower cadence grades, but increased as higher ranked targets naturally received lower and lower cadence grades as they accumulated more and more observations. If we still could not fit in all planned observations after lowering the cadence grades of lower ranked targets, we eliminated targets one by one from this observing run to save more time, starting from the bottom of the ranking.

Besides limited telescope time, weather and seeing conditions add more realistic constraints to the execution of our designed observing strategies. If we could not observe a target at its designated window in our cadence design due to weather or bad seeing conditions, we simply skip this observation without compensation. This is equivalent to adding another random factor in our cadence design that is decoupled from planetary properties or human intervention. For Las Campanas, the typical weather loss (non-spectroscopic nights) is around 20--25\% and the 25/50/75\% seeing percentiles are 0.51/0.62/0.79\arcsec\ \citep{thomas-osip2010}. In seeing worse than 0.8\arcsec\ we try not to conduct observations on targets fainter than $V=11$ mag (as the photon noise approaches the read noise level of the CCD). Since designing the observing cadence of any star in a given observing run depends on its coverage in previous runs, we record each star's observation cadence according to the \textit{actual} execution of the observations, instead of the originally designated cadence. For example, if a star was originally planned for high cadence observations according to the cadence design flowchart in Figure~\ref{fig:cadence_flowchart}, but the actual cadence was only medium due to weather or seeing, then we record ``medium cadence'' for this star in this run.

For element (3) in our observation design, the exposure time and number of visits\footnote{We use the term ``visit'' instead of exposure because a visit could consist of multiple consecutive exposures to limit the maximum length of an exposure to 20 minutes to limit the number of cosmic rays in a single frame.} per night are determined as follows:
\begin{itemize}
    \item All targets receive at least 10 minutes total exposure time per visit. This is to average over RV jitter induced by stellar oscillation in F, G, and early K dwarfs. We keep this rule for later type stars for simplicity.
    \item Exposure time is set to achieve $<$2 m~s$^{-1}$ photon limited precision per visit.\footnote{This means the exposure time could change according to seeing conditions, as PFS is a slit-fed spectrograph.} If the target is too faint to achieve 2 m~s$^{-1}$  photon limited precision within an hour, the combined visits per night should aim for a 2 m/s precision (i.e., the nightly binned precision) unless the target is not visible for more than 1--2 hours.
    \item If the target is observable for more than 2--3 hours during a night, it should receive two visits separated as much time as possible. This is to average down RV jitter caused by stellar granulation in the case of F, G and early K dwarfs (e.g., \citealt{dumusque2011}). We keep this strategy for other types of stars for simplicity (which also helps boost the nightly binned precision in general).
    \item If the target has $P < 2$ days, then the number of visits within a night should be: HC -- as many as possible, separated by 2--3 hours; MC and LC -- 2, with a higher priority to carry out multiple visits than other MC or LC targets; Trailing -- one visit.
\end{itemize}

\section{Uniform Determination of Planet Radii}\label{sec:radii}

Now we move on to our own analysis of the MTS sample. First, we discuss here the uniform determination of the planet/planet candidate's radii from the TESS light curves. In Secton \ref{sec:mass} we then describe our analysis of the RV data and uniform derivation of the planet/planet candidate's masses.

    \subsection{TESS Light Curve Generation}\label{subsec:tess_lc}

The TESS 2-min cadence images of our MTS sample were first reduced through the Science Processing Operations Center\ (SPOC) pipeline \citep{Jenkins:2016}, which was developed at NASA Ames Research Center mainly based on Kepler mission's science pipeline. Transit Planet Search \citep[TPS;][]{Jenkins2002,Jenkins2017,Twicken:DVdiagnostics2018, Li:DVmodelFit2019} was then performed to look for transit-like periodic signals. All of the targets in the MTS sample were initially identified as planet candidates in the TESS Object of Interest catalog (TOI), and alerted on the MIT TESS Alerts portal\footnote{\url{https://tess.mit.edu/alerts/}}. For this paper, we only used year 1 TESS data, but we plan to employ all available TESS data for our selected targets in future work.

We downloaded all TESS Presearch Data Conditioning Simple Aperture Photometry \citep[PDCSAP;][]{twicken2010,morris2020,Stumpe2012,Smith2012,Stumpe2014} light curves of our MTS sample from the Mikulski Archive for Space Telescopes\ (MAST\footnote{\url{http://archive.stsci.edu/tess/}}), which have been corrected for the instrumental and systematic errors as well as light dilution effects from nearby stars. 
We removed all measurements where the SPOC quality flag was set in order to reduce the effect of stray light from Earth or Moon in camera field of view. 

We retrieved the orbital periods (i.e., $P_{1},P_{2}...P_{i}$), mid-transit times (i.e., $t_{0,1},t_{0,2}...t_{0,i}$) and duration times (i.e., $\delta_{1},\delta_{2}...\delta_{i}$) of each TOI from ExoFOP\footnote{\url{https://exofop.ipac.caltech.edu}}, iteratively masking out all in-transits of all planets/planet candidates and leaving the out-of-transit (OOT) part of the total light curve for each system through phase-selection:

\begin{equation}
    {\Phi_{i}} \geq \frac{\delta_{i}}{2\times P_{i}}+\Delta,
\end{equation}
where $\Phi_{i}$ represents the orbital phase of the $i^{th}$ planet/planet candidate. 
We accounted for the uncertainties on $P_{i}$, $t_{0,i}$ and $\delta_{i}$ by including an additional factor $\Delta$, which was set to 0.01. We then employed the \texttt{celerite} package \citep{Foreman-Mackey2017} to fit a Gaussian Process (GP) model to the OOT light curve with a simple Matern32 kernel formulated as:
\begin{equation}
    k_{i,j}(\tau) = \sigma^{2}\left(1+\frac{\sqrt{3}\tau}{\rho}\right){\rm exp}\left(\frac{\sqrt{3}\tau}{\rho}\right),
\end{equation}
where $\tau$ is the time-lag, $\sigma$ and $\rho$ are the covariance amplitude and the correlation timescale of the GP, respectively. 

We detrended the PDCSAP light curves using the best-fit GP models. We note that the TESS light curve of TOI-186 used in this work is the same as the one in \cite{Dragomir2019} and \cite{Gan2021}, which was extracted from the publicly available target pixel stamps using a simple photometric aperture with a customized pipeline and has not been corrected for light dilution or systematic signals. We detrended this light curve by using a median filter as the GP model failed to deal with the sharp downturn between BJD = 2458420 and BJD = 2458424. We used these reprocessed light curves in our further transit fitting analysis. 

    \subsection{Transit Fitting}\label{subsec:tess_fitting}

We utilized the \texttt{juliet} package \citep{Espinoza2019} to perform the light curve fit. The transit is modelled by \texttt{batman} \citep{Kreidberg2015}. We applied the dynamic nested sampling approach to determine the posterior probability distribution of the system parameters based on the public package \texttt{dynesty} \citep{Higson2019,Speagle2018}.

As the PDCSAP light curves have already been corrected for light contamination,\footnote{We note that the PDCSAP light curves were updated with better contamination estimates after the submission of this paper, and as a result, the transit depths reported in this work are subject to a very small correction. However, the effects of these corrections are negligible given the typical uncertainties on the planetary radius. For example, the largest change in the transit depths occurred for TOI-134.01, which has a decrease in $R_{p}/R_{\star}$ of $0.0224\cdot \sqrt{0.0128)} \sim0.002$, i.e., $R_{p}/R_{\star}$ decreased from 0.0224 to 0.0204, about 1.8\%. The typical uncertainty on $R_{\star}$ is significantly larger than 1.8\% (typically 0.04 $R_\odot$ in TICv8), so the updates on the PDCSAP light curves would not alter our results or conclusion in any considerable way.} we fixed the dilution factors $D_{\rm TESS}$ equal to 1 in all our runs\footnote{As noted in Section \ref{subsec:tess_lc}, the light curve of TOI-186 was not corrected for light contamination, thus we fitted the dilution effect by placing a Gaussian prior on $D_{\rm TESS}$, centering at 0.86, the value reported by TIC v8, with a width of 0.01.}. For every planet/planet candidate in each system, we placed Gaussian prior constraints on the periods $P_{i}$ and times of transit $t_{0,i}$, centered at the values reported in the TESS validation reports with a $1\sigma$ width of 0.1 day. In addition, instead of directly fitting for the planet-to-star radius ratio ($R_{p}/R_{\star}$), orbital inclination ($i$) and scaled separation distance ($a/R_{\star}$), we applied the new parameterizations $r_{1}$ and $r_{2}$ to keep all values physically plausible \citep{Espinoza2018}, both of which we adopted uniform priors. A quadratic limb-darkening law was adopted for the TESS photometry, as parameterized by \cite{Kipping2013}. We fitted circular orbits for all planets with Gaussian priors set on the stellar densities, taken from TICv8. Finally, we also considered the instrument offset and the photometric jitter, accounting for the white noise. 

We list our general prior settings along with detailed descriptions in Table \ref{transit_fit_prior}. The final main results of each fit are presented in Table \ref{trasit_fit_posteriors}. Other related best-fit parameters are listed in Appendix \ref{sec:appendix_tess_lc_fits} in Table \ref{trasit_fit_posteriors_2}. We present the TESS lightcurves and our detrending models and residuals for all the stars, as well as phase-folded lightcurves for all the planets in this paper, in Appendix \ref{sec:appendix_tess_lc_detrends} and \ref{sec:appendix_tess_lc_fits}. The complete figure set is available in the online journal.

We note that the advantage of our transit analysis is that it is uniform across our sample in terms of the fitting procedure, the choices of priors and the underlying assumptions behind the LC modeling, as well as the source for the stellar parameters. However, other works typically employ more precise stellar parameters than TICv8, from SED fittings and/or spectral analyses, especially works that focus on validating or confirming individual planetary systems. They may also include ground-based photometry in addition to the TESS light curves, which could yield better estimates for the planet radii than our work. Adopting the published radii is unlikely to change our population analysis in any significant way, as our estimates for the planet radii all agree with the previously published values (if available) within 1$\sigma$.

\begin{table*}[htbp]
    \centering
    \caption{Model parameters and prior settings for the TESS light curves of each MTS planetary system.}
    \begin{tabular}{lcr}
        \hline\hline
        Parameter       &Prior     &Description\\\hline
        \it{Planetary parameters}\\
        $P_{i}$ (days)  &$\mathcal{N}^{[1]}$ ($P^{\ast [2]}_{i}$\ ,\ $0.1^{2}$)
        &Orbital period of the $i^{\rm th}$ planet in the system.\\
        $t_{0,i}$ (BJD-2457000)    &$\mathcal{N}$ ($t^{\ast}_{0,i}$\ ,\ $0.1^{2}$) 
        &Mid-transit time of the $i^{\rm th}$ planet in the system.\\
        $r_{1,i}$    &$\mathcal{U}^{[3]}$ (0\ ,\ 1)
        &Parametrisation for {\it p} and {\it b} of the $i^{\rm th}$ planet in the system.\\
        $r_{2,i}$    &$\mathcal{U}$ (0\ ,\ 1)
        &Parametrisation for {\it p} and {\it b} of the $i^{\rm th}$ planet in the system.\\
        $e_{i}$   &$0$ (Fixed)  &Orbital eccentricity of the $i^{\rm th}$ planet in the system.\\
        $\omega_{i}$ (deg)  &$90$ (Fixed)  &Argument of periapsis of the $i^{\rm th}$ planet in the system.\\
        \\
        \it{TESS photometry parameters}\\
        $D_{\rm TESS}$  &$1$ (Fixed)      &TESS photometric dilution factor\\
        $M_{\rm TESS}$ (ppm)    &$\mathcal{N}$ (0\ ,\ $0.1^{2}$)      &Mean out-of-transit flux of TESS photometry.\\
        $\sigma_{\rm TESS}$ (ppm) &$\mathcal{J}^{[4]}$ ($10^{-6}$\ ,\ $10^{6}$)      &TESS additive photometric jitter term.\\
        $q_{1}$        &$\mathcal{U}$ (0\ ,\ 1)  &Quadratic limb darkening coefficient.\\
        $q_{2}$        &$\mathcal{U}$ (0\ ,\ 1)  &Quadratic limb darkening coefficient.\\
        \\
        \it{Stellar parameters}\\
        ${\rho}_{\star}$ ($\rm kg\ m^{-3}$)  &$\mathcal{N}$ ($\rho^{\ast [5]}_{\star}$\ ,\ $\rm \sigma_{\rho^{\ast}_{\star}}^{2}$) &Stellar density.\\
        \hline\hline 
    \end{tabular}
    \begin{tablenotes}
    \item[1]  [1]\ $\mathcal{N}$($\mu\ ,\ \sigma^{2}$) means a normal prior with mean $\mu$ and standard deviation $\sigma$. 
    \item[2]  [2]\ The priors are centered at the values taken from TESS validation reports.
    \item[3]  [3]\ $\mathcal{U}$(a\ , \ b) stands for a uniform prior ranging from a to b.
    \item[4]  [4]\ $\mathcal{J}$(a\ , \ b) stands for a Jeffrey's prior ranging from a to b.
    \item[5]  [5]\ The prior on stellar density is set based on the estimates from TESS Input Catalog (TIC) V8 \citep{Stassun2017tic,Stassun2019tic}.
    \end{tablenotes}
    \label{transit_fit_prior}
\end{table*}

\begin{table*}[htbp]
    \centering
    \caption{Median and the 68\% credibility of the posterior distribution for the transit fit of each system along with the final derived planet radius.}
    \begin{tabular}{ccccccc}
        \hline\hline
        TOI &$P$ (d) &$t_{0}$ (BJD-2457000) &$R_{p}/R_{\star}$ &$a/R_{\star}$ &$i$ (deg) &$R_{p}^{[1]}$ (R$_{\oplus}$) \\\hline
        
        $134.01$\tablenotemark{[a]} &$1.401459\pm0.000124$ &$1326.0332\pm0.0013$ &$0.0224\pm0.0010$ &$7.39\pm0.12$ &$85.56\pm0.61$ &$1.46\pm0.11$ \\
    
        $141.01$ &$1.007907\pm0.000076$ &$1325.5388\pm0.0010$ &$0.0144\pm0.0004$ &$3.83\pm0.24$ &$79.52\pm1.53$ &$1.78\pm0.13$ \\
        
        $144.01$ &$6.267843\pm0.000018$ &$1325.5037\pm0.0007$ &$0.0163\pm0.0003$ &$13.07\pm0.87$ &$87.35\pm0.50$ &$2.05\pm0.12$ \\
        
        $174.01$ &$17.667062\pm0.000354$ &$1325.3291\pm0.0017$ &$0.0273\pm0.0008$ &$36.09\pm0.93$ &$88.89\pm0.06$ &$2.06\pm0.19$ \\
        
        $174.02$ &$29.797264\pm0.000766$ &$1340.3038\pm0.0021$ &$0.0260\pm0.0010$ &$51.13\pm1.31$ &$89.16\pm0.04$ &$1.96\pm0.20$ \\
        
        $174.03$ &$12.163131\pm0.000335$ &$1335.7401\pm0.0043$ &$0.0161\pm0.0007$ &$28.14\pm0.72$ &$89.48\pm0.42$ &$1.22\pm0.13$ \\
        
        $175.01$ &$3.690644\pm0.000005$ &$1356.2032\pm0.0002$ &$0.0402\pm0.0004$ &$21.24\pm0.14$ &$88.76\pm0.06$ &$1.38\pm0.05$ \\
        
        $175.02$ &$7.450802\pm0.000020$ &$1355.2874\pm0.0006$ &$0.0447\pm0.0009$ &$33.93\pm0.22$ &$88.46\pm0.01$ &$1.53\pm0.08$ \\
        
        $175.03$\tablenotemark{[b]} &$2.253114\pm0.000005$ &$1354.9050\pm0.0004$ &$0.0250\pm0.0004$ &$15.28\pm0.10$ &$87.94\pm0.08$ &$0.86\pm0.04$ \\
        
        $186.01$ &$35.613301\pm0.000561$ &$1350.3116\pm0.0007$ &$0.0382\pm0.0012$ &$63.36\pm3.62$ &$89.37\pm0.08$ &$2.94\pm0.33^{[2]}$ \\
        
        $186.02$\tablenotemark{[b]} &$7.789163\pm0.000307$ &$1332.2825\pm0.0028$ &$0.0141\pm0.0005$ &$23.00\pm1.31$ &$89.25\pm0.47$ &$1.08\pm0.09$ \\
        
        $260.01$ &$13.478048\pm0.005188$ &$1392.2946\pm0.0029$ &$0.0261\pm0.0017$ &$33.92\pm3.88$ &$88.75\pm0.30$ &$1.76\pm0.30$ \\
        
        $402.01$ &$4.756415\pm0.000195$ &$1411.4619\pm0.0009$ &$0.0182\pm0.0006$ &$14.01\pm0.81$ &$88.77\pm0.71$ &$1.67\pm0.14$ \\
        
        $402.02$ &$17.178210\pm0.001417$ &$1414.5500\pm0.0011$ &$0.0268\pm0.0011$ &$32.97\pm1.91$ &$88.49\pm0.12$ &$2.47\pm0.23$ \\
        
        $431.01$\tablenotemark{[b]} &$12.461062\pm0.000788$ &$1440.6301\pm0.0006$ &$0.0410\pm0.0005$ &$29.52\pm0.70$ &$89.72\pm0.25$ &$3.26\pm0.23$ \\
        
        $431.02$ &$0.490068\pm0.000009$ &$1438.3795\pm0.0006$ &$0.0161\pm0.0003$ &$3.41\pm0.08$ &$85.66\pm1.98$ &$1.28\pm0.10$ \\
        
        $455.01$ &$5.358971\pm0.000334$ &$1423.4263\pm0.0004$ &$0.0442\pm0.0013$ &$29.40\pm0.08$ &$89.36\pm0.16$ &$1.33\pm0.08$ \\
        
        $562.01$ &$3.930638\pm0.000152$ &$1517.9997\pm0.0006$ &$0.0328\pm0.0008$ &$20.44\pm0.19$ &$88.54\pm0.10$ &$1.29\pm0.07$ \\
        
        $652.01$ &$3.984832\pm0.000428$ &$1543.2539\pm0.0017$ &$0.0186\pm0.0005$ &$10.67\pm0.63$ &$86.95\pm0.58$ &$2.10\pm0.14$ \\
        
        $784.01$ &$2.797179\pm0.000158$ &$1570.2232\pm0.0019$ &$0.0196\pm0.0006$ &$8.93\pm0.43$ &$88.75\pm1.20$ &$1.92\pm0.17$ \\
        
        $836.01$ &$8.592004\pm0.002328$ &$1599.7675\pm0.0036$ &$0.0333\pm0.0013$ &$24.02\pm1.76$ &$88.96\pm0.41$ &$2.51\pm0.33$ \\
        
        $836.02$ &$3.816514\pm0.000757$ &$1599.9956\pm0.0034$ &$0.0240\pm0.0014$ &$13.99\pm1.03$ &$87.95\pm0.83$ &$1.81\pm0.27$ \\
        
        $1233.01$\tablenotemark{[a]} &$14.176416\pm0.000981$ &$1571.3364\pm0.0013$ &$0.0280\pm0.0005$ &$29.74\pm0.44$ &$89.65\pm0.26$ &$2.64\pm0.17$ \\
        
        $1233.02$\tablenotemark{[b]} &$19.592851\pm0.001619$ &$1586.5663\pm0.0013$ &$0.0305\pm0.0006$ &$36.90\pm0.54$ &$89.46\pm0.13$ &$2.88\pm0.18$ \\
        
        $1233.03$ &$6.204033\pm0.000581$ &$1572.3938\pm0.0016$ &$0.0211\pm0.0004$ &$17.14\pm0.25$ &$89.54\pm0.41$ &$1.99\pm0.13$ \\
        
        $1233.04$ &$3.795184\pm0.000289$ &$1572.1137\pm0.0022$ &$0.0165\pm0.0006$ &$12.35\pm0.18$ &$88.17\pm0.56$ &$1.55\pm0.13$ \\
        
        $1233.05$\tablenotemark{[b]} &$29.529048\pm0.00928$ &$1586.4927\pm0.0069$ &$0.0181\pm0.0008$ &$48.50\pm0.71$ &$89.98\pm0.22$ &$1.71\pm0.15$ \\
        \hline\hline 
    \end{tabular}
    \begin{tablenotes}
    \item [1] [a] Median metric rank $>30$.
    \item [1] [b] Does not constrain the mass-radius relationship.
    \item[2]  [1]\ We adopted the stellar radius from TESS Input Catalog (TIC) V8
    \citep{Stassun2017tic,Stassun2019tic}.
    \item[2]  [2]\ The radius we derive for 186.01 or HD 21749b is larger than that reported in the literature (\citealt{Dragomir2019} and \citealt{Gan2021}), although it is still consistent within error bars, mostly due to a difference in stellar parameters (TICv8 reports a larger stellar radius). 
    \end{tablenotes}
    \label{trasit_fit_posteriors}
\end{table*}


    \section{Uniform Determination of Planet Masses \label{sec:mass}}
    
We describe here our RV data and analyses to uniformly derive the masses for the planets in our sample in this section, including a comparison with the literature results and the caveats in our work. For clarity, we present our figures of RV fits for all the targets in Appendix \ref{sec:appendix_rv_fits}.
    
    \subsection{Radial Velocity Data \label{subsec:rv_data}}
    
 The focus of this paper is on the analysis of observations and modeling of the derived masses from the Planet Finder Spectrograph \citep[PFS;][]{crane2006, crane2008,crane2010}, a precision optical radial velocity spectrograph calibrated by an iodine cell on the Magellan II telescope at Las Campanas Observatory. PFS was commissioned in October 2009 and has been used for a long-term monitoring campaign to search for new planets, as well as following up transiting planet candidates from HAT-South \citep{Bakos2004,Bakos2009}, WASP \citep{Pollacco2006}, and recently TESS \citep[e.g.,][]{Dragomir2019,Teske2020}.

 In January 2018 the PFS detector was upgraded to a new larger (10k$\times$10k) CCD with smaller (9 $\mu$m) pixels, and the default slit width used for science observations was changed from 0.5$\arcsec$ to 0.3$\arcsec$, increasing the resolving power from $\sim$80,000 to $\sim$130,000. We refer to pre- and post-upgrade PFS data as ``PFS1'' and ``PFS2''. All PFS spectra are reduced and analyzed using a custom IDL pipeline based on \citet{butler1996} that regularly delivers sub-1 m~s$^{-1}$ photon-limited precision and, especially with the higher resolution data, is capable of delivering near-0.5 m~s$^{-1}$ photon-limited precision. The scatter from RV standard stars measured with PFS post-upgrade is typically 1.5-2 m~s$^{-1}$. The PFS radial velocities used in our analysis are listed in Table \ref{tab:pfsrvs} by TOI number.
 
 We also perform the same mass-radius modeling using a set of masses from the literature, some of which also include a large fraction or all of the PFS data in this paper. For TOI-836 and TOI-260 we perform our own updated fit, including observations from additional spectrographs; no previously published mass measurements exist for these TOIs. Winters et al. (in prep) and Osborn et al. (submitted) will provide updated mass measurements (including more data) of TOI-455 and TOI-431, respectively. There are no additional public data for TOI-784 or TOI-1233. 
 
 Additional data for TOI-836 come from the HARPS and iSHELL precision RV spectrographs. During the period between 16th and 23rd March 2020, 15 spectra of TOI-836 were obtained with the High Accuracy Radial velocity Planet Searcher \citep[HARPS,][]{Pepe:2002,2003Msngr.114...20M} on the ESO 3.6m telescope at La Silla, Chile. Spectra were typically taken following a cadence of two per night, with a separation of 2--3 hours. These spectra were obtained as part of the \textit{NCORES} programme (PI:Armstrong, 1102.C-0249) designed to study the internal structure of hot worlds. We used the high-accuracy mode of HARPS with a $1\arcsec$ science fibre on the star and a second on-sky fibre monitoring the background flux during exposure. The nominal exposure time was 1500 seconds, resulting in a typical SNR per pixel of 75 and an RV error of 1.5 m~s$^{-1}$. Spectra and RV information were extracted using the offline HARPS data reduction pipeline hosted at Geneva Observatory. 
We use a flux template matching a K5 star to correct the continuum-slope in each echelle order. The spectra were cross correlated with a K5 mask to derive the cross correlation function (CCF) \citep{1996A&AS..119..373B}, on which we fit a Gaussian function to obtain RVs, FWHM and contrast. 
Additionally, we compute the bisector-span \citep{2001A&A...379..279Q} of the CCF and spectral indices tracing chromospheric activity \citep{2011A&A...534A..30G,2009A&A...495..959B}. No correlation between the RVs and any activity indicator was observed.
 
We obtained 55 spectra of TOI-836 during 10 nights with the iSHELL spectrometer on the NASA Infrared Telescope Facility \citep{2016SPIE.9908E..84R}, spanning 43 days from January -- March 2020. The exposure times were 5 minutes, repeated 3--10 times within a night to reach a cumulative photon signal-to-noise ratio per spectral pixel varying from 134--241 at $\sim$2.4 $\mu$m (the approximate center of the blaze for the middle order). This achieves a per-night RV percision of 2--6 ms$^{-1}$ with a median of 3.4 ms$^{-1}$. Spectra were reduced and RVs extracted using updated methods to those outlined in \cite{2019AJ....158..170C}. These observations were carried out under IRTF programs 2019B047 and 2020A088 (PI: Cale).

 Additional data for TOI-260 come from Keck I/HIRES \citep{Vogt1994} and PFS archival data. The 37 epochs of Keck I/HIRES observations span September 2008 to January 2014, and were published in \cite{Butler2017} and corrected for small systematic effects in \cite{Tal-Or2019}. Typical internal errors of the HIRES TOI-260 data are $\sim$2 m~s$^{-1}$. The 28 epochs of archival PFS observations (PFS1) span March 2010 to August 2016, and were taken with the 0.5$\arcsec$ slit providing lower resolving power than the MTS-specific observations. Typical internal errors of the PFS1 TOI-260 RVs are $\sim$1.5 m~s$^{-1}$, and these measurements are included in Table \ref{tab:pfsrvs}. 
 
        \begin{table}
	        \centering
	        \caption{PFS Radial Velocities }
	        \begin{tabular}{lcccc} 
	        TOI & JD-2450000 	&	RV & $\sigma_{\rm RV}$  &Note\\
	        \hline
	               & (- 2450000)& (\mps) & (\mps)&\\
	        \hline 
           134 &8406.61644  &  1.83& 1.01 &PFS2\\
           134 &8407.61934 &  -3.66& 0.79& PFS2   \\
          134 &8408.6066   &  1.48 &0.85 &PFS2  \\ 
           134 &8409.60288  &  4.47& 0.73 &PFS2  \\ 
            134 &8410.57667 &  -4.24 &0.77 &PFS2  \\ 
            134 &8411.60363 &  -4.49& 0.66 &PFS2   \\	
            ...&...&...&...&...\\
	        \hline
	        \end{tabular}
	    \tablecomments{This table is published in its entirety in the machine-readable format. A portion is shown here for guidance regarding its form and content.}
        \label{tab:pfsrvs}
        \end{table}
    
    \subsection{Radial Velocity Modeling using Only PFS Data}\label{subsec:pfs_rv_fitting}
        \subsubsection{RadVel}\label{subsubsec:Radvel}
     
We began our RV modeling using \texttt{RadVel} v.1.3.2 \citep{Fulton2018a} and the nightly-binned PFS radial velocity observations, with observations acquired during planet transits removed. We used the \texttt{`per tc secosw sesinw k'} basis for fitting and posterior sampling, and fixed the period and time of inferior conjunction to the values listed in ExoFOP-TESS\footnote{https://exofop.ipac.caltech.edu/tess/} from the TESS project.\footnote{The period and time of inferior conjunction for each planet are consistent between our transit fits and the released values on ExoFOP-TESS. The differences are smaller than their respective error bars, which are on timescales too short for RV observations to make any significant impact in the RV fitting.} For the systems in which there has been a published detection of additional non-transiting planet(s), we also included those planets in our model but allowed their $P$ and $t_c$ values (in days) to vary within 2$\times$ the errors of the published values (that is, we set a Gaussian prior with $\sigma=2\times$ the published error value). For our initial fit, we also fixed `secosw', `sesinw' ($\sqrt{e}\cdot$sin($\omega$) and $\sqrt{e} \cdot$cos($\omega$)), slope ($\dot{\gamma}$,m~s$^{-1}$~day$^{-1}$), and curvature ($\ddot{\gamma}$, m~s$^{-1}$~day$^{-2}$) to zero, and initialized each planet's semi-amplitude $K$ value in m~s$^{-1}$ to the value that we predicted for our ranking metric (\S\ref{sec:survey}). We set a wide prior on the PFS white noise jitter term ($\mathcal{U}$(0,15) m~s$^{-1}$), and a wide prior on $K$ ($\mathcal{U}$(-20,20) m~s$^{-1}$). We allowed negative $K$ values to avoid biasing our mass measurements toward too-high values and use these instances as an indicator that we are only measuring an upper limit. Section \S\ref{sec:hbm} describes how we incorporated negative masses into our mass-radius relation modeling.

We used the `fit' command in \texttt{RadVel} to perform a fit to maximize the posterior probability (maximum posteriori optimization, MAP), which in \texttt{RadVel} is found using Powell's method (Powell 1964) implemented via \texttt{scipy.optimize.minimize}. To estimate uncertainties on the fitted parameters, \texttt{RadVel} implements an Affine Invariant Markov Chain Monte Carlo (MCMC) sampler via \texttt{emcee} \citep{foreman-mackey2013}, and computes the Gelman-Rubin Statistic \citep{gelman2003} to check for convergence; values close to one indicate the chains have converged. The default approach in \texttt{RadVel} is to (1) run an initial set of MCMC chains until the Gelman-Rubin statistic is $<1.03$ for all free parameters,  (2) discard these burn-in chains and launch new chains from the last position, (3) run the MCMC until the Gelman-Rubin statistic is $<1.01$ and the number of independent samples ($T_z$, \citealt{ford2006}) is $>1000$ for all free parameters for at least five consecutive checks (which happen every 50 steps). \texttt{RadVel} runs eight independent ensembles of samplers in parallel with 50 walkers per ensemble, for up to a maximum of 10,000 steps per walker or until convergence is achieved. The default initial step sizes are 10\% of the values of all free parameters, except period, which is set to 0.001\% of the value. 

We systematically tested more complicated fits in \texttt{RadVel}, leaving more parameters -- slope ($\dot{\gamma}$), curvature ($\ddot{\gamma}$), and `secosw' and `sesinw' -- free to vary, and including a correlated noise model via Gaussian process (GP) model with a quasi-periodic kernel as implemented by \texttt{celerite} \citep{Foreman-Mackey2017}:

\begin{equation}
 C_{ij} = \frac{B}{2+C} \times e^{-|t_i - t_j|/L} \times (\rm{cos}(\frac{2\pi|t_i-t_j|}{P_{rot}}) + (1+C)). 
\end{equation}

\noindent The addition of a GP term is motivated by previous works that have successfully modeled stellar activity effects in RV data sets to help tease out the true planetary signals \citep[e.g.,][]{Haywood2014,Grunblatt2015,Lopez-Morales2016,Barragan2019}. To standardize our approach, we used the same broad priors on the varied parameters for every target as listed in Table \ref{tab:rv_priors}. We ran our fits with different combinations of these free parameters and with or without a GP component. Finally, we computed the Akaike Information Criterion (AIC) to evaluate if adding complexity to our fits improved the results significantly, with lower AIC values corresponding to a more favored model. A general rule of thumb is that $\Delta$AIC values $<2$ indicate the models are indistinguishable, where larger $\Delta$AIC values start to indicate the model with the lower AIC value is favored. In the case of a small $\Delta$AIC values between models, we chose the model with the least number of free parameters as the ``best fit''. The $K$ values, $\Delta$AIC, and the corresponding planet masses from the ``best fit'' \texttt{RadVel} models are compiled in Table \ref{tab:pfs_K}. All other ancillary parameters from the fits are included in the appendix in Table \ref{tab:radvel_ancilliary}.

            \subsubsection{juliet}

 While \texttt{RadVel} provides the Bayesian Information Criterion (BIC) and AIC to evaluate differences between models, these criteria depend only on point estimates of the posterior and are less robust than other model comparison techniques which consider the full posterior. To complement \texttt{RadVel}, we also implement the planet modeling code \texttt{juliet}, as in \S\ref{sec:radii}, which provides a possibly more robust estimate, via nested sampling, of the evidence of different models, so they can be compared and selected or combined. Briefly, \texttt{juliet} (1) makes use of \texttt{RadVel} to compute the Keplerian component of the probabilistic model for the RV data, (2) gives several different options for posterior sampling -- classical or importance nested sampling via \texttt{MultiNest} (\citealt{Feroz2009,Feroz2019,Buchner2014}), or dynamic nested sampling via \texttt{dynesty} \citealt{Speagle2018} -- that, unlike MCMC algorithms, do not require initial guesses of the parameters, and then (3) estimates not only the posterior distribution of parameters $\theta$ given the data $\mathcal{D}$, but also the Bayesian evidences $Z_i = p(\mathcal{D}$|$M_i$) for model $M_i$ via the posterior odds ratio, $p(M_i$|$\mathcal{D}$)/$p(M_j$|$\mathcal{D}$) = ($Z_i$/$Z_j$)($p(M_i)$/$p(M_j)$). Following \cite{Espinoza2019}, who follow \cite{Trotta2008}, we consider $\Delta$ln$Z=2$ as moderate evidence of a model being preferred, and $\Delta$ln$Z=5$ as strong evidence. Similarly to our \texttt{RadVel} model selection, in the case of a small $\Delta$ln$Z$ values between models, we chose the model with the least number of free parameters as the ``best fit''. 
 
 For our analysis we used \texttt{juliet} v.2.0 from the command line, and approached RV fitting in a similar systematic fashion as with \texttt{RadVel}, using the nightly-binned PFS radial velocity observations with those acquired during planet transits not included in the fit (but included in Table \ref{tab:pfsrvs}. For each planet, in our initial fit we again fixed the period and time of inferior conjuction (in days) to the values listed in ExoFOP-TESS from the TESS project, fixed $\sqrt{e}\times$ sin($\omega\times 180$/ $\pi$) and $\sqrt{e} \times$ cos($\omega\times 180$/ $\pi$) to zero, and did not include any trends in the data. \texttt{juliet} takes priors on each parameter as input, so we started with wide uniform priors on $K$ $\mathcal{U}(-20,20)$ m~s$^{-1}$, based loosely on our expected planet masses. As with \texttt{RadVel}, we set a wide prior on the PFS white noise jitter term ($\mathcal{U}$(0,15) m~s$^{-1}$) and a narrower prior on the PFS velocity zero-offset term $\mu$ ($\mathcal{U}$(-2,2) m~s$^{-1}$), since PFS velocities were offset to have a median of 0 m/s out of the RV extraction pipeline prior to fitting. 
 The priors are listed in Table \ref{tab:rv_priors}. Also, similar to our \texttt{RadVel} fits, for the systems in which there has been a published detection of additional non-transiting planet(s) we also included those planets in our model but allowed their $P$ and $t_c$ values to vary within 2$\sigma$ of the published values.
 
 We then proceeded to increase the complexity of the \texttt{juliet} model by allowing a linear trend, quadratic trend, non-zero eccentricity and the argument of periastron, and finally by including a Gaussian process correlated noise model. In keeping consistent with our \texttt{RadVel} approach, we used the same quasi-periodic kernel as implemented by \texttt{celerite} \citep{Foreman-Mackey2017}. For each model,  We used $\Delta$ln$Z$ for each model to evaluate which $K$ value to report as a our ``best fit'' for each system; these $K$ values, the model evidences, and the corresponding planet masses are compiled in Table \ref{tab:pfs_K}. All other ancillary parameters from the fits are included in the appendix in Table \ref{tab:juliet_ancilliary}.

    \begin{table*}
        \centering
        \footnotesize
        \caption{Priors used in fits \texttt{juliet} and \texttt{RadVel}.}
        \label{tab:rv_priors}
        \begin{tabular}{lcc}
            \hline
            \hline
            \noalign{\smallskip}
            Parameter & Prior & Description \\
            \noalign{\smallskip}
            \hline
        $P$    & fixed to TESS project value & planet period (d) \\
        $t_0$  & fixed to TESS project value & center-of-transit time (d) \\
        $\sqrt{e} \times \mathrm{sin} ~\omega$ &  $\mathcal{N}(0,1)$ & parameterization of $e$ and $\omega$ \\
        $\sqrt{e} \times \mathrm{cos}~ \omega$ &  $\mathcal{N}(0,1)$ & parameterization of $e$ and $\omega$ \\
        $K$ &  $\mathcal{U}$(-20,20) & RV semi-amplitude (m~s$^{-1}$) \\
        $\gamma_{PFS}$ & $\mathcal{U}(-2,2)$ & velocity zero point (all PFS RVs are normalized before fitting)\\
        $\sigma_{PFS}$ & $\mathcal{U}(0,15)$ & PFS jitter (m~s$^{-1}$)\\
        GP$_{B}$ & $\mathcal{U}(0,\infty)$ & GP variability amplitude (m~s$^{-1}$) \\
        GP$_{C}$ & $\mathcal{U}(0,\infty)$ & GP periodic characteristic length  \\
        GP$_{L}$ &  $\mathcal{U}(0,\infty)$ &  GP non-periodic characteristic length (d) \\
        GP$_{Prot}$ &  $\mathcal{N}$(30,20) & GP variability period (d) \\
        \hline
        \end{tabular}
    \end{table*}

\subsection{Radial Velocity Modeling to Derive Masses to Compare to Literature Values}\label{subsec:all_rv_fitting}

Our survey is designed to be self-contained, reliant only on PFS data collected by our team according to the prioritization and cadence schemes outlined in \S\ref{sec:survey}. This enables us to better account for target selection and observational decisions in our hierarchical Bayesian modeling (\S\ref{sec:hbm}). However, limiting our mass derivation to only PFS data could, and in some of the targets likely does, limit the precision of the masses. So, we are also interested in whether inferences about the population change after including additional RV data in the planet mass derivation. As an exercise, we wish to compare the mass-radius relation derived from ``MTS-only'' masses to that derived from masses taken from a set of disparate sources. 

First, we searched for masses from the literature, which are listed in the right-most columns of Table \ref{tab:pfs_K}. For most of our sample, there are recent publications that include an analysis of all (or all at the time) available RV data to derive the planet masses, as described in Appendix \ref{sec:individual_systems}. In cases where no prior literature mass exists, we tried to collect additional data and combine with the MTS data to derive a mass analogous to a literature value (TOI-836 and TOI-260); this was not possible in all cases. To create our ``literature value'' sample, we then chose the mass with best precision for each TOI. In cases where there were no available additional data and not (yet) a literature value (TOI-431, TOI-455, TOI-784, TOI-1233), we included our MTS mass.

For TOI-836 (HIP 73472) and TOI-260 (HIP 1532), we combine the available public or contributed RV data and the PFS data to derive a mass using the most complete data set possible. We follow the same systematic procedure that is outlined above for the ``PFS data only'' fits. The results of these ``additional data'' fits for TOI-260 and TOI-836 are given in the last three columns of Table \ref{tab:pfs_K}. In the case of TOI-836 our fit includes iSHELL data taken in $K$-band, and in this paper we ignore that these data will have a different (smaller) activity amplitude than the visible wavelength spectrographs; analyzing the effects of different wavelengths of RV data is beyond the scope of this work.  
 
        \subsection{Caveats to RV fitting}

While our goal was to systematically step through increasing complexity in each system with \texttt{RadVel} and \texttt{juliet}, there were some special cases where this was not feasible. First, fitting TOI-144 and 652, we input and/or used as priors the most recently published information on the long-period, RV-detected giant planets that accompany the inner transiting planets. Our MTS RV data does not cover a long enough time span to place good constraints on the long-period planets, but they still influence the final fit to the system, so we need to account for them. For TOI-144, we use the long-period planet $P$ and $t_c$ values from \cite{Damasso2020}, and for TOI-652, we use the long-period planet $P$ and $t_c$ values from \cite{Teske2020}. For TOI-141, \cite{Espinoza2020} detected a second non-transiting planet at 4.79 d, and we use the parameters from that paper for the non-transiting planet in our model for TOI-141. For TOI-562, \cite{Luque2019} detected two non-transiting planets at 9.12 d and 55.7 d, and we use the parameters from that paper for the non-transiting planets in our model for TOI-141.

Second, the number of epochs of RV observations we have in hand ranges from system to system, and is sometimes as few as nine. For TOI-455 (10 MTS RVs), we do not attempt any of the more complex fits, and just report the ``best fit'' as that in which only $K$ varies. For TOI-1233 (33 MTS RVs) and TOI-431 (9 MTS RVs) we do not attempt a fit with the correlated noise (GP) component, given the number of free parameters in a four-planet and two-planet fit, respectively. For TOI-455 (12 MTS RVs), we only test and report the simplest fit where $K$ alone is free.

Third, in the case of TOI-174, TOI-431, and TOI-1233, our RV modeling consistently produced $K \lesssim 0$ values for at least one of the planets in the system. In each of these negative $K$ cases, we refit the system without the negative $K$ planet, and used the RMS (root-mean-square) residuals of that fit as the upper limit on the removed planet's $K$ value. We also tested whether removing the negative $K$ planets from the fits significantly changed the $K$ values of the other planets. For TOI-174.03 and TOI-1233.01, we found no significant difference within 1$\sigma$ errors. For TOI-431, our RV modeling produced a negative $K$ value for 431.02, but removing this planet from the fit more significantly changed the $K$ value of 431.01 -- from 1.34$\pm$1.7 m~s$^{-1}$ with 431.02 included to 2.51$\pm$1.10 m~s$^{-1}$ without it, using \texttt{RadVel}. Although TOI-431.01 is too large to be in our sample, we report both the 2-planet and 1-planet fit masses for this planet in Table \ref{tab:pfs_K}. 
 
We also note that, while for many of the TOIs the favored models of \texttt{RadVel} and \texttt{juliet} were the same, this was not always the case. In the case of TOI-174, in which the favored \texttt{RadVel} model includes a GP and the favored \texttt{juliet} model does not, the resulting upper limits on TOI 174.03 (at 12.2d) are significantly different, such that the \texttt{RadVel} limit is much lower than that from \texttt{juliet}. 

\begin{deluxetable*}{lccccccccccc}
\setlength{\tabcolsep}{1.5pt} 
\renewcommand{\arraystretch}{0.95} 
\rotate
\tablecolumns{12}
\tabletypesize{\scriptsize}
 \tablecaption{Results of RV fitting of PFS data \label{tab:pfs_K}}  
 \tablehead{
            \colhead{TOI}	& \colhead{\texttt{RadVel}} & \colhead{\texttt{RadVel} $| \Delta $AIC$|$} &	\colhead{\texttt{RadVel} $K$ }& \colhead{\texttt{RadVel} mass } &	\colhead{\texttt{juliet}} & \colhead{\texttt{juliet} $| \Delta  $ln($Z$)$|$}	& \colhead{\texttt{juliet} $K$} &\colhead{\texttt{juliet} mass}	& \colhead{Published or } & \colhead{Published or } & \colhead{Published or}\\[-0.3cm]
            \colhead{}	& \colhead{Model} & \colhead{(vs. next} &	\colhead{ }&	\colhead{} & \colhead{Model}	& \colhead{(vs. next}	& \colhead{} & \colhead{} & \colhead{all-data model} & \colhead{add.-data $K$} & \colhead{add.-data mass}\\[-0.3cm]
            \colhead{}	& \colhead{} & \colhead{best model)} &	\colhead{m~s$^{-1}$  }&	\colhead{M$_{\oplus}$} & \colhead{}	& \colhead{best model)}	& \colhead{m~s$^{-1}$} & \colhead{ M$_{\oplus}$} & \colhead{ } & \colhead{ m~s$^{-1}$} & \colhead{M$_{\oplus}$ }\\[-0.3cm]
            }
\startdata
    134.01\tablenotemark{a} &  $K$  & 3.20 &	2.94$\pm$1.18 & 3.61$\pm$1.43 &	$K$  & 0.24	& 2.90$\pm$1.17 & 3.55$\pm$1.44 & $K+e+$GP & 3.66$^{+0.47}_{-0.46}$ & 4.60$\pm$0.56 \\ 
    141.01 & $K+$trend  & 5.03 &	4.06$\pm$0.94& 6.53$\pm$1.62 & $K$  &8.77	& 3.09$\pm$1.01& 4.96$\pm$1.68	& $K+$GP & 5.30$\pm$0.39 & 8.83$^{+0.66}_{-0.65}$\\ 
    non-transiting 4.78d & $K$+trend  & 5.03 &	6.50$\pm$0.67 & 17.52$\pm$2.32 &$K$  &8.77	& 6.35$\pm$0.93 & 17.16$\pm$2.88 & $K+$GP & 7.26$^{+0.48}_{-0.47}$ & 19.95$^{+1.38}_{-1.36}$\tablenotemark{c} \\
    \hspace{4pt} planet in TOI-141\tablenotemark{b} &  &  &	&  & &	& &	&  & \\
    144.01  & $K+e_b$  & 11.14 &	1.42$\pm$0.47 & 4.35$\pm$1.48 &	$K+e_b+$GP$+$trend & 0.21 & 1.42$\pm$0.41 & 4.36$\pm$1.32 & $K+e_b$ & 1.50$\pm$0.20 & 4.30$\pm$0.70 \\ 
    non-transiting 2088d & $K+e_b$  & 11.14 &	196.43$\pm$0.55 & 4171.17$\pm$357.76\tablenotemark{c} &	$K+e_b+$GP$+$trend & 0.21 &196.47$\pm$0.47 &4172.10$\pm$357.77\tablenotemark{c}	&  $K+e_b$ & 196.10$\pm$0.70 & 9.89$\pm$0.25\tablenotemark{c} \\
    \hspace{4pt} planet in TOI-144\tablenotemark{b} &  &  &	&  & &	& &	&  & \\
    174.01 & $K+$GP & 6.31 &	2.99$\pm$0.39 & 10.32$\pm$1.56 &	$K$  & 7.03	& 3.39$\pm$0.27 & 11.70$\pm$1.30	& $K$ & 5.33$^{+0.67}_{-4.20}$& 17.92$^{1.41}_{-14.00}$ \\ 
    174.02 &  $K+$GP & 6.31  &	$1.39\pm 0.48$ & 5.68$\pm$2.00 &	$K$  & 7.03	& 1.18$\pm$0.30 & 4.85$\pm$1.28	& $K$ & 4.29$^{0.26}_{-3.44}$& 17.18$^{+1.07}_{-13.77}$\\ 
    174.03 & $K+$GP & 6.31  &	$<$0.51 & $<$1.55 &	$K$  & 7.03 & $<$1.49 & $<$4.53 & n.a. & n.a. & n.a.\\
    175.01 &  $K+$trend  & 12.53 & 1.97$\pm$0.66 & 2.11$\pm$0.72 & $K$  & 5.73& 1.62$\pm$1.15 & 1.72$\pm$1.24	& $K+e+$GP & 2.21$\pm$0.28 & 2.42$^{+0.35}_{-0.34}$\\ 
    175.02 &  $K+$trend   & 12.53 & 0.91$\pm$0.70 & 1.24$\pm$0.94 &	$K$   & 5.73	& 0.82$\pm$1.28 & 1.12$\pm$1.73	& $K+e+$GP &1.67$\pm$0.31 & 2.31$^{+0.46}_{-0.45}$ \\ 
    175.03\tablenotemark{b} &  $K+$trend   & 12.53 &	1.46$\pm$0.81 &1.32$\pm$0.73 &	$K$   & 5.73	& 0.94$\pm$1.44 &  0.84$\pm$1.29	& $K+e+$GP & $<1.06$ &$<1.01$ \\ 
    186.01 &  $K+e_{0.01}+$GP & 3.51 & 5.83$\pm$2.23 & 24.36$\pm$9.14 &	$K+$GP & 1.23	& 4.54$\pm$1.01 &18.91$\pm$4.48	& $K+$GP & 4.86$\pm$0.61 & 20.0$\pm$2.7 \\ 
    186.02\tablenotemark{b} &  $K+e_{0.01}+$GP & 3.51 &	0.55$\pm$0.35 & 1.38$\pm$0.88 &	$K+$GP & 1.23& 0.54$\pm$0.34 & 1.34$\pm$0.87  & $K$  & $<1.43$ & $<3.70$ \\ 
    260.01 &  $K+$GP & 7.77 & 1.09$\pm$1.56 & 2.93$\pm$4.26 &	$K+e$  & 0.95	& 1.03$\pm$3.81 & 2.75$\pm$10.39	& $K+$GP & 1.47$^{+0.85}_{-0.83}$\tablenotemark{d}  & 3.97$^{+2.39}_{-2.23}$\tablenotemark{d} \\
    402.01 &  $K+$GP & 12.16 & 2.85$\pm$0.31 & 6.88$\pm$0.97 &	$K$  & 2.14	& 2.84$\pm$0.49 & 6.87$\pm$1.34	& $K+e+$trend$+$GP & 3.11$\pm$0.35 &  7.20$\pm$0.81  \\ 
    402.02 &  $K+$GP &  12.16 &	1.81$\pm$0.78 & 6.76$\pm$2.94 &	$K$  & 2.14 & 1.96$\pm$0.50 & 7.25$\pm$1.94 &  $K+e+$trend$+$GP  & 2.48$\pm$0.47  & 8.79$\pm$1.68  \\ 
    431.01 (2pl)\tablenotemark{b} &  $K$  & 13.13 &	1.34$\pm$1.70 & 4.14$\pm$5.33 &	$K$  & 13.34	& 2.04$\pm$1.37 & 6.37$\pm$4.28	& n.a. & n.a. & n.a. \\ 
    431.01 (1pl)\tablenotemark{b} &  $K$  & 21.31 &	2.51$\pm$1.10 & 7.83$\pm$3.38 &	$K$  & 28.83	& 2.47$\pm$0.96 &7.70$\pm$3.03 	& n.a. & n.a.  & n.a. \\ 
    431.02 &  $K$  & 21.31 & $<$1.52 & $<$1.50 & $K$  & 28.83	& $<$1.42 & $<$1.50	& n.a. & n.a. & n.a \\
    455.01 & $K$  & n.a. &	2.40$\pm$1.09 & 2.61$\pm$1.19 &	$K$  & n.a.	& 2.38$\pm$1.07 & 2.57$\pm$1.17	& n.a & n.a.  & n.a. \\ 
    562.01 & $K$  & 5.26 &	2.28$\pm$0.55 & 2.78$\pm$0.69 & $K$  & 0.77	& 2.30$\pm$0.51 & 2.8$\pm$0.63	& $K+$GP & 1.52$\pm$0.25 & 1.84$\pm$0.31 \\ 
    non-transiting 9.12d &  $K$  & 5.26 &	1.81$\pm$0.45 & 2.91$\pm$0.72\tablenotemark{c} &	$K$  & 0.77	&1.80$\pm$0.41 & 2.90$\pm$0.66\tablenotemark{d}	& $K+$GP & 2.13$\pm$0.28 & 3.40$\pm$0.46\tablenotemark{c} \\
   \hspace{4pt} planet in TOI-562\tablenotemark{b} &  &  &	&  & &	& &	&  & \\
    non-transiting 55.96d &  $K$  & 5.26 &	2.76$\pm$0.47& 8.12$\pm$1.42\tablenotemark{c} &	$K$ & 0.77	& 2.76$\pm$0.43 & 8.11$\pm$1.31\tablenotemark{c}	& $K+$GP & 2.09$^{+0.34}_{-0.35}$ & 6.10$\pm$1.0\tablenotemark{c} \\
   \hspace{4pt} planet in TOI-562\tablenotemark{b} &  &  &	&  & &	& &	&  & \\
    652.01 &  $K+e_b$ & 1.17 & 3.30$\pm$0.53 & 8.65$\pm$1.58 & $K+e_b$  & 0.49 & 3.08$\pm$0.50 & 8.07$\pm$1.47	& $K+e_b$ & 2.89$^{+0.46}_{-0.43}$& 7.25$^{+1.19}_{-1.12}$\\ 
     non-transiting 1651d &  $K+e_b$  & 1.17 & 7.09$\pm$0.61 & 138.31$\pm$16.46\tablenotemark{c} &	$K+e_b$  & 0.49	& 6.25$\pm$0.53 & 121.68$\pm$14.26\tablenotemark{d}	& $K+e_b$ & 7.74$^{+0.69}_{-0.70}$ & 143.00$^{+12.71}_{-15.89}$\tablenotemark{c} \\
   \hspace{4pt} planet in TOI-652\tablenotemark{b} &  &  &	&  & &	& &	&  & \\
    784.01 &  $K+$trend & 8.35 &	4.52$\pm$0.52 & 9.71$\pm$1.38& $K$  & 0.82	& 4.83$\pm$1.31 & 10.36$\pm$2.96	& n.a. & n.a. & n.a. \\
    836.01 &  $K+$GP & 6.40 &	6.13$\pm$1.42 & 14.94$\pm$3.67 &	$K+$GP & 0.89& 5.95$\pm$0.81 & 14.44$\pm$2.28	& $K+$GP & 6.07$\pm$0.77\tablenotemark{d}  & 14.66$^{+2.33}_{-2.11}$\tablenotemark{d} \\
    836.02 &  $K+$GP & 6.40  &	3.09$\pm$0.56 & 5.73$\pm$1.12 &	$K+$GP& 0.89	& 3.22$\pm$0.53 & 5.97$\pm$1.09	& $K+$GP & 3.21$\pm$0.49\tablenotemark{d}  & 5.91$^{+1.04}_{-0.97}$\tablenotemark{d}  \\
    1233.01\tablenotemark{a} &  $K+e_{0.02}$ & 12.23 & $<$1.89 & $<$7.25 &	$K+e$  & 2.96	& $<$2.02 &$<$7.75	& n.a. & n.a. & n.a. \\
    1233.02\tablenotemark{b}& $K+e_{0.02}$  &  12.23 &	4.92$\pm$0.78 &20.99$\pm$3.84 &	$K+e$  & 2.96	& 4.48$\pm$0.83 & 19.10$\pm$3.91	& n.a. & n.a. & n.a. \\
    1233.03 & $K+e_{0.02}$  &  12.23 &	1.33$\pm$0.57 & 3.88$\pm$1.71 &	$K+e$  & 2.96	& 1.36$\pm$0.60 & 3.97$\pm$1.77	& n.a.& n.a. & n.a. \\
    1233.04 & $K+e_{0.02}$  &  12.23 &	1.64$\pm$0.59 & 4.06$\pm$1.51 &	$K+e$   & 2.96	& 1.75$\pm$0.59 & 4.33$\pm$1.54	& n.a. & n.a. & n.a. \\
    1233.05\tablenotemark{b} & $K+e_{0.02}$  &  12.23 &	4.59$\pm$1.33 & 22.42$\pm$6.90 &	$K+e$   & 2.96	& 2.20$\pm$1.12 & 10.85$\pm$5.55	& n.a. & n.a. & n.a. \\
    \enddata
        \begin{flushleft}
        \tablenotetext{a}{Median metric rank $>30$.}
         \vspace{-8pt}
       \tablenotetext{b}{Does not constrain the mass-radius relationship.}
       \vspace{-8pt}
     \tablenotetext{c}{Minimum mass of non-transiting planet.}
     \vspace{-8pt}
     \tablenotetext{d}{Fit using additional RV data as described in the text.}
     \vspace{-8pt}
      \tablecomments{Published Mass References: TOI-134 \cite{Astudillo-Defru2020}, TOI-141 \cite{Espinoza2020}, TOI-144 \cite{Damasso2020}, TOI-174 \cite{Trifonov2019}, TOI-175 \cite{Cloutier2019}, TOI-186 \cite{Gan2021} \& \cite{Dragomir2019}, TOI-402 \cite{Dumusque2019} but see also \cite{Gandolfi2019} (respectively), TOI-562 \cite{Luque2019}, TOI-652 \cite{Teske2020}. The masses of TOI-455.01 and TOI-431.01 \& 431.02  will be measured from a more extensive RV data sets and published by Winters et al. (in prep) and Osborn et al. (submitted), respectively. In the case of the all-data models, we quote our \texttt{RadVel} results; these are all consistent with our \texttt{juliet} results.}
        \end{flushleft}
    \end{deluxetable*}

\section{Population Analysis of the Mass-Radius Relation}\label{sec:hbm}

The MTS has been designed from the outset to enable population inferences. As this big-picture view drives our science, we choose to break with common practice in the RV community and present preliminary population-wide results before the survey is complete.  The purpose of this decision is to provide a published, transparent record of how the mass-radius relation can change as the data to which it is fit grows, and to examine how our survey design might affect the resulting M-R relation.  

Since we are starting with a small sample, we do not believe that a complex M-R model is either warranted or wise(for a direct demonstration of this, see \S \ref{subsec:broken_pl}).  We acknowledge that a population model that more directly ties the mass-radius space to planetary compositions can provide more intuitive interpretation of this data.  However, such composition-focused analyses are heavily dependent on the utilized interior structure models and the range of compositions that they are allowed to take.  Furthermore, an important goal of this study is to test the effect of observational choices on the M-R relation, which is more straightforward to assess with observed quantities like mass and radius.  As a result, we implement a simple model (Eqn \ref{eq:simpleMR}) that can provide apples-to-apples comparisons with prior work.

\subsection{Framework: Hierarchical Bayesian Model}\label{subsec:HBM}

There are non-negligible measurement uncertainties on small exoplanet masses and radii, and some masses that are effectively only upper limits. Given our goal of quantifying how much astrophysical scatter there is in planet masses around the radius gap, we must use an analysis method that can account for multiple sources of scatter in the data and incorporate uncertainties on both the individual and population-wide level.  Hierarchical Bayesian modeling (HBM) provides just this framework.  HBM has been described in detail in previous papers \citep{Hogg2010, Wolfgang2015} and used in several papers on the exoplanet mass-radius relation \citep{wolfgang2016,chenkipping2017,Neil&Rogers2020}; we defer to these sources for a detailed description of this method and its advantages.  For our purposes here, we note that hierarchical Bayesian modeling 
involves a likelihood where parameters on both the individual planet and population-wide levels are allowed to vary.  The probabilistic relationships between parameters on multiple levels also allows multiple sources of uncertainty to be propagated through the posteriors of all of the parameters.  In what follows, we focus on the posteriors for the population-wide parameters, hereafter referred to as ``hyperparameters''.

The model that we use for this analysis consists of a power-law  with some intrinsic dispersion at the population level.  Mathematically it can be summarized as
\begin{align}\label{eq:simpleMR}
\mu_M &= C\Big(\frac{R_{p}}{\text{R}_\oplus}\Big)^\gamma \nonumber\\
    \frac{M_{p}}{\text{M}_\oplus} &\sim \text{Normal}(\mu_M,\sigma)
\end{align}
where the hyperparameters $\gamma$ and $C$ define the ``mean'' power law between mass and radius, and $\sigma$ is the scatter of the individual mass measurements around this power law ($\sim$ means ``drawn from'' and indicates that the function on the right is a probability distribution).  Collectively, these hyperparameters define the mass-radius relation, and are the parameters whose posteriors are displayed in Figures \ref{fig:posteriors} and \ref{fig:posteriors_others}.  Often $\sigma$ is interpreted as representing the astrophysical diversity in planet compositions at a given radius, although the observed scatter in planet masses at similar radii could have other origins (see \S \ref{subsec:litresults}).

The full hierarchical Bayesian model consists of the likelihood $\mathscr{L}$ of the data ${D}$ (the observed masses ${\hat{M}}$ and radii ${\hat{R}}$) given the model parameters ${\theta}$ (the unobserved, infinitely precise ``true" masses ${M_t}$ and radii ${R_t}$ and the above hyperparameters of the mass-radius relation), scaled by the priors $\pi$ for all parameters:

\begin{align}\label{eq:HBMlik}
\mathscr{L}&({D}|{\theta})\pi({\theta}) \nonumber\\
&= \mathscr{L}({\hat{M}},{\hat{R}}|{M_t},{R_t},C,\gamma,\sigma) \pi({M_t},{R_t},C,\gamma,\sigma) \nonumber\\ 
&= p({\hat{M}}|{M_t})p({\hat{R}}|{R_t})p({M_t}|{R_t},C,\gamma,\sigma)p({R_t})p(C,\gamma,\sigma) \nonumber\\
&= p(C)p(\gamma)p(\sigma)\prod^N_{i=1} p(\hat{M}^{(i)}|M_t^{(i)}) p(\hat{R}^{(i)}|R_t^{(i)}) \nonumber\\
&\mspace{160mu} \times p(M_t^{(i)}|R_t^{(i)},C,\gamma,\sigma)p(R_t^{(i)})
\end{align}
where 1) we have assumed that the measurement uncertainty distributions for the observed masses and radii are independent, as is appropriate given their separate measurement techniques; 2) we have defined the mass-radius relation as a mass conditional on a radius, as in Eqn \ref{eq:simpleMR}; and 3) all masses and radii have units of M$_\oplus$ and R$_\oplus$, respectively, so that the hyperparameters are unitless.  The last step in defining the full hierarchical model is specifying each $p()$; this is given below.  In general, distinguishing between conditional probabilities ($p(x|y)$), joint probabilities ($p(x,y)$), and marginal probabilities ($p(x)$ and $p(y)$) is important for correct construction of hierarchical models, so we note that we have only dropped the ``$|$'' below for brevity.
\begin{align}\label{eq:HBM}
p(\gamma)&=\mathcal{N}(1,1)\nonumber\\
p(\text{ln}(C))&=\mathcal{U}(-3,3)\nonumber\\
p(\text{log}(\sigma^2))&=\mathcal{U}(-8,3)\nonumber\\
p(R_t^{(i)})&=\mathcal{U}(0.3,20)\nonumber\\
p(M_t^{(i)}) &=\mathcal{N}\Big(C(R_t^{(i)})^\gamma,\sigma\Big) \nonumber\\
p(\hat{R}^{(i)})&=\mathcal{N}(R_t^{(i)},\hat{\sigma}_R^{(i)})\nonumber\\
p(\hat{M}^{(i)}|\hat{M}^{(i)}>0) &= \hat{p}_{30}^{(i)}\mspace{5mu}\mathcal{N}(M_t^{(i)},\hat{\sigma}_M^{(i)}) \nonumber\\ 
& \mspace{30mu} + (1-\hat{p}_{30}^{(i)}) \mspace{3mu} \mathcal{U}(0,30)\nonumber\\ 
P(\hat{M}^{(i)}<\hat{M}_{lim}^{(i)}|\hat{M}^{(i)}<0) &= \hat{p}_{30}^{(i)} \int_0^{\hat{M}_{lim}^{(i)}}\mathcal{N}(M_t^{(i)},\frac{\hat{M}_{lim}^{(i)}}{2})dM_t^{(i)} \nonumber\\ & \mspace{30mu} + (1-\hat{p}_{30}^{(i)})  \int_0^{\hat{M}_{lim}^{(i)}}\mathcal{U}(0,30)dM_t^{(i)}
\end{align}

In the rest of this section, we describe the above hierarchical Bayesian model in detail and note some practical implementation requirements which have influenced these details.  To provide context to this discussion, we note now that we have fit this model with STAN \citep{STAN}, a platform for statistical modeling that produces full Bayesian statistical inference with Markov chain Monte Carlo (MCMC) sampling.  The specific algorithm used by STAN is Hamiltonian Monte Carlo (HMC) with a No-U-Turn sampler (NUTS).  This MCMC method uses the derivatives of the specified likelihood to perform efficient sampling of the posterior.  In particular, it numerically integrates to first order a Hamiltonian-like set of coupled differential equations describing the position and ``momentum'' (determined by the ``potential energy'', i.e. the posterior probability) of a location in parameter space to find a new location that would, if the posterior does not have significant curvature, result in a higher probability.  That step is either accepted or rejected with the standard Metropolis acceptance probability, which compares the actual probability of the new location with that of the current location.  The No-U-turn sampler avoids trajectories in parameter space that double back on themselves.

To return to Eqn \ref{eq:HBM}, the hyperparameters $\gamma$, $C$, and $\sigma$ are drawn from arbitrary but wide prior distributions, such as the uniform distribution denoted by $\mathcal{U}$(lower bound, upper bound) and the normal distribution denoted by $\mathcal{N}$(mean,standard deviation) where the standard deviation is large.  These distributions are largely uninformative, meaning that the data do not need to contain much information before the posteriors are pulled away from the prior distributions.  In other words, as long as the priors on the hyperparameters are sufficiently uninformative, the resulting M-R relation does not change with changes in the prior distribution.  In particular, the prior we use for $\sigma$ is a Jeffreys prior for the standard deviation of a normal distribution (note that $\sigma \sim 1/\sigma$ transforms to a uniform distribution on ln$(\sigma)$), which is a common uninformative prior for scale parameters. 

Below the hyperparameters, the parameters for the individual planets, denoted by ``$^{(i)}$'', are drawn from distributions that depend on the hyperparameters -- this multi-level probabilistic structure is what makes this Bayesian analysis hierarchical.  In particular, the ``true'' masses $M_t$ and ``true'' radii $R_t$ of the planets are defined as latent, unobserved variables -- in contrast to the observed quantities denoted by ``$\mspace{5mu}\hat{ }\mspace{5mu}$'' -- where each planet's $M_t$ is drawn from the M-R relation given its $R_t$.  Within the hierarchical model we require that $M_t>0$, as no planet can have a negative mass in nature\footnote{While STAN allows bounds on parameters to be formally restricted at declaration, these restrictions can impose hard boundaries in parameter space which STAN's HMC algorithm struggles to efficiently explore.  Specifically, sharp boundaries can cause divergences in the MCMC chains, for which there is concern that the accepted posterior samples may not reflect the actual underlying posterior probability.  In practice we found that the formal $M_t > 0$ bound produced a high ($>60$\%) fraction of divergences, so we impose a softer boundary using a sigmoid function which allows only slightly negative ($\lesssim|10^{-2}|$) $M_t$ values.  Fortunately in our case, both treatments of the $M_t > 0$ bound produced very similar posteriors.}.  However, we allow the \textit{observed} mass to possibly be below zero due to measurement uncertainty, which we approximate with a usual Gaussian distribution whose standard deviation is set to the standard deviation of the posterior samples ($\hat{\sigma}_M$). In general, the posteriors computed in \S \ref{sec:mass} may not be Gaussian, but in practice, a normal distribution is a close approximation of the posterior shape for nearly all constrained mass measurements (see Figures \ref{fig:MRresult_Radvel} and \ref{fig:MRresult_others} for contours computed directly from the individual planets' posterior masses and radii; note that the x-axis is in log scale).  We investigated using a more flexible approximation for the individual planet mass posteriors through the quantile-defined functions developed by \citet{Hadlock2019}\footnote{STAN requires that all probability distributions are analytically defined so that it can compute the derivatives of the likelihood that are required for the HMC algorithm.  Accordingly, we could not sample directly from the \texttt{RadVel} or \texttt{juliet} posteriors while using STAN.}, but found that the computational tradeoffs were not worth the change.  In particular, the computational time increased, mixing within chains decreased with higher autocorrelation between samples, and in the end the hyperparameter posterior did not significantly differ from the posterior that assumed normal distributions for the measurement uncertainties.

Returning to Eqn \ref{eq:HBM}, this mass measurement uncertainty is incorporated into the last two lines of the model.  The penultimate line provides the likelihood that we observed the specific value $\hat{M}^{(i)}$ for the planet's mass (in practice, this is the mean of the \texttt{RadVel} or \texttt{juliet} posteriors) when the mass posterior provides an acceptable constraint (which we define as $\hat{M}^{(i)}>0$).  $p(\hat{M}^{(i)})$ depends on: 1) the planet's true mass $M_t$, which in turn depends on the mass-radius relation; and 2) the probability $\hat{p}_{30}^{(i)}$ that the planet is ranked in the top 30 TOIs via the merit function of Eqn \ref{eqn:merit}, and thus is in the MTS sample.  As a result, this likelihood, and the corresponding posterior constraints on the M-R relation, include the quantitative selection function that is a salient feature of this survey (see discussion in \S \ref{subsubsec:whyquantselect}).  Specifically, $p(\hat{M}^{(i)})$ is a mixture of two components: One in which the planet is ranked highly enough to be included in the MTS sample, and thus has an informative mass measurement (first term), and one in which the planet falls out of the MTS sample and so does not contribute an informative mass measurement to the M-R relation (second term).   $\hat{p}_{30}^{(i)}$ is calculated by simulating the TOI parent sample (the list of TOIs that meet our eligibility criteria; see \S \ref{subsubsec:elimination}) several thousand times -- we sample from the measurement uncertainties of each planet's insolation flux and radius, including the uncertainties in the host star's $T_{eff}$, $M_\star$, and $R_\star$, evaluate the merit function, and rank the TOIs based on that merit for each of these several thousand instances.  The fraction of simulation instances for which a planet is ranked in the top 30 TOIs yields $\hat{p}_{30}^{(i)}$, which is treated as data (versus a parameter that can vary) in the model.\footnote{While it is possible with STAN to calculate the merit function and the associated ranking at each step of the MCMC chain, in practice the convergence of the chains is very poor when the instantaneous in/out-of-MTS-sample state is used.  Because STAN calculates derivatives of the likelihood, step functions in likelihood parameters prove to be particularly problematic.  This in/out-of-sample variable is a discrete, binary parameter which is set by such a step function: it equaled 1 if the planet's ranking was $\leq 30$, or 0 if not.  We tried substituting this step function with a differentiable-yet-still-steep sigmoid function; in practice, fast switching between the informative term and the uninformative term in the mixture still throttled convergence. The above mixture model formulation sidesteps these computational difficulties by effectively marginalizing over this instantaneous in/out state variable with the introduction of $\hat{p}_{30}^{(i)}$, which is calculated outside of the MCMC chains.}

The last line of Eqn \ref{eq:HBM} incorporates the information about the mass-radius relation that is provided by mass upper limits.  This information can be valuable because even though upper limits don't show where a planet is located on the mass-radius plane, they \emph{do} show where that planet is \emph{not} located.  With an analysis that incorporates upper limits, M-R relations that pass through that portion of mass-radius space are ruled out; meanwhile, an analysis that discards upper limits erroneously considers those M-R relations as possibilities. Therefore, analyses that do not consider upper limits are potentially arriving at incorrect conclusions, depending on the proportion of constrained to unconstrained measurements at crucial locations in the mass-radius space, such as at small radii.  The standard way to incorporate upper or lower limits in Bayesian models is to marginalize over the unobserved portion of the likelihood; this is accomplished via the integrals in the last line of Eqn \ref{eq:HBM}.  We note that this probability also incorporates the MTS selection function through the mixture model weighted with $\hat{p_{30}}^{(i)}$.  We identify which planets effectively have only an upper limit on its mass when the \texttt{RadVel} or \texttt{juliet} posterior has a negative mean value ($\hat{M}^{(i)}<0$); in these cases we do not consider the mass posterior to be reliable.  Therefore we do not use the posterior to inform the upper limit in the M-R modeling; instead, we set $\hat{M}_{lim}^{(i)}$ to the RMS of the residual RV time series, from which all planets with constrained mass measurements (i.e. $\hat{M}^{(i)}>0$) have been removed.  Because this mass upper limit is the only information we retain about the measurement uncertainty distribution, we had to choose a standard deviation that was reasonable given that upper limit.  We generated residual RV time series with differing true RV semiamplitudes hidden in white noise scaled to these RMS values and determined that a standard deviation equal to half of the upper limit was a conservative estimate of the width of this unobserved measurement uncertainty distribution: if $M_t^{(i)}$ was close to zero, then $\hat{M}_{lim}^{(i)}$ would be a 2-$\sigma$ upper limit, but given the more likely scenario that $M_t^{(i)}$ is at least 1 M$_\oplus$, $\hat{M}_{lim}^{(i)}$ as implemented above is closer to a 1-$\sigma$ upper limit.

\begin{equation}
p(M_{ob}^{(i)}|I^{(i)}=0,M_t^{(i)},...) = \int_0^{M_{lim}^{(i)}}\text{N}(M_t^{(i)},\sigma_{M})dM_t^{(i)} \nonumber
\end{equation}

This model borrows very heavily from \citet{wolfgang2016} with some minor adjustments, including the explicit integration of mass upper limits in the likelihood, the inclusion of the MTS selection function, and the removal of the theoretical upper mass limit that was used in \citet{wolfgang2016}.  While there has been significant developments in studying the mass-radius relation since then (i.e., \citealt{chenkipping2017, Ning2018, Neil&Rogers2020}), \citet{Ning2018} found that a power-law was a sufficient descriptor of the mass-radius relation for planets with radii $<5~R_\oplus$.  Since we are starting with a smaller sample and testing the effect of observational choices on the population result, we have decided to start with a simple model that can provide straightforward apples-to-apples comparisons.  Nevertheless, this is a clear area for future improvement (see \S \ref{subsec:futurework}).

Convergence of the posterior is assessed in the standard ways, by 1) plotting the parameter value as a function of the position in the chain to ensure no discontinuous jumps to new values have occurred after warm-up; 2) comparing the variance of parameter values within a chain to the variance between chains via the \^{R} value; and 3) assessing the autocorrelation from step to step within a chain.  All of the posteriors whose results are presented in the following sections were sampled by 8 chains for 1,000,000 steps apiece, with the first 500,000 discarded as warm-up and the value of every 100th step saved as output.  The average effective (i.e. uncorrelated) number of posterior samples ($n_{eff}$) per hyperparameter over the five fits we display below is roughly equal to the number of saved samples, pointing to low autocorrelation lengths, a positive indicator of convergence. \^{R} is universally equal to 1.0, indicating that the chains were well mixed.

\subsection{Preliminary M-R Relation with Mid-Survey Masses}\label{subsec:prelimMR}

\begin{figure}[htbp]
\centering
\includegraphics[width=\columnwidth]{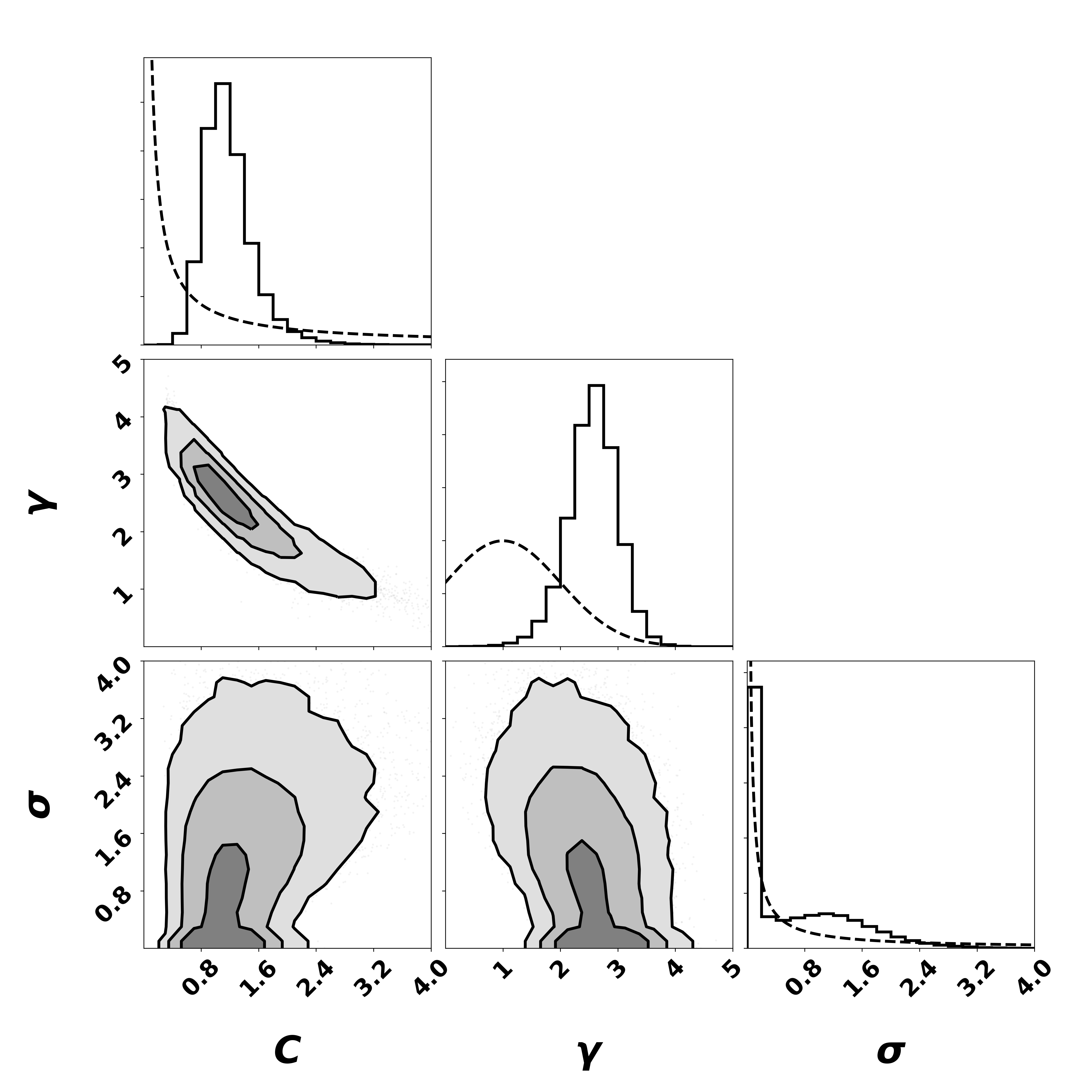}
\caption{The posteriors of the empirical power-law mass-radius relation (see Eqn \ref{eq:HBM}, which draws heavily from the model presented in \citealt{wolfgang2016}) fit to the 22 planets with non-zero $\hat{p}_{30}$ (see Eqn \ref{eq:HBM}) and mass measurements or upper limits derived using \texttt{Radvel}; these results will likely change as the survey grows.  The contours show the 68\%, 95\%, and 99.7\% credible regions of various 2-D projections of the joint 3-D hyperparameter posterior, which consists of $C$, the normalization constant of the power law, $\gamma$, the power-law index, and $\sigma$, the astrophysical scatter around the mean relation.  The histograms on the diagonal show the marginal 1-D posteriors for those hyperparameters; the dashed lines denote the corresponding prior distribution. Note that the power-law normalization constant $C$ is close to 1, and that there is currently no statistically significant astrophysical scatter in this dataset, as the $\sigma$ posterior mostly follows the prior distribution and is still consistent with 0; these preliminary results contrast with prior published work.  See \S \ref{subsec:mrcompare} for a full discussion.}
\label{fig:posteriors}%
\end{figure}

\begin{figure*}[htbp]%
\centering
\plottwo{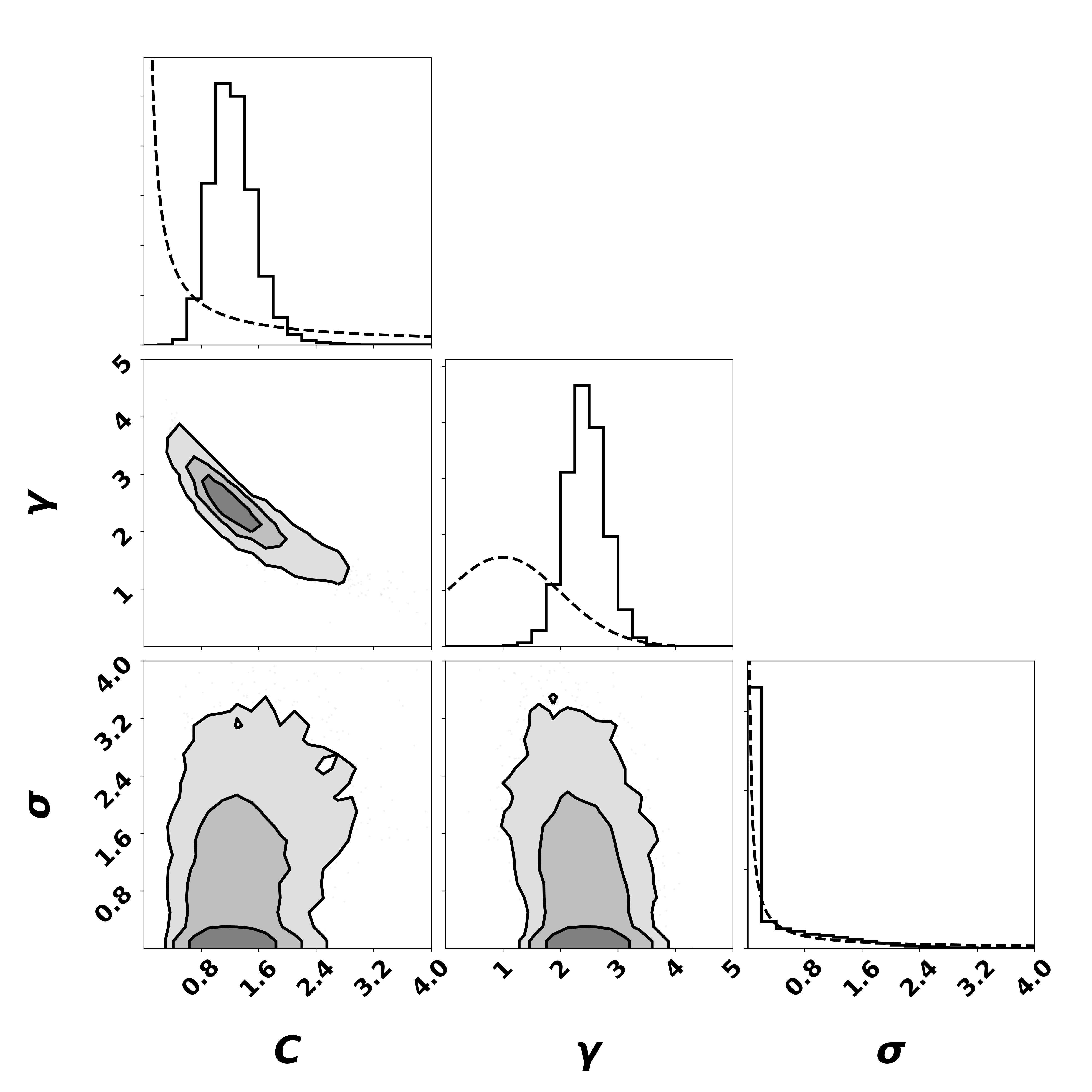}{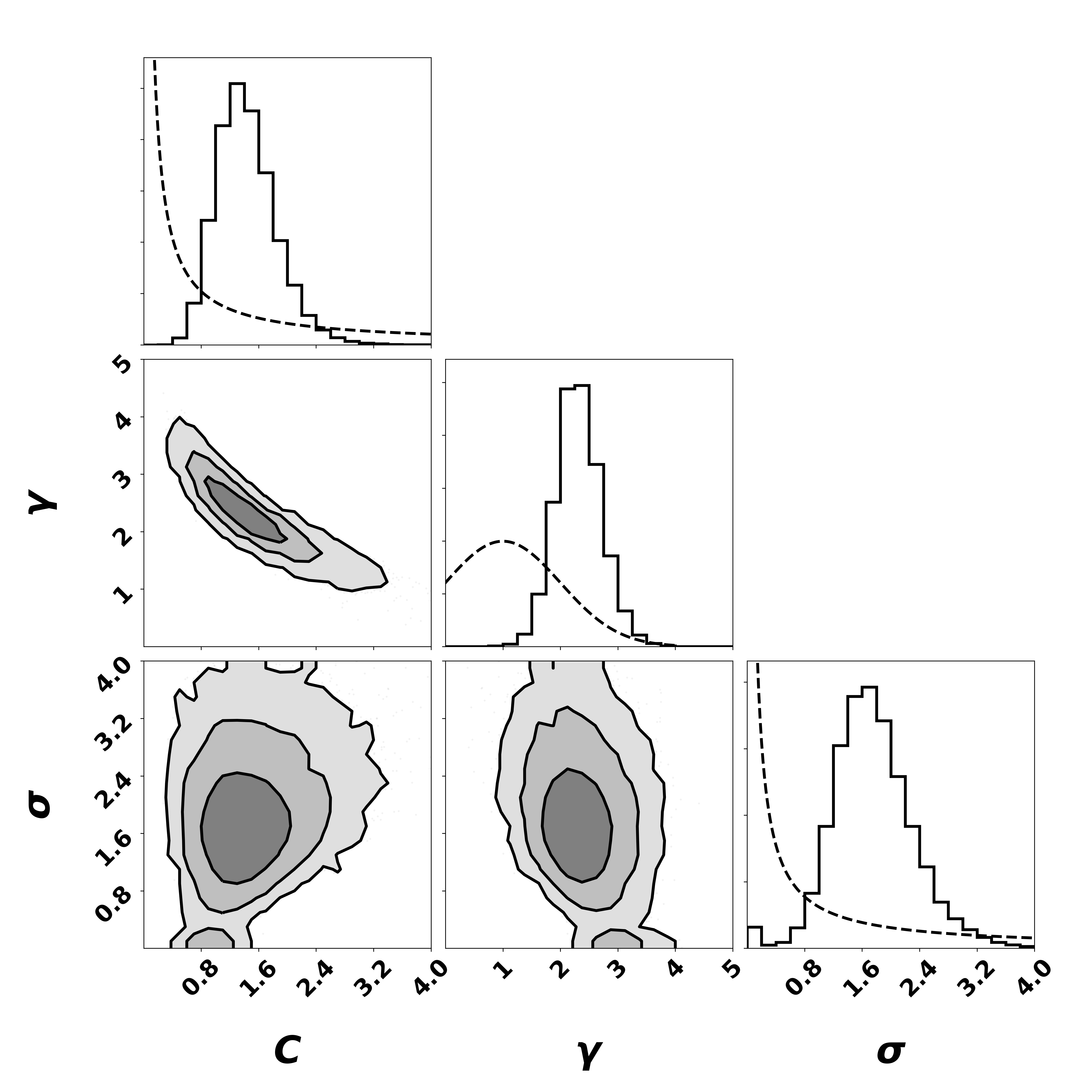}
\caption{Same as Figure \ref{fig:posteriors}, but using the \texttt{juliet}-derived masses (left) and the heterogeneous masses pulled from the literature, where available (right).  Note that all axes are on the same scale as Figure \ref{fig:posteriors}.  The hyperparameter posteriors for the \texttt{juliet}-derived M-R relation are very similar to the hyperparameter posteriors for the \texttt{Radvel}-derived M-R relation; both relations use the exact same radial velocity data from PFS.  On the other hand, the hyperparameter posteriors for the literature-derived M-R relation is noticeably different from both of these for $\sigma$, wherein there is significant probability away from zero; while the same planets were included in this relation as the \texttt{Radvel}-derived and \texttt{juliet}-derived ones, the underlying radial velocity data differs and is drawn from heterogenous sources.  Part of the difference in $\sigma$ may be due to the fact that overall the literature masses are currently more precise (see discussion in \S \ref{subsec:limitations}).  If the final MTS mass error bars become similar in size to those in the literature, and if the MTS $\sigma$ remains consistent with zero, then the non-zero intrinsic scatter fit by previous M-R studies to heterogeneous mass data may not actually astrophysical but may instead be driven by differences in how different RV teams acquire, reduce, and analyze their data.  See \S \ref{subsec:mrcompare} for a full discussion.}%
\label{fig:posteriors_others}%
\end{figure*}

In Figure \ref{fig:posteriors} we display the posteriors that result from fitting the M-R relation described in Eqn \ref{eq:HBM} to the 22 planets with non-zero $\hat{p}_{30}$ (see Eqn \ref{eq:HBM}) and with mass measurements or upper limits obtained with \texttt{RadVel}, as described in \S \ref{subsubsec:Radvel}. The posterior medians and central 68\% credible intervals for the three hyperparameters are: ${C=1.13_{-0.26}^{+0.35},\gamma=2.57_{-0.42}^{+0.39},\sigma=0.28_{-0.28}^{+1.19}}$.  The left subplot of Figure~\ref{fig:posteriors_others} shows the posteriors from our M-R relation fits using masses and upper limits derived with \texttt{juliet}; these results are consistent with the \texttt{RadVel} results in Figure~\ref{fig:posteriors}: ${C=1.22_{-0.27}^{+0.31},\gamma=2.43_{-0.33}^{+0.34},\sigma=0.03_{-0.03}^{+0.89}}$.  This consistency is expected, as the two sets of mass measurements were derived using the same dataset, and our \texttt{RadVel} and \texttt{juliet} masses agree with each other within error bars.

\begin{figure*}[htbp]
\centering
\includegraphics[width=\textwidth]{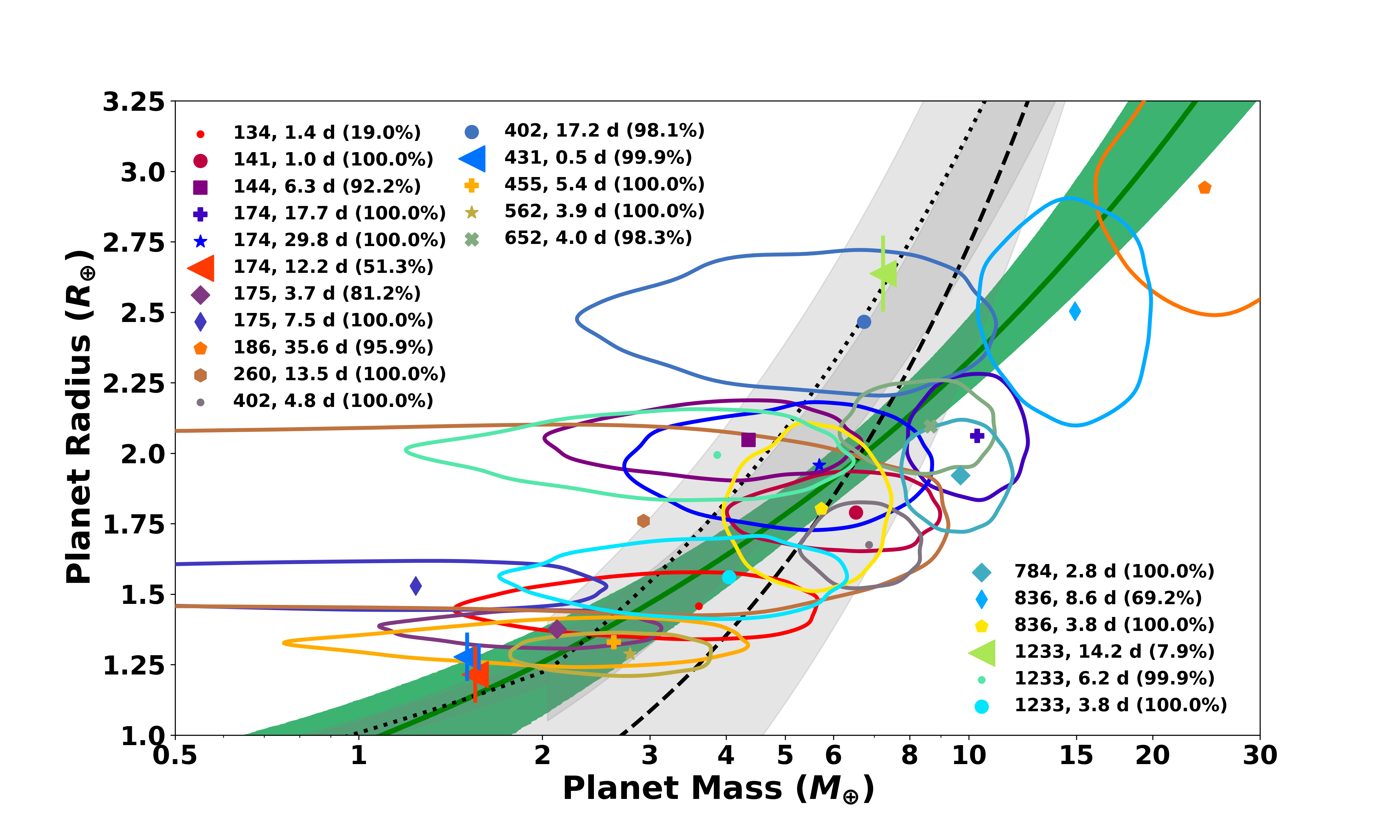}
\caption{The empirical mass-radius relation of Eqn \ref{eq:HBM} (green solid line) with the 22 planets with mass constraints and non-zero $\hat{p}_{30}$ (see Eqn \ref{eq:HBM}).  Mass measurements are represented by the colored contours (only the central 68\% of the joint mass, radius uncertainty distribution is shown) with the mean values denoted with a symbol; upper limits are denoted with left-pointing triangles, which correspond to rms of the residual RV time series for that planetary system.  The legend contains the TOI host star, planetary orbital period, and $\hat{p}_{30}$, the probability that the TOI is ranked in the top 30 by our merit function (Eqn \ref{eqn:merit}).  The green band corresponds to the posterior predictive M-R relation, meaning that it was produced by ranging over the hyperparameter posteriors (see Figure \ref{fig:posteriors}).  Since the $\sigma$ for this dataset is consistent with 0, the width of this band is dominated by the uncertainty in the mean mass-radius relation, not by any intrinsic variation in these planets' compositions. Specifically, the band represents the central 68\% credible interval of the posterior predictive masses at a given radius.  The gray bands denote previous mass-radius relations from the literature, fit to datasets that differ significantly than the one used here: the dashed line centered in the wider gray band corresponds to the \citet{wolfgang2016} mass-radius relation that was fit to the RV-only data using the model with a radius-dependent astrophysical scatter; and the dotted line centered in the narrower gray band corresponds to the \citet{chenkipping2017} relation.  The M-R relation fit here has some preliminary quantitative differences with this prior work (discussed in \S \ref{subsec:mrcompare}) that will be monitored as we acquire more data.}
\label{fig:MRresult_Radvel}
\end{figure*}

\begin{figure*}[htbp]
\centering
\plottwo{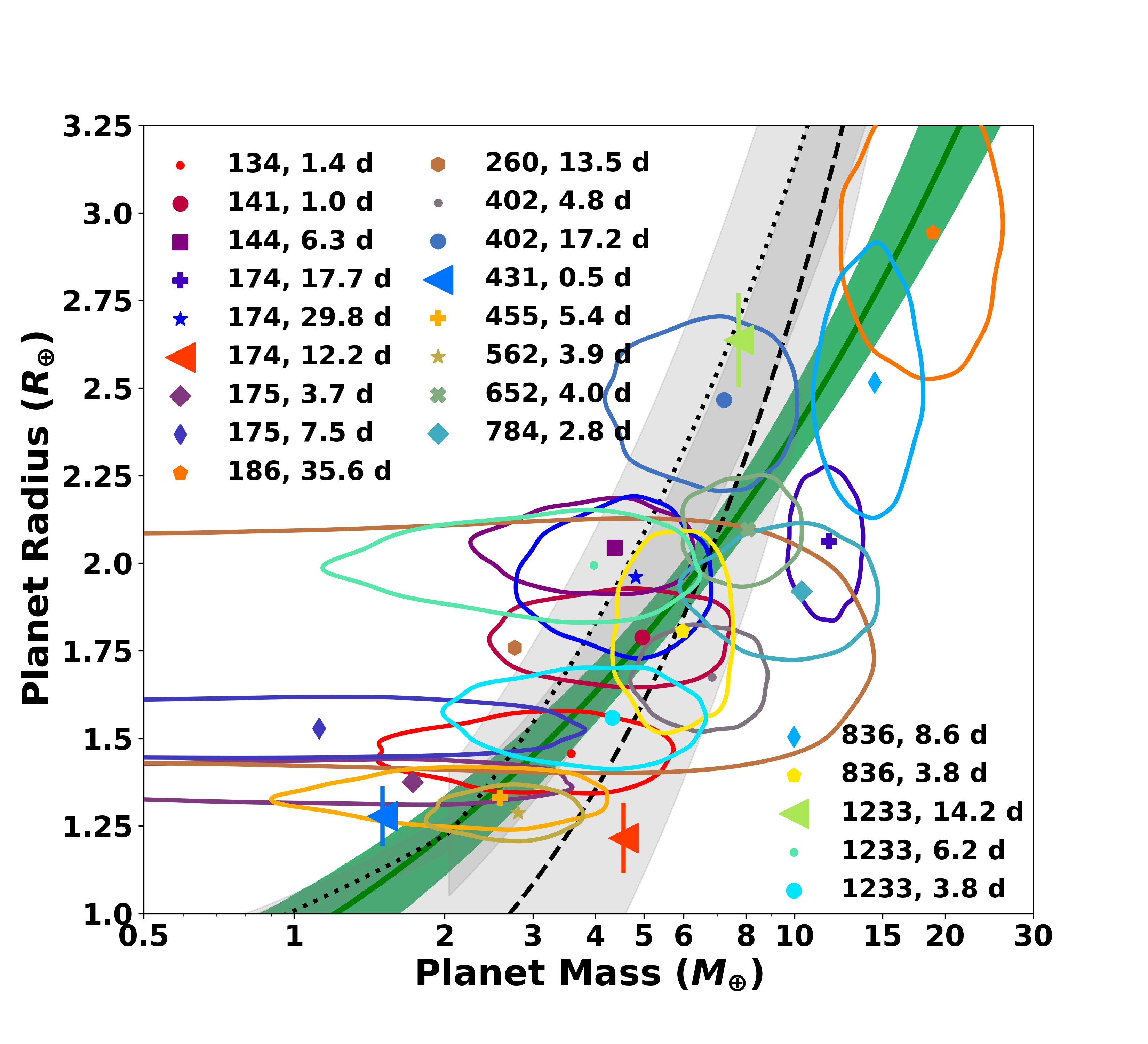}{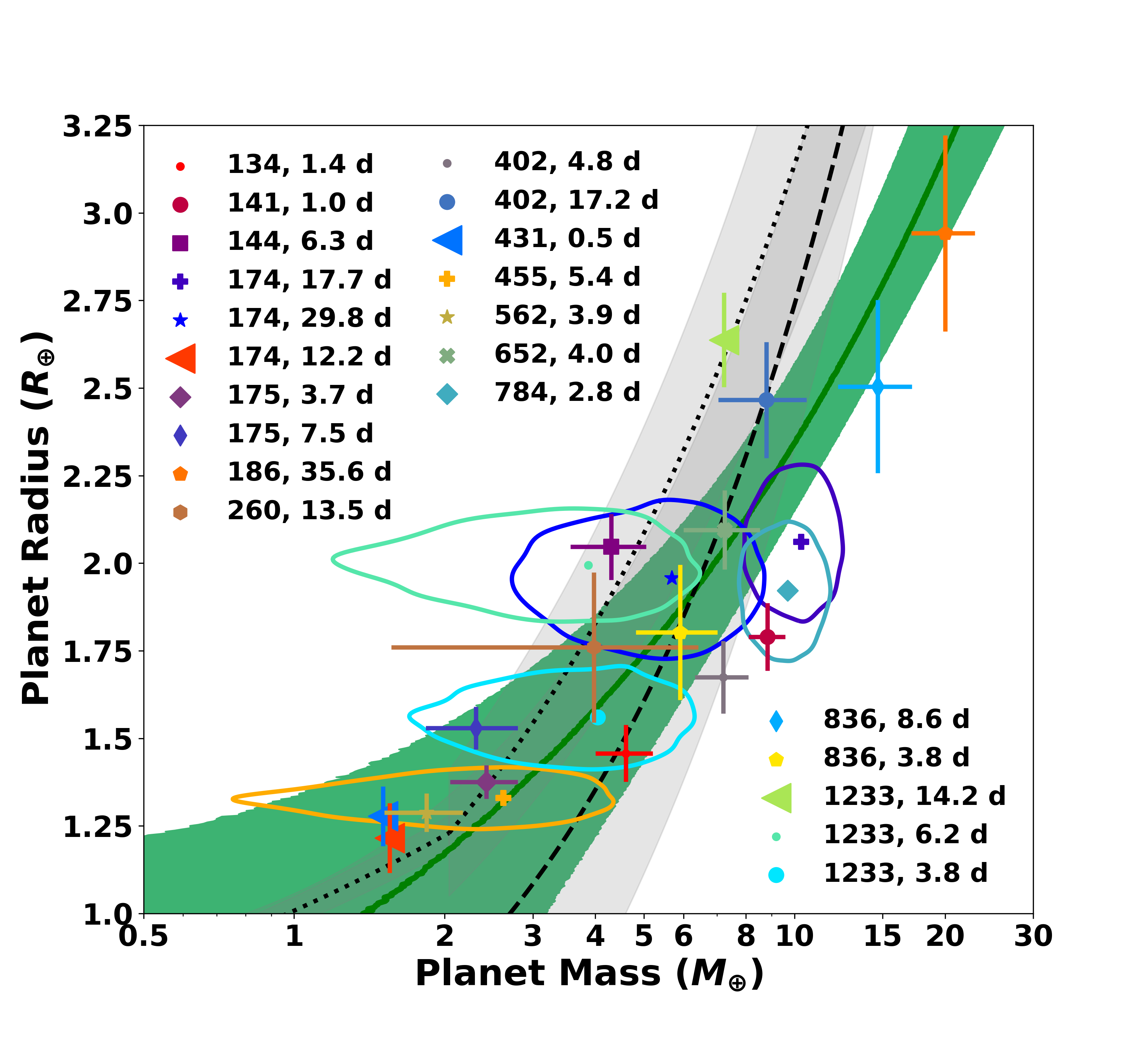}
\caption{Same as Figure \ref{fig:MRresult_Radvel}, but for the \texttt{juliet}-derived masses (left) and for the masses from the literature, where available (right, see \S \ref{subsec:litresults}).  On the right, the data from the literature is denoted by the crosses; for planets without previously published masses or whose published masses were less precise than the ones computed here, we use our \texttt{RadVel} mass estimates and denote those planets' mass and radius uncertainties with the same contours as Figure \ref{fig:MRresult_Radvel}.  The \texttt{juliet} M-R relation is very similar to the \texttt{RadVel} relation, which reflects the fact that the majority of the individual planet \texttt{Radvel} and \texttt{juliet} masses are consistent within 1-sigma.  On the other hand, the M-R relation fit to the literature values is quantitatively different than that fit to the masses calculated here, in both the width of the relation and in its mean predicted mass at $1 R_\oplus$.  For full discussion, see \S \ref{subsec:litresults}.}
\label{fig:MRresult_others}
\end{figure*}

Figures \ref{fig:MRresult_Radvel} and \ref{fig:MRresult_others} display these results projected on the mass-radius plane.  In particular, in green we plot the posterior predictive mass-radius relation, which means that it was produced by ranging over the above hyperparameter posteriors rather than just using the ``best fit'' values.  As a result, the width of the band includes the uncertainty on the M-R relation itself.  Figure \ref{fig:MRresult_Radvel} displays the M-R relation that was fit to the \texttt{RadVel} masses; given that the intrinsic scatter for this fit is consistent with zero, the uncertainty in the M-R relation hyperparameters dominates the width of the band.  The left panel of Figure~\ref{fig:MRresult_others} displays the M-R relation that was fit to the \texttt{juliet} masses; again, it looks very similar to the fit to the \texttt{RadVel} masses. The colored contours or upper limits (triangles) are the mass and radius constraints for the 22 planets with non-zero $\hat{p}_{30}$, color-coded by individual planets; the mean mass and radius for each planet is demarcated by a symbol. The gray lines and regions are M-R relations from \cite{wolfgang2016} (dashed line) and \cite{chenkipping2017} (dotted line). We highlight some features of our M-R relation results in comparison with these previous ones and discuss the implications in Section~\ref{subsec:mrcompare}.

\subsection{Analysis Including Masses Derived from Literature Values}\label{subsec:litresults}

As described in \S\ref{subsec:cadence_choice}, in the MTS we use a detailed RV observation cadence scheme to determine when to observe our targets and their relative priority within an observing run, with the ultimate goal of producing a relatively homogeneous sample in terms of observational coverage and precision on the mass measurement. Studies that combine mass and radius measurements from different teams would likely produce a more heterogeneous dataset, although the exact effects of such heterogeneity are difficult to quantify without a homogeneous control sample.  Given the uniformity of our approach, we are able to shed light on this issue.  Specifically, we compare the MTS M-R relation results from \S \ref{subsec:prelimMR} to those fit to a heterogeneous sample that includes literature values where they are available (see \S \ref{subsec:all_rv_fitting} for details). As the planet sample itself is not changing, any differences between the MTS- and literature-based M-R results would not arise from differences in target selection, but instead may be due to heterogeneous analysis techniques or to heterogeneous data acquisition processes (modulo typical statistical variance).  Performing this test therefore allows us to distinguish how much the differences between the MTS M-R relation and previous M-R relations in the literature are due to our selection function versus other analysis differences. 

Analogous to the results presented in \S \ref{subsec:prelimMR}, the right panel of Figure~\ref{fig:posteriors_others} shows the posterior distributions for the M-R relation hyperparameters from our HBM modeling of the heterogenous sample.  The posterior medians and central 68\% credible intervals for the three hyperparameters are: ${C=1.39_{-0.36}^{+0.44},\gamma=2.30_{-0.36}^{+0.40},\sigma=1.71_{-0.47}^{+0.57}}$.
Figure~\ref{fig:MRresult_others}, right panel, shows the posterior predictive M-R relation derived from the heterogenous sample; the literature radius and mass measurements are shown in crosses to distinguish them from the MTS-only measurements, which are used when there is no independent paper yet published on the individual planets or when the MTS-only \texttt{RadVel} mass measurements are more precise.

While the heterogeneous literature sample results in posterior median values for $C$ and $\gamma$ that are consistent at the 1-$\sigma$ level with the \texttt{RadVel} median values, the posterior median value for $\sigma$ is only marginally consistent with the \texttt{RadVel} values, and would be  considered inconsistent if not for the long tail in the \texttt{RadVel} $\sigma$ posterior. In Section~\ref{subsec:mrcompare} we further discuss the implications of differences between the M-R relations of the MTS-only data and the literature data.

\subsection{Beyond the Single Power Law M-R relation \label{subsec:broken_pl}}

 The basic power law M-R relation is useful as an apples-to-apples comparison with prior work and as a tool to explore the impact that target selection and data acquisition choices can have on these population results.  However, it is limited in its flexibility as an M-R model and does not necessarily reflect our evolving \emph{a priori} expectations for the densities of small planets. In particular, it does not allow different populations of planets in this $1-3$ R$_\oplus$ range to be distinguished, which is a key motivation of the MTS.  In this paper we have attempted to balance transparency through sharing in-progress results with realism about what can be concluded from preliminary analyses; that said, we acknowledge that there is keen interest, both amongst ourselves (see \S \ref{subsec:futurework}) and in the community to fit more physically interpretable population models to datasets like these.  To provide a window into these possibilities, especially with respect to distinguishing between multiple populations, we offer some preliminary results from a model that is a small step up in complexity from the basic single power law.  In doing so, we demonstrate that more complex modeling on this dataset as it currently stands would need to be carefully interpreted.

A broken power law is one such model that is only slightly more complex than the single power law (so as to maximize the information contained in this dataset per hyperparameter), yet is able to distinguish between different planetary populations.  Mathematically, this M-R relation is:
\begin{align}\label{eq:brokenMR}
\mu_M &=
C\Big[g(\Delta_R)\Big(\frac{R_{p}}{\text{R}_\oplus}\Big)^{\gamma_1} + g(-\Delta_R)*R_{trans}^{(\gamma_1-\gamma_2)}\Big(\frac{R_{p}}{\text{R}_\oplus}\Big)^{\gamma_2}\Big] \nonumber\\
\frac{M_{pl}}{\text{M}_\oplus} &\sim \text{Normal}(\mu_M,\sigma)
\end{align}
where the $R_{trans}^{(\gamma_1-\gamma_2)}$ factor forces the two segments to be continuous at the transition point, and where $g(\Delta_R)=(1+e^{-100(R_{trans} -R_p)})^{-1}$ is a sigmoid function that allows the broken power law to be differentiable (required for convergence with STAN) while maintaining a fast transition between the two segments.  Note that the intrinsic scatter, $\sigma$, is the same for both power-law segments, as it was already consistent with zero for the single power law and as we aim to limit the number of additional free parameters for ease of comparison.  To assess differences in the M-R relation on either side of the super-Earth/sub-Neptune radius gap, we set $R_{trans} = 1.7$.  With $R_{trans}$ fixed, this broken, mean-continuous power-law HBM has a total of four free hyperparameters ($C,\gamma_1,\gamma_2,\sigma$), which is one more than the HBM displayed in Eqn \ref{eq:HBM}; other than the above change in the mean power law, all other details of the HBM remain the same between the two models ($p(\gamma_1) = p(\gamma_2)= \mathcal{N}(1,1)$).

\begin{figure}[htbp]
\centering
\includegraphics[width=\columnwidth]{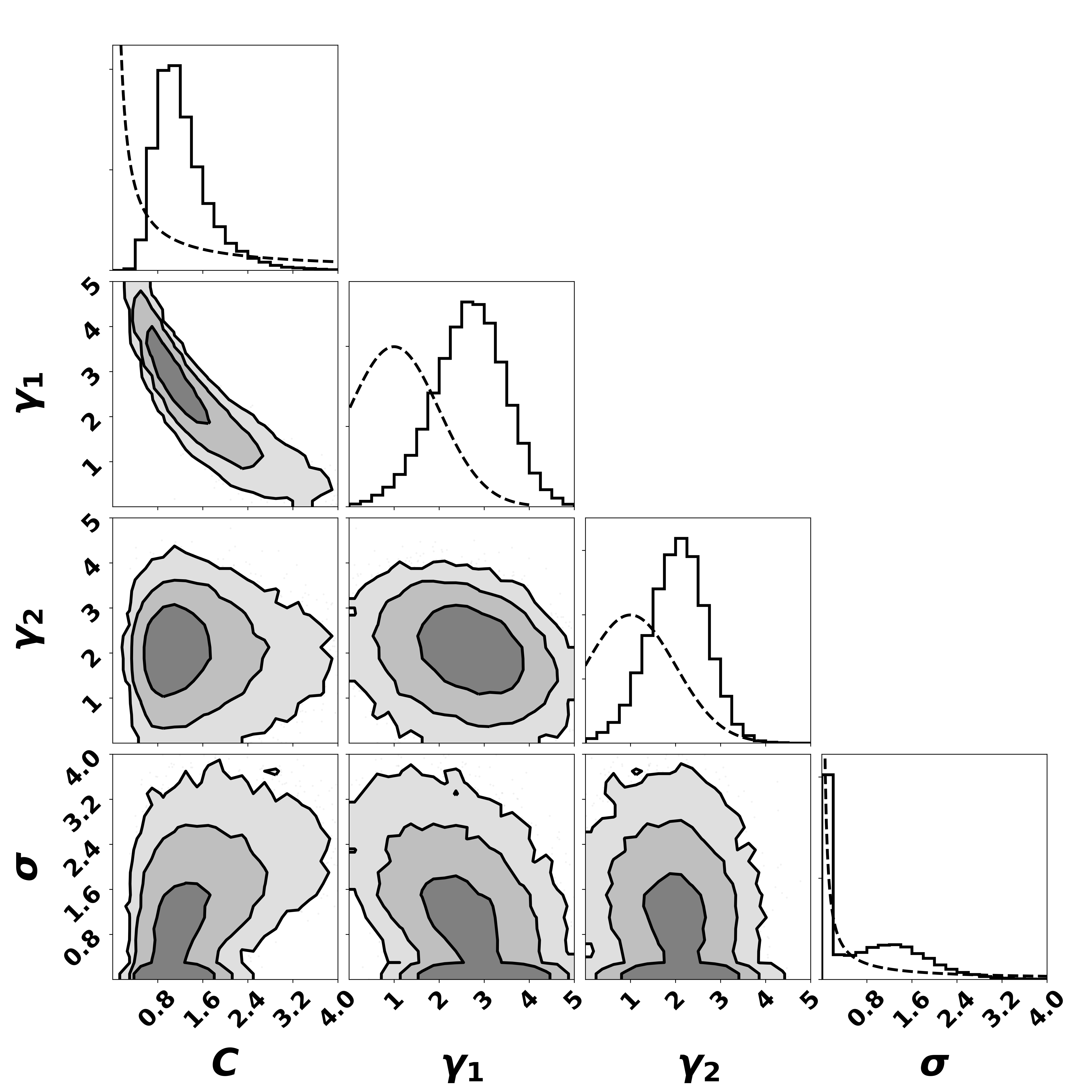} 
\caption{The posteriors of the broken power-law mass-radius relation (see Eqn \ref{eq:brokenMR}), fit to the 22 planets with mass constraints and non-zero $\hat{p}_{30}$ (see Eqn \ref{eq:HBM}).  The contours show the 68\%, 95\%, and 99.7\% credible regions of various 2-D projections of the joint 4-D hyperparameter posterior, which consists of: $C$, the normalization constant of the power law; $\gamma_1$, the power-law index of the segment with $R_p<R_{trans}=1.7\ \text{R}_\oplus$; $\gamma_2$, the power-law index of the segment with $R_p>R_{trans}=1.7\ \text{R}_\oplus$; and $\sigma$, the astrophysical scatter around the mean relation, which is the same for both power-law segments.  The histograms on the diagonal show the marginal 1-D posteriors for those hyperparameters; the dashed lines denote the corresponding prior distribution. Note that there are weaker constraints on $\gamma_2$ than $\gamma_1$, and that they are consistent with each other at the 1-$\sigma$ level.  This indicates that the population-distinguishing power in the current dataset is weak.} 
\label{fig:posteriors_brokenfix}%
\end{figure}

Figure \ref{fig:posteriors_brokenfix} displays the posterior of these four hyperparameters, on the same scale as Figures \ref{fig:posteriors} and \ref{fig:posteriors_others}.  The posterior medians and central 68\% credible intervals for the hyperparameters are: $C=1.14_{-0.33}^{+0.54},\gamma_1=2.70_{-0.81}^{+0.75}$,$\gamma_2=2.06_{-0.65}^{+0.60},\sigma=0.60_{-0.60}^{+1.08}$.  These constraints are looser than that for the single power law, as would be expected when an additional free parameter is introduced to the fit.  Note also that there are weaker constraints on $\gamma_2$ than $\gamma_1$, and that they are consistent with each other at the 1-$\sigma$ level.  Together, these indicate that the population-distinguishing power in the current dataset is weak, and that a single power law is a more informative fit to the current dataset.

\begin{figure}[htbp]
\centering
\includegraphics[width=\columnwidth]{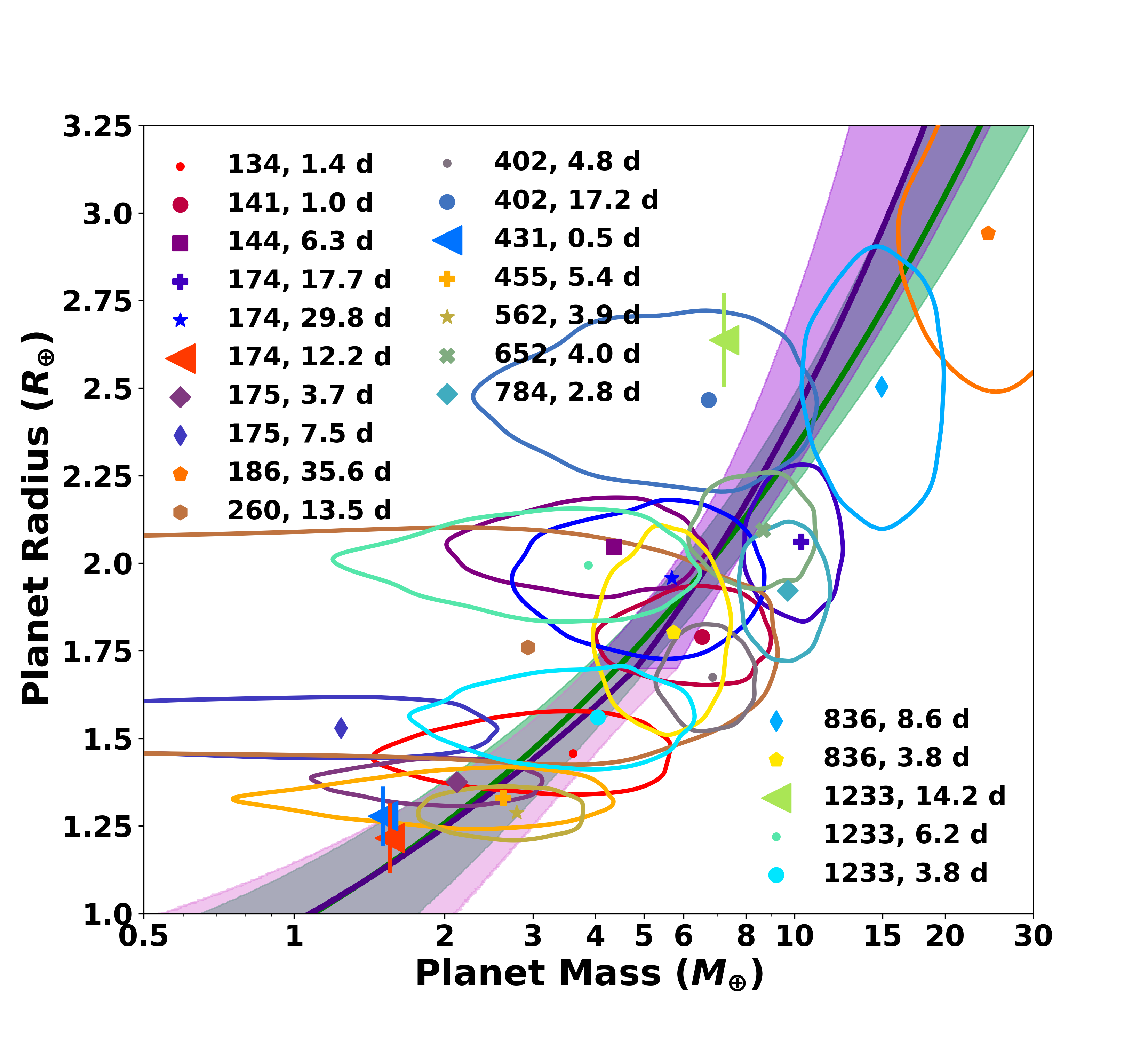} 
\caption{The single power law mass-radius relation (in green) and a broken, mean-continuous power law (in purple) with a transition point fixed at 1.7 R$_\oplus$, fit to the 22 planets with mass constraints and non-zero $\hat{p}_{30}$ (see Eqn \ref{eq:HBM}).  These are posterior predictive distributions; because the intrinsic scatter terms for these fits are near zero, the widths of the relation dominated by the uncertainty in the hyperparameters.  Contours denoting individual planets' measurement uncertainties and the single power law M-R relation are the same as in Figure \ref{fig:MRresult_Radvel}.  The small-radius segment (lighter purple) closely follows the single power law, and the large-radius segment (darker purple) is consistent with the single power law within 1-$\sigma$.  This suggests that there is currently little evidence for more than one population in this mid-survey dataset.  As the survey grows and the mass measurements become more precise, this result may change.}
\label{fig:MRresult_brokencont}%
\end{figure}

Figure \ref{fig:MRresult_brokencont} shows the broken power law fit to the 22 MTS planets with \texttt{RadVel} mass constraints and non-zero $\hat{p}_{30}$.  The posterior predictive M-R relation of Figure \ref{fig:MRresult_Radvel} is plotted in green for ready comparison with the posterior predictive broken power law in purple. The small-radius segment (lighter purple) closely follows the single power law, and the large-radius segment (darker purple) is consistent with the single power law within 1-$\sigma$.  Given the non-negligible uncertainties in both sets of hyperparameters, this suggests that: 1) there is currently little evidence for more than one population in this mid-survey dataset, at least when assuming a sharp transition in radius; 2) the single power law is actually a reasonable model given the current size of both the sample and the measurement uncertainties; and 3) future analyses of more complex M-R models will need to carefully consider the full width of the hyperparameter posteriors before drawing conclusions about the presence of different planet populations and their respective M-R relations.

\begin{figure}[htbp]
\centering
\includegraphics[width=\columnwidth]{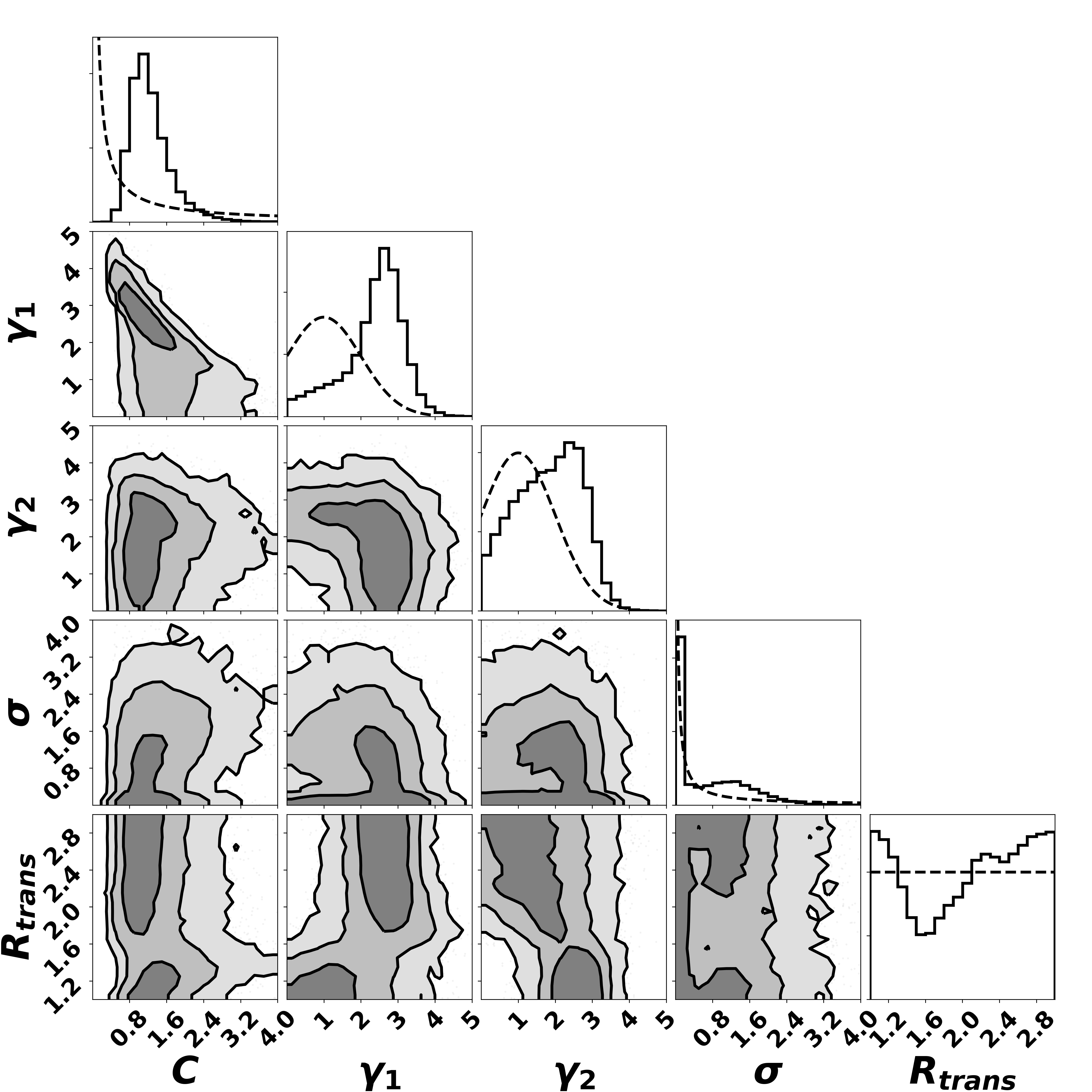} 
\caption{The posteriors of the broken power-law mass-radius relation (see Eqn \ref{eq:brokenMR}) with the transition point $R_{trans}$ allowed to vary.  The contours show the 68\%, 95\%, and 99.7\% credible regions of various 2-D projections of the joint 5-D hyperparameter posterior.  The histograms on the diagonal show the marginal 1-D posteriors for those hyperparameters; the dashed lines denote the corresponding prior distribution. Note the marginal posterior for $R_{trans}$ in the rightmost panel: the more probable transition points occur at the edges of the allowed range ($\sim$ 1 and $\sim 2.5-3$ R$_\oplus$), which indicates that a single power law is preferred.  Taken with Figure \ref{fig:MRresult_brokencont}, this demonstrates that there is little evidence for more than one population in the current dataset.  As the survey grows and the mass measurements become more precise, this result may change.} 
\label{fig:posteriors_brokenvar}%
\end{figure}

What if we had allowed the transition radius $R_{trans}$ to vary?  Would the HBM have been able to find a break point that better emphasizes the difference in the two segments?  As displayed in Figure \ref{fig:posteriors_brokenvar}, the answer is no: The marginal posterior for $R_{trans}$ (the rightmost bottom panel) illustrates that the more probable transition points occur at the edges of the allowed range ($\sim$ 1 and $\sim 2.5-3$ R$_\oplus$; the prior on $R_{trans}$ was a uniform distribution between 1 and 3 R$_\oplus$).  In fact, a transition at 1.6-1.7 R$_\oplus$ that is physically motivated given prior work is actually the least probable break point given this current dataset. This indicates that the MCMC algorithm prefers placing the data into one rather than two segments. This conclusion is further supported by the disjoint, multimodal $R_{trans}$ vs.\ $\gamma_1$ and $R_{trans}$ vs.\ $\gamma_2$ projections of the five-dimensional hyperparameter posterior, as the most probable values for $\gamma_2$ (the power-law index for the larger radius segment) when $R_{trans}$ is small is very similar to that for $\gamma_1$ (the power-law index for the smaller radius segment) when $R_{trans}$ is large.  Together with Figure \ref{fig:MRresult_brokencont}, this demonstrates that there is little evidence for more than one population in the current dataset.  We emphasize, however, that this result may change as the survey grows and the mass measurements become more precise.

\section{Discussion}\label{sec:discussion}

\subsection{Comparison of Mass-Radius Relations} \label{subsec:mrcompare}

First, we caution readers that any comparisons between mass-radius relations at this point should only be considered in a qualitative fashion and should not be over-interpreted on a quantitative level.  The results from our data are only preliminary, as we are only halfway through our survey.  
As the MTS acquires more RV data, these mass contours will shrink, and more planet masses will be added to Figure \ref{fig:MRresult_Radvel}, which could change our results on a quantitative level (see discussion in \S \ref{subsec:limitations} and \S \ref{subsec:futurework}).  Therefore we save an in-detail discussion of the MTS M-R relation as compared to others in the literature (e.g. \citealt{chenkipping2017,Ning2018,Kanodia2019,Neil&Rogers2020}) for a future analysis. For now, since we aimed to begin with a simple model that can provide apples-to-apples comparisons, we limit ourselves to more straightforward comparisons.  The model presented here (a single power law with some intrinsic scatter across our radius range of 1--3$R_\oplus$) is most similar to that in \citet{wolfgang2016}; therefore we restrict our discussion to the comparison between this prior work and the MTS-derived (\S \ref{subsec:prelimMR}) and  literature-derived (\S \ref{subsec:litresults}) results.

With these caveats, we note a few interesting differences as compared to \citet{wolfgang2016}, which are visually apparent in Figure \ref{fig:MRresult_Radvel}. First, in our MTS-only relation, the power-law normalization constant $C$ is close to 1.  This means that we predict 1 R$_\oplus$ planets to have masses of $\sim 1$ M$_\oplus$.  This is noteable because \citet{wolfgang2016} and other prior work predicted masses of $> 2-3$ M$_\oplus$ for 1 R$_\oplus$ planets.  While our sample does not contain planets at exactly 1 R$_\oplus$, the smallest planets in our sample are $1.2-1.3$ R$_\oplus$.  Therefore very little extrapolation of our mass-radius relation is required to make a mass prediction at 1 R$_\oplus$, especially compared to the datasets used by \citet{wolfgang2016} and others, which contained a larger gap between 1 R$_\oplus$ and the smallest exoplanet. 

What causes our M-R relation to predict $\sim 1$ M$_\oplus$ for 1 R$_\oplus$ planets while previous ones didn't? Studies such as \citet{Burt2018} have shown that imposing a threshold on the statistical significance of a planet's mass measurement before it is published creates a bias toward higher mass/density planets, which manifests in the empirical M-R relation as higher predicted masses for 1 R$_\oplus$ planets. Our M-R relation appears to remove much of this bias, as it passes closer to (1 R$_{\oplus}$, 1 M$_{\oplus}$) without needing be anchored to the solar system terrestrial planets.  Initially one might attribute this to the combination of the selection function we used to define our sample and our inclusion of mass upper limits for the smallest radius planets. Together, these survey design choices address the population bias identified by \citet{Burt2018}, as we are not discarding the low-significance mass measurements that are more likely to occur for small planets.

However, the comparison between the MTS-only fit and the fit with the literature values (see \S \ref{subsec:litresults}) indicates that the origin of this difference is more complex, as both the planet sample and the two low-radius upper limits are the same in both datasets.  The literature M-R relation predicts masses closer to 1 M$_\oplus$ for 1 R$_\oplus$ planets versus \citet{wolfgang2016}, indicating that the usage of a selection function with upper limits does have some mitigating effect, but it is still higher than that for the MTS-only relation.  Looking at the posteriors in Figure \ref{fig:posteriors_others}, we see a slight correlation between $\sigma$ and $C$, so that if the data contains enough information to constrain $\sigma$ away from zero, then $C$ will also shift away from 1. Indeed, $\sigma=0$ and $C=1$ is consistent with the literature M-R relation at the $2\sigma$ level, considering the slightly bimodal posterior.  Given that the masses in the literature dataset are overall more precise than those in the MTS-only dataset, this could be evidence that our simple M-R relation is a less sufficient fit for this more mass-precise dataset. In particular, the assumption that $\sigma$ is constant over the entire M-R relation, or that there is only one M-R relation rather than a mixture of several, may need to be relaxed in future analyses with more precise mass measurements.  

A second difference between \citet{wolfgang2016} and the M-R relations presented here is that there is currently no statistically significant evidence for intrinsic scatter in the MTS-only M-R relation as the $\sigma$ posterior closely follows the prior distribution. This is likely due to our small sample size and larger mass measurement uncertainty -- using a larger sample, \citet{wolfgang2016} found a non-zero intrinsic scatter that was statistically significant, and using the same sample with slightly different and somewhat more precise mass measurements on the per-planet level, the literature relation presented here has a mostly non-zero $\sigma$ posterior. It will be important to perform this comparison again once our survey is completed and the size of the MTS mass measurement errors shrink, which will allow the model to more easily distinguish the intrinsic scatter around the mean M-R relation from ordinary measurement uncertainty (see \S \ref{subsec:limitations} for a more detailed discussion of why $\sigma$ will likely change as the survey grows).  If the final MTS mass error bars are similar in size to those in the literature and if a consistent-with-zero $\sigma$ persists, then there will be evidence that the non-zero intrinsic scatter fit to heterogeneous mass data is not actually astrophysical but is instead driven by differences in how different RV teams acquire, reduce, and analyze their data, even if the differences in masses are statistically consistent with each other on the individual planet level.  

We note a seemingly contradictory observation about the M-R relation fit to our putatively less biased sample: it predicts higher masses than other relations at the 3 R$_{\oplus}$ end.  First, we acknowledge that our survey and results are not completely free of limitations and biases; we detail these in Section~\ref{subsec:limitations}.  We do not believe this to be a true bias, but a manifestation of statistical variance due to a small sample size, and expect that this aspect of our relation will change as our survey continues.  To explain, we note that the majority of the data to which the \citet{wolfgang2016} M-R relation was fit lies in a slightly different mass and radius regime than the sample presented here. 
In particular, our sample is less biased toward higher mass planets for planets with radii < 2 R$_{\oplus}$, as we are especially interested in the compositions around the radius valley and thus target planets within that radius range (which is also reflected in our selection metric described in \S \ref{subsubsec:ranking}). However, the trade-off to this choice is that we have fewer planets in our sample above $\sim$2.2 R$_{\oplus}$, and so the M-R relation toward the larger planets is more subject to small number statistics.  By chance, the larger planets currently in our sample have mass measurements that are higher than other $\sim$2.5-3 R$_{\oplus}$ planets not in our sample.  This happenstance, along with the fact that we do not fit the relation to any planets above 3 R$_{\oplus}$, causes our relation to be shallower and therefore to predict higher masses at higher radii.  This observation serves as a valuable reminder that these results are preliminary.  Overall, we expect a clearer understanding of these apparent discrepancies with previous results in future analyses with more sophisticated and realistic M-R models (see \S \ref{subsec:futurework}), as appropriate given the ultimate size of our sample.

\subsection{Implications of Planet Compositions from this Study \label{subsec:compositions}}

The M-R relation derived above is an empirical one obtained from the observed data. In this section we compare this empirical relation to theoretical M-R relationships derived from specific, constant planetary compositions. The goal is to explore what our sample implies for the variety of compositions produced by nature. This exploration is preliminary since the masses reported here will likely change and/or increase in precision as additional RV measurements are collected, but it is still a useful exercise to see if there are any current trends which we can monitor as the survey grows.

\begin{figure*}[htbp]
\centering
\includegraphics[width=\textwidth]{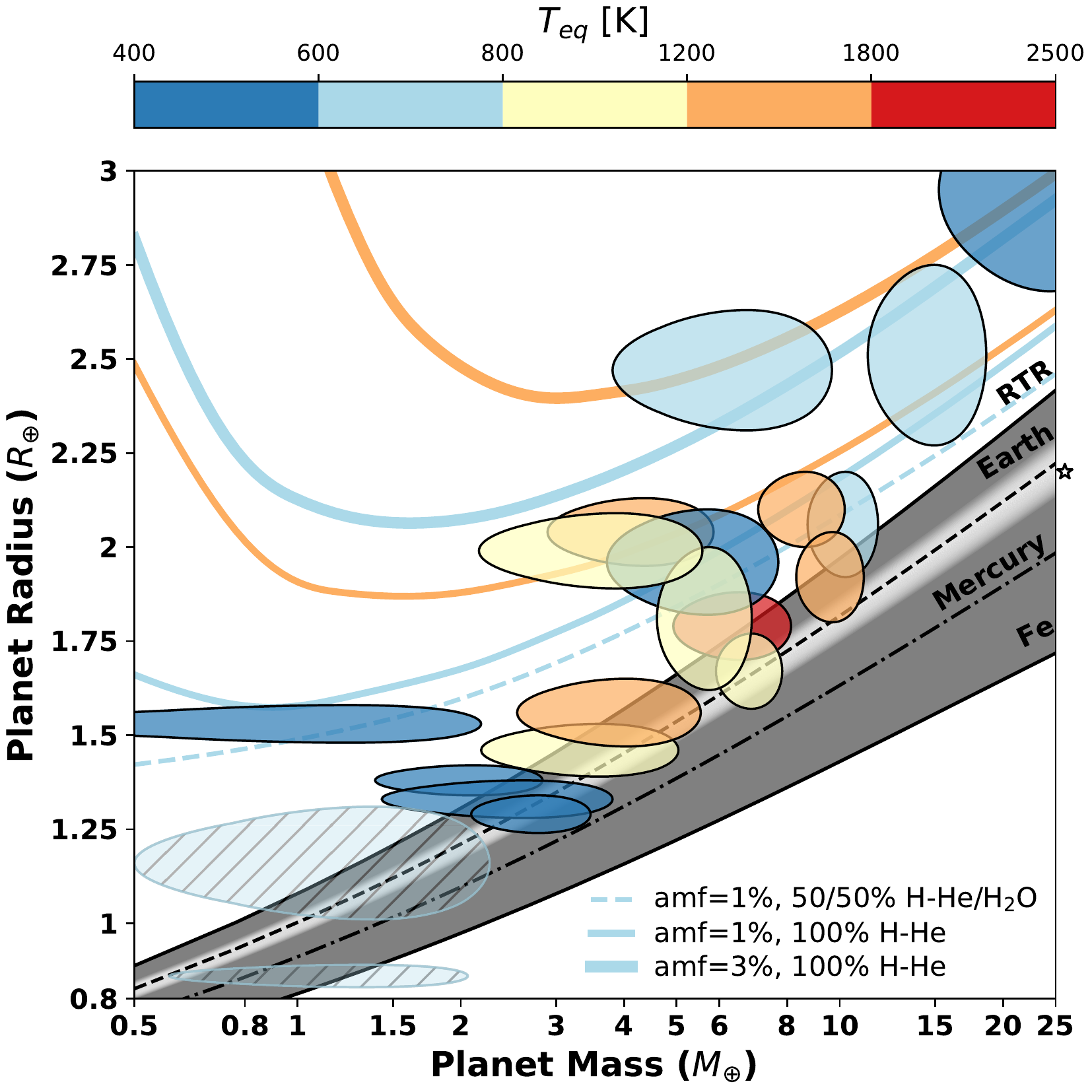}
\caption{Planets in our sample compared to mass-radius relationships for different compositions. The shaded grey region corresponds to rocky compositions:  pure iron-nickel (solid black line at the bottom), Mercury's composition (core-mass fraction of 0.67, dotted-dashed black line), Earth's composition (core-mass fraction of 0.33, dashed black line), Mg-Si rock with all relevant phases (RTR, rocky threshold radius).  Planets with the same refractory ratios as that of stars would lie within the white shaded region (marked with a small star symbol). Planets above the RTR require volatiles.  The M-R relationships for mini-Neptunes correspond to an Earth-like rocky interior overlaid by an atmospheric composition of pure H-He (solid lines) or an even mixture of H-He and water vapor (dashed line). We show two different atmospheric-mass fractions (amf) -- 1\% (thin lines) and 3\% (thick lines) -- for two different temperatures temperatures -- 800K (light blue) and 1600 K (orange).  Planet masses shown in this figure were derived using RadVel. Planets are color coded according to their equilibrium temperatures (calculated with A$_b=0.3$ and a redistribution factor of 4) and the contours show their one-sigma values.  The planet with hatch symbol did not rank highly in our sample and thus was excluded from analysis, but a mass measurement was obtained anyway because it is in a system with planets that are highly ranked. }
\label{fig:MR_chi}
\end{figure*}

Figure \ref{fig:MR_chi} shows where these planets fall on a M-R diagram compared to different likely compositions.   The M-R constant composition curves are obtained with interior structure code based on \cite{Valencia2006} and \cite{Valencia2013} with updated equations of state for the rocky minerals \citep{Plotnykov&Valencia2020}. We have also included the mass-radius space of planets that have the same refractory ratios as planet-hosting stars in general \citep{Hinkel2014} (though not necessarily our specific host stars, see \S\ref{subsec:futurework}).  That is, we have translated the rocky component of planet-hosting stars to constant composition curves.  Given that stars have a distribution of Fe/Mg and Fe/Si refractory ratios, we have shown the one-sigma values of the probability density distribution of stars translated to composition curves as a white strip within the grey shaded rocky region for comparison \footnote{For a repository that converts stars composition to rocky M-R relationships see \href{https://github.com/mplotnyko/SuperEarth.py}{https://github.com/mplotnyko/SuperEarth.py}}. Interestingly, most planets that fall within the rocky region (shaded grey) have masses less than 8 M$_{\oplus}$ and also overlap with the composition of the general sample of planet-hosting stars, suggesting these planets could have a primordial origin (i.e. they did not suffer any compositional processing during formation and would have a similar composition as the host star). This trend is in agreement with previous assumptions that aim at breaking the degeneracy in low-mass exoplanets by using the refractory ratios of the star to constrain the rocky part of the planets \citep{dorn2015}, although see also the more recent work of \cite{adibekyan2021} showing the a correlation that is not one-to-one between the iron contents in stars and their orbiting planets. Within the data included in this paper, the only outlier to this trend is TOI-175.03, with a compact radius of $0.86\pm 0.04$ R$_{\oplus}$ and a small (preliminary) mass of $1.32\pm 0.73$  M$_{\oplus}$. While  TOI-175.03 is excluded from the empirical relation analysis above based on its rank in our metric, we have mass constraints since its ``siblings'' were highly ranked. 
We note that with a larger RV data set, \cite{Cloutier2019} puts a limit on the mass of TOI-175.03 to be $<1.01$ M$_{\oplus}$, which would place it within the rocky region and thus consistent with a similar composition to host stars. 
 
The Rocky Threshold Radius (RTR), shown in Fig. \ref{fig:MR_chi}), is the largest size a rocky planet can be (\citealt{Plotnykov&Valencia2020} and can be used to investigate the transition between planets that could be rocky to those that have volatiles. Within our data, this transition is not entirely clear, but appears to happens near 1.7  R$_{\oplus}$. Again, given the preliminary nature of our mass measurements, it will be interesting to monitor how this transition changes with additional RV data and thus better mass constraints. 
 
More than a third of planets in our sample, all with different equilibrium temperatures,  cluster around a composition that is just above or barely intersects the RTR at masses between 5-10 M$_{\oplus}$. Should they have volatiles, these planets would have less than 1\% atmosphere by mass. Instead, should they be rocky, they would be depleted in iron with respect to Earth -- a puzzling scenario as we do not have a theory that can form iron-poor massive rocky planets \citep{Scora2020}. As we continue to gather RV data on these planets our hope is to uncover useful constraints on their composition.

\subsection{Limitations of this Study \label{subsec:limitations}}

There are several salient limitations to this work that we wish to explicitly delineate. Our survey was designed to explore a population rather than thoroughly analyze thirty individual planets, and so we prefer a uniform analysis as much as possible versus a customized one that could ultimately produce lower uncertainties on the planets' masses. Furthermore, the purpose of this mid-survey paper is to describe our motivation, survey structure, and analysis process -- the results presented here will change as we add more data and will be updated in our next MTS paper. The sample presented here is mostly comprised of vetted TOIs, but a few are still planet candidates. We did not attempt to detail the full suite of follow-up observations that have or will be conducted for these TOIs, as that is beyond the scope of this work, but single-system papers on at least TOI-260, TOI-784, and TOI-836 are warranted.  

One set of limitations of our study comes from the analysis of the radial velocity data.  In many cases our PFS data-in-hand are not large in number, so provide only limited information about the planet masses (although our HBM analysis enables us to incorporate this information regardless of how loose the resulting constraints are). In each TOI system we tested a series of increasingly complex models to fit the RV data, but certainly more variations in models could be tested and may provide improvements on the resulting ``goodness of fit'' metrics. Furthermore, we acknowledge that the RV noise from stellar activity is often not well behaved, and adding more data to a given time series can shift the mode of $K$ posterior, and thus the mass measurement of the planet, away from initial values.  That said, we do test Gaussian process models for the stellar noise, and we suspect that for quiet stars the improvements provided by more complex RV models would not be large given only the data presented here. For example, although we used the computationally efficient \texttt{celerite} approximation for the quasi-periodic Gaussian process kernel, we repeated our \texttt{RadVel} fits with a traditional quasi-periodic kernel and found no difference in the resulting $\Delta K$ values greater than 0.6 m~s$^{-1}$, which in many cases is on par with or lower than our quoted $K$ errors.  As for the overall activity of our stellar sample, 
we have endeavored to identify a quieter stellar sample by removing stars with photometric rotational periods less than 15 days and with $R_\star > 1.2 R_\odot$.

We also present some cautionary notes about the population results presented in this work. As the survey is not yet complete, these population results should be interpreted as a status report that we wish to have a record of in the literature.  In particular, more data means that more of our top-ranked planets will be observed and that the mass uncertainties will shrink. One may expect more planets and/or smaller mass uncertainties to simply shrink the M-R relation hyperparameter posteriors, but this may not necessarily be the case.
If the underlying distribution of planet masses and radii follow a well-behaved symmetric distribution, such as the normal distribution that we have assumed in our power-law model, then we would expect that the mean power law of the final MTS M-R relation would be consistent with these preliminary results. (The intrinsic scatter term, however, will almost certainly change; see discussion below.)  However, we do not actually know whether the underlying distribution of masses and radii are symmetric.  If it is asymmetric, then the planets whose masses we have yet to observe may tend to be lower or higher than the current sample, which would change the power-law index and its normalization constant, and we would need to reassess the symmetry assumed in the M-R relation of Eqn \ref{eq:simpleMR} (see \citet{Ning2018} for an analysis of this assumption given the \emph{Kepler} M,R dataset).  Having the mean power law shift away from these preliminary ``best-fit" values (instead of the posteriors shrinking in around them) is even more of a possibility given that our dataset is relatively small.  In the final MTS paper we plan to perform tests on how robust the final M-R result is to the presence/absence of one to a few planets, but these tests are out of the scope of this current work.  

Even if the underlying distribution of planet masses and radii are symmetric, the intrinsic scatter term in the M-R relation will almost certainly change as more data are added to this analysis.  In order to understand why, it is important to realize that hierarchical modeling enables different sources of uncertainty/scatter in the data to be separated.  In this model, there are two different ways that a planet’s mass is able to be “scattered” around the mean mass-radius relation (i.e., the power law by itself).  First, there is measurement uncertainty, which scatters the measured mass around its “true” value due to uncertainty in the observations and mass measurement process.  Second, there is the intrinsic scatter term, (i.e., $\sigma$), which is intended to represent the compositional diversity of extrasolar planets at a single radius value.\footnote{In truth, $\sigma$ includes this astrophysical diversity plus the sources of measurement uncertainty/systematic error that are not accurately captured in the reported mass measurement error bar.}  There is a trade-off between these two sources of scatter as fit by the hierarchical model.  When the measurement uncertainty is large, then more of the error contours intersect with the mean mass-radius relation, and there is no need for any astrophysical scatter --- the observed variation around the power law can be completely explained by measurement uncertainty, as is the case for these preliminary results.  However, as the mass measurement uncertainty shrinks as we acquire more data --- even if the error contours shrink toward their current means --- then fewer of the planets’ error contours intersect with the mean power law, and measurement uncertainty may not be enough to explain the observed dispersion.  This is when the intrinsic/astrophysical scatter term, $\sigma$, starts to be nonzero and thus qualitatively different from our current results, a situation which may or may not arise at the end of the survey.

Our hierarchical Bayesian modeling also does not account for biases introduced in the original vetting and reporting of the TOIs by the TESS team, which are likely sector-dependent and will affect our sample. We apply our selection function only to those TOIs announced by the TESS project and use only their reported parameters in the ranking process; we do not include any community TOIs in our parent sample. As discussed in \S \ref{subsubsec:whyquantselect}, explicitly incorporating a selection function is particularly important for occurrence rate studies that aim to extract underlying frequencies of planets with different compositions.  While occurrence rate studies are a natural extension of this work, they require understanding the completeness and reliability of the TESS detections, which is outside the scope of this paper.  Instead, the primary utility of the selection function for the population results presented here is simply ensuring -- in a way that is easy to reproduce -- that we are measuring masses of a wide variety of sub-Neptune-sized planets and eliminating some of the more pernicious publication biases present in the current exoplanet mass, radius dataset collated from the literature. 

Finally, our sample has a limited radius range and has fewer planets toward the sub-Neptune regime. As described in \S\ref{subsec:mrcompare}, this limits the predictive power of our M-R relation above $\sim$2.2~R$_{\oplus}$. We will monitor this effect as we add more RV data and planets to our sample, but note that it should not significantly impact our ability to address the science questions motivating the survey, which our selection metric was designed to address.

\subsection{Future Work \label{subsec:futurework}}

The full results and interpretation of the MTS will come in future analyses. The preliminary masses and densities presented here will likely change and/or become more precise as we continue the MTS and add to our RV data quantity and time baseline, which will affect our mass-radius model and inferences about the planet compositions. We will also add TOIs to our sample to reach our stated goal of 30 $R_p \leq 3$ R$_{\oplus}$ planet mass measurements.  

In this work we have not started to explore how the planet densities depend on the three parameters listed in \S\ref{subsec:motivation} -- insolation flux, host star composition, and system architecture. Given the lack of intrinsic scatter in our preliminary measurements (\S\ref{sec:hbm}), adding an insolation flux dependence is not yet warranted. Also given the current varying quantities and baseline of our RV data, we leave to our end-of-survey paper an examination of the residuals of our RV fits and searches for long term trends that might be indicative of non-transiting/longer-period planet companions. Finally, host star compositions will be investigated in the future via a uniform analysis (using an equivalent width, line-by-line differential approach, similar to \citealt{teske2019}) of our high-resolution, high-signal-to-noise, iodine-free PFS template spectra to measure a suite of elemental abundances (O, C, Si, Mg, Ni, Ti, etc.). These could be used jointly with the planet masses and radii to help break degeneracies in the bulk composition and constrain the planetary refractory content (e.g., \citealt{dorn2015}, however see also \citealt{Plotnykov&Valencia2020} for evidence calling this assumption into question). 

There are areas of improvement for the hierarchical model implemented in this paper.  In particular, we plan to:
\begin{enumerate}
    \item Test the dependence of the M-R relation on other planetary and stellar properties, once/if a nonzero $\sigma$ indicates that the dataset contains astrophysical scatter whose origins can be explored.
    \item Model the planet transit depth and RV semi-amplitude uncertainty separately from the stellar parameter uncertainties, instead of combining measurement uncertainty of the planet and star parameters into the masses and radii that are inputted into the hierarchical model.  This will allow us to account for correlated errors between the stellar masses and radii, and therefore the planet masses and radii.
    \item Assess the impact of different assumptions for the underlying mass-radius relation.  \citet{Ning2018} relaxed the assumption of the power-law for the entire \textit{Kepler} sample and found that it still described well the M-R relationship for planets $< 5 R_\oplus$, which spans the entire MTS sample.  That said, there have been a number of studies since \citet{wolfgang2016}, from which we draw heavily here, which have improved on this early M-R model.  As the survey grows, we intend to incorporate some of these improvements, most notably those in \citet{Neil&Rogers2020}, to provide closer comparisons with the state of the art in M-R modeling.
\end{enumerate}

Looking beyond just planet bulk density, a natural extension of the MTS is a campaign to collect atmospheric observations of our targets with JWST. All of the TOIs in our sample orbit bright, slowly rotating stars that are mostly ``well behaved'' (not very jittery), and fall in a parameter space of atmospheric studies that remain virtually unexplored. The planets in our sample span a wide range of densities and equilibrium temperatures, and could host a variety of primary or secondary (out-gassed) atmospheres measurable by JWST. As seen in Figure \ref{fig:MR_chi}, based on our preliminary results, many of our targets could straddle the RTR, so we may only be able to know their volatile fraction through atmospheric observations. All of the planets in our sample have mass constraints (albeit some only upper limits, which will improve with more time and data), which \cite{batalha2019} showed is an important ingredient for accurately and precisely interpreting atmosphere transmission observations. If the targets for a future atmospheric composition survey were also selected from MTS with a quantitative metric, this would provide a natural framework for a population-level inference study of small exoplanet atmosphere. Indeed, several of the TOIs in this MTS Year 1.5 paper will be observed in the first year of JWST operations -- 134.01, 175.01 \& .02, 260.01, 402.01 \& .02, 455.01, 562.01, 836.01 \& .02 (see proposals GO 2512 from \citealt{2021jwst.prop.2512B}, GO 1981 from \citealt{2021jwst.prop.1981S}, GO 2708 from \citealt{2021jwst.prop.2708B}, and GTO 1224 from \citealt{2017jwst.prop.1224B}).

\section{Summary \label{sec:summary}}

Here we introduce the Magellan-TESS Survey, designed to broadly explore whether super-Earths and sub-Neptunes have similar or divergent formation pathways by searching for trends between planet density and insolation flux, host star composition, and the presence of additional planets. The MTS was conceived as a three-year survey with the Planet Finder Spectrograph on Magellan II to measure the RV masses of 30 $R_p \leq 3 R_{\oplus}$ transiting planets detected by TESS in its first year scanning the southern hemisphere. In this work representing approximately the mid-point of our survey, we detail the survey motivation and structure, including target selection and the RV observation strategy. Most of the 27 TOIs presented here have already been validated in separate papers, but some of them are still planet candidates; we provide a summary of what was known about all the TOIs in our sample prior to this work in Appendix \ref{sec:individual_systems}. 

The main contributions from this paper are as follows:
\begin{itemize}
    \item We uniformly measure the transit depths of our targets from the TESS light curves and combine these with the TICv8 stellar radii to derive planet radii (\S\ref{sec:radii} and Table \ref{trasit_fit_posteriors}).
    
    \item We systematically test how well increasingly complex radial velocity models fit our MTS PFS data, and report the best fits from both \texttt{RadVel} and \texttt{juliet}, which have different approaches to posterior sampling and determining ``goodness of fit''. The best fit values are then combined with TICv8 stellar masses to derive planet masses (\S\ref{sec:mass} and Table \ref{tab:pfs_K}). 
    
    \item We analyze the $R_p$ and $M_p$ measurements computed in this work with a single power law hierarchical Bayesian model of the mass-radius relation that, for the first time, is based on quantitative target selection and an observational decision tree designed to decrease bias (\S\ref{sec:hbm}). 
    
    \item Our preliminary empirical mass-radius relation (e.g., Figure \ref{fig:MRresult_Radvel}) shows two noteworthy differences versus prior works: (1) the power law constant $C$ is close to 1, meaning the relation passes through 1~$R_{\oplus}$, 1~$M_{\oplus}$ without anchoring to solar system bodies and potentially indicating our survey strategy has mitigated some previous bias; and (2) the intrinsic scatter parameter $\sigma$ closely follows the prior distribution (which peaks at zero) (Figure \ref{fig:posteriors}). The second point likely reflects our smaller sample size and still-large measurement uncertainties, but may indicate  that the non-zero $\sigma$ in previously published mass-radius relations was driven by differences in RV observation decisions rather than by astrophysical diversity. We will assess these differences further once our survey is complete.
    
    \item We also explore a continuous broken power law mass-radius relation (\S\ref{subsec:broken_pl} and Figure \ref{fig:MRresult_brokencont}) and find it closely follows the single power law at small radii and is consistent at $\sim 1 \sigma$ at large radii. This indicates that there is little evidence for more than one population in this mid-survey dataset.
    
    \item We compare our $R_p$ and $M_p$ values to curves representing specific planetary compositions to estimate the compositional diversity in our sample (\S\ref{subsec:compositions} and Figure \ref{fig:MR_chi}).
    
\end{itemize}

The $R_p$ and especially $M_p$ results in this paper are preliminary and will likely change as we add more RV data in the second half of our survey, and/or add TESS transits from the Extended Mission; for a full discussion see \S\ref{subsec:limitations} and \S \ref{subsec:futurework}. Thus the reader is advised to avoid over-interpreting or relying very heavily on our HBM mass-radius relation and the comparison of our measurements to composition curves. Sharing a preliminary data release and analysis prior to survey completion is not the ``norm'' in the radial velocity community, but given the population-level goals of this survey, we have decided that there is utility in producing a record of how our results can evolve as more data are added. Instead of final masses or a final mass-radius relation, the goal of this paper is to detail the framework and process of MTS, and demonstrate that it is possible to design and conduct an RV survey of transiting planets that allows for robust population inference. This type of approach may become increasingly necessary as we move deeper into the era of exoplanet characterization, especially in the small planet regime where observational constraints on both masses and atmospheres are more technically challenging and expensive.

\acknowledgements

Funding for the \textit{TESS} mission is provided by NASA's Science Mission directorate. We acknowledge the use of public \textit{TESS} Alert data from pipelines at the \textit{TESS} Science Office and at the \textit{TESS} Science Processing Operations Center. Resources supporting this work were provided by the NASA High-End Computing (HEC) Program through the NASA Advanced Supercomputing (NAS) Division at Ames Research Center for the production of the SPOC data products. This research has made use of the Exoplanet Follow-up Observation Program website, which is operated by the California Institute of Technology, under contract with the National Aeronautics and Space Administration under the Exoplanet Exploration Program. This paper includes data collected by the \textit{TESS} mission, which are publicly available from the Mikulski Archive for Space Telescopes (MAST). Support for this work was provided by NASA through grant 18-XRP18\_2-0048. Support for this work was also provided by NASA through Hubble Fellowship grant HST-HF2-51399.001 awarded by the Space Telescope Science Institute, which is operated by the Association of Universities for Research in Astronomy, Inc., for NASA, under contract NAS5-26555. Part of this research was carried out at the Jet Propulsion Laboratory, California Institute of Technology, under a contract with the National Aeronautics and Space Administration (NASA). This work makes use of observations from the LCOGT network. MEarth is funded by the David and Lucile Packard Fellowship for Science and Engineering, the National Science Foundation under grants AST-0807690, AST-1109468, AST-1004488 (Alan T. Waterman Award) and AST-1616624, and the John Templeton Foundation.  This publication was made possible through the support of a grant from the John Templeton Foundation.  The opinions expressed in this publication are those of the authors and do not necessarily reflect the views of the John Templeton Foundation. The Center for Exoplanets and Habitable Worlds is supported by the Pennsylvania State University, the Eberly College of Science, and the Pennsylvania Space Grant Consortium. This work is partly supported by JSPS KAKENHI Grant Numbers JP18H01265 and JP18H05439, and JST PRESTO Grant Number JPMJPR1775. DJA acknowledges support from the STFC via an Ernest Rutherford Fellowship (ST/R00384X/1). SH acknowledge support by the fellowships PD/BD/128119/2016 funded by FCT (Portugal). This work was supported by FCT - Fundacao para a Ciencia e a Tecnologia through national funds and by FEDER through COMPETE2020 - Programa Operacional Competitividade e Internacionalizacao by these grants: UID/FIS/04434/2019; UIDB/04434/2020; UIDP/04434/2020; PTDC/FIS-AST/32113/2017 \& POCI-01-0145-FEDER- 032113; PTDC/FIS-AST/28953/2017 \& POCI-01-0145-FEDER-028953; PTDC/FIS-AST/28987/2017 \& POCI-01-0145-FEDER-028987. S.C.C.B acknowledges support from FCT through FCT contracts nr. IF/01312/2014/CP1215/CT0004.
DJAB acknowledges support from the UK Space Agency. O.D.S.D. is supported in the form of work contract (DL 57/2016/CP1364/CT0004) funded by FCT.	D.B. has been funded by the Spanish State Research Agency (AEI) Projects No.PID2019-107061GB-C61 and No. MDM-2017-0737 Unidad de Excelencia "Mar\'ia de Maeztu"- Centro de Astrobiolog\'ia (INTA-CSIC). S.G.S acknowledges the support from FCT through Investigador FCT contract nr. CEECIND/00826/2018 and POPH/FSE (EC). J.L-B has been funded by the Spanish State Research Agency (AEI) Projects No.ESP2017-87676-C5-1-R and No. MDM-2017-0737 Unidad de Excelencia "Mar\'ia de Maeztu"- Centro de Astrobiolog\'ia (INTA-CSIC). V. A. acknowledges the support from FCT through Investigador FCT contract no. IF/00650/2015/CP1273/CT0001. The authors wish to recognize and acknowledge the very significant cultural role and reverence that the summit of Maunakea has always had within the indigenous Hawaiian community.  We are most fortunate to have the opportunity to conduct observations from this mountain. J.K.T. acknowledges Dr. Thomas Connor for his prompt help with LaTeX formatting, and of all of the staff at Las Campanas Observatory for their continued support. We thank Joseph D. Twicken for collecting updated background estimates for some of our TOIs. We also thank the anonymous referee for their comments that helped improve the clarity and quality of this paper.

\facilities{TESS, Magellan:Clay (PFS), \\
Keck:I (HIRES),ESO:3.6m (HARPS),IRTF (iSHELL), Las Cumbres Observatory Global Telescope (LCOGT)}
\software{astropy \citep{astropy}, emcee \citep{emcee}, Lightkurve \citep{Lightkurve}, STAN \citep{STAN}, celerite \citep{Foreman-Mackey2017}, radvel \citep{Fulton2018a}, juliet \citep{Espinoza2019}, batman \citep{Kreidberg2015}, dynesty \citep{Higson2019,Speagle2018}, SuperEarth \citep{Plotnykov&Valencia2020} }
\bibliographystyle{aasjournal}
\bibliography{references}

\begin{thebibliography}{}
\expandafter\ifx\csname natexlab\endcsname\relax\def\natexlab#1{#1}\fi
\providecommand{\url}[1]{\href{#1}{#1}}

\bibitem[{{Adibekyan} {et~al.}(2021){Adibekyan}, {Dorn}, {Sousa}, {Santos},
  {Bitsch}, {Israelian}, {Mordasini}, {Barros}, {Delgado Mena}, {Demangeon},
  {Faria}, {Figueira}, {Hakobyan}, {Oshagh}, {Kunitomo}, {Takeda}, {Jofr{\'e}},
  {Petrucci}, \& {Martioli}}]{adibekyan2021}
{Adibekyan}, V., {Dorn}, C., {Sousa}, S.~G., {et~al.} 2021, arXiv e-prints,
  arXiv:2102.12444

\bibitem[{{Arriagada}(2011)}]{arriagada2011}
{Arriagada}, P. 2011, \apj, 734, 70

\bibitem[{{Arriagada} {et~al.}(2010){Arriagada}, {Butler}, {Minniti},
  {L{\'o}pez-Morales}, {Shectman}, {Adams}, {Boss}, \&
  {Chambers}}]{arriagada2010}
{Arriagada}, P., {Butler}, R.~P., {Minniti}, D., {et~al.} 2010, \apj, 711, 1229

\bibitem[{{Astudillo-Defru} {et~al.}(2020){Astudillo-Defru}, {Cloutier},
  {Wang}, {Teske}, {Brahm}, {Hellier}, {Ricker}, {Vand erspek}, {Latham},
  {Seager}, {Winn}, {Jenkins}, {Collins}, {Stassun}, {Ziegler}, {Almenara},
  {Anderson}, {Artigau}, {Bonfils}, {Bouchy}, {Brice{\~n}o}, {Butler},
  {Charbonneau}, {Conti}, {Crane}, {Crossfield}, {Davies}, {Delfosse},
  {D{\'\i}az}, {Doyon}, {Dragomir}, {Eastman}, {Espinoza}, {Essack}, {Feng},
  {Figueira}, {Forveille}, {Gan}, {Glidden}, {Guerrero}, {Hart}, {Henning},
  {Horch}, {Isopi}, {Jenkins}, {Jord{\'a}n}, {Kielkopf}, {Law}, {Lovis},
  {Mallia}, {Mann}, {de Medeiros}, {Melo}, {Mennickent}, {Mignon}, {Murgas},
  {Nusdeo}, {Pepe}, {Relles}, {Rose}, {Santos}, {S{\'e}gransan}, {Shectman},
  {Shporer}, {Smith}, {Torres}, {Udry}, {Villasenor}, {Winters}, \&
  {Zhou}}]{Astudillo-Defru2020}
{Astudillo-Defru}, N., {Cloutier}, R., {Wang}, S.~X., {et~al.} 2020, \aap, 636,
  A58

\bibitem[{{Bae} {et~al.}(2016){Bae}, {Nelson}, \& {Hartmann}}]{bae2016}
{Bae}, J., {Nelson}, R.~P., \& {Hartmann}, L. 2016, \apj, 833, 126

\bibitem[{{Bakos} {et~al.}(2004){Bakos}, {Noyes}, {Kov{\'a}cs}, {Stanek},
  {Sasselov}, \& {Domsa}}]{Bakos2004}
{Bakos}, G., {Noyes}, R.~W., {Kov{\'a}cs}, G., {et~al.} 2004, \pasp, 116, 266

\bibitem[{{Bakos} {et~al.}(2009){Bakos}, {Afonso}, {Henning}, {Jord{\'a}n},
  {Holman}, {Noyes}, {Sackett}, {Sasselov}, {Kov{\'a}cs}, {Csubry}, \&
  {P{\'a}l}}]{Bakos2009}
{Bakos}, G., {Afonso}, C., {Henning}, T., {et~al.} 2009, in Transiting Planets,
  ed. F.~{Pont}, D.~{Sasselov}, \& M.~J. {Holman}, Vol. 253, 354--357

\bibitem[{{Baranne} {et~al.}(1996){Baranne}, {Queloz}, {Mayor}, {Adrianzyk},
  {Knispel}, {Kohler}, {Lacroix}, {Meunier}, {Rimbaud}, \&
  {Vin}}]{1996A&AS..119..373B}
{Baranne}, A., {Queloz}, D., {Mayor}, M., {et~al.} 1996, \aaps, 119, 373

\bibitem[{{Barclay} {et~al.}(2018){Barclay}, {Pepper}, \&
  {Quintana}}]{barclay2018}
{Barclay}, T., {Pepper}, J., \& {Quintana}, E.~V. 2018, \apjs, 239, 2

\bibitem[{{Barnes}(2007)}]{Barnes2007}
{Barnes}, S.~A. 2007, \apj, 669, 1167

\bibitem[{{Barrag{\'a}n} {et~al.}(2019){Barrag{\'a}n}, {Aigrain}, {Kubyshkina},
  {Gand olfi}, {Livingston}, {Fridlund}, {Fossati}, {Korth}, {Parviainen},
  {Malavolta}, {Palle}, {Deeg}, {Nowak}, {Rajpaul}, {Zicher}, {Antoniciello},
  {Narita}, {Albrecht}, {Bedin}, {Cabrera}, {Cochran}, {de Leon},
  {Eigm{\"u}ller}, {Fukui}, {Granata}, {Grziwa}, {Guenther}, {Hatzes},
  {Kusakabe}, {Latham}, {Libralato}, {Luque},
  {Monta{\~n}{\'e}s-Rodr{\'\i}guez}, {Murgas}, {Nardiello}, {Pagano}, {Piotto},
  {Persson}, {Redfield}, \& {Tamura}}]{Barragan2019}
{Barrag{\'a}n}, O., {Aigrain}, S., {Kubyshkina}, D., {et~al.} 2019, \mnras,
  490, 698

\bibitem[{{Batalha} {et~al.}(2021){Batalha}, {Teske}, {Alam}, {Alderson},
  {Batalha}, {Gao}, {Lopez-Morales}, {Marley}, {Shahar}, {Wakeford}, \&
  {Wolfgang}}]{2021jwst.prop.2512B}
{Batalha}, N., {Teske}, J., {Alam}, M., {et~al.} 2021, {Seeing the Forest and
  the Trees: Unveiling Small Planet Atmospheres with a Population-Level
  Framework}, JWST Proposal. Cycle 1, ,

\bibitem[{{Batalha} {et~al.}(2019){Batalha}, {Lewis}, {Fortney}, {Batalha},
  {Kempton}, {Lewis}, \& {Line}}]{batalha2019}
{Batalha}, N.~E., {Lewis}, T., {Fortney}, J.~J., {et~al.} 2019, \apjl, 885, L25

\bibitem[{{Berta-Thompson} {et~al.}(2021){Berta-Thompson}, {Diamond-Lowe}, \&
  {Winters}}]{2021jwst.prop.2708B}
{Berta-Thompson}, Z.~K., {Diamond-Lowe}, H., \& {Winters}, J. 2021, {The
  Thermal Emission Spectrum of the Closest M Dwarf Transiting Rocky Planet},
  JWST Proposal. Cycle 1, ,

\bibitem[{{Birkmann} {et~al.}(2017){Birkmann}, {Ferruit}, \&
  {Valenti}}]{2017jwst.prop.1224B}
{Birkmann}, S., {Ferruit}, P., \& {Valenti}, J.~A. 2017, {Transiting exoplanet
  characterization with JWST/NIRSPEC}, JWST Proposal. Cycle 1, ,

\bibitem[{{Bitsch}(2019)}]{bitsch2019}
{Bitsch}, B. 2019, \aap, 630, A51

\bibitem[{{Boisse} {et~al.}(2009){Boisse}, {Moutou}, {Vidal-Madjar}, {Bouchy},
  {Pont}, {H{\'e}brard}, {Bonfils}, {Croll}, {Delfosse}, {Desort}, {Forveille},
  {Lagrange}, {Loeillet}, {Lovis}, {Matthews}, {Mayor}, {Pepe}, {Perrier},
  {Queloz}, {Rowe}, {Santos}, {S{\'e}gransan}, \& {Udry}}]{2009A&A...495..959B}
{Boisse}, I., {Moutou}, C., {Vidal-Madjar}, A., {et~al.} 2009, \aap, 495, 959

\bibitem[{{Bonfanti} {et~al.}(2021){Bonfanti}, {Delrez}, {Hooton}, {Wilson},
  {Fossati}, {Alibert}, {Hoyer}, {Mustill}, {Osborn}, {Adibekyan}, {Gandolfi},
  {Salmon}, {Sousa}, {Tuson}, {Van Grootel}, {Cabrera}, {Nascimbeni}, {Maxted},
  {Barros}, {Billot}, {Bonfils}, {Borsato}, {Broeg}, {Davies}, {Deleuil},
  {Demangeon}, {Fridlund}, {Lacedelli}, {Lendl}, {Persson}, {Santos},
  {Scandariato}, {Szab{\'o}}, {Collier Cameron}, {Udry}, {Benz}, {Beck},
  {Ehrenreich}, {Fortier}, {Isaak}, {Queloz}, {Alonso}, {Asquier}, {Bandy},
  {B{\'a}rczy}, {Barrado}, {Barrag{\'a}n}, {Baumjohann}, {Beck}, {Bekkelien},
  {Bergomi}, {Brandeker}, {Busch}, {Cessa}, {Charnoz}, {Chazelas}, {Corral Van
  Damme}, {Demory}, {Erikson}, {Farinato}, {Futyan}, {Garcia Mu{\~n}oz},
  {Gillon}, {Guedel}, {Guterman}, {Hasiba}, {Heng}, {Hernandez}, {Kiss},
  {Kuntzer}, {Laskar}, {Lecavelier des Etangs}, {Lovis}, {Magrin}, {Malvasio},
  {Marafatto}, {Michaelis}, {Munari}, {Olofsson}, {Ottacher}, {Ottensamer},
  {Pagano}, {Pall{\'e}}, {Peter}, {Piazza}, {Piotto}, {Pollacco}, {Ragazzoni},
  {Rando}, {Ratti}, {Rauer}, {Ribas}, {Rieder}, {Rohlfs}, {Safa}, {Salatti},
  {S{\'e}gransan}, {Simon}, {Smith}, {Sordet}, {Steller}, {Thomas},
  {Tschentscher}, {Van Eylen}, {Viotto}, {Walter}, {Walton}, {Wildi}, \&
  {Wolter}}]{Bonfanti2021}
{Bonfanti}, A., {Delrez}, L., {Hooton}, M.~J., {et~al.} 2021, \aap, 646, A157

\bibitem[{{Bonomo} {et~al.}(2019){Bonomo}, {Zeng}, {Damasso}, {Leinhardt},
  {Justesen}, {Lopez}, {Lund}, {Malavolta}, {Silva Aguirre}, {Buchhave},
  {Corsaro}, {Denman}, {Lopez-Morales}, {Mills}, {Mortier}, {Rice}, {Sozzetti},
  {Vanderburg}, {Affer}, {Arentoft}, {Benbakoura}, {Bouchy},
  {Christensen-Dalsgaard}, {Collier Cameron}, {Cosentino}, {Dressing},
  {Dumusque}, {Figueira}, {Fiorenzano}, {Garc{\'\i}a}, {Hand berg},
  {Harutyunyan}, {Johnson}, {Kjeldsen}, {Latham}, {Lovis}, {Lundkvist},
  {Mathur}, {Mayor}, {Micela}, {Molinari}, {Motalebi}, {Nascimbeni}, {Nava},
  {Pepe}, {Phillips}, {Piotto}, {Poretti}, {Sasselov}, {S{\'e}gransan}, {Udry},
  \& {Watson}}]{Bonomo2019}
{Bonomo}, A.~S., {Zeng}, L., {Damasso}, M., {et~al.} 2019, Nature Astronomy, 3,
  416

\bibitem[{{Brewer} \& {Fischer}(2018)}]{brewer&fischer2018}
{Brewer}, J.~M., \& {Fischer}, D.~A. 2018, \apjs, 237, 38

\bibitem[{{Bryan} {et~al.}(2019){Bryan}, {Knutson}, {Lee}, {Fulton}, {Batygin},
  {Ngo}, \& {Meshkat}}]{bryan2019}
{Bryan}, M.~L., {Knutson}, H.~A., {Lee}, E.~J., {et~al.} 2019, \aj, 157, 52

\bibitem[{{Buchner} {et~al.}(2014){Buchner}, {Georgakakis}, {Nandra}, {Hsu},
  {Rangel}, {Brightman}, {Merloni}, {Salvato}, {Donley}, \&
  {Kocevski}}]{Buchner2014}
{Buchner}, J., {Georgakakis}, A., {Nandra}, K., {et~al.} 2014, \aap, 564, A125

\bibitem[{{Burke} \& {Catanzarite}(2017)}]{Burke2017&Catanzarite2017}
{Burke}, C.~J., \& {Catanzarite}, J. 2017, {Planet Detection Metrics:
  Per-Target Detection Contours for Data Release 25}, Kepler Science Document
  KSCI-19111-002, ,

\bibitem[{{Burke} {et~al.}(2015){Burke}, {Christiansen}, {Mullally}, {Seader},
  {Huber}, {Rowe}, {Coughlin}, {Thompson}, {Catanzarite}, {Clarke}, {Morton},
  {Caldwell}, {Bryson}, {Haas}, {Batalha}, {Jenkins}, {Tenenbaum}, {Twicken},
  {Li}, {Quintana}, {Barclay}, {Henze}, {Borucki}, {Howell}, \&
  {Still}}]{burke2015}
{Burke}, C.~J., {Christiansen}, J.~L., {Mullally}, F., {et~al.} 2015, \apj,
  809, 8

\bibitem[{{Burt} {et~al.}(2018){Burt}, {Holden}, {Wolfgang}, \&
  {Bouma}}]{Burt2018}
{Burt}, J., {Holden}, B., {Wolfgang}, A., \& {Bouma}, L.~G. 2018, \aj, 156, 255

\bibitem[{{Butler} {et~al.}(1996){Butler}, {Marcy}, {Williams}, {McCarthy},
  {Dosanjh}, \& {Vogt}}]{butler1996}
{Butler}, R.~P., {Marcy}, G.~W., {Williams}, E., {et~al.} 1996, \pasp, 108, 500

\bibitem[{{Butler} {et~al.}(2017){Butler}, {Vogt}, {Laughlin}, {Burt},
  {Rivera}, {Tuomi}, {Teske}, {Arriagada}, {Diaz}, {Holden}, \&
  {Keiser}}]{Butler2017}
{Butler}, R.~P., {Vogt}, S.~S., {Laughlin}, G., {et~al.} 2017, \aj, 153, 208

\bibitem[{{Cale} {et~al.}(2019){Cale}, {Plavchan}, {LeBrun}, {Gagn{\'e}},
  {Gao}, {Tanner}, {Beichman}, {Xuesong Wang}, {Gaidos}, {Teske}, {Ciardi},
  {Vasisht}, {Kane}, \& {von Braun}}]{2019AJ....158..170C}
{Cale}, B., {Plavchan}, P., {LeBrun}, D., {et~al.} 2019, \aj, 158, 170

\bibitem[{{Chen} \& {Rogers}(2016)}]{chen&rogers2016}
{Chen}, H., \& {Rogers}, L.~A. 2016, \apj, 831, 180

\bibitem[{{Chen} \& {Kipping}(2017)}]{chenkipping2017}
{Chen}, J., \& {Kipping}, D. 2017, \apj, 834, 17

\bibitem[{{Chiang} \& {Laughlin}(2013)}]{chiang&laughlin2013}
{Chiang}, E., \& {Laughlin}, G. 2013, \mnras, 431, 3444

\bibitem[{{Christiansen}(2017)}]{Christiansen2017}
{Christiansen}, J.~L. 2017, {Planet Detection Metrics: Pixel-Level Transit
  Injection Tests of Pipeline Detection Efficiency for Data Release 25}, Kepler
  Science Document KSCI-19110-001, ,

\bibitem[{{Christiansen} {et~al.}(2020){Christiansen}, {Clarke}, {Burke},
  {Jenkins}, {Bryson}, {Coughlin}, {Mullally}, {Twicken}, {Batalha},
  {Catanzarite}, {Uddin}, {Zamudio}, {Smith}, {Henze}, \&
  {Campbell}}]{Christiansen2020}
{Christiansen}, J.~L., {Clarke}, B.~D., {Burke}, C.~J., {et~al.} 2020, \aj,
  160, 159

\bibitem[{{Cloutier} {et~al.}(2019){Cloutier}, {Astudillo-Defru}, {Bonfils},
  {Jenkins}, {Berdi{\~n}as}, {Ricker}, {Vand erspek}, {Latham}, {Seager},
  {Winn}, {Jenkins}, {Almenara}, {Bouchy}, {Delfosse}, {D{\'\i}az},
  {D{\'\i}az}, {Doyon}, {Figueira}, {Forveille}, {Kurtovic}, {Lovis}, {Mayor},
  {Menou}, {Morgan}, {Morris}, {Muirhead}, {Murgas}, {Pepe}, {Santos},
  {S{\'e}gransan}, {Smith}, {Tenenbaum}, {Torres}, {Udry}, {Vezie}, \&
  {Villasenor}}]{Cloutier2019}
{Cloutier}, R., {Astudillo-Defru}, N., {Bonfils}, X., {et~al.} 2019, \aap, 629,
  A111

\bibitem[{{Cossou} {et~al.}(2014){Cossou}, {Raymond}, {Hersant}, \&
  {Pierens}}]{cossou2014}
{Cossou}, C., {Raymond}, S.~N., {Hersant}, F., \& {Pierens}, A. 2014, \aap,
  569, A56

\bibitem[{{Crane} {et~al.}(2006){Crane}, {Shectman}, \& {Butler}}]{crane2006}
{Crane}, J.~D., {Shectman}, S.~A., \& {Butler}, R.~P. 2006, in \procspie, Vol.
  6269, Society of Photo-Optical Instrumentation Engineers (SPIE) Conference
  Series, 626931

\bibitem[{{Crane} {et~al.}(2010){Crane}, {Shectman}, {Butler}, {Thompson},
  {Birk}, {Jones}, \& {Burley}}]{crane2010}
{Crane}, J.~D., {Shectman}, S.~A., {Butler}, R.~P., {et~al.} 2010, in
  \procspie, Vol. 7735, Ground-based and Airborne Instrumentation for Astronomy
  III, 773553

\bibitem[{{Crane} {et~al.}(2008){Crane}, {Shectman}, {Butler}, {Thompson}, \&
  {Burley}}]{crane2008}
{Crane}, J.~D., {Shectman}, S.~A., {Butler}, R.~P., {Thompson}, I.~B., \&
  {Burley}, G.~S. 2008, in \procspie, Vol. 7014, Ground-based and Airborne
  Instrumentation for Astronomy II, 701479

\bibitem[{{Dai} {et~al.}(2019){Dai}, {Masuda}, {Winn}, \& {Zeng}}]{Dai2019}
{Dai}, F., {Masuda}, K., {Winn}, J.~N., \& {Zeng}, L. 2019, \apj, 883, 79

\bibitem[{{Damasso} {et~al.}(2020){Damasso}, {Sozzetti}, {Lovis}, {Barros},
  {Sousa}, {Demangeon}, {Faria}, {Lillo-Box}, {Cristiani}, {Pepe}, {Rebolo},
  {Santos}, {Zapatero Osorio}, {Gonz{\'a}lez Hern{\'a}ndez}, {Amate},
  {Pasquini}, {Zerbi}, {Adibekyan}, {Abreu}, {Affolter}, {Alibert}, {Aliverti},
  {Allart}, {Allende Prieto}, {{\'A}lvarez}, {Alves}, {Avila}, {Baldini},
  {Bandy}, {Benz}, {Bianco}, {Borsa}, {Bossini}, {Bourrier}, {Bouchy}, {Broeg},
  {Cabral}, {Calderone}, {Cirami}, {Coelho}, {Conconi}, {Coretti}, {Cumani},
  {Cupani}, {D'Odorico}, {Deiries}, {Dekker}, {Delabre}, {Di Marcantonio},
  {Dumusque}, {Ehrenreich}, {Figueira}, {Fragoso}, {Genolet}, {Genoni},
  {G{\'e}nova Santos}, {Hughes}, {Iwert}, {Kerber}, {Knudstrup}, {Landoni},
  {Lavie}, {Lizon}, {Lo Curto}, {Maire}, {Martins}, {M{\'e}gevand}, {Mehner},
  {Micela}, {Modigliani}, {Molaro}, {Monteiro}, {Monteiro}, {Moschetti},
  {Mueller}, {Murphy}, {Nunes}, {Oggioni}, {Oliveira}, {Oshagh}, {Pall{\'e}},
  {Pariani}, {Poretti}, {Rasilla}, {Rebord{\~a}o}, {Redaelli}, {Riva}, {Santana
  Tschudi}, {Santin}, {Santos}, {S{\'e}gransan}, {Schmidt}, {Segovia},
  {Sosnowska}, {Span{\`o}}, {Su{\'a}rez Mascare{\~n}o}, {Tabernero}, {Tenegi},
  {Udry}, \& {Zanutta}}]{Damasso2020}
{Damasso}, M., {Sozzetti}, A., {Lovis}, C., {et~al.} 2020, arXiv e-prints,
  arXiv:2007.06410

\bibitem[{{Daylan} {et~al.}(2020){Daylan}, {Pingle}, {Wright}, {Guenther},
  {Stassun}, {Kane}, {Vanderburg}, {Jontof-Hutter}, {Rodriguez}, {Shporer},
  {Huang}, {Mikal-Evans}, {Badenas-Agusti}, {Collins}, {Rackham}, {Quinn},
  {Cloutier}, {Collins}, {Guerra}, {Jensen}, {Kielkopf}, {Massey}, {Schwarz},
  {Charbonneau}, {Lissauer}, {Irwin}, {Basturk}, {Fulton}, {Soubkiou},
  {Howell}, {Ziegler}, {Briceno}, {Law}, {Mann}, {Scott}, {Furlan}, {Ciardi},
  {Matson}, {Hellier}, {Anderson}, {Butler}, {Crane}, {Teske}, {Schectman},
  {Ricker}, {Vanderspek}, {Seager}, {Winn}, {Jenkins}, {Berta Thompson},
  {Bouma}, {Fong}, {Furesz}, {Henze}, {Morgan}, {Quintana}, {Ting}, \&
  {Twicken}}]{Daylan2020}
{Daylan}, T., {Pingle}, K., {Wright}, J., {et~al.} 2020, arXiv e-prints,
  arXiv:2004.11314

\bibitem[{{Delgado Mena} {et~al.}(2015){Delgado Mena}, {Bertr{\'a}n de Lis},
  {Adibekyan}, {Sousa}, {Figueira}, {Mortier}, {Gonz{\'a}lez Hern{\'a}ndez},
  {Tsantaki}, {Israelian}, \& {Santos}}]{DelgadoMena2015}
{Delgado Mena}, E., {Bertr{\'a}n de Lis}, S., {Adibekyan}, V.~Z., {et~al.}
  2015, \aap, 576, A69

\bibitem[{{Dorn} {et~al.}(2015){Dorn}, {Khan}, {Heng}, {Connolly}, {Alibert},
  {Benz}, \& {Tackley}}]{dorn2015}
{Dorn}, C., {Khan}, A., {Heng}, K., {et~al.} 2015, \aap, 577, A83

\bibitem[{{Dragomir} {et~al.}(2019){Dragomir}, {Teske}, {G{\"u}nther},
  {S{\'e}gransan}, {Burt}, {Huang}, {Vanderburg}, {Matthews}, {Dumusque},
  {Stassun}, {Pepper}, {Ricker}, {Vanderspek}, {Latham}, {Seager}, {Winn},
  {Jenkins}, {Beatty}, {Bouchy}, {Brown}, {Butler}, {Ciardi}, {Crane},
  {Eastman}, {Fossati}, {Francis}, {Fulton}, {Gaudi}, {Goeke}, {James},
  {Klaus}, {Kuhn}, {Lovis}, {Lund}, {McDermott}, {Paegert}, {Pepe},
  {Rodriguez}, {Sha}, {Shectman}, {Shporer}, {Siverd}, {Garcia Soto},
  {Stevens}, {Twicken}, {Udry}, {Villanueva}, {Wang}, {Wohler}, {Yao}, \&
  {Zhan}}]{Dragomir2019}
{Dragomir}, D., {Teske}, J., {G{\"u}nther}, M.~N., {et~al.} 2019, \apjl, 875,
  L7

\bibitem[{{Dressing} \& {Charbonneau}(2013)}]{dressing&charbonneau2013}
{Dressing}, C.~D., \& {Charbonneau}, D. 2013, \apj, 767, 95

\bibitem[{{Dressing} {et~al.}(2015){Dressing}, {Charbonneau}, {Dumusque},
  {Gettel}, {Pepe}, {Collier Cameron}, {Latham}, {Molinari}, {Udry}, {Affer},
  {Bonomo}, {Buchhave}, {Cosentino}, {Figueira}, {Fiorenzano}, {Harutyunyan},
  {Haywood}, {Johnson}, {Lopez-Morales}, {Lovis}, {Malavolta}, {Mayor},
  {Micela}, {Motalebi}, {Nascimbeni}, {Phillips}, {Piotto}, {Pollacco},
  {Queloz}, {Rice}, {Sasselov}, {S{\'e}gransan}, {Sozzetti}, {Szentgyorgyi}, \&
  {Watson}}]{Dressing2015}
{Dressing}, C.~D., {Charbonneau}, D., {Dumusque}, X., {et~al.} 2015, \apj, 800,
  135

\bibitem[{{Dumusque} {et~al.}(2011){Dumusque}, {Udry}, {Lovis}, {Santos}, \&
  {Monteiro}}]{dumusque2011}
{Dumusque}, X., {Udry}, S., {Lovis}, C., {Santos}, N.~C., \& {Monteiro},
  M.~J.~P.~F.~G. 2011, \aap, 525, A140

\bibitem[{{Dumusque} {et~al.}(2017){Dumusque}, {Borsa}, {Damasso},
  {D{\'{\i}}az}, {Gregory}, {Hara}, {Hatzes}, {Rajpaul}, {Tuomi}, {Aigrain},
  {Anglada-Escud{\'e}}, {Bonomo}, {Bou{\'e}}, {Dauvergne}, {Frustagli},
  {Giacobbe}, {Haywood}, {Jones}, {Laskar}, {Pinamonti}, {Poretti}, {Rainer},
  {S{\'e}gransan}, {Sozzetti}, \& {Udry}}]{dumusque2017}
{Dumusque}, X., {Borsa}, F., {Damasso}, M., {et~al.} 2017, \aap, 598, A133

\bibitem[{{Dumusque} {et~al.}(2019){Dumusque}, {Turner}, {Dorn}, {Eastman},
  {Allart}, {Adibekyan}, {Sousa}, {Santos}, {Mordasini}, {Bourrier}, {Bouchy},
  {Coffinet}, {Davies}, {D{\'\i}az}, {Fausnaugh}, {Glidden}, {Guerrero},
  {Henze}, {Jenkins}, {Latham}, {Lovis}, {Mayor}, {Pepe}, {Quintana}, {Ricker},
  {Rowden}, {Segransan}, {Mascare{\~n}o}, {Seager}, {Twicken}, {Udry}, {Vand
  erspek}, \& {Winn}}]{Dumusque2019}
{Dumusque}, X., {Turner}, O., {Dorn}, C., {et~al.} 2019, \aap, 627, A43

\bibitem[{{Espinoza}(2018)}]{Espinoza2018}
{Espinoza}, N. 2018, Research Notes of the American Astronomical Society, 2,
  209

\bibitem[{{Espinoza} {et~al.}(2019){Espinoza}, {Kossakowski}, \&
  {Brahm}}]{Espinoza2019}
{Espinoza}, N., {Kossakowski}, D., \& {Brahm}, R. 2019, \mnras, 490, 2262

\bibitem[{{Espinoza} {et~al.}(2020){Espinoza}, {Brahm}, {Henning},
  {Jord{\'a}n}, {Dorn}, {Rojas}, {Sarkis}, {Kossakowski}, {Schlecker},
  {D{\'\i}az}, {Jenkins}, {Aguilera-Gomez}, {Jenkins}, {Twicken}, {Collins},
  {Lissauer}, {Armstrong}, {Adibekyan}, {Barrado}, {Barros}, {Battley},
  {Bayliss}, {Bouchy}, {Bryant}, {Cooke}, {Demangeon}, {Dumusque}, {Figueira},
  {Giles}, {Lillo-Box}, {Lovis}, {Nielsen}, {Pepe}, {Pollacco}, {Santos},
  {Sousa}, {Udry}, {Wheatley}, {Turner}, {Marmier}, {S{\'e}gransan}, {Ricker},
  {Latham}, {Seager}, {Winn}, {Kielkopf}, {Hart}, {Wingham}, {Jensen},
  {He{\l}miniak}, {Tokovinin}, {Brice{\~n}o}, {Ziegler}, {Law}, {Mann},
  {Daylan}, {Doty}, {Guerrero}, {Boyd}, {Crossfield}, {Morris}, {Henze}, \&
  {Chacon}}]{Espinoza2020}
{Espinoza}, N., {Brahm}, R., {Henning}, T., {et~al.} 2020, \mnras, 491, 2982

\bibitem[{{Fabrycky} {et~al.}(2014){Fabrycky}, {Lissauer}, {Ragozzine}, {Rowe},
  {Steffen}, {Agol}, {Barclay}, {Batalha}, {Borucki}, {Ciardi}, {Ford},
  {Gautier}, {Geary}, {Holman}, {Jenkins}, {Li}, {Morehead}, {Morris},
  {Shporer}, {Smith}, {Still}, \& {Van Cleve}}]{fabrycky2014}
{Fabrycky}, D.~C., {Lissauer}, J.~J., {Ragozzine}, D., {et~al.} 2014, \apj,
  790, 146

\bibitem[{{Feroz} {et~al.}(2009){Feroz}, {Hobson}, \& {Bridges}}]{Feroz2009}
{Feroz}, F., {Hobson}, M.~P., \& {Bridges}, M. 2009, \mnras, 398, 1601

\bibitem[{{Feroz} {et~al.}(2019){Feroz}, {Hobson}, {Cameron}, \&
  {Pettitt}}]{Feroz2019}
{Feroz}, F., {Hobson}, M.~P., {Cameron}, E., \& {Pettitt}, A.~N. 2019, The Open
  Journal of Astrophysics, 2, 10

\bibitem[{{Ford}(2006)}]{ford2006}
{Ford}, E.~B. 2006, \apj, 642, 505

\bibitem[{{Foreman-Mackey} {et~al.}(2017){Foreman-Mackey}, {Agol},
  {Ambikasaran}, \& {Angus}}]{Foreman-Mackey2017}
{Foreman-Mackey}, D., {Agol}, E., {Ambikasaran}, S., \& {Angus}, R. 2017, \aj,
  154, 220

\bibitem[{{Foreman-Mackey} {et~al.}(2013{\natexlab{a}}){Foreman-Mackey},
  {Hogg}, {Lang}, \& {Goodman}}]{foreman-mackey2013}
{Foreman-Mackey}, D., {Hogg}, D.~W., {Lang}, D., \& {Goodman}, J.
  2013{\natexlab{a}}, \pasp, 125, 306

\bibitem[{{Foreman-Mackey} {et~al.}(2013{\natexlab{b}}){Foreman-Mackey},
  {Hogg}, {Lang}, \& {Goodman}}]{emcee}
---. 2013{\natexlab{b}}, \pasp, 125, 306

\bibitem[{{Fressin} {et~al.}(2013){Fressin}, {Torres}, {Charbonneau}, {Bryson},
  {Christiansen}, {Dressing}, {Jenkins}, {Walkowicz}, \&
  {Batalha}}]{fressin2013}
{Fressin}, F., {Torres}, G., {Charbonneau}, D., {et~al.} 2013, \apj, 766, 81

\bibitem[{{Fulton} \& {Petigura}(2018)}]{fulton&petigura2018}
{Fulton}, B.~J., \& {Petigura}, E.~A. 2018, ArXiv e-prints, arXiv:1805.01453

\bibitem[{{Fulton} {et~al.}(2018){Fulton}, {Petigura}, {Blunt}, \&
  {Sinukoff}}]{Fulton2018a}
{Fulton}, B.~J., {Petigura}, E.~A., {Blunt}, S., \& {Sinukoff}, E. 2018, \pasp,
  130, 044504

\bibitem[{{Fulton} {et~al.}(2017){Fulton}, {Petigura}, {Howard}, {Isaacson},
  {Marcy}, {Cargile}, {Hebb}, {Weiss}, {Johnson}, {Morton}, {Sinukoff},
  {Crossfield}, \& {Hirsch}}]{Fulton2017}
{Fulton}, B.~J., {Petigura}, E.~A., {Howard}, A.~W., {et~al.} 2017, \aj, 154,
  109

\bibitem[{{Gaia Collab.} {et~al.}(2018){Gaia Collab.}, {Brown}, {Vallenari},
  {Prusti}, {de Bruijne}, {Babusiaux}, {Bailer-Jones}, {Biermann}, {Evans},
  {Eyer}, \& et~al.}]{Gaia:2018}
{Gaia Collab.}, {Brown}, A.~G.~A., {Vallenari}, A., {et~al.} 2018, \aap, 616,
  A1

\bibitem[{{Gan} {et~al.}(2021){Gan}, {Wang}, {Teske}, {Mao}, {Howard}, {Law},
  {Batalha}, {Vanderburg}, {Dragomir}, {Huang}, {Feng}, {Butler}, {Crane},
  {Shectman}, {Beletsky}, {Shporer}, {Montet}, {Burt}, {Feinstein}, {Flowers},
  {Nandakumar}, {Barbieri}, {Corbett}, {Ratzloff}, {Galliher}, {Chavez},
  {Vasquez}, {Glazier}, \& {Haislip}}]{Gan2021}
{Gan}, T., {Wang}, S.~X., {Teske}, J.~K., {et~al.} 2021, \mnras, 501, 6042

\bibitem[{{Gandolfi} {et~al.}(2018){Gandolfi}, {Barrag{\'a}n}, {Livingston},
  {Fridlund}, {Justesen}, {Redfield}, {Fossati}, {Mathur}, {Grziwa}, {Cabrera},
  {Garc{\'\i}a}, {Persson}, {Van Eylen}, {Hatzes}, {Hidalgo}, {Albrecht},
  {Bugnet}, {Cochran}, {Csizmadia}, {Deeg}, {Eigm{\"u}ller}, {Endl}, {Erikson},
  {Esposito}, {Guenther}, {Korth}, {Luque}, {Monta{\~n}es Rodr{\'\i}guez},
  {Nespral}, {Nowak}, {P{\"a}tzold}, \& {Prieto-Arranz}}]{Gandolfi2018}
{Gandolfi}, D., {Barrag{\'a}n}, O., {Livingston}, J.~H., {et~al.} 2018, \aap,
  619, L10

\bibitem[{{Gandolfi} {et~al.}(2019){Gandolfi}, {Fossati}, {Livingston},
  {Stassun}, {Grziwa}, {Barrag{\'a}n}, {Fridlund}, {Kubyshkina}, {Persson},
  {Dai}, {Lam}, {Albrecht}, {Batalha}, {Beck}, {Justesen}, {Cabrera},
  {Cartwright}, {Cochran}, {Csizmadia}, {Davies}, {Deeg}, {Eigm{\"u}ller},
  {Endl}, {Erikson}, {Esposito}, {Garc{\'\i}a}, {Goeke}, {Gonz{\'a}lez-Cuesta},
  {Guenther}, {Hatzes}, {Hidalgo}, {Hirano}, {Hjorth}, {Kabath}, {Knudstrup},
  {Korth}, {Li}, {Luque}, {Mathur}, {Monta{\~n}es Rodr{\'\i}guez}, {Narita},
  {Nespral}, {Niraula}, {Nowak}, {Palle}, {P{\"a}tzold}, {Prieto-Arranz},
  {Rauer}, {Redfield}, {Ribas}, {Skarka}, {Smith}, {Rowden}, {Torres}, {Van
  Eylen}, \& {Vezie}}]{Gandolfi2019}
{Gandolfi}, D., {Fossati}, L., {Livingston}, J.~H., {et~al.} 2019, \apjl, 876,
  L24

\bibitem[{{Garc{\'\i}a Mu{\~n}oz} {et~al.}(2020){Garc{\'\i}a Mu{\~n}oz},
  {Youngblood}, {Fossati}, {Gandolfi}, {Cabrera}, \&
  {Rauer}}]{Garcia_Munoz2020}
{Garc{\'\i}a Mu{\~n}oz}, A., {Youngblood}, A., {Fossati}, L., {et~al.} 2020,
  \apjl, 888, L21

\bibitem[{Gelman {et~al.}(2003)Gelman, Carlin, Stern, \& Rubin}]{gelman2003}
Gelman, A., Carlin, J.~B., Stern, H.~S., \& Rubin, D.~B. 2003, Bayesian Data
  Analysis (Chapman and Hall/CRC)

\bibitem[{{Ginzburg} \& {Sari}(2017)}]{ginzburg&sari2017}
{Ginzburg}, S., \& {Sari}, R. 2017, \mnras, 464, 3937

\bibitem[{{Ginzburg} {et~al.}(2018){Ginzburg}, {Schlichting}, \&
  {Sari}}]{Ginzburg2018}
{Ginzburg}, S., {Schlichting}, H.~E., \& {Sari}, R. 2018, \mnras, 476, 759

\bibitem[{{Gomes da Silva} {et~al.}(2011){Gomes da Silva}, {Santos}, {Bonfils},
  {Delfosse}, {Forveille}, \& {Udry}}]{2011A&A...534A..30G}
{Gomes da Silva}, J., {Santos}, N.~C., {Bonfils}, X., {et~al.} 2011, \aap, 534,
  A30

\bibitem[{{Gray} {et~al.}(2006){Gray}, {Corbally}, {Garrison}, {McFadden},
  {Bubar}, {McGahee}, {O'Donoghue}, \& {Knox}}]{Gray2006}
{Gray}, R.~O., {Corbally}, C.~J., {Garrison}, R.~F., {et~al.} 2006, \aj, 132,
  161

\bibitem[{{Grunblatt} {et~al.}(2015){Grunblatt}, {Howard}, \&
  {Haywood}}]{Grunblatt2015}
{Grunblatt}, S.~K., {Howard}, A.~W., \& {Haywood}, R.~D. 2015, \apj, 808, 127

\bibitem[{{Gupta} \& {Schlichting}(2019)}]{Gupta&Schlichting2019}
{Gupta}, A., \& {Schlichting}, H.~E. 2019, \mnras, 487, 24

\bibitem[{{Gupta} \& {Schlichting}(2020)}]{Gupta&Schlichting2020}
---. 2020, \mnras, 493, 792

\bibitem[{{Hadlock} \& {Bickel}(2019)}]{Hadlock2019}
{Hadlock}, C.~C., \& {Bickel}, J.~E. 2019, Decision Analysis, 16, 85

\bibitem[{{Hansen} \& {Murray}(2012)}]{hansen&murray2012}
{Hansen}, B.~M.~S., \& {Murray}, N. 2012, \apj, 751, 158

\bibitem[{{Hansen} \& {Murray}(2013)}]{hansen&murray2013}
---. 2013, \apj, 775, 53

\bibitem[{{Haywood} {et~al.}(2014){Haywood}, {Collier Cameron}, {Queloz},
  {Barros}, {Deleuil}, {Fares}, {Gillon}, {Lanza}, {Lovis}, {Moutou}, {Pepe},
  {Pollacco}, {Santerne}, {S{\'e}gransan}, \& {Unruh}}]{Haywood2014}
{Haywood}, R.~D., {Collier Cameron}, A., {Queloz}, D., {et~al.} 2014, \mnras,
  443, 2517

\bibitem[{{Higson} {et~al.}(2019){Higson}, {Handley}, {Hobson}, \&
  {Lasenby}}]{Higson2019}
{Higson}, E., {Handley}, W., {Hobson}, M., \& {Lasenby}, A. 2019, Statistics
  and Computing, 29, 891

\bibitem[{Hinkel {et~al.}(2014)Hinkel, Timmes, Young, Pagano, \&
  Turnbull}]{Hinkel2014}
Hinkel, N.~R., Timmes, F.~X., Young, P.~A., Pagano, M.~D., \& Turnbull, M.~C.
  2014, The Astronomical Journal, 148, 54

\bibitem[{{Hogg} {et~al.}(2010){Hogg}, {Myers}, \& {Bovy}}]{Hogg2010}
{Hogg}, D.~W., {Myers}, A.~D., \& {Bovy}, J. 2010, \apj, 725, 2166

\bibitem[{{Howard} {et~al.}(2012){Howard}, {Marcy}, {Bryson}, {Jenkins},
  {Rowe}, {Batalha}, {Borucki}, {Koch}, {Dunham}, {Gautier}, {Van Cleve},
  {Cochran}, {Latham}, {Lissauer}, {Torres}, {Brown}, {Gilliland}, {Buchhave},
  {Caldwell}, {Christensen-Dalsgaard}, {Ciardi}, {Fressin}, {Haas}, {Howell},
  {Kjeldsen}, {Seager}, {Rogers}, {Sasselov}, {Steffen}, {Basri},
  {Charbonneau}, {Christiansen}, {Clarke}, {Dupree}, {Fabrycky}, {Fischer},
  {Ford}, {Fortney}, {Tarter}, {Girouard}, {Holman}, {Johnson}, {Klaus},
  {Machalek}, {Moorhead}, {Morehead}, {Ragozzine}, {Tenenbaum}, {Twicken},
  {Quinn}, {Isaacson}, {Shporer}, {Lucas}, {Walkowicz}, {Welsh}, {Boss},
  {Devore}, {Gould}, {Smith}, {Morris}, {Prsa}, {Morton}, {Still}, {Thompson},
  {Mullally}, {Endl}, \& {MacQueen}}]{howard2012}
{Howard}, A.~W., {Marcy}, G.~W., {Bryson}, S.~T., {et~al.} 2012, \apjs, 201, 15

\bibitem[{{Huang} {et~al.}(2018){Huang}, {Burt}, {Vanderburg}, {G{\"u}nther},
  {Shporer}, {Dittmann}, {Winn}, {Wittenmyer}, {Sha}, {Kane}, {Ricker}, {Vand
  erspek}, {Latham}, {Seager}, {Jenkins}, {Caldwell}, {Collins}, {Guerrero},
  {Smith}, {Quinn}, {Udry}, {Pepe}, {Bouchy}, {S{\'e}gransan}, {Lovis},
  {Ehrenreich}, {Marmier}, {Mayor}, {Wohler}, {Haworth}, {Morgan}, {Fausnaugh},
  {Ciardi}, {Christiansen}, {Charbonneau}, {Dragomir}, {Deming}, {Glidden},
  {Levine}, {McCullough}, {Yu}, {Narita}, {Nguyen}, {Morton}, {Pepper},
  {P{\'a}l}, {Rodriguez}, {Stassun}, {Torres}, {Sozzetti}, {Doty},
  {Christensen-Dalsgaard}, {Laughlin}, {Clampin}, {Bean}, {Buchhave}, {Bakos},
  {Sato}, {Ida}, {Kaltenegger}, {Palle}, {Sasselov}, {Butler}, {Lissauer},
  {Ge}, \& {Rinehart}}]{Huang2018}
{Huang}, C.~X., {Burt}, J., {Vanderburg}, A., {et~al.} 2018, \apjl, 868, L39

\bibitem[{{Inamdar} \& {Schlichting}(2015)}]{inamdar&schlichting2015}
{Inamdar}, N.~K., \& {Schlichting}, H.~E. 2015, \mnras, 448, 1751

\bibitem[{{Inamdar} \& {Schlichting}(2016)}]{inamdar&schlichting2016}
---. 2016, \apjl, 817, L13

\bibitem[{{Isaacson} \& {Fischer}(2010)}]{Isaacson&Fischer2010}
{Isaacson}, H., \& {Fischer}, D. 2010, \apj, 725, 875

\bibitem[{{Izidoro} {et~al.}(2017){Izidoro}, {Ogihara}, {Raymond},
  {Morbidelli}, {Pierens}, {Bitsch}, {Cossou}, \& {Hersant}}]{izidoro2017}
{Izidoro}, A., {Ogihara}, M., {Raymond}, S.~N., {et~al.} 2017, \mnras, 470,
  1750

\bibitem[{{Jenkins}(2002)}]{Jenkins2002}
{Jenkins}, J.~M. 2002, \apj, 575, 493

\bibitem[{{Jenkins} {et~al.}(2017){Jenkins}, {Tenenbaum}, {Seader}, {Burke},
  {McCauliff}, {Smith}, {Twicken}, \& {Chandrasekaran}}]{Jenkins2017}
{Jenkins}, J.~M., {Tenenbaum}, P., {Seader}, S., {et~al.} 2017, {Kepler Data
  Processing Handbook: Transiting Planet Search}, Tech. rep.

\bibitem[{{Jenkins} {et~al.}(2016){Jenkins}, {Twicken}, {McCauliff},
  {Campbell}, {Sanderfer}, {Lung}, {Mansouri-Samani}, {Girouard}, {Tenenbaum},
  {Klaus}, {Smith}, {Caldwell}, {Chacon}, {Henze}, {Heiges}, {Latham},
  {Morgan}, {Swade}, {Rinehart}, \& {Vanderspek}}]{Jenkins:2016}
{Jenkins}, J.~M., {Twicken}, J.~D., {McCauliff}, S., {et~al.} 2016, in Society
  of Photo-Optical Instrumentation Engineers (SPIE) Conference Series, Vol.
  9913, \procspie, 99133E

\bibitem[{{Jin} \& {Mordasini}(2018)}]{jin&mordasini2018}
{Jin}, S., \& {Mordasini}, C. 2018, \apj, 853, 163

\bibitem[{{Jin} {et~al.}(2014){Jin}, {Mordasini}, {Parmentier}, {van Boekel},
  {Henning}, \& {Ji}}]{jin2014}
{Jin}, S., {Mordasini}, C., {Parmentier}, V., {et~al.} 2014, \apj, 795, 65

\bibitem[{{Jones} {et~al.}(2002){Jones}, {Paul Butler}, {Tinney}, {Marcy},
  {Penny}, {McCarthy}, {Carter}, \& {Pourbaix}}]{Jones2002}
{Jones}, H. R.~A., {Paul Butler}, R., {Tinney}, C.~G., {et~al.} 2002, \mnras,
  333, 871

\bibitem[{{Kanodia} {et~al.}(2019){Kanodia}, {Wolfgang}, {Stefansson}, {Ning},
  \& {Mahadevan}}]{Kanodia2019}
{Kanodia}, S., {Wolfgang}, A., {Stefansson}, G.~K., {Ning}, B., \& {Mahadevan},
  S. 2019, \apj, 882, 38

\bibitem[{{Kipping}(2013)}]{Kipping2013}
{Kipping}, D.~M. 2013, \mnras, 435, 2152

\bibitem[{{Kostov} {et~al.}(2019){Kostov}, {Schlieder}, {Barclay}, {Quintana},
  {Col{\'o}n}, {Brand e}, {Collins}, {Feinstein}, {Hadden}, {Kane},
  {Kreidberg}, {Kruse}, {Lam}, {Matthews}, {Montet}, {Pozuelos}, {Stassun},
  {Winters}, {Ricker}, {Vanderspek}, {Latham}, {Seager}, {Winn}, {Jenkins},
  {Afanasev}, {Armstrong}, {Arney}, {Boyd}, {Barentsen}, {Barkaoui}, {Batalha},
  {Beichman}, {Bayliss}, {Burke}, {Burdanov}, {Cacciapuoti}, {Carson},
  {Charbonneau}, {Christiansen}, {Ciardi}, {Clampin}, {Collins}, {Conti},
  {Coughlin}, {Covone}, {Crossfield}, {Delrez}, {Domagal-Goldman}, {Dressing},
  {Ducrot}, {Essack}, {Everett}, {Fauchez}, {Foreman-Mackey}, {Gan}, {Gilbert},
  {Gillon}, {Gonzales}, {Hamann}, {Hedges}, {Hocutt}, {Hoffman}, {Horch},
  {Horne}, {Howell}, {Hynes}, {Ireland }, {Irwin}, {Isopi}, {Jensen}, {Jehin},
  {Kaltenegger}, {Kielkopf}, {Kopparapu}, {Lewis}, {Lopez}, {Lissauer}, {Mann},
  {Mallia}, {Mandell}, {Matson}, {Mazeh}, {Monsue}, {Moran}, {Moran}, {Morley},
  {Morris}, {Muirhead}, {Mukai}, {Mullally}, {Mullally}, {Murray}, {Narita},
  {Palle}, {Pidhorodetska}, {Quinn}, {Relles}, {Rinehart}, {Ritsko},
  {Rodriguez}, {Rowden}, {Rowe}, {Sebastian}, {Sefako}, {Shahaf}, {Shporer},
  {Ta{\~n}{\'o}n Reyes}, {Tenenbaum}, {Ting}, {Twicken}, {van Belle}, {Vega},
  {Volosin}, {Walkowicz}, \& {Youngblood}}]{Kostov2019}
{Kostov}, V.~B., {Schlieder}, J.~E., {Barclay}, T., {et~al.} 2019, \aj, 158, 32

\bibitem[{{Kreidberg}(2015)}]{Kreidberg2015}
{Kreidberg}, L. 2015, \pasp, 127, 1161

\bibitem[{{Lee} \& {Chiang}(2015)}]{lee&chiang2015}
{Lee}, E.~J., \& {Chiang}, E. 2015, \apj, 811, 41

\bibitem[{{Lee} \& {Chiang}(2016)}]{lee&chiang2016}
---. 2016, \apj, 817, 90

\bibitem[{{Lee} {et~al.}(2014){Lee}, {Chiang}, \& {Ormel}}]{lee2014}
{Lee}, E.~J., {Chiang}, E., \& {Ormel}, C.~W. 2014, \apj, 797, 95

\bibitem[{{Lee} \& {Connors}(2020)}]{Lee&Connors2020}
{Lee}, E.~J., \& {Connors}, N.~J. 2020, arXiv e-prints, arXiv:2008.01105

\bibitem[{{Li} {et~al.}(2019){Li}, {Tenenbaum}, {Twicken}, {Burke}, {Jenkins},
  {Quintana}, {Rowe}, \& {Seader}}]{Li:DVmodelFit2019}
{Li}, J., {Tenenbaum}, P., {Twicken}, J.~D., {et~al.} 2019, \pasp, 131, 024506

\bibitem[{{Lightkurve Collaboration} {et~al.}(2018){Lightkurve Collaboration},
  {Cardoso}, {Hedges}, {Gully-Santiago}, {Saunders}, {Cody}, {Barclay}, {Hall},
  {Sagear}, {Turtelboom}, {Zhang}, {Tzanidakis}, {Mighell}, {Coughlin}, {Bell},
  {Berta-Thompson}, {Williams}, {Dotson}, \& {Barentsen}}]{Lightkurve}
{Lightkurve Collaboration}, {Cardoso}, J.~V.~d.~M., {Hedges}, C., {et~al.}
  2018, {Lightkurve: Kepler and TESS time series analysis in Python},
  Astrophysics Source Code Library, , , ascl:1812.013

\bibitem[{{Lissauer} {et~al.}(2011){Lissauer}, {Ragozzine}, {Fabrycky},
  {Steffen}, {Ford}, {Jenkins}, {Shporer}, {Holman}, {Rowe}, {Quintana},
  {Batalha}, {Borucki}, {Bryson}, {Caldwell}, {Carter}, {Ciardi}, {Dunham},
  {Fortney}, {Gautier}, {Howell}, {Koch}, {Latham}, {Marcy}, {Morehead}, \&
  {Sasselov}}]{lissauer2011}
{Lissauer}, J.~J., {Ragozzine}, D., {Fabrycky}, D.~C., {et~al.} 2011, \apjs,
  197, 8

\bibitem[{{Liu} {et~al.}(2015){Liu}, {Hori}, {Lin}, \& {Asphaug}}]{liu2015}
{Liu}, S.-F., {Hori}, Y., {Lin}, D.~N.~C., \& {Asphaug}, E. 2015, \apj, 812,
  164

\bibitem[{{Lopez} \& {Fortney}(2013{\natexlab{a}})}]{lopezfortney2013}
{Lopez}, E.~D., \& {Fortney}, J.~J. 2013{\natexlab{a}}, \apj, 776, 2

\bibitem[{{Lopez} \& {Fortney}(2013{\natexlab{b}})}]{lopez&fortney2013}
---. 2013{\natexlab{b}}, \apj, 776, 2

\bibitem[{{Lopez} {et~al.}(2012){Lopez}, {Fortney}, \& {Miller}}]{lopez2012}
{Lopez}, E.~D., {Fortney}, J.~J., \& {Miller}, N. 2012, \apj, 761, 59

\bibitem[{{Lopez} \& {Rice}(2018)}]{Lopez&Rice2018}
{Lopez}, E.~D., \& {Rice}, K. 2018, \mnras, 479, 5303

\bibitem[{{L{\'o}pez-Morales} {et~al.}(2016){L{\'o}pez-Morales}, {Haywood},
  {Coughlin}, {Zeng}, {Buchhave}, {Giles}, {Affer}, {Bonomo}, {Charbonneau},
  {Collier Cameron}, {Consentino}, {Dressing}, {Dumusque}, {Figueira},
  {Fiorenzano}, {Harutyunyan}, {Johnson}, {Latham}, {Lopez}, {Lovis},
  {Malavolta}, {Mayor}, {Micela}, {Molinari}, {Mortier}, {Motalebi},
  {Nascimbeni}, {Pepe}, {Phillips}, {Piotto}, {Pollacco}, {Queloz}, {Rice},
  {Sasselov}, {Segransan}, {Sozzetti}, {Udry}, {Vanderburg}, \&
  {Watson}}]{Lopez-Morales2016}
{L{\'o}pez-Morales}, M., {Haywood}, R.~D., {Coughlin}, J.~L., {et~al.} 2016,
  \aj, 152, 204

\bibitem[{{Luhn} {et~al.}(2020){Luhn}, {Wright}, \& {Isaacson}}]{Luhn2020}
{Luhn}, J.~K., {Wright}, J.~T., \& {Isaacson}, H. 2020, \aj, 159, 236

\bibitem[{{Luque} {et~al.}(2019){Luque}, {Pall{\'e}}, {Kossakowski},
  {Dreizler}, {Kemmer}, {Espinoza}, {Burt}, {Anglada-Escud{\'e}}, {B{\'e}jar},
  {Caballero}, {Collins}, {Collins}, {Cort{\'e}s-Contreras},
  {D{\'\i}ez-Alonso}, {Feng}, {Hatzes}, {Hellier}, {Henning}, {Jeffers},
  {Kaltenegger}, {K{\"u}rster}, {Madden}, {Molaverdikhani}, {Montes}, {Narita},
  {Nowak}, {Ofir}, {Oshagh}, {Parviainen}, {Quirrenbach}, {Reffert}, {Reiners},
  {Rodr{\'\i}guez-L{\'o}pez}, {Schlecker}, {Stock}, {Trifonov}, {Winn},
  {Zapatero Osorio}, {Zechmeister}, {Amado}, {Anderson}, {Batalha}, {Bauer},
  {Bluhm}, {Burke}, {Butler}, {Caldwell}, {Chen}, {Crane}, {Dragomir},
  {Dressing}, {Dynes}, {Jenkins}, {Kaminski}, {Klahr}, {Kotani}, {Lafarga},
  {Latham}, {Lewin}, {McDermott}, {Monta{\~n}{\'e}s-Rodr{\'\i}guez}, {Morales},
  {Murgas}, {Nagel}, {Pedraz}, {Ribas}, {Ricker}, {Rowden}, {Seager},
  {Shectman}, {Tamura}, {Teske}, {Twicken}, {Vanderspeck}, {Wang}, \&
  {Wohler}}]{Luque2019}
{Luque}, R., {Pall{\'e}}, E., {Kossakowski}, D., {et~al.} 2019, \aap, 628, A39

\bibitem[{{Mamajek} \& {Hillenbrand}(2008)}]{Mamajek2008}
{Mamajek}, E.~E., \& {Hillenbrand}, L.~A. 2008, \apj, 687, 1264

\bibitem[{{Marcy} {et~al.}(2014){Marcy}, {Isaacson}, {Howard}, {Rowe},
  {Jenkins}, {Bryson}, {Latham}, {Howell}, {Gautier}, {Batalha}, {Rogers},
  {Ciardi}, {Fischer}, {Gilliland}, {Kjeldsen}, {Christensen-Dalsgaard},
  {Huber}, {Chaplin}, {Basu}, {Buchhave}, {Quinn}, {Borucki}, {Koch}, {Hunter},
  {Caldwell}, {Van Cleve}, {Kolbl}, {Weiss}, {Petigura}, {Seager}, {Morton},
  {Johnson}, {Ballard}, {Burke}, {Cochran}, {Endl}, {MacQueen}, {Everett},
  {Lissauer}, {Ford}, {Torres}, {Fressin}, {Brown}, {Steffen}, {Charbonneau},
  {Basri}, {Sasselov}, {Winn}, {Sanchis-Ojeda}, {Christiansen}, {Adams},
  {Henze}, {Dupree}, {Fabrycky}, {Fortney}, {Tarter}, {Holman}, {Tenenbaum},
  {Shporer}, {Lucas}, {Welsh}, {Orosz}, {Bedding}, {Campante}, {Davies},
  {Elsworth}, {Handberg}, {Hekker}, {Karoff}, {Kawaler}, {Lund}, {Lundkvist},
  {Metcalfe}, {Miglio}, {Silva Aguirre}, {Stello}, {White}, {Boss}, {Devore},
  {Gould}, {Prsa}, {Agol}, {Barclay}, {Coughlin}, {Brugamyer}, {Mullally},
  {Quintana}, {Still}, {Thompson}, {Morrison}, {Twicken}, {D{\'e}sert},
  {Carter}, {Crepp}, {H{\'e}brard}, {Santerne}, {Moutou}, {Sobeck}, {Hudgins},
  {Haas}, {Robertson}, {Lillo-Box}, \& {Barrado}}]{marcy2014}
{Marcy}, G.~W., {Isaacson}, H., {Howard}, A.~W., {et~al.} 2014, \apjs, 210, 20

\bibitem[{{Marmier} {et~al.}(2013){Marmier}, {S{\'e}gransan}, {Udry}, {Mayor},
  {Pepe}, {Queloz}, {Lovis}, {Naef}, {Santos}, {Alonso}, {Alves}, {Berthet},
  {Chazelas}, {Demory}, {Dumusque}, {Eggenberger}, {Figueira}, {Gillon},
  {Hagelberg}, {Lendl}, {Mardling}, {M{\'e}gevand }, {Neveu}, {Sahlmann},
  {Sosnowska}, {Tewes}, \& {Triaud}}]{marmier2013}
{Marmier}, M., {S{\'e}gransan}, D., {Udry}, S., {et~al.} 2013, \aap, 551, A90

\bibitem[{{Martinez} {et~al.}(2019){Martinez}, {Cunha}, {Ghezzi}, \&
  {Smith}}]{Martinez2019}
{Martinez}, C.~F., {Cunha}, K., {Ghezzi}, L., \& {Smith}, V.~V. 2019, \apj,
  875, 29

\bibitem[{{Mayor} {et~al.}(2003){Mayor}, {Pepe}, {Queloz}, {Bouchy},
  {Rupprecht}, {Lo Curto}, {Avila}, {Benz}, {Bertaux}, {Bonfils}, {Dall},
  {Dekker}, {Delabre}, {Eckert}, {Fleury}, {Gilliotte}, {Gojak}, {Guzman},
  {Kohler}, {Lizon}, {Longinotti}, {Lovis}, {Megevand}, {Pasquini}, {Reyes},
  {Sivan}, {Sosnowska}, {Soto}, {Udry}, {van Kesteren}, {Weber}, \&
  {Weilenmann}}]{2003Msngr.114...20M}
{Mayor}, M., {Pepe}, F., {Queloz}, D., {et~al.} 2003, The Messenger, 114, 20

\bibitem[{{Misener} \& {Schlichting}(2021)}]{Misener&Schlichting2021}
{Misener}, W., \& {Schlichting}, H.~E. 2021, arXiv e-prints, arXiv:2103.09212

\bibitem[{{Montet}(2018)}]{montet2018}
{Montet}, B.~T. 2018, Research Notes of the American Astronomical Society, 2,
  28

\bibitem[{{Morris} {et~al.}(2020){Morris}, {Twicken}, {Smith}, {Clarke},
  {Jenkins}, {Bryson}, {Girouard}, \& {Klaus}}]{morris2020}
{Morris}, R.~L., {Twicken}, J.~D., {Smith}, J.~C., {et~al.} 2020, {Kepler Data
  Processing Handbook: Photometric Analysis}, Kepler Data Processing Handbook
  (KSCI-19081-003), ,

\bibitem[{{Naoz}(2016)}]{naoz2016}
{Naoz}, S. 2016, \araa, 54, 441

\bibitem[{{Neil} \& {Rogers}(2020)}]{Neil&Rogers2020}
{Neil}, A.~R., \& {Rogers}, L.~A. 2020, \apj, 891, 12

\bibitem[{{Ning} {et~al.}(2018){Ning}, {Wolfgang}, \& {Ghosh}}]{Ning2018}
{Ning}, B., {Wolfgang}, A., \& {Ghosh}, S. 2018, \apj, 869, 5

\bibitem[{{Owen} \& {Wu}(2013)}]{owen&wu2013}
{Owen}, J.~E., \& {Wu}, Y. 2013, \apj, 775, 105

\bibitem[{{Owen} \& {Wu}(2017)}]{owen&wu2017}
---. 2017, ArXiv e-prints, arXiv:1705.10810

\bibitem[{{Pepe} {et~al.}(2002){Pepe}, {Mayor}, {Rupprecht}, {Avila},
  {Ballester}, {Beckers}, {Benz}, {Bertaux}, {Bouchy}, {Buzzoni}, {Cavadore},
  {Deiries}, {Dekker}, {Delabre}, {D'Odorico}, {Eckert}, {Fischer}, {Fleury},
  {George}, {Gilliotte}, {Gojak}, {Guzman}, {Koch}, {Kohler}, {Kotzlowski},
  {Lacroix}, {Le Merrer}, {Lizon}, {Lo Curto}, {Longinotti}, {Megevand},
  {Pasquini}, {Petitpas}, {Pichard}, {Queloz}, {Reyes}, {Richaud}, {Sivan},
  {Sosnowska}, {Soto}, {Udry}, {Ureta}, {van Kesteren}, {Weber}, {Weilenmann},
  {Wicenec}, {Wieland}, {Christensen-Dalsgaard}, {Dravins}, {Hatzes},
  {K{\"u}rster}, {Paresce}, \& {Penny}}]{Pepe:2002}
{Pepe}, F., {Mayor}, M., {Rupprecht}, G., {et~al.} 2002, The Messenger, 110, 9

\bibitem[{{Petigura} {et~al.}(2013){Petigura}, {Howard}, \&
  {Marcy}}]{petigura2013}
{Petigura}, E.~A., {Howard}, A.~W., \& {Marcy}, G.~W. 2013, Proceedings of the
  National Academy of Science, 110, 19273

\bibitem[{{Plotnykov} \& {Valencia}(2020)}]{Plotnykov&Valencia2020}
{Plotnykov}, M., \& {Valencia}, D. 2020, \mnras, 499, 932

\bibitem[{{Pollacco} {et~al.}(2006){Pollacco}, {Skillen}, {Collier Cameron},
  {Christian}, {Hellier}, {Irwin}, {Lister}, {Street}, {West}, {Anderson},
  {Clarkson}, {Deeg}, {Enoch}, {Evans}, {Fitzsimmons}, {Haswell}, {Hodgkin},
  {Horne}, {Kane}, {Keenan}, {Maxted}, {Norton}, {Osborne}, {Parley}, {Ryans},
  {Smalley}, {Wheatley}, \& {Wilson}}]{Pollacco2006}
{Pollacco}, D.~L., {Skillen}, I., {Collier Cameron}, A., {et~al.} 2006, \pasp,
  118, 1407

\bibitem[{{Queloz} {et~al.}(2001){Queloz}, {Henry}, {Sivan}, {Baliunas},
  {Beuzit}, {Donahue}, {Mayor}, {Naef}, {Perrier}, \&
  {Udry}}]{2001A&A...379..279Q}
{Queloz}, D., {Henry}, G.~W., {Sivan}, J.~P., {et~al.} 2001, \aap, 379, 279

\bibitem[{{Rajpaul} {et~al.}(2015){Rajpaul}, {Aigrain}, {Osborne}, {Reece}, \&
  {Roberts}}]{Rajpaul2015}
{Rajpaul}, V., {Aigrain}, S., {Osborne}, M.~A., {Reece}, S., \& {Roberts}, S.
  2015, \mnras, 452, 2269

\bibitem[{{Raymond} {et~al.}(2008){Raymond}, {Barnes}, \&
  {Mandell}}]{raymond2008}
{Raymond}, S.~N., {Barnes}, R., \& {Mandell}, A.~M. 2008, \mnras, 384, 663

\bibitem[{{Raymond} \& {Cossou}(2014)}]{raymond&cossou2014}
{Raymond}, S.~N., \& {Cossou}, C. 2014, \mnras, 440, L11

\bibitem[{{Raymond} {et~al.}(2014){Raymond}, {Kokubo}, {Morbidelli},
  {Morishima}, \& {Walsh}}]{raymond2014}
{Raymond}, S.~N., {Kokubo}, E., {Morbidelli}, A., {Morishima}, R., \& {Walsh},
  K.~J. 2014, Protostars and Planets VI, 595

\bibitem[{{Rayner} {et~al.}(2016){Rayner}, {Tokunaga}, {Jaffe}, {Bonnet},
  {Ching}, {Connelley}, {Kokubun}, {Lockhart}, \&
  {Warmbier}}]{2016SPIE.9908E..84R}
{Rayner}, J., {Tokunaga}, A., {Jaffe}, D., {et~al.} 2016, in \procspie, Vol.
  9908, Ground-based and Airborne Instrumentation for Astronomy VI, 990884

\bibitem[{{Ricker} {et~al.}(2015){Ricker}, {Winn}, {Vanderspek}, {Latham},
  {Bakos}, {Bean}, {Berta-Thompson}, {Brown}, {Buchhave}, {Butler}, {Butler},
  {Chaplin}, {Charbonneau}, {Christensen-Dalsgaard}, {Clampin}, {Deming},
  {Doty}, {De Lee}, {Dressing}, {Dunham}, {Endl}, {Fressin}, {Ge}, {Henning},
  {Holman}, {Howard}, {Ida}, {Jenkins}, {Jernigan}, {Johnson}, {Kaltenegger},
  {Kawai}, {Kjeldsen}, {Laughlin}, {Levine}, {Lin}, {Lissauer}, {MacQueen},
  {Marcy}, {McCullough}, {Morton}, {Narita}, {Paegert}, {Palle}, {Pepe},
  {Pepper}, {Quirrenbach}, {Rinehart}, {Sasselov}, {Sato}, {Seager},
  {Sozzetti}, {Stassun}, {Sullivan}, {Szentgyorgyi}, {Torres}, {Udry}, \&
  {Villasenor}}]{ricker2015}
{Ricker}, G.~R., {Winn}, J.~N., {Vanderspek}, R., {et~al.} 2015, Journal of
  Astronomical Telescopes, Instruments, and Systems, 1, 014003

\bibitem[{{Rogers} \& {Owen}(2020)}]{Rogers&Owen2020}
{Rogers}, J.~G., \& {Owen}, J.~E. 2020, arXiv e-prints, arXiv:2007.11006

\bibitem[{{Schlecker} {et~al.}(2020){Schlecker}, {Mordasini}, {Emsenhuber},
  {Klahr}, {Henning}, {Burn}, {Alibert}, \& {Benz}}]{Schlecker2020}
{Schlecker}, M., {Mordasini}, C., {Emsenhuber}, A., {et~al.} 2020, arXiv
  e-prints, arXiv:2007.05563

\bibitem[{{Schlichting}(2014)}]{schlichting2014}
{Schlichting}, H.~E. 2014, \apjl, 795, L15

\bibitem[{{Scora} {et~al.}(2020){Scora}, {Valencia}, {Morbidelli}, \&
  {Jacobson}}]{Scora2020}
{Scora}, J., {Valencia}, D., {Morbidelli}, A., \& {Jacobson}, S. 2020, \mnras,
  493, 4910

\bibitem[{{Shaikhislamov} {et~al.}(2020){Shaikhislamov}, {Fossati},
  {Khodachenko}, {Lammer}, {Garc{\'\i}a Mu{\~n}oz}, {Youngblood}, {Dwivedi}, \&
  {Rumenskikh}}]{Shaikhislamov2020}
{Shaikhislamov}, I.~F., {Fossati}, L., {Khodachenko}, M.~L., {et~al.} 2020,
  \aap, 639, A109

\bibitem[{{Shporer} {et~al.}(2020){Shporer}, {Collins}, {Astudillo-Defru},
  {Irwin}, {Bonfils}, {Collins}, {Matthews}, {Winters}, {Anderson},
  {Armstrong}, {Charbonneau}, {Cloutier}, {Daylan}, {Gan}, {G{\"u}nther},
  {Hellier}, {Horne}, {Huang}, {Jensen}, {Kielkopf}, {Palle}, {Sefako},
  {Stassun}, {Tan}, {Vanderburg}, {Ricker}, {Latham}, {Vanderspek}, {Seager},
  {Winn}, {Jenkins}, {Colon}, {Dressing}, {L{\'e}epine}, {Muirhead}, {Rose},
  {Twicken}, \& {Villasenor}}]{Shporer2020}
{Shporer}, A., {Collins}, K.~A., {Astudillo-Defru}, N., {et~al.} 2020, \apjl,
  890, L7

\bibitem[{{Smith} {et~al.}(2012){Smith}, {Stumpe}, {Van Cleve}, {Jenkins},
  {Barclay}, {Fanelli}, {Girouard}, {Kolodziejczak}, {McCauliff}, {Morris}, \&
  {Twicken}}]{Smith2012}
{Smith}, J.~C., {Stumpe}, M.~C., {Van Cleve}, J.~E., {et~al.} 2012, \pasp, 124,
  1000

\bibitem[{{Sousa} {et~al.}(2011){Sousa}, {Santos}, {Israelian}, {Mayor}, \&
  {Udry}}]{Sousa2011}
{Sousa}, S.~G., {Santos}, N.~C., {Israelian}, G., {Mayor}, M., \& {Udry}, S.
  2011, \aap, 533, A141

\bibitem[{{Speagle} \& {Barbary}(2018)}]{Speagle2018}
{Speagle}, J.~S., \& {Barbary}, K. 2018, {dynesty: Dynamic Nested Sampling
  package}, Astrophysics Source Code Library, , , ascl:1809.013

\bibitem[{{Stan Development Team}(2019)}]{STAN}
{Stan Development Team}. 2019, Users Guide and Reference Manual, v2.25,
  https://mc-stan.org

\bibitem[{{Stassun} {et~al.}(2018){Stassun}, {Oelkers}, {Pepper}, {Paegert},
  {De Lee}, {Torres}, {Latham}, {Charpinet}, {Dressing}, {Huber}, {Kane},
  {L{\'e}pine}, {Mann}, {Muirhead}, {Rojas-Ayala}, {Silvotti}, {Fleming},
  {Levine}, \& {Plavchan}}]{Stassun2017tic}
{Stassun}, K.~G., {Oelkers}, R.~J., {Pepper}, J., {et~al.} 2018, \aj, 156, 102

\bibitem[{{Stassun} {et~al.}(2019){Stassun}, {Oelkers}, {Paegert}, {Torres},
  {Pepper}, {De Lee}, {Collins}, {Latham}, {Muirhead}, {Chittidi},
  {Rojas-Ayala}, {Fleming}, {Rose}, {Tenenbaum}, {Ting}, {Kane}, {Barclay},
  {Bean}, {Brassuer}, {Charbonneau}, {Ge}, {Lissauer}, {Mann}, {McLean},
  {Mullally}, {Narita}, {Plavchan}, {Ricker}, {Sasselov}, {Seager}, {Sharma},
  {Shiao}, {Sozzetti}, {Stello}, {Vanderspek}, {Wallace}, \&
  {Winn}}]{Stassun2019tic}
{Stassun}, K.~G., {Oelkers}, R.~J., {Paegert}, M., {et~al.} 2019, \aj, 158, 138

\bibitem[{{Stevenson} {et~al.}(2021){Stevenson}, {Lustig-Yaeger}, {Alam},
  {Batalha}, {Lopez-Morales}, {Lothringer}, {MacDonald}, {May}, {Moran},
  {Peacock}, {Rustamkulov}, {Sing}, {Sotzen}, {Valenti}, \&
  {Wakeford}}]{2021jwst.prop.1981S}
{Stevenson}, K., {Lustig-Yaeger}, J., {Alam}, M., {et~al.} 2021, {Tell Me How
  I?m Supposed To Breathe With No Air: Measuring the Prevalence and Diversity
  of M-Dwarf Planet Atmospheres}, JWST Proposal. Cycle 1, ,

\bibitem[{{Stumpe} {et~al.}(2014){Stumpe}, {Smith}, {Catanzarite}, {Van Cleve},
  {Jenkins}, {Twicken}, \& {Girouard}}]{Stumpe2014}
{Stumpe}, M.~C., {Smith}, J.~C., {Catanzarite}, J.~H., {et~al.} 2014, \pasp,
  126, 100

\bibitem[{{Stumpe} {et~al.}(2012){Stumpe}, {Smith}, {Van Cleve}, {Twicken},
  {Barclay}, {Fanelli}, {Girouard}, {Jenkins}, {Kolodziejczak}, {McCauliff}, \&
  {Morris}}]{Stumpe2012}
{Stumpe}, M.~C., {Smith}, J.~C., {Van Cleve}, J.~E., {et~al.} 2012, \pasp, 124,
  985

\bibitem[{{Sullivan} {et~al.}(2015){Sullivan}, {Winn}, {Berta-Thompson},
  {Charbonneau}, {Deming}, {Dressing}, {Latham}, {Levine}, {McCullough},
  {Morton}, {Ricker}, {Vanderspek}, \& {Woods}}]{sullivan2015}
{Sullivan}, P.~W., {Winn}, J.~N., {Berta-Thompson}, Z.~K., {et~al.} 2015, \apj,
  809, 77

\bibitem[{{Tal-Or} {et~al.}(2019){Tal-Or}, {Trifonov}, {Zucker}, {Mazeh}, \&
  {Zechmeister}}]{Tal-Or2019}
{Tal-Or}, L., {Trifonov}, T., {Zucker}, S., {Mazeh}, T., \& {Zechmeister}, M.
  2019, \mnras, 484, L8

\bibitem[{{Teske} {et~al.}(2020){Teske}, {D{\'\i}az}, {Luque}, {Mo{\v{c}}nik},
  {Seidel}, {Otegi}, {Feng}, {Jenkins}, {Pall{\`e}}, {S{\'e}gransan}, {Udry},
  {Collins}, {Eastman}, {Ricker}, {Vanderspek}, {Latham}, {Seager}, {Winn},
  {Jenkins}, {Anderson}, {Barclay}, {Bouchy}, {Burt}, {Butler}, {Caldwell},
  {Collins}, {Crane}, {Dorn}, {Flowers}, {Haldemann}, {Helled}, {Hellier},
  {Jensen}, {Kane}, {Law}, {Lissauer}, {Mann}, {Marmier}, {Nielsen}, {Rose},
  {Shectman}, {Shporer}, {Torres}, {Wang}, {Wolfgang}, {Wong}, \&
  {Ziegler}}]{Teske2020}
{Teske}, J., {D{\'\i}az}, M.~R., {Luque}, R., {et~al.} 2020, \aj, 160, 96

\bibitem[{{Teske} {et~al.}(2018){Teske}, {Ciardi}, {Howell}, {Hirsch}, \&
  {Johnson}}]{teske2018b}
{Teske}, J.~K., {Ciardi}, D.~R., {Howell}, S.~B., {Hirsch}, L.~A., \&
  {Johnson}, R.~A. 2018, ArXiv e-prints, arXiv:1804.10170

\bibitem[{{Teske} {et~al.}(2019){Teske}, {Thorngren}, {Fortney}, {Hinkel}, \&
  {Brewer}}]{teske2019}
{Teske}, J.~K., {Thorngren}, D., {Fortney}, J.~J., {Hinkel}, N., \& {Brewer},
  J.~M. 2019, \aj, 158, 239

\bibitem[{{The Astropy Collaboration} {et~al.}(2018){The Astropy
  Collaboration}, {Price-Whelan}, {Sip{\H o}cz}, {G{\"u}nther}, {Lim},
  {Crawford}, {Conseil}, {Shupe}, {Craig}, {Dencheva}, {Ginsburg},
  {VanderPlas}, {Bradley}, {P{\'e}rez-Su{\'a}rez}, {de Val-Borro}, {Aldcroft},
  {Cruz}, {Robitaille}, {Tollerud}, {Ardelean}, {Babej}, {Bachetti}, {Bakanov},
  {Bamford}, {Barentsen}, {Barmby}, {Baumbach}, {Berry}, {Biscani}, {Boquien},
  {Bostroem}, {Bouma}, {Brammer}, {Bray}, {Breytenbach}, {Buddelmeijer},
  {Burke}, {Calderone}, {Cano Rodr{\'{\i}}guez}, {Cara}, {Cardoso},
  {Cheedella}, {Copin}, {Crichton}, {D{\'A}vella}, {Deil}, {Depagne},
  {Dietrich}, {Donath}, {Droettboom}, {Earl}, {Erben}, {Fabbro}, {Ferreira},
  {Finethy}, {Fox}, {Garrison}, {Gibbons}, {Goldstein}, {Gommers}, {Greco},
  {Greenfield}, {Groener}, {Grollier}, {Hagen}, {Hirst}, {Homeier}, {Horton},
  {Hosseinzadeh}, {Hu}, {Hunkeler}, {Ivezi{\'c}}, {Jain}, {Jenness}, {Kanarek},
  {Kendrew}, {Kern}, {Kerzendorf}, {Khvalko}, {King}, {Kirkby}, {Kulkarni},
  {Kumar}, {Lee}, {Lenz}, {Littlefair}, {Ma}, {Macleod}, {Mastropietro},
  {McCully}, {Montagnac}, {Morris}, {Mueller}, {Mumford}, {Muna}, {Murphy},
  {Nelson}, {Nguyen}, {Ninan}, {N{\"o}the}, {Ogaz}, {Oh}, {Parejko}, {Parley},
  {Pascual}, {Patil}, {Patil}, {Plunkett}, {Prochaska}, {Rastogi}, {Reddy
  Janga}, {Sabater}, {Sakurikar}, {Seifert}, {Sherbert}, {Sherwood-Taylor},
  {Shih}, {Sick}, {Silbiger}, {Singanamalla}, {Singer}, {Sladen}, {Sooley},
  {Sornarajah}, {Streicher}, {Teuben}, {Thomas}, {Tremblay}, {Turner},
  {Terr{\'o}n}, {van Kerkwijk}, {de la Vega}, {Watkins}, {Weaver}, {Whitmore},
  {Woillez}, \& {Zabalza}}]{astropy}
{The Astropy Collaboration}, {Price-Whelan}, A.~M., {Sip{\H o}cz}, B.~M.,
  {et~al.} 2018, ArXiv e-prints, arXiv:1801.02634

\bibitem[{{Thomas-Osip} {et~al.}(2010){Thomas-Osip}, {McCarthy}, {Prieto},
  {Phillips}, \& {Johns}}]{thomas-osip2010}
{Thomas-Osip}, J.~E., {McCarthy}, P., {Prieto}, G., {Phillips}, M.~M., \&
  {Johns}, M. 2010, in Society of Photo-Optical Instrumentation Engineers
  (SPIE) Conference Series, Vol. 7733, Ground-based and Airborne Telescopes
  III, ed. L.~M. {Stepp}, R.~{Gilmozzi}, \& H.~J. {Hall}, 77331L

\bibitem[{{Thompson} {et~al.}(2018){Thompson}, {Coughlin}, {Hoffman},
  {Mullally}, {Christiansen}, {Burke}, {Bryson}, {Batalha}, {Haas},
  {Catanzarite}, {Rowe}, {Barentsen}, {Caldwell}, {Clarke}, {Jenkins}, {Li},
  {Latham}, {Lissauer}, {Mathur}, {Morris}, {Seader}, {Smith}, {Klaus},
  {Twicken}, {Van Cleve}, {Wohler}, {Akeson}, {Ciardi}, {Cochran}, {Henze},
  {Howell}, {Huber}, {Pr{\v{s}}a}, {Ram{\'\i}rez}, {Morton}, {Barclay},
  {Campbell}, {Chaplin}, {Charbonneau}, {Christensen-Dalsgaard}, {Dotson},
  {Doyle}, {Dunham}, {Dupree}, {Ford}, {Geary}, {Girouard}, {Isaacson},
  {Kjeldsen}, {Quintana}, {Ragozzine}, {Shabram}, {Shporer}, {Silva Aguirre},
  {Steffen}, {Still}, {Tenenbaum}, {Welsh}, {Wolfgang}, {Zamudio}, {Koch}, \&
  {Borucki}}]{Thompson2018}
{Thompson}, S.~E., {Coughlin}, J.~L., {Hoffman}, K., {et~al.} 2018, \apjs, 235,
  38

\bibitem[{{Trifonov} {et~al.}(2019){Trifonov}, {Rybizki}, \&
  {K{\"u}rster}}]{Trifonov2019}
{Trifonov}, T., {Rybizki}, J., \& {K{\"u}rster}, M. 2019, \aap, 622, L7

\bibitem[{{Trotta}(2008)}]{Trotta2008}
{Trotta}, R. 2008, Contemporary Physics, 49, 71

\bibitem[{{Twicken} {et~al.}(2010){Twicken}, {Clarke}, {Bryson}, {Tenenbaum},
  {Wu}, {Jenkins}, {Girouard}, \& {Klaus}}]{twicken2010}
{Twicken}, J.~D., {Clarke}, B.~D., {Bryson}, S.~T., {et~al.} 2010, in
  \procspie, Vol. 7740, Software and Cyberinfrastructure for Astronomy, 774023

\bibitem[{{Twicken} {et~al.}(2018){Twicken}, {Catanzarite}, {Clarke},
  {Girouard}, {Jenkins}, {Klaus}, {Li}, {McCauliff}, {Seader}, {Tenenbaum},
  {Wohler}, {Bryson}, {Burke}, {Caldwell}, {Haas}, {Henze}, \&
  {Sanderfer}}]{Twicken:DVdiagnostics2018}
{Twicken}, J.~D., {Catanzarite}, J.~H., {Clarke}, B.~D., {et~al.} 2018, \pasp,
  130, 064502

\bibitem[{{Valencia} {et~al.}(2013){Valencia}, {Guillot}, {Parmentier}, \&
  {Freedman}}]{Valencia2013}
{Valencia}, D., {Guillot}, T., {Parmentier}, V., \& {Freedman}, R.~S. 2013,
  \apj, 775, 10

\bibitem[{{Valencia} {et~al.}(2006){Valencia}, {O'Connell}, \&
  {Sasselov}}]{Valencia2006}
{Valencia}, D., {O'Connell}, R.~J., \& {Sasselov}, D. 2006, \icarus, 181, 545

\bibitem[{{Van Eylen} {et~al.}(2018){Van Eylen}, {Agentoft}, {Lundkvist},
  {Kjeldsen}, {Owen}, {Fulton}, {Petigura}, \& {Snellen}}]{vaneylen2018}
{Van Eylen}, V., {Agentoft}, C., {Lundkvist}, M.~S., {et~al.} 2018, \mnras,
  479, 4786

\bibitem[{{Vogt} {et~al.}(1994){Vogt}, {Allen}, {Bigelow}, {Bresee}, {Brown},
  {Cantrall}, {Conrad}, {Couture}, {Delaney}, {Epps}, {Hilyard}, {Hilyard},
  {Horn}, {Jern}, {Kanto}, {Keane}, {Kibrick}, {Lewis}, {Osborne},
  {Pardeilhan}, {Pfister}, {Ricketts}, {Robinson}, {Stover}, {Tucker}, {Ward},
  \& {Wei}}]{Vogt1994}
{Vogt}, S.~S., {Allen}, S.~L., {Bigelow}, B.~C., {et~al.} 1994, in \procspie,
  Vol. 2198, Instrumentation in Astronomy VIII, ed. D.~L. {Crawford} \& E.~R.
  {Craine}, 362

\bibitem[{{Weiss} \& {Marcy}(2014)}]{weiss&marcy2014}
{Weiss}, L.~M., \& {Marcy}, G.~W. 2014, \apjl, 783, L6

\bibitem[{{Winters} {et~al.}(2019){Winters}, {Medina}, {Irwin}, {Charbonneau},
  {Astudillo-Defru}, {Horch}, {Eastman}, {Vrijmoet}, {Henry}, {Diamond-Lowe},
  {Winston}, {Barclay}, {Bonfils}, {Ricker}, {Vanderspek}, {Latham}, {Seager},
  {Winn}, {Jenkins}, {Udry}, {Twicken}, {Teske}, {Tenenbaum}, {Pepe}, {Murgas},
  {Muirhead}, {Mink}, {Lovis}, {Levine}, {L{\'e}pine}, {Jao}, {Henze},
  {Fur{\'e}sz}, {Forveille}, {Figueira}, {Esquerdo}, {Dressing}, {D{\'\i}az},
  {Delfosse}, {Burke}, {Bouchy}, {Berlind}, \& {Almenara}}]{Winters2019}
{Winters}, J.~G., {Medina}, A.~A., {Irwin}, J.~M., {et~al.} 2019, \aj, 158, 152

\bibitem[{{Wolfgang} \& {Lopez}(2015)}]{Wolfgang2015}
{Wolfgang}, A., \& {Lopez}, E. 2015, \apj, 806, 183

\bibitem[{{Wolfgang} {et~al.}(2016){Wolfgang}, {Rogers}, \&
  {Ford}}]{wolfgang2016}
{Wolfgang}, A., {Rogers}, L.~A., \& {Ford}, E.~B. 2016, \apj, 825, 19

\bibitem[{{Xuan} \& {Wyatt}(2020)}]{Xuan&Wyatt2020}
{Xuan}, J.~W., \& {Wyatt}, M.~C. 2020, \mnras, 497, 2096

\bibitem[{{Zechmeister} {et~al.}(2018){Zechmeister}, {Reiners}, {Amado},
  {Azzaro}, {Bauer}, {B{\'e}jar}, {Caballero}, {Guenther}, {Hagen}, {Jeffers},
  {Kaminski}, {K{\"u}rster}, {Launhardt}, {Montes}, {Morales}, {Quirrenbach},
  {Reffert}, {Ribas}, {Seifert}, {Tal-Or}, \& {Wolthoff}}]{Zechmeister2018}
{Zechmeister}, M., {Reiners}, A., {Amado}, P.~J., {et~al.} 2018, \aap, 609, A12

\bibitem[{{Zhu} \& {Wu}(2018)}]{zhu&wu2018}
{Zhu}, W., \& {Wu}, Y. 2018, \aj, 156, 92

\end{thebibliography}
\newpage

\appendix

\section{Summary of Prior Knowledge of Individual Systems in MTS \label{sec:individual_systems}}
 
For TOIs with no previously published validation, we briefly summarize the ground-based photometric follow-up observations in Table \ref{table:SG1-observations} that have been made by the TFOP SG1 using the facilities listed in Table \ref{table:sg1-observatories}. These observations attempt to rule out false positives from neighboring stars that may be blended with the target in the TESS data.

    \subsection{TOI-134.01}

L 168-9 b (TOI-134.01) is  a 1.39$\pm$0.09 R$_{\oplus}$ planet in a 1.4-d orbit around a bright ($V=11$, $d=25$ pc) M1 host star, detected by TESS in Sector 1 and published by \cite{Astudillo-Defru2020}. The authors measured the mass of the planet using HARPS and PFS RV observations to be 4.60$\pm$0.56 M$_{\oplus}$, and put an upper limit on the orbital eccentricity of 0.21 (95\% confidence). Using additional photometry of L 168-9 from WASP collected between 2010 and 2014, the authors measured the rotation period ($P_{\rm{rot}}$) of the star using a Gaussian process regression model with a quasi-periodic covariance kernel to be 29.8$\pm$1.3 days. A similar Gaussian process component was included in the author's model of the RV data, with $P_{\rm{RV}} = P_{\rm{rot}}$. The mass and radius of TOI-134.01 make it one of the densest planets above 4 M$_{\oplus}$, and it is one of the top rocky planet targets for emission spectroscopy with JWST.

    \subsection{TOI-141.01}

HD 213885 b (TOI-141.01) is an ultrashort 1.008-d period, 1.75$\pm$0.05 R$_{\oplus}$ planet detected in TESS Sector 1 around a bright $V=7.9$ ($d$=48 pc) G dwarf slightly hotter than the Sun, and published by \cite{Espinoza2020}. The authors measured the planet's mass as 8.8$\pm$0.6 M$_{\oplus}$ using radial velocity observations from FEROS, HARPS, and CORALIE. The authors also searched for additional Keplerian signals in the RV data, and found their modeling to strongly favor a two-planet model, with a second planet (HD 213885 c or TOI-141c) with a period of 4.785 days and a semi-amplitude of 7.26$^{+0.48}_{-0.47}$ m~s$^{-1}$, translating to a minimum mass of 19.95$^{+1.38}_{-1.36}$ M$_{\oplus}$ (assuming M$_{*}$ of 1.068 M$_{\odot}$ from their own stellar parameter analysis). Both planets' orbits were fit assuming $e=0$, although the authors give 95\% upper confidence interval limits on the eccentricities of 0.24 and 0.16 for the short and long period planets, respectively. The authors searched for transits of TOI-141c in the TESS data but did not find evidence of this. The authors also found the model that was favored (highest evidence) included a Gaussian process component with a squared-exponential kernel to model correlated noise in the RV data. Overall, TOI-141.01 presents an interesting prospect for atmospheric characterization -- the star is the second brightest known to host an USP exoplanet, and the planet is a very close ``twin'' to the well-known USP super-Earth 55 Cancri e.  

\begin{deluxetable}{ccccc}
\tabletypesize{\footnotesize}
\tablewidth{\columnwidth}
 \tablecaption{TFOP SG1 Photometric Observations
 \label{table:SG1-observations}}
 \tablehead{ 
 \colhead{TOI} & \colhead{TIC} & \colhead{Telescope} & \colhead{Date} & \colhead{Filter}
 }
\startdata 
\multirow{6}{*}{174.01} & \multirow{6}{*}{425997655} &  IRSF/Sirius  &  2018-11-08  &  $JHK_s$ \\
                           &                          &  LCO 1m  &  2018-12-14  &  $i^\prime$ \\
                           &                          &  LCO 0.4m  &  2019-01-18  &  $i^\prime$ \\
                           &                          &  iTel. T17  &  2019-01-18  &  clear \\
                           &                          &  LCO 1m  &  2019-10-28  &  $z_s$ \\
                           &                          &  LCO 1m  &  2019-12-20  &  $z_s$ \\
\cmidrule{1-5}
174.02 & 425997655 & LCO 1m  &  2018-11-07  &  $r^\prime$ \\
\cmidrule{1-5}
\multirow{2}{*}{260.01} & \multirow{2}{*}{37749396} &  LCO 1m  &  2019-08-19  &  y \\
                           &                          &  LCO 1m  &  2020-10-10  &  $z_s$ \\
\cmidrule{1-5}
784.01 & 460984940 & LCO 1m  &  2020-02-26  &  $z_s$ \\
\cmidrule{1-5}
\multirow{6}{*}{836.01} & \multirow{6}{*}{440887364} &  MEarth-South  &  2019-07-04  &  RG715 \\
                           &                          &  NGTS  &  2019-08-15  &  NGTS\tablenotemark{[1]} \\
                           &                          &  LCO 1m  &  2020-02-29  &  y \\
                           &                          &  LCO 1m  &  2020-04-12  &  $z_s$ \\
                           &                          &  PEST  &  2020-04-29  &  $R_c$ \\
                           &                          &  LCO 1m  &  2020-05-16  &  $z_s$ \\
\cmidrule{1-5}
\multirow{9}{*}{836.02} & \multirow{9}{*}{440887364} &  MEarth-South  &  2019-07-04  &  RG715 \\
                           &                          &  LCO 1m  &  2019-07-04  &  $i^\prime$ \\
                           &                          &  LCO 1m  &  2020-03-08  &  y \\
                           &                          &  LCO 1m  &  2020-03-20  &  y \\
                           &                          &  LCO 1m  &  2020-05-04  &  $z_s$ \\
                           &                          &  LCO 1m  &  2020-05-15  &  $z_s$ \\
                           &                          &  LCO 1m  &  2020-05-19  &  $z_s$ \\
                           &                          &  LCO 1m  &  2020-06-15  &  $z_s$ \\
                           &                          &  LCO 1m  &  2020-07-08  &  $z_s$ \\
\enddata
\begin{tablenotes}
\item[1] [1] The custom NGTS filter covers 520--890~nm.
\end{tablenotes}
\end{deluxetable}
\begin{deluxetable*}{llDDc}
\decimals
\tablecaption{Facilities used for TFOP SG1 followup \label{table:sg1-observatories}}
\tablehead{\colhead{Facility or Instrument} & \colhead{Location} & \twocolhead{Aperture} & \twocolhead{Pixel scale} & \colhead{FOV}\\[-2mm]
 & & \twocolhead{(m)} & \twocolhead{(arcsec)} & \colhead{(arcmin)}
}
\startdata
CTIO / MEarth-South & La Serena, Chile & 0.4 & 0.84 & $29 \times 29$ \\
iTelescope T17 & Siding Spring, Australia & 0.43 & 0.92 & $15.5 \times 15.5$ \\
IRSF/Sirius & Sutherland, South Africa & 1.4 & 0.45 & $7.7 \times 7.7$ \\
LCOGT 0.4m &  (various) & 0.4 & 0.57 & $29.2 \times 19.5$ \\
LCOGT 1m & (various) & 1 & 0.39 & $26.5 \times 26.5$ \\
NGTS & Paranal, Chile & 0.2 & 4.97 &  $170 \times 170$ \\
Perth Exoplanet Survey Tel.\ (PEST) & Perth, Australia & 0.3 & 1.2 & $31 \times 21$
\enddata
\end{deluxetable*}

    \subsection{TOI-144.01}

HD 39091 c ($\pi$ Men c, TOI-144.01) is a $\sim$2 R$_{\oplus}$ planet orbiting a naked-eye ($V=5.65$, $d=18.27$ pc) star in a 6.27 day orbit. The planet was detected in TESS Sector 1 data, and was the first published TESS planet \citep{Huang2018,Gandolfi2018}; the G0V star was also previously revealed with RV observations to host a long-period, eccentric sub-stellar companion \citep[HD 39091 b,][]{Jones2002}. The authors of both papers used archival RV data from the Anglo-Australian Telescope (AAT) University College London Echelle Spectrograph (UCLES) and HARPS (both pre- and post-fiber upgrade) to derive the mass of the inner TESS-detected planet, finding that the 6.27-d period signal was fit with an RV semi-amplitude $K$ of 1.55$\pm0.27$ \citep{Huang2018} and 1.58$^{+0.26}_{-0.28}$ m~s$^{-1}$ \citep{Gandolfi2018}. Both papers fixed the orbital eccentricity of HD 39091 c to 0, and included separate systemic velocity and white noise (jitter) terms for the three data sets (UCLES, HARPS-pre, HARPS-post), but no correlated noise model. Given the slightly different stellar masses derived by the different authors (1.094$\pm$0.039 M$_{\odot}$ by \citealt{Huang2018} and 1.02$\pm$0.03 by M$_{\odot}$ \citealt{Gandolfi2018}), the derived masses for HD 39091 c were 4.82$^{+0.84}_{-0.86}$ and 4.52$\pm$0.81 M$_{\oplus}$, respectively. An updated analysis incorporating new TESS observations, previously-unreleased CORALIE RV observations, and new VLT/ESPRESSO radial velocity observations (with typical nightly binned uncertainties of 10 cm~s$^{-1}$) result in a $K$ value for HD 39091 c of 1.5$\pm$0.2 m~s$^{-1}$, again using a white noise model and assuming $e=0$ \citep{Damasso2020}. The author's analysis of their ESPRESSO spectra result in a derived M$_{*} = 1.07\pm0.04$ M$_{\odot}$ and $M_{\rm{p}} = 4.3\pm0.7$ M$_{\oplus}$. Both \cite{Damasso2020} and \cite{Xuan&Wyatt2020} find evidence for a large mutual inclination between HD 39091 b and c by combining the RVs with astrometry from \textit{Hipparcos} and \textit{Gaia} DR2, suggestive of a significant dynamical evolution history in the system. 

The brightness of the host star makes HD 39091 c still one of the best small planet atmospheric targets (not accounting for saturation issues), despite the larger size of the host star relative to other planets with predicted strong atmospheric signals. In addition, given the differences in stellar activity over time between G and M dwarfs, the atmospheric properties of the planet could be different from similar planets around lower mass stars; the planet also sits near the radius gap of small, close-in planets \citep{Fulton2017,vaneylen2018} so could have an atmosphere in transition. \cite{Garcia_Munoz2020} used HST/STIS HI Ly$\alpha$ observations to search for hydrogen from photodissociating H$_2$ or H$_2$O, but could only place an upper limit. \cite{Garcia_Munoz2020} and later \cite{Shaikhislamov2020} suggest that this non-detection could be due to a high mean-molecular weight (heavier than H) atmosphere in HD 39091 c.

    \subsection{TOI-174.01, 174.02, \& 174.03}
    
HD 23472 (TOI-174) is a fairly bright ($V=9.7$, $d=39.08$ pc \citealt{Gaia:2018}) mid-K dwarf of somewhat low [Fe/H] \citep[$\sim -0.2$,][]{Sousa2011}, around which TESS detected multiple planet candidates in sectors 1-4: TOI-174.01 with $R_{\rm{p}}= 1.89~R_{\oplus}$ and $P=17.67$ d, TOI-174.02 with $R_{\rm{p}} = 1.73 R_{\oplus}$ and $P=29.80$ d, TOI-174.03 with $R_{\rm{p}} = 1.25~R_{\oplus}$ and $P=12.61$ d, and TOI-174.04 with $R_{\rm{p}} = 0.93~R_{\oplus}$ and $P=3.98$ d. We do not detect the latter TOI in our TESS light curve analysis (\S\ref{sec:radii}), so it is not considered in our ranking. \cite{Trifonov2019} reprocessed 14 archival HARPS RVs, taken between February 2004 and February 2013, with the SpEctrum Radial Velocity AnaLyser \citep[SERVAL,][]{Zechmeister2018} pipeline and modeled the first two TOI-174 planets. The authors included the planet periods and phases as free parameters in their modeling to allow for small deviations of the orbit over the 14 year span of RV observations. Indeed, they find the exact period estimates from TESS give a poor fit, but also caution that the small amount of RV makes it challenging to derive firm constraints on the planet masses. Nonetheless, they report $K$ values (assuming circular orbits) of 5.33$^{+0.67}_{-4.20}$ m~s$^{-1}$ and 4.29$^{+0.26}_{-3.44}$ m~s$^{-1}$ for TOI-174.01 and 174.02, respectively, resulting in masses of 17.92$^{+1.41}_{-14.00}$ and 17.18$^{+1.07}_{-13.77}$ M$_{\oplus}$ assuming M$_*$ = 0.75$^{+0.04}_{-0.03}$ M$_{\odot}$. No further papers validating the TOI-174 planet candidates as true planets have been published, although there is a paper in preparation (Burt et al.). TFOP SG1 photometry at the TESS ephemeris timing of TOI-174.01 and TOI-174.02 (Table \ref{table:SG1-observations}) rules out all neighboring stars bright enough to cause the transit seen in the TESS data; one observation of TOI-174.01 shows a tentative on-source detection of a $\sim$0.6 ppt transit.

    \subsection{TOI-175.01 \& 175.02}   
    
L 98-59 (TOI-174) is a nearby M3 dwarf ($V=11.685$, $d=10.6$ pc) around which TESS first detected three small planet candidates in Sector 2 (and subsequently observed in sectors 5, 8, \& 9-12): TOI-175.01 with a period of 3.69 d and a radius of $1.375~R_{\oplus}$, TOI-175.02 with a period 7.45 d and a radius of $1.476~R_{\oplus}$, and TOI-175.03 with a period of 2.25 d and a radius of $0.861~R_{\oplus}$. Given the small expected signal $K$ of TOI-175.03, it is not highly-ranked by our metric and thus not included in our sample. \cite{Kostov2019} presented the discovery of the system via a thorough analysis of the TESS photometry and extensive suite of follow-up observations to confirm the host star and planet properties of this TOI. Following that work, \cite{Cloutier2019} set out to measure the masses of the planets and acquired 164 HARPS spectra of L 98-59 between October 2018 and April 2019. They used a semi-parametric Gaussian process (GP) regression model (with a quasi-periodic kernel) of the RV stellar activity, as measured through H$\alpha$, and simultaneously modeled the activity and RV planetary signals. The authors only detect the outer two planets with robust $K>0$ m~s$^{-1}$ values, with $K=2.21\pm0.28$ m~s$^{-1}$ for the 3.69 d planet and $K=1.67\pm0.31$ m~s$^{-1}$ for the 7.45 d planet; the inner-most planet at 2.25 d has a $K$ upper limits of 1.06 m~s$^{-1}$. Assuming a $M_* = 0.312\pm0.031$, this results in mass derivations of 2.42$^{+0.35}_{-0.34}$, 2.31$^{+0.46}_{-0.45}$, and $<1.01$ M$_{\oplus}$ for TOI-175.01, .02, and .03, respectively. The authors also constrain the planet's orbital eccentricity values through a stability analysis to be $<$0.19, $<$0.31, and $<$0.53 for TOI-175.01, .02, and 0.03, respectively. This analysis confirms the TOI-175 system is similar to other compact systems with low eccentricities. Overall, TOI-175 is an interesting laboratory for studying terrestrial planets with the host star parameters held constant and is a prime target for transmission spectroscopy with JWST (TSM values between $\sim$15 and 210; \citealt{Cloutier2019}).

    \subsection{TOI-186.01}
    
HD 21749 b (TOI-186.01) is a 2.85$\pm$0.2 R$_{\oplus}$ planet orbiting at 35 days around a bright ($V=8.1$) K dwarf detected by TESS in sectors 2, 3, and 4 \citep{Gan2021}. The planet was originally alerted as a single-transit candidate. The first publication of this planet was by \cite{Trifonov2019}, who used archival RV data from HARPS to ``Doppler validate'' the planet, finding a period of 35.6 d, a moderately eccentric orbit with $e=0.33$, and $K=7.6$ m~s$^{-1}$, resulting in a derived planet mass of 30.63$^{+2.63}_{-2.67}$ M$_{\oplus}$ (assuming M$_{*}=0.76^{+0.03}_{-0.02}$ M$_{\odot}$). \cite{Dragomir2019} were the first to present an analysis of the TESS data on this system. After being alerted to the single transit in sector 2, the authors analyzed archival HARPS RVs of the star, which showed a clear signal at 35.57 days. Additional TESS observations in sectors 3 and 4 confirmed the RV signal was indeed TOI-186.01. Furthermore, the fully-combined (sectors 1-4) TESS observations revealed a second planet candidate with a period of 7.9 days (TOI-186.02) and a radius of 0.892$^{+0.064}_{-0.058}$ R$_{\oplus}$. \cite{Dragomir2019} combined the archival HARPS RVs with archival and new PFS RVs (sporadic cadence from January 2010 to October 2018, and then high cadence in December 2018) to measure  $K= 1.51^{+0.41}_{-0.33}$ m~s$^{-1}$ for HD 21749 b, resulting in a derived mass $M_{\rm{p}} = 22.7^{+2.2}_{-1.9}$ M$_{\oplus}$ (assuming M$_{*}=0.73\pm0.07$ M$_{\odot}$). The authors used a white noise model (jitter term) only for the RVs. They were also able to put an upper limit on the mass of HD 21749 c of 3.70 M$_{\oplus}$ (from $K < 1.43$ m~s$^{-1}$).

However, HD 21749 is a particularly challenging system because the TESS photometry as well as KELT photometry (published in \citealt{Dragomir2019}) indicate a stellar rotation period at almost the same value as the outer planet's orbit period. Stellar activity like that manifesting as stellar rotation signals can induce RV variations that interfere with precise planet mass measurements. This motivated \cite{Gan2021} to reanalyze the observations included in \citeauthor{Dragomir2019}, and also 53 new PFS RV epochs and new Evryscope-South photometry, to determine to what extent the stellar rotation signal from HD 21749 was influencing the derived planet masses. The authors incorporated a Gaussian process in their model to fit the RV data, with hyperparameter priors informed by the Gaussian process model (with the same kernel) applied to the TESS photometry. The GP+Keplerian fit results in $K=4.92\pm0.66$ m~s$^{-1}$ and $e=0.177\pm0.065$, which combined with the assumed stellar mass of 0.73$\pm$0.07 M$_{\odot}$ results in a planet mass of 20.2$\pm$2.7 M$_{\oplus}$. The authors also updated the radius of HD 21749 b with the new TICv8 dilution factor, which was not available at the time \cite{Dragomir2019} was published.

    \subsection{TOI-260.01} 

HIP 1532 (TOI-260) is a nearby ($V=9.9$, $d=20.197$ pc, \citealt{Gaia:2018}) early M star observed by TESS in Sector 3. The TESS SPOC pipeline detected one transiting planet candidate, TOI-260.01 with $R_{\rm{p}}$ = 1.60 R$_{\oplus}$ and $P=13.47$ days. This star was monitored as part of the California Planet Search program and Lick-Carnegie Exoplanet Survey \citep{Isaacson&Fischer2010,Butler2017}, and as part of the Magellan Planet Search Program \citep[when it was using Magellan II/MIKE and PFS][]{arriagada2011}, but no prior planet detections have been published. There has been no publication validating this TOI as a true planet.  TFOP SG1 photometric observations (Table \ref{table:SG1-observations}) have ruled out any nearby stars bright enough to be the source of the 0.7 ppt transit seen in the TESS data but have not conclusively detected the transit on-source, suggesting that the SPOC ephemeris may be slightly off. Two papers in prep further validate this planet, Beard et al. and Lavie et al.

    \subsection{TOI-402.01 \& 402.02}

HD 15337 (TOI-402) was announced as the host of two transiting planets, TOI-402.01 and .02, in January 2019 based on TESS sector 3 and 4 data. The host star is a bright ($V=9.1$, $d=44.86$ pc \citealt{Gaia:2018}) K1 dwarf star with solar metallicity. The system was validated in two separate publications, \cite{Dumusque2019} and \cite{Gandolfi2019}, which both made use of HARPS radial velocity measurements. In \cite{Dumusque2019}, their analysis of the TESS light curves results in radii and periods of $R_{\rm{p}}= 1.699^{+0.062}_{-0.059}~R_{\oplus}$ and $P=4.756$ d and $R_{\rm{p}}= 2.522^{+0.106}_{-0.102}~R_{\oplus}$ and $P=17.178$ d for TOI-402.01 and 402.02, respectively, assuming R$_{*}=0.839^{+0.018}_{-0.016}$ R$_{\odot}$. The authors re-reduced 85 HARPS spectra using the latest Data Reduction Software (DRS) that includes a better wavelength solution, which results in significant differences in the data gathered after the change of HARPS fibers in June 2015. These authors model the RVs using a Gaussian process model with a squared exponential kernel to account for stellar activity, after examining the log($R'_{\rm{HK}}$) measurements for the star. \cite{Dumusque2019} find through their analyses of the HARPS RVs $K$ semi-amplitudes of 3.1079$\pm$0.3515 m~s$^{-1}$ and 2.4819$\pm$0.4679 m~s$^{-1}$ for TOI-402.01 and 402.02, respectively, which translate into masses of 7.20$\pm$0.81 $M_{\oplus}$ and 8.79$\pm$1.68 $M_{\oplus}$, assuming M$_{*}=0.851^{+0.042}_{-0.034}$ M$_{\odot}$. \cite{Dumusque2019} also find non-zero eccentricities of the planets to be $e_1=0.17\pm0.09$ and $e_2=0.19\pm0.10$. 

\cite{Gandolfi2019} use the same HARPS data but a slightly different approach to deriving the masses of the TOI-402 planets. They rely on a joint analysis of the TESS light curves and RV data, and include in their RV model a coherent sine-like curve with a period prior constrained by the HARPS FWHM measurements of the cross-correlation function (CCF) and uniform priors on the phase and amplitude. The authors tested swapping out the coherent signal with a GP model with a quasi-periodic kernel as described by \cite{Rajpaul2015}, and found similar results. \cite{Gandolfi2019} report $K$ semi-amplitudes of $3.08^{+0.44}_{-0.41}$ m~s$^{-1}$ and 2.16$^{+0.48}_{-0.45}$ m~s$^{-1}$ for TOI-402.01 and 402.02, respectively, which translate into masses of $7.51^{+1.09}_{-1.01}~ M_{\oplus}$ and 8.11$^{+1.82}_{-1.69} ~M_{\oplus}$, assuming M$_{*}=0.90\pm0.03$ M$_{\odot}$. \cite{Gandolfi2019} also find a possible non-zero eccentricity of TOI-402.01, $e_1=0.09\pm0.05$, but that the eccentricity of TOI-402.02 is consistent with 0, $e_2=0.05^{+0.06}_{-0.04}$. Their analysis of the TESS light curves results in radii and periods of $R_{\rm{p}}= 1.64\pm0.06 ~R_{\oplus}$ and $P=4.756$ d and $R_{\rm{p}}= 2.39\pm0.12 ~R_{\oplus}$ and $P=17.178$ d for TOI-402.01 and 402.02, respectively, assuming R$_{*}=0.856\pm0.017$ R$_{\odot}$.

The TOI-402 system will be particularly interesting for further studies of atmospheric composition. Based on its density and consideration of likely atmospheric evolution in the presence of stellar irradiation, TOI-402.01 should not have a H/He atmosphere, but could have a secondary atmosphere, whereas the less irradiated TOI-402.02 should have a primordial H/He atmosphere \citep{Dumusque2019}. Placing limits on the current atmospheric compositions of these planets will help elucidate their past evolution and will represent comparative planetology across the radius gap.

  \subsection{TOI-431.02}

HIP 26013 (TOI-431) is a nearby ($V=9.13$, $\sim$32.6 pc \citealt{Gaia:2018}) K3 dwarf of moderate activity \citep{Gray2006} and solar metallicity \citep{DelgadoMena2015} around which TESS detected two transiting planets, TOI-431.01 at $R_p = 3.61831\pm0.230095$~R$_{\oplus}$ and $P=12.45941\pm0.0008$ d and TOI-431.02 at $R_p =1.419494\pm0.697035$~R$_{\oplus}$ and $P=0.490058\pm0.00001$ d. Due to the radius of TOI-431.01, it is not included in our population analysis but we analyze the entire system for mass constraints and provide those results here. There are not yet any publications validating this TOI as a true planet, but Osborn et al. (submitted) will present observations that both validate the planets in this system, and provide additional RV data with which they will derive updated mass estimates.

    \subsection{TOI-455.01}
    
    GJ 3193 b (TOI-455.01, LTT 1445Ab) is 1.38$^{+0.13}_{-0.12}$ R$_{\oplus}$ planet orbiting at 5.36 days around one mid-M dwarf (A) in a hierarchical triple composed of three mid-to-late M dwarfs (B and C also orbit each other) observed by TESS in Sector 4. The discovery of this system was published in \cite{Winters2019}, and the authors also presented a detailed characterization of the very nearby (6.9 pc) host star system using high-resolution astrometry from the Fourth Interferometric Catalog, speckle imaging data from WIYN/DSSI, high-resolution spectroscopy from FLWO/TRES, along with MEarth and TESS photometry. This characterization was especially important in this system in which all three stars were blended in one  21\arcsec square TESS pixel. Using archival and new HARPS data, the authors place a 3$\sigma$ upper limit on the planet's mass of 8.4 M$_{\oplus}$. GJ 3193 b is the nearest planet known to transit an M dwarf and a prime target for atmosphere studies due to the relatively large transit depth (0.2\%) and bright host star ($V=11.22$, $K=6.50$); the blended BC pair is also a potentially valuable calibration source with the same spectral type as the host star. Winters et al. (in prep) present an updated analysis of the system with new RV data from multiple instruments, including the PFS data analyzed in this work, from which they derive a precise mass measurement for GJ 3193 b.

    \subsection{TOI-562.01}
    
    GJ 357 b (TOI-562.01) is a 1.217$\pm$0.085 R$_{\oplus}$, 3.93 d planet orbiting a nearby ($d=9.4$ pc) M2.5V star, detected by TESS in Sector 8 and published by \cite{Luque2019}. The authors combined RV data from HIRES, HARPS, PFS, and CARMENES to measure the mass of the planet to be 1.84$\pm$0.31 M$_{\oplus}$. They also performed a search via generalized Lomb-Scargle peridogram and found significant evidence for additional Keplerian signals, which they tested with more extensive modeling of the RVs. The model that best explains the data has three planets at 3.93, 9.1, and 55.7 days, and a Gaussian process component for correlated noise (using an exponential kernel). The minimum masses of the two outer planets are 3.40$\pm$0.45 M$_{\oplus}$ (9.1 days, GJ 357 c) and 6.1$\pm$1.0 M$_{\oplus}$ (55.7 days; GJ 357 d), and they do not show transits in the TESS data. The authors further ruled out possible transits of GJ 357 c by comparing a transiting vs. non-transiting model for the planet, finding the non-transiting model preferred. However, based on the predicted transit epoch from the RV fits, in its 28-day observing window TESS would have missed the transit of GJ 357 d. The \textit{a priori} transit probability of GJ 357 d is only 0.8\%, but the authors note that further RVs could improve the transit predictions and allow for a more precise transit search. The authors were also able to determine the host star's rotation period $P_{\rm{rot}} = 77.8^{+2.1}_{-2.0}$ by fitting photometry from ASAS, NSVS, ASAS-SN, and WASP with a quasi-periodic Gaussian process model. Overall, GJ 357 b is one of the best targets for atmospheric characterization with both JWST and ground-based ELTs.

    \subsection{TOI-652.01}

HD 86226 c (TOI-652.01) is a short-period sub-Neptune planet orbiting a bright ($V=7.9$) solar-type host star detected by TESS in Sector 9, and published by \cite{Teske2020}. From the TESS data, the authors derived a radius $R{\rm{p}} = 2.16\pm 0.08$ R$_{\oplus}$ and period $P=3.98$ days for HD 86226 c. From CORALIE, MIKE, and PFS RV data the authors derived a mass for HD 86226 c of $M_{\rm{p}}= 7.25^{+1.19}_{-1.12}$ $M_{\oplus}$, and fit for $e$ but found it consistent with zero ($e=0.075^{+0.065}_{-0.048}$). The MIKE and most of the CORALIE RV observations were used previously to detect a long-period, eccentric giant planet \citep[HD 86226 b;][]{arriagada2010,marmier2013}, the parameters of which Teske et al. updated to $M_{\rm{p}}$~sin$i = 0.45^{+0.04}_{-0.05}$ $M_{\rm{Jup}}$, $P=1628^{+22}_{-21}$, and $e=0.059^{+0.062}_{-0.039}$, decreasing both the mass and eccentricity significantly from Marmier et al. HD 86226 c fits the prediction of \citet{zhu&wu2018} and \citet{bryan2019} that long-period giant planets are very likely to have short-period small planet companions, and its location close to but not within the hot Neptune desert makes it an interesting target for better understanding the processes that influence small exoplanet sizes.

   \subsection{TOI-784.01} 
    HD 307842 is a $V=9.33$ solar-type mid-G star observed by TESS in Sectors 10 and 11. The TESS SPOC pipeline detected one transiting planet candidate, TOI-784.01 with $R_{\rm{p}} = 1.86$ R$_{\oplus}$ and $P=2.80$ days. There have been no publications validating this TOI as a true planet. TFOP SG1 photometry  (Table \ref{table:SG1-observations}) rules out all neighboring stars bright enough to cause the transit seen in the TESS data and shows a tentative on-source detection of the 0.4 ppt transit.

    \subsection{TOI-836.01 \& 836.02}

HIP 73427 (TOI-836) is a fairly bright ($V=9.9$, $d=27.5$ pc, \citealt{Gaia:2018}) late K dwarf star observed by TESS in Sector 11. The TESS SPOC pipeline \citep{Jenkins:2016} detected two transiting planet candidates, TOI-836.01 with $R_{\rm{p}} = 2.52 ~R_{\oplus}$ and $P=8.59$ days and TOI-836.02 with $R_{\rm{p}} = 1.76~R_{\oplus}$ and $P=3.81$ days. This star was monitored as part of the California Planet Search program and Lick-Carnegie Exoplanet Survey \citep{Isaacson&Fischer2010,Butler2017}, and as part of the Magellan Planet Search Program \citep[when it was using Magellan II/MIKE;][]{arriagada2011}, but no prior planet detections have been published. There have been no publications validating these TOIs as true planets. TFOP SG1 photometric observations (Table \ref{table:SG1-observations}) have ruled out blends from neighboring stars as the source of the TESS transit signal. Transits of both TOIs have been detected in SG1 observations in multiple bands.  There are some suggestions of TTVs in the SG1 data.

    \subsection{TOI-1233.01, 1233.03, \& 1233.04}

HD 108236 (TOI-1233) hosts four exoplanets detected by TESS in sectors 10 and 11, with periods of 3.8, 6.2, 14.2, and 19.6 days and radii of 1.586$\pm$0.098, 2.068$^{+0.10}_{-0.091}$, 2.72$\pm$0.11, and 3.12$^{+0.13}_{-0.12}$ R$_{\oplus}$, respectively \citep{Daylan2020}. The TOIs were detected in a different order from their periods -- TOI-1233.04 is a 3.8 days, TOI-1233.03 is at 6.2 days, TOI-1233.01 is at 14.2 days, and finally TOI-1233.02 is at 19.6 days. \cite{Daylan2020} validated the planets using photmetry from TESS, LCOGT,  MEarth-South, and WASP; reconnaissance spectroscopy from LCO/NRES and SMARTS/CHIRON; high-resolution speckle imaging from SOAR/HRCAM and Gemini/Zorro; and a subset of the precision RV data from PFS that is presented here.
A fifth planet was detected by CHEOPS high-precision photometric observations; from a combined analysis of CHEOP and TESS data, \cite{Bonfanti2021} report that this planet (TOI-1233.05) has a period of 29.5 days and a radius of 2.017$^{+0.052}_{-0.057}$ R$_{\oplus}$. The authors also update the radii of the previously-detected planets (.01 - .04) to be 2.539$^{+0.062}_{-0.065}$, 3.083$\pm$ 0.052, 2.017$\pm$0.052, and 1.615$\pm$0.051, respectively. HD 108236 is a bright (V$=9.2$, $d= 64$ pc), Sun-like star (R$_{*}=0.888\pm0.017$ R$_{\odot}$, T$_{\rm{eff}} = 5730\pm50$ K; \citealt{Daylan2020}), and the third system brighter than $V=10$ to host more than four transiting planets.  While the system does consist of closely-packed planets, none of them are in or near a strong mean motion resonance, however non-resonant (chopping) TTVs are feasible and could be used to measure the masses of these planets. No other papers have been published on the TOI-1233 system, so this work presents the first planet mass measurements. Overall, TOI-1233 stands out as one of the highest-multiplicity TESS systems yet detected, offering a controlled experiment for studying small planet evolution.

\section{TESS Light Curve Detrends \label{sec:appendix_tess_lc_detrends}}

\figsetstart
\figsetnum{11}
\figsettitle{TESS SAP, PDCSAP and detreded PDCSAP light curves of each target.}

\figsetgrpstart
\figsetgrpnum{11.1}
\figsetgrptitle{TOI-134}
\figsetplot{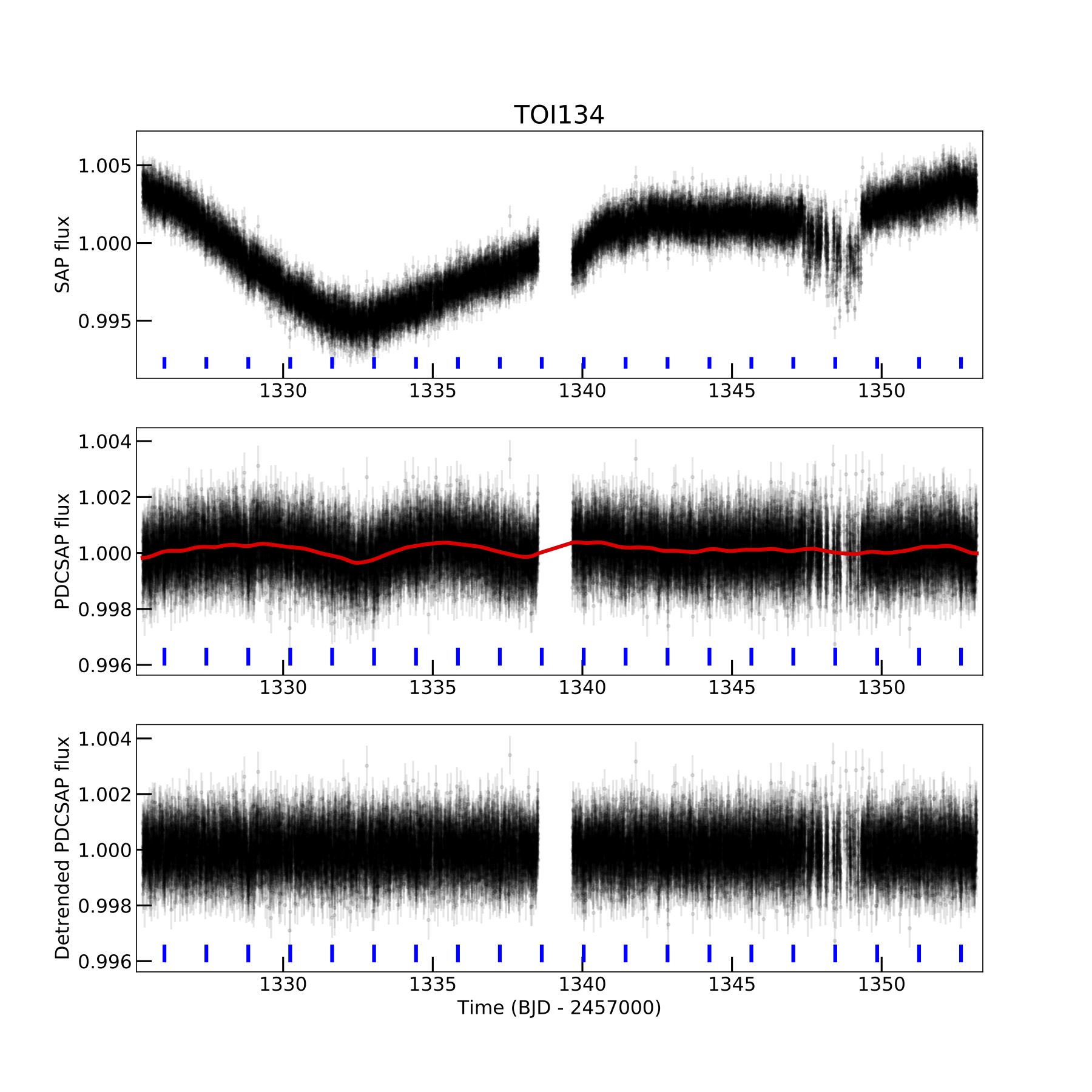}
\figsetgrpnote{The SAP, PDCSAP and detrended PDCSAP light curves of TESS. The best-fit GP model is shown as a red solid line. The vertical colored ticks represent the expected transits of each planet.}
\figsetgrpend

\figsetgrpstart
\figsetgrpnum{11.2}
\figsetgrptitle{TOI-1411}
\figsetplot{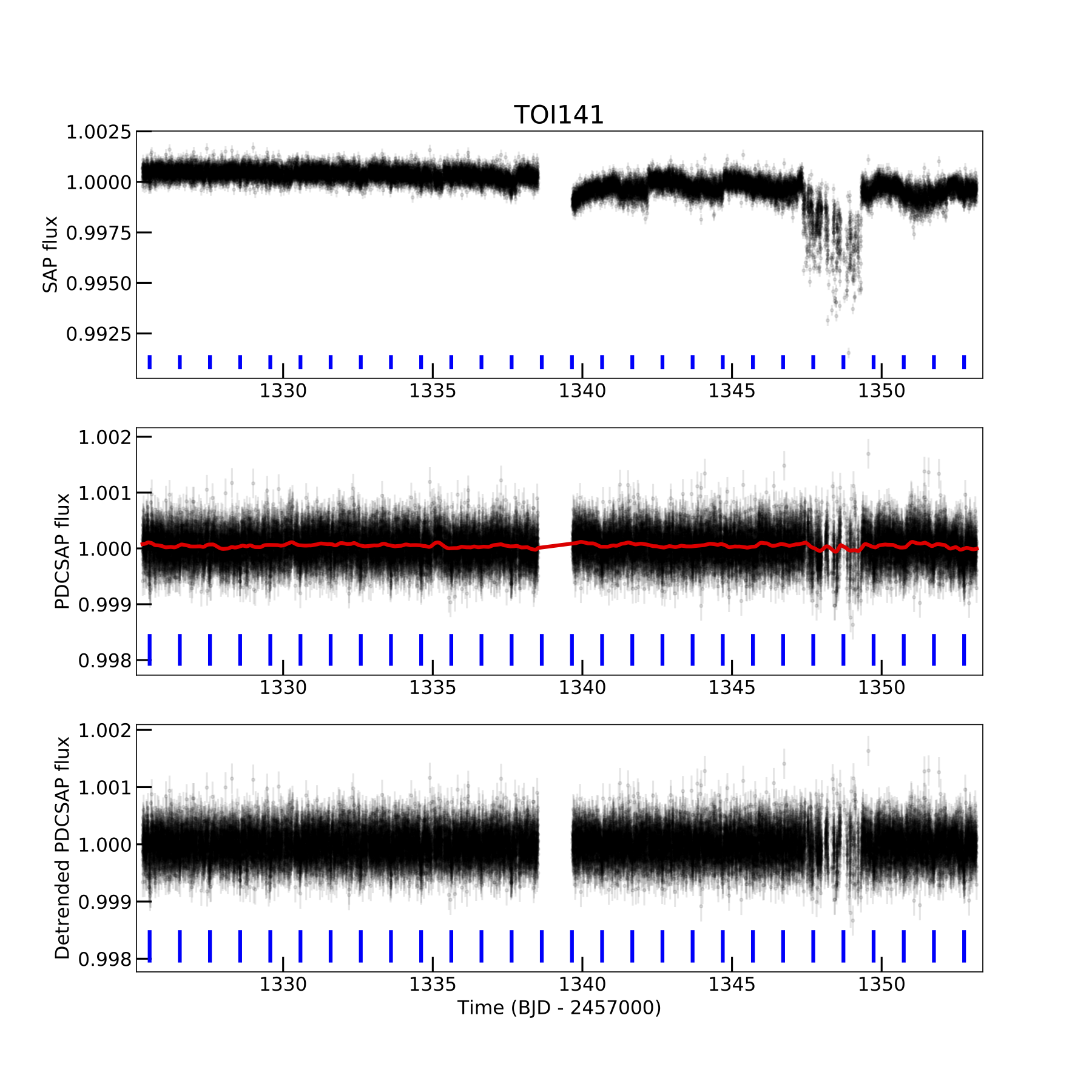}
\figsetgrpnote{The SAP, PDCSAP and detrended PDCSAP light curves of TESS. The best-fit GP model is shown as a red solid line. The vertical colored ticks represent the expected transits of each planet.}
\figsetgrpend

\figsetgrpstart
\figsetgrpnum{11.3}
\figsetgrptitle{TOI-144}
\figsetplot{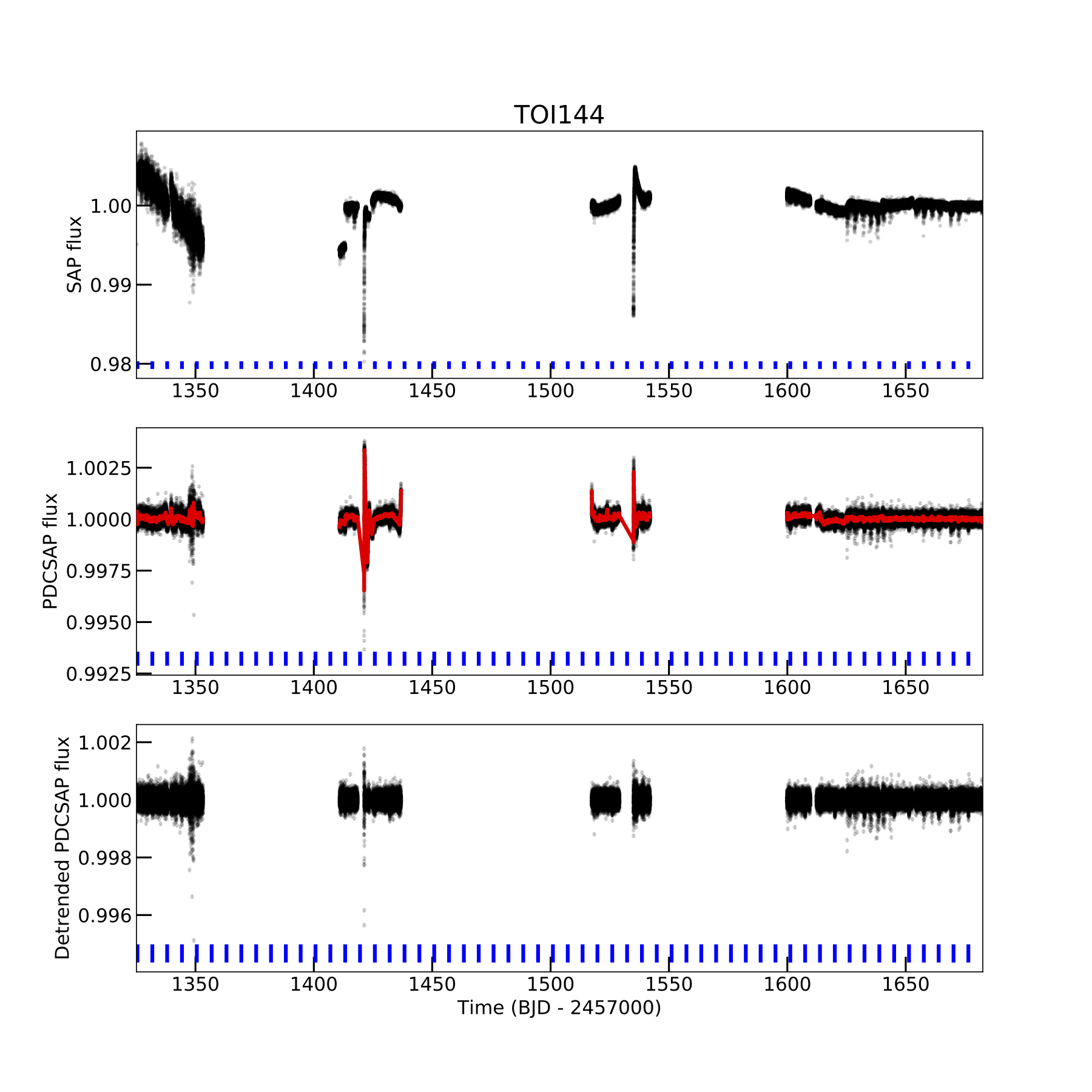}
\figsetgrpnote{The SAP, PDCSAP and detrended PDCSAP light curves of TESS. The best-fit GP model is shown as a red solid line. The vertical colored ticks represent the expected transits of each planet.}
\figsetgrpend

\figsetgrpstart
\figsetgrpnum{11.4}
\figsetgrptitle{TOI-174}
\figsetplot{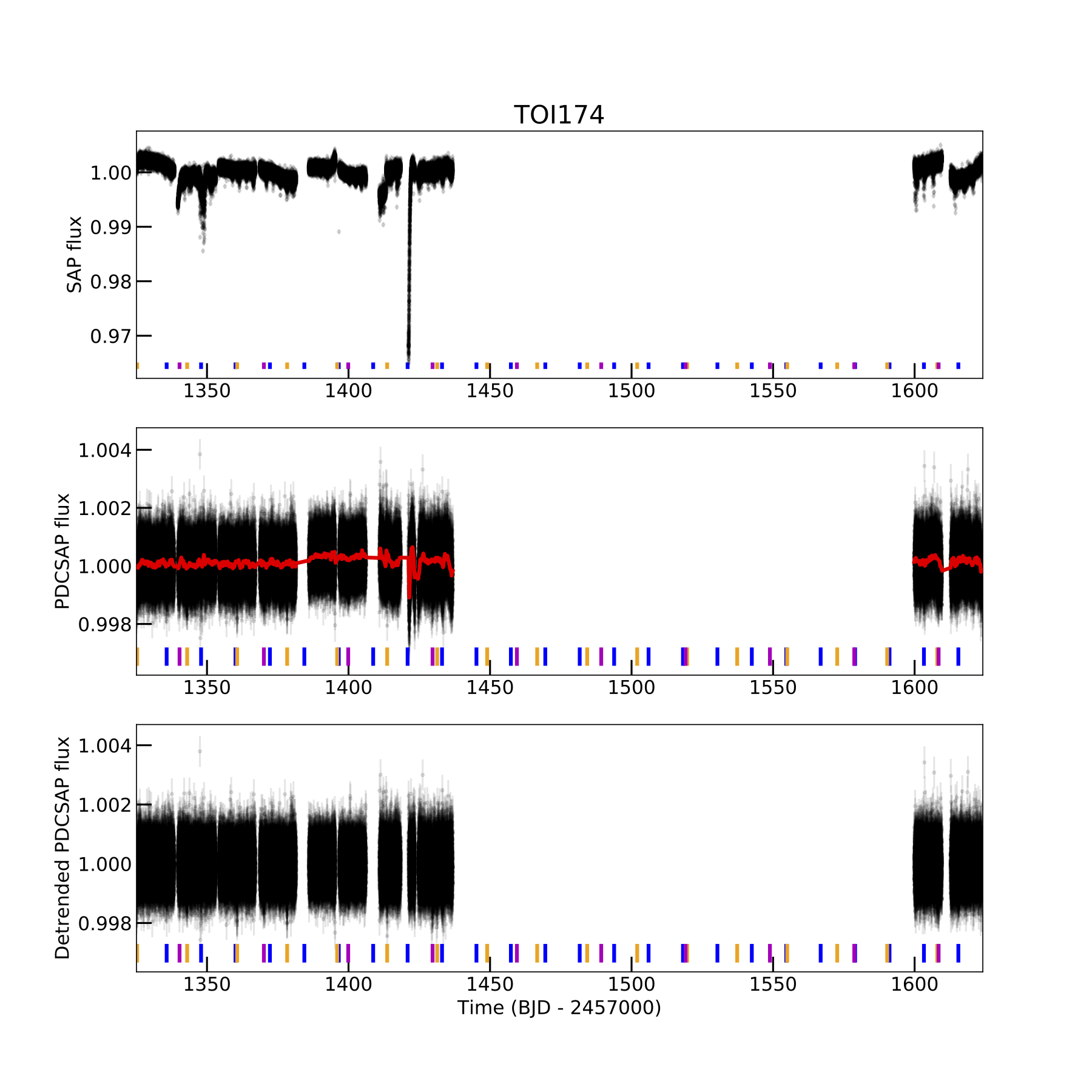}
\figsetgrpnote{The SAP, PDCSAP and detrended PDCSAP light curves of TESS. The best-fit GP model is shown as a red solid line. The vertical colored ticks represent the expected transits of each planet.}
\figsetgrpend

\figsetgrpstart
\figsetgrpnum{11.5}
\figsetgrptitle{TOI-175}
\figsetplot{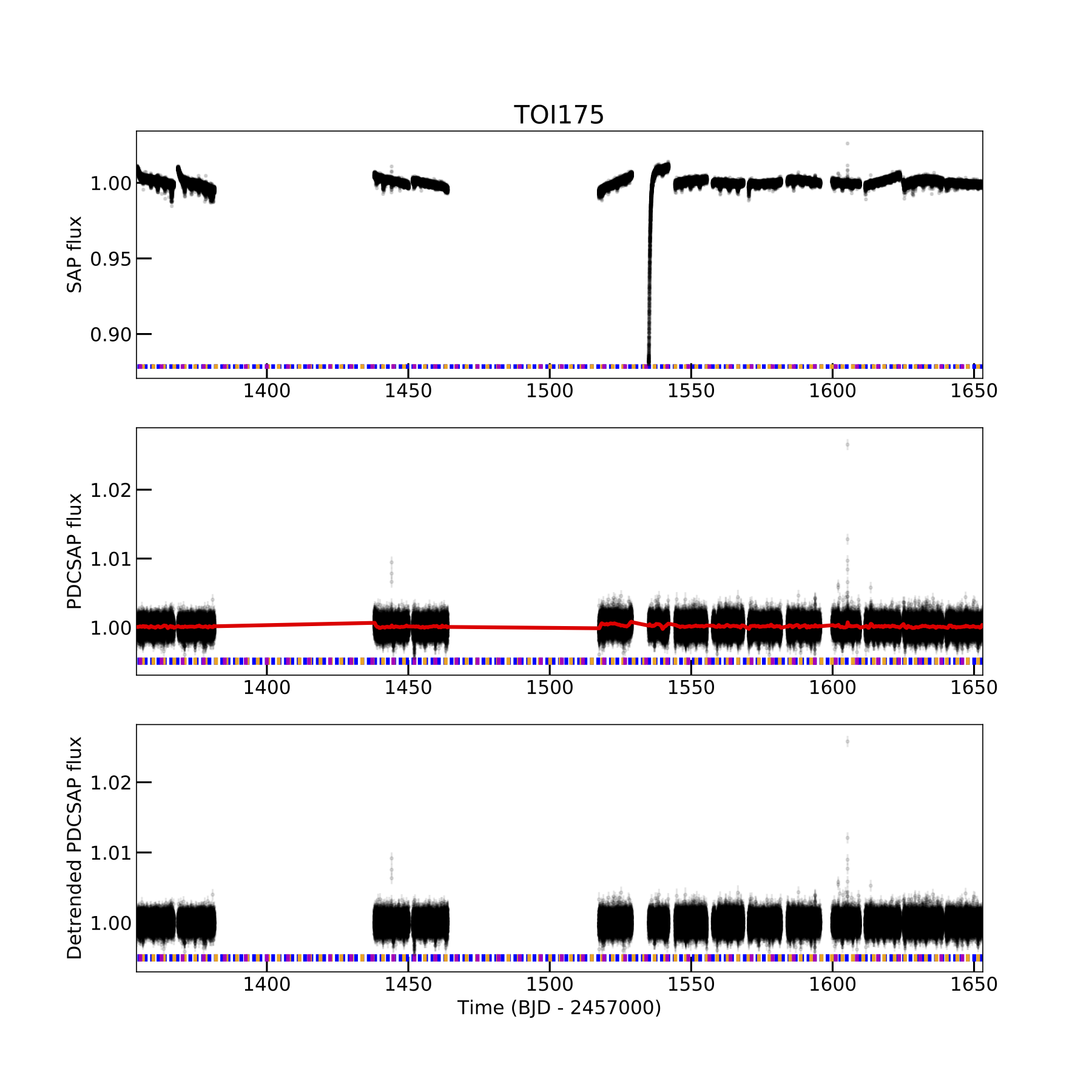}
\figsetgrpnote{The SAP, PDCSAP and detrended PDCSAP light curves of TESS. The best-fit GP model is shown as a red solid line. The vertical colored ticks represent the expected transits of each planet.}
\figsetgrpend

\figsetgrpstart
\figsetgrpnum{11.6}
\figsetgrptitle{TOI-186}
\figsetplot{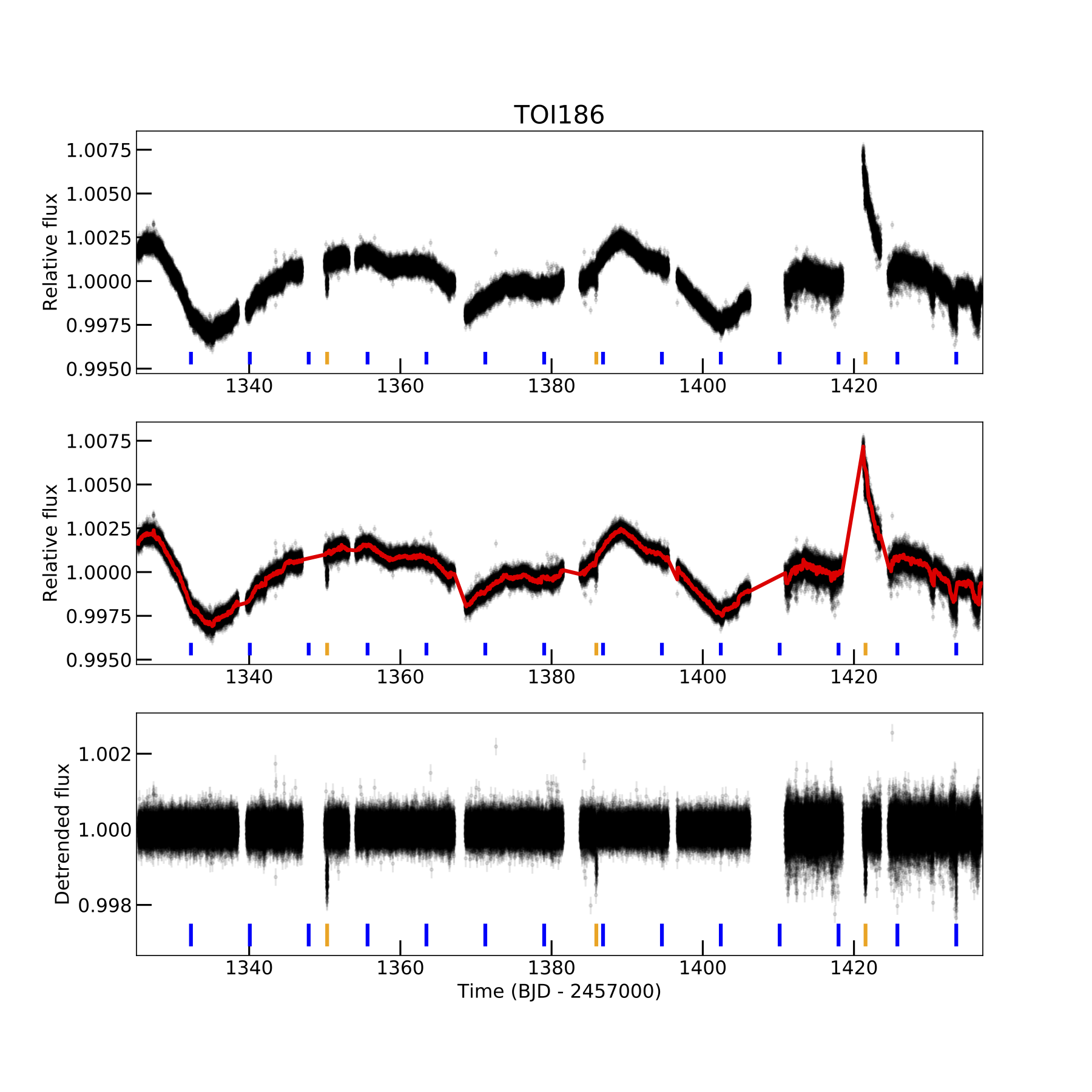}
\figsetgrpnote{The SAP, PDCSAP and detrended PDCSAP light curves of TESS. The best-fit GP model is shown as a red solid line. The vertical colored ticks represent the expected transits of each planet.}
\figsetgrpend

\figsetgrpstart
\figsetgrpnum{11.7}
\figsetgrptitle{TOI-260}
\figsetplot{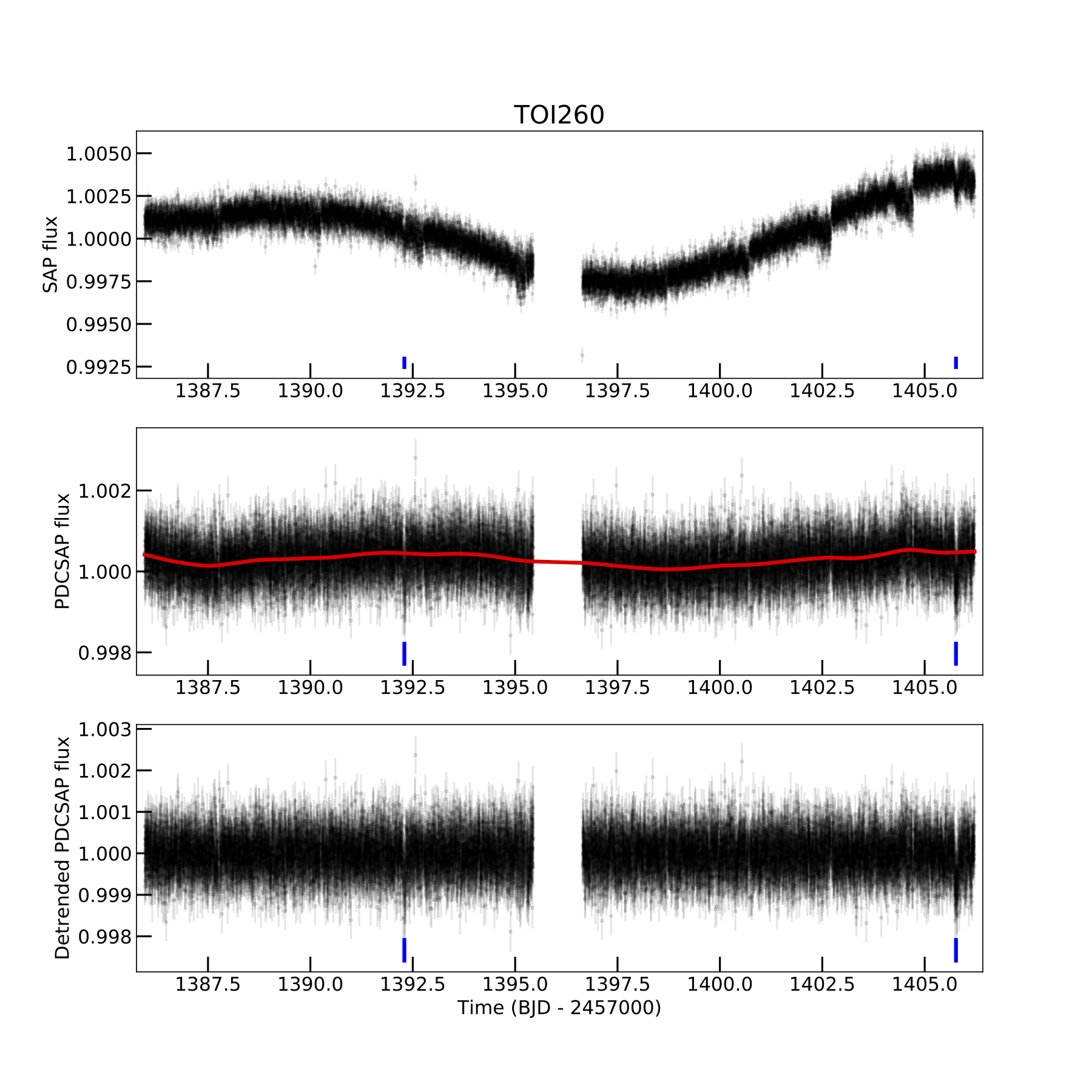}
\figsetgrpnote{The SAP, PDCSAP and detrended PDCSAP light curves of TESS. The best-fit GP model is shown as a red solid line. The vertical colored ticks represent the expected transits of each planet.}
\figsetgrpend

\figsetgrpstart
\figsetgrpnum{11.8}
\figsetgrptitle{TOI-402}
\figsetplot{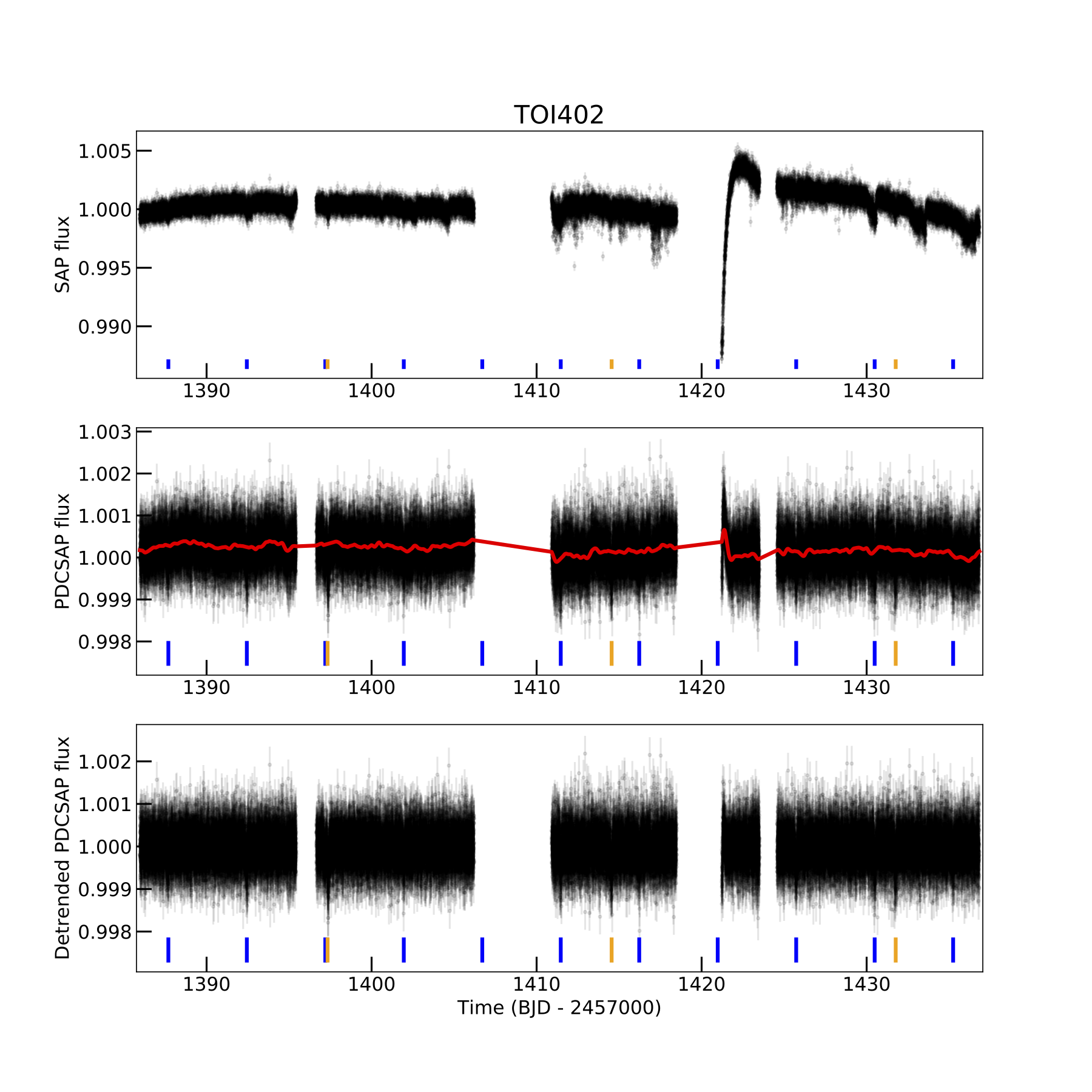}
\figsetgrpnote{The SAP, PDCSAP and detrended PDCSAP light curves of TESS. The best-fit GP model is shown as a red solid line. The vertical colored ticks represent the expected transits of each planet.}
\figsetgrpend

\figsetgrpstart
\figsetgrpnum{11.9}
\figsetgrptitle{TOI-431}
\figsetplot{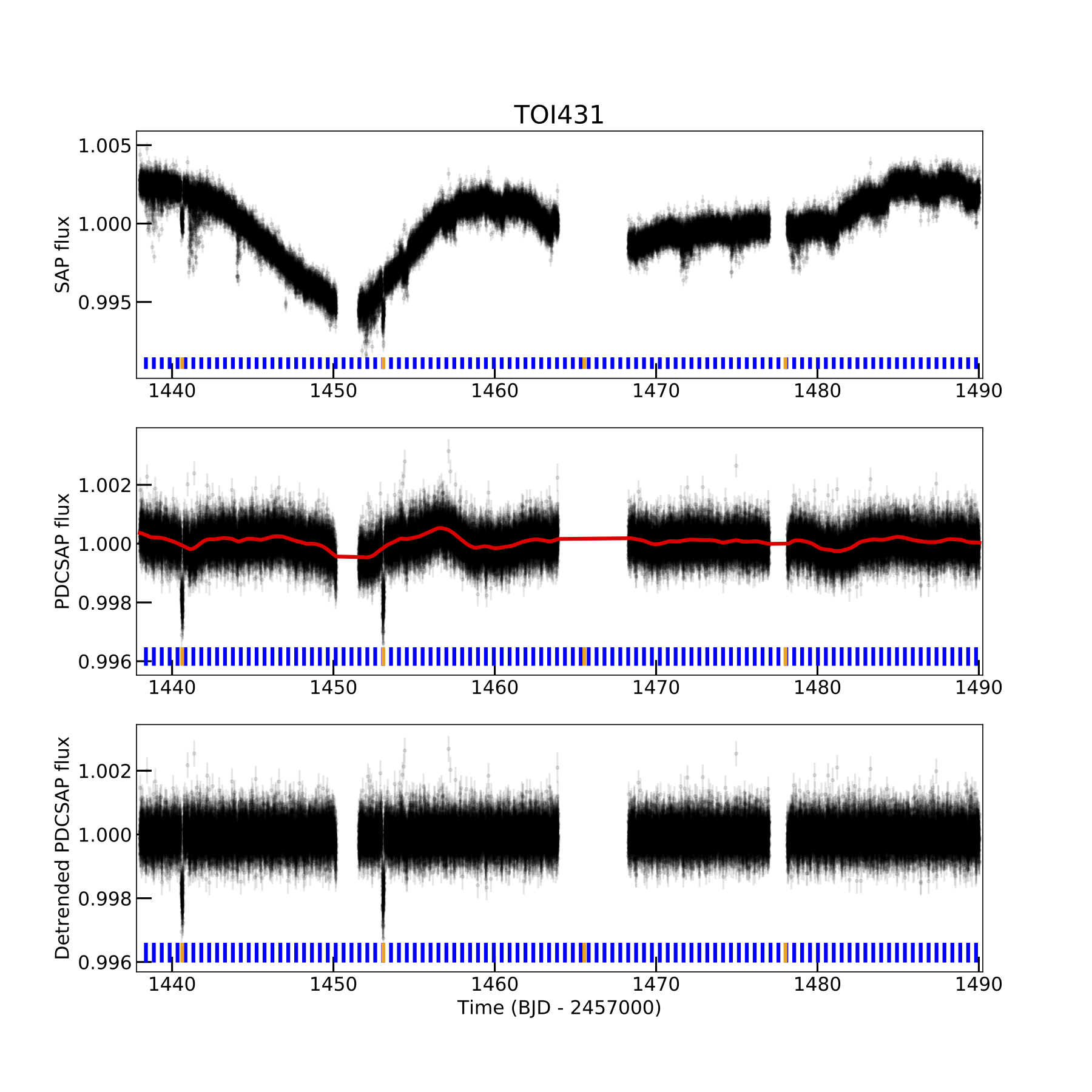}
\figsetgrpnote{The SAP, PDCSAP and detrended PDCSAP light curves of TESS. The best-fit GP model is shown as a red solid line. The vertical colored ticks represent the expected transits of each planet.}
\figsetgrpend

\figsetgrpstart
\figsetgrpnum{11.10}
\figsetgrptitle{TOI-455}
\figsetplot{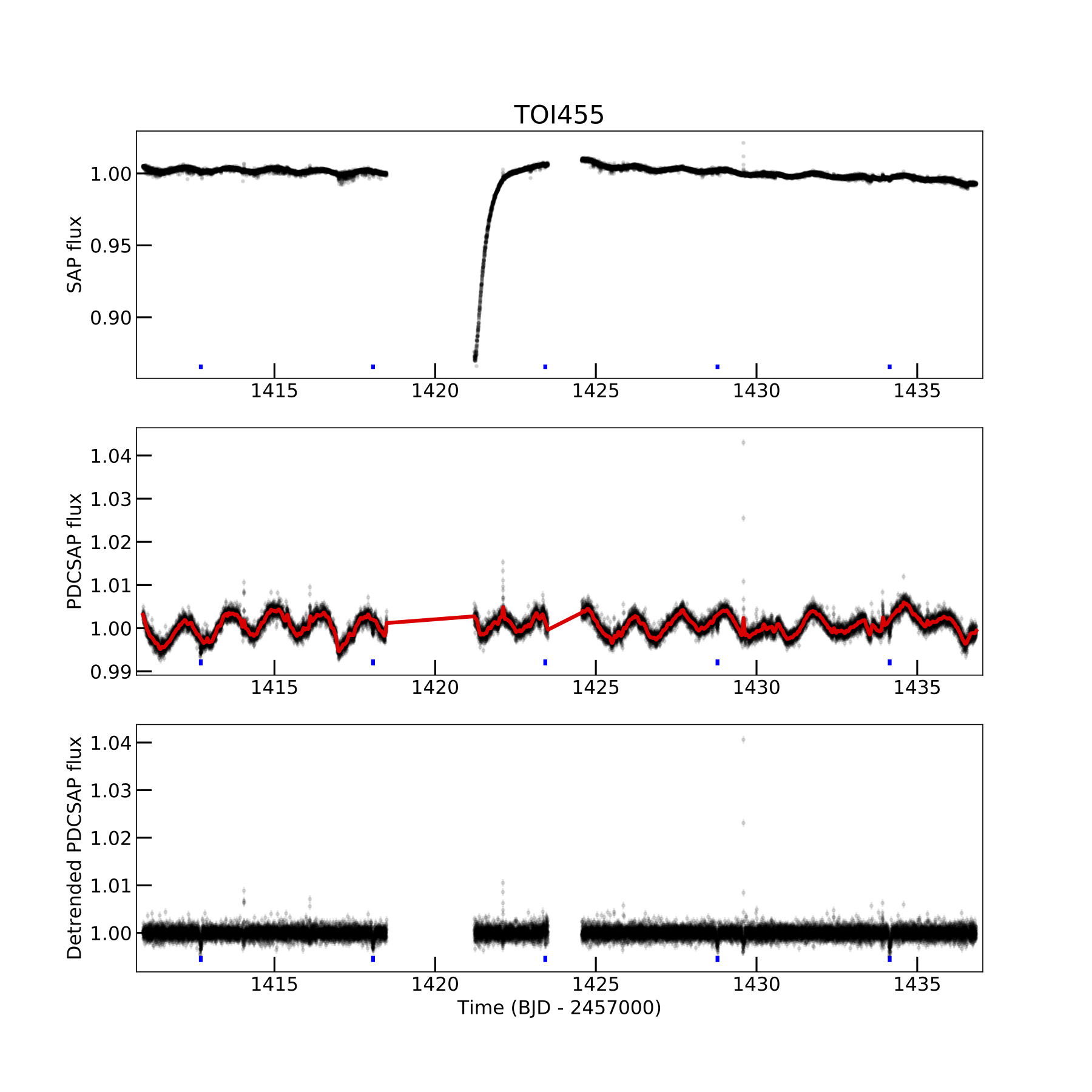}
\figsetgrpnote{The SAP, PDCSAP and detrended PDCSAP light curves of TESS. The best-fit GP model is shown as a red solid line. The vertical colored ticks represent the expected transits of each planet.}
\figsetgrpend

\figsetgrpstart
\figsetgrpnum{11.11}
\figsetgrptitle{TOI-562}
\figsetplot{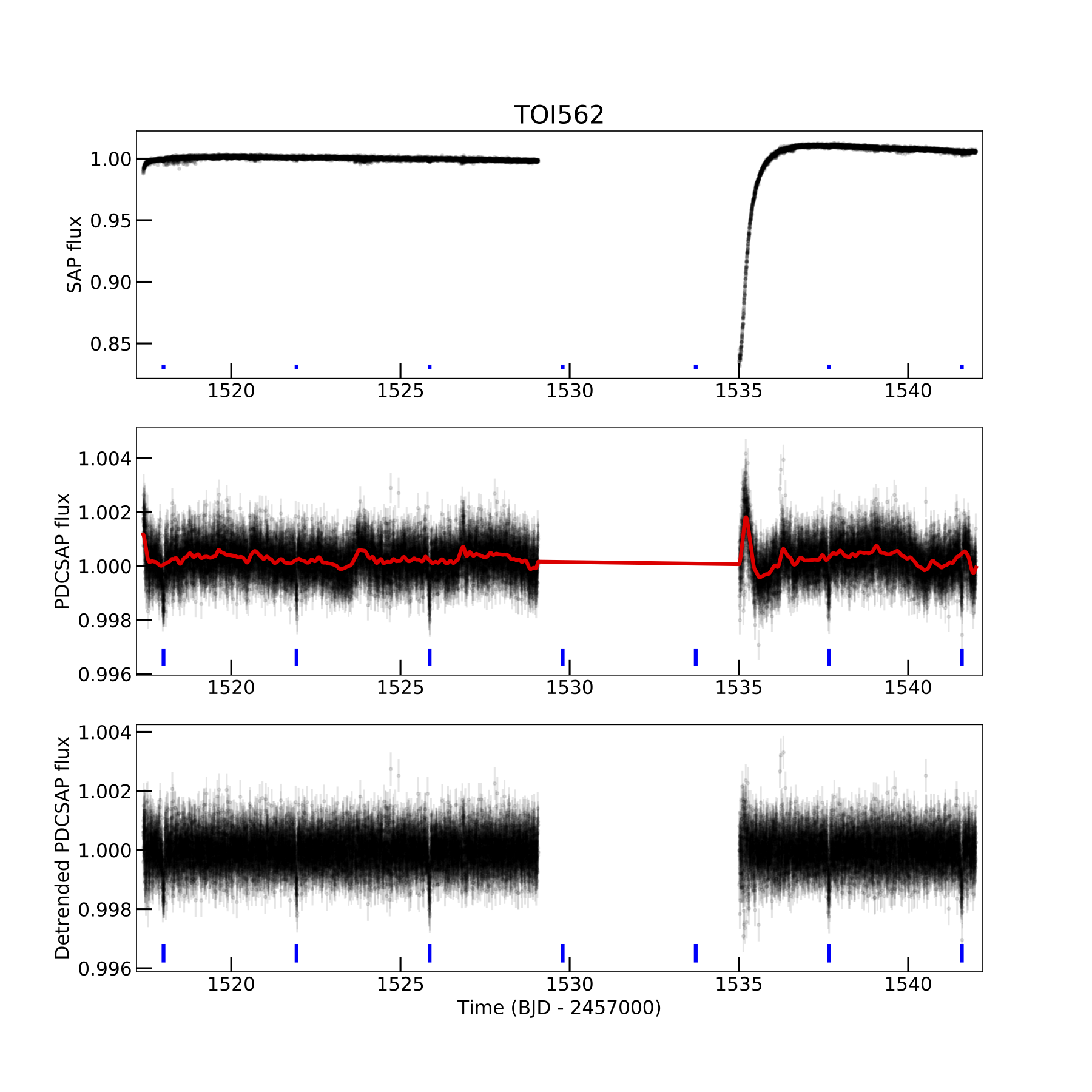}
\figsetgrpnote{The SAP, PDCSAP and detrended PDCSAP light curves of TESS. The best-fit GP model is shown as a red solid line. The vertical colored ticks represent the expected transits of each planet.}
\figsetgrpend

\figsetgrpstart
\figsetgrpnum{11.12}
\figsetgrptitle{TOI-652}
\figsetplot{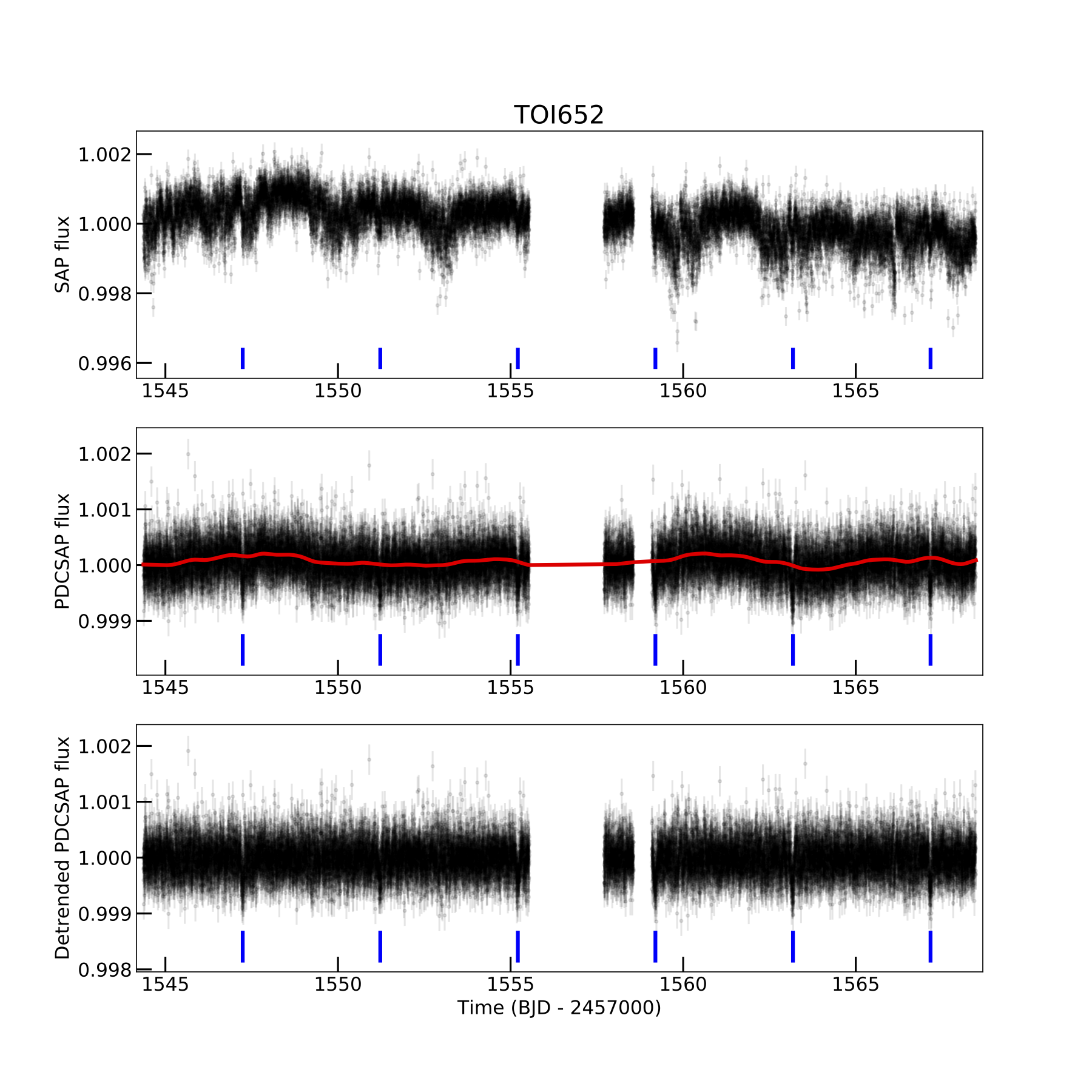}
\figsetgrpnote{The SAP, PDCSAP and detrended PDCSAP light curves of TESS. The best-fit GP model is shown as a red solid line. The vertical colored ticks represent the expected transits of each planet.}
\figsetgrpend

\figsetgrpstart
\figsetgrpnum{11.13}
\figsetgrptitle{TOI-784}
\figsetplot{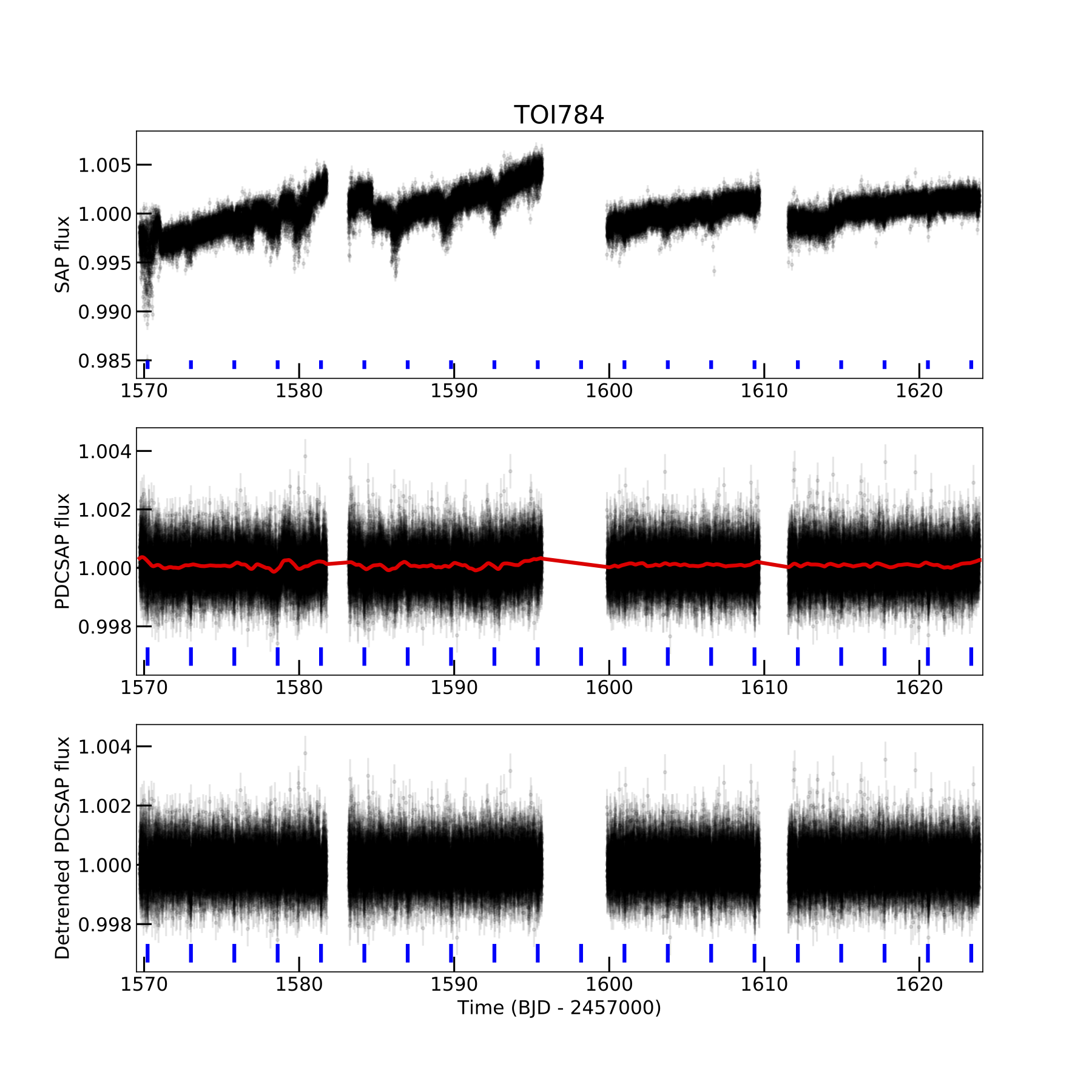}
\figsetgrpnote{The SAP, PDCSAP and detrended PDCSAP light curves of TESS. The best-fit GP model is shown as a red solid line. The vertical colored ticks represent the expected transits of each planet.}
\figsetgrpend

\figsetgrpstart
\figsetgrpnum{11.14}
\figsetgrptitle{TOI-836}
\figsetplot{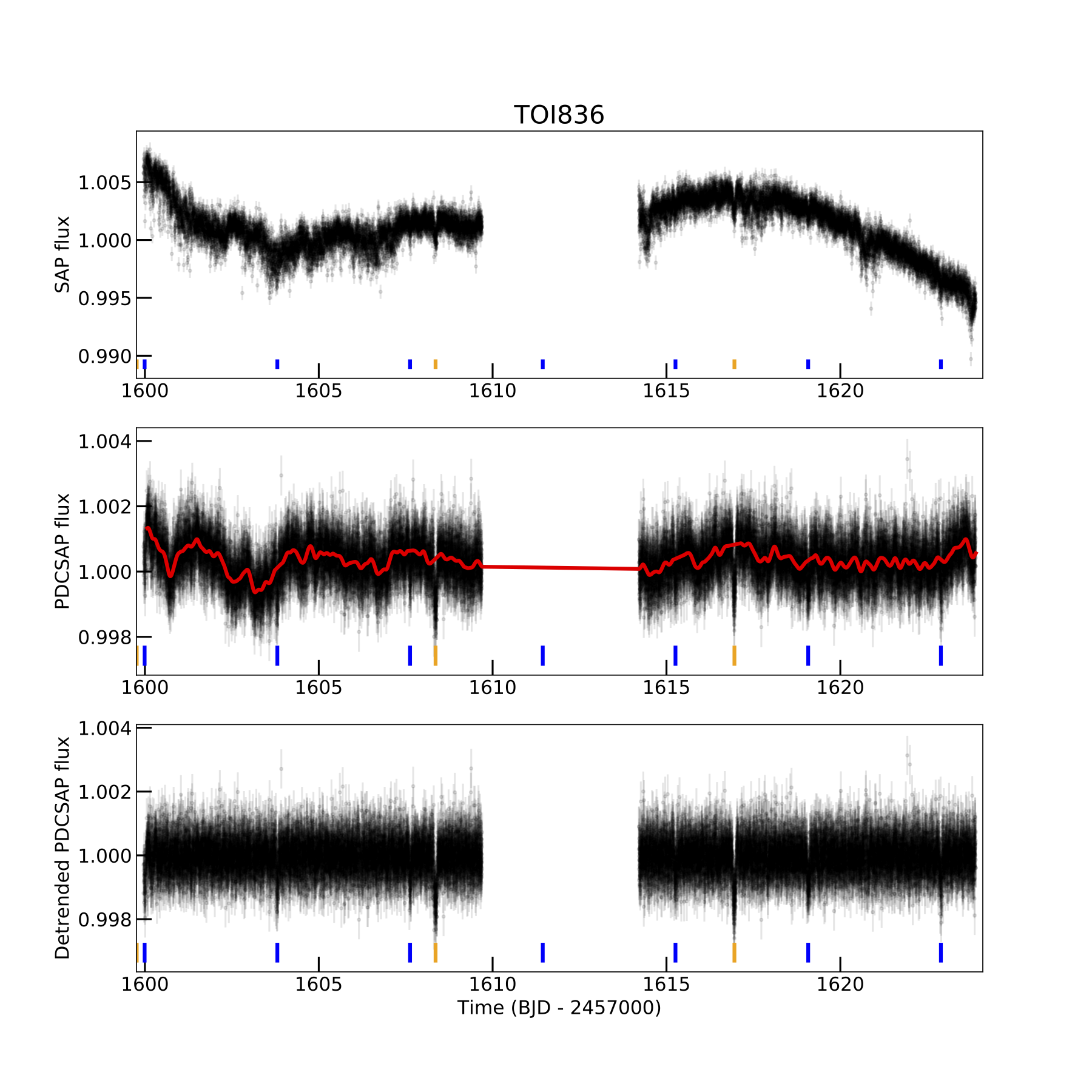}
\figsetgrpnote{The SAP, PDCSAP and detrended PDCSAP light curves of TESS. The best-fit GP model is shown as a red solid line. The vertical colored ticks represent the expected transits of each planet.}
\figsetgrpend

\figsetgrpstart
\figsetgrpnum{11.15}
\figsetgrptitle{TOI-1233}
\figsetplot{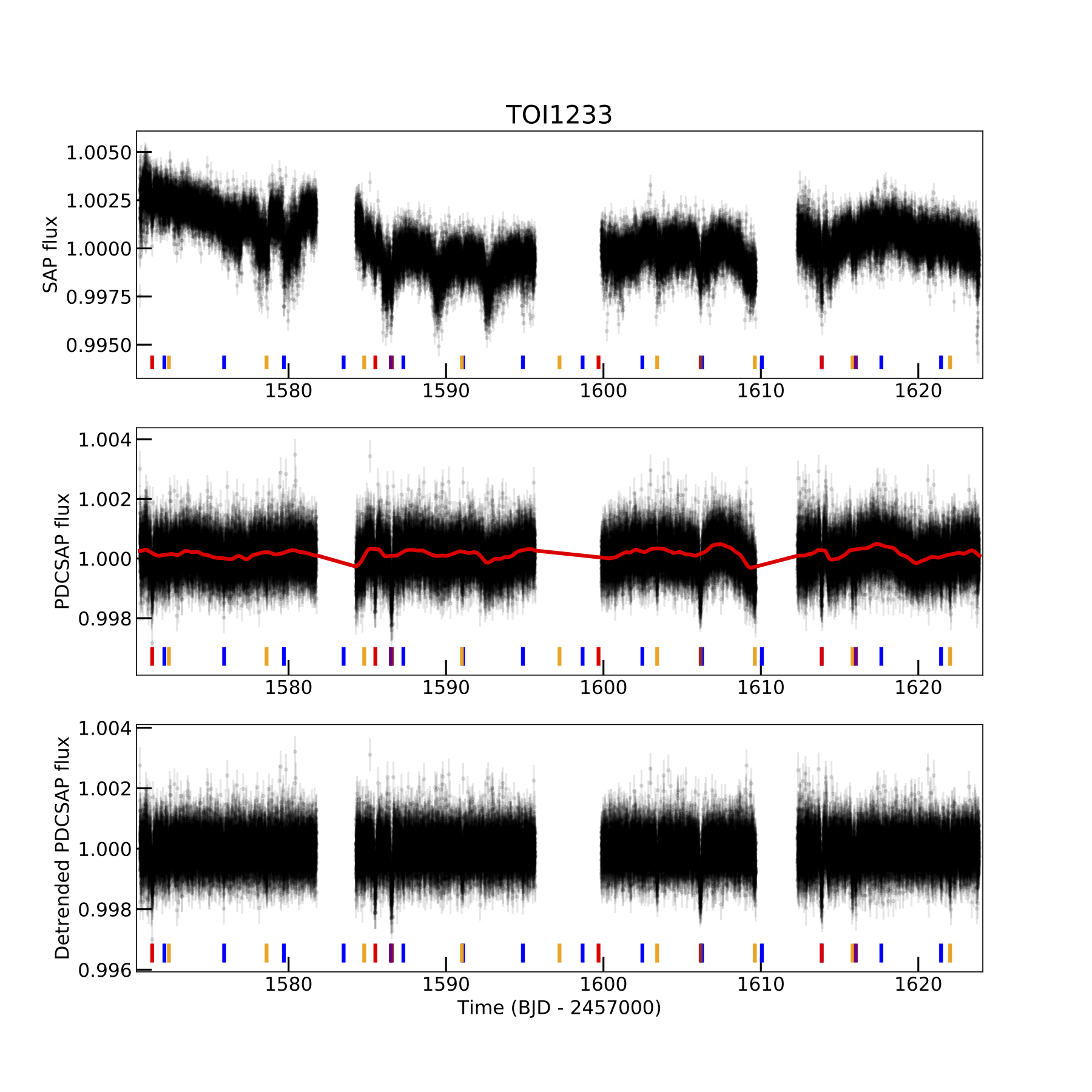}
\figsetgrpnote{The SAP, PDCSAP and detrended PDCSAP light curves of TESS. The best-fit GP model is shown as a red solid line. The vertical colored ticks represent the expected transits of each planet.}
\figsetgrpend

\figsetend

\begin{figure*}
\figurenum{11}
\plotone{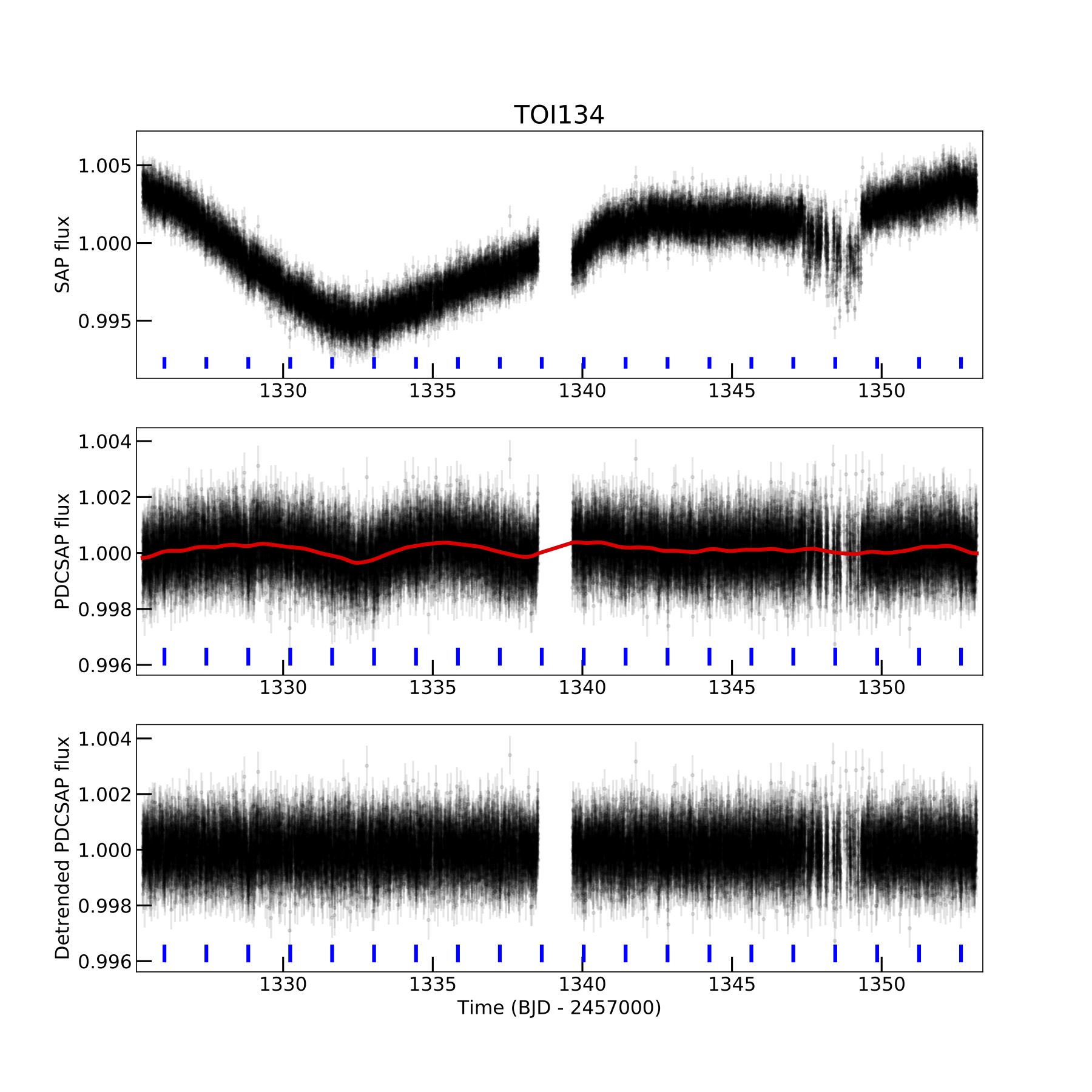}
\caption{The SAP, PDCSAP and detrended PDCSAP light curves of TESS. The best-fit GP model is shown as a red solid line. The vertical colored ticks represent the expected transits of each planet.}
\end{figure*}

\newpage

\section{TESS Light Curve Fits \label{sec:appendix_tess_lc_fits}}

\figsetstart
\figsetnum{12}
\figsettitle{TESS phase-folded curves of each TOI.}

\figsetgrpstart
\figsetgrpnum{12.1}
\figsetgrptitle{TOI-134.01}
\figsetplot{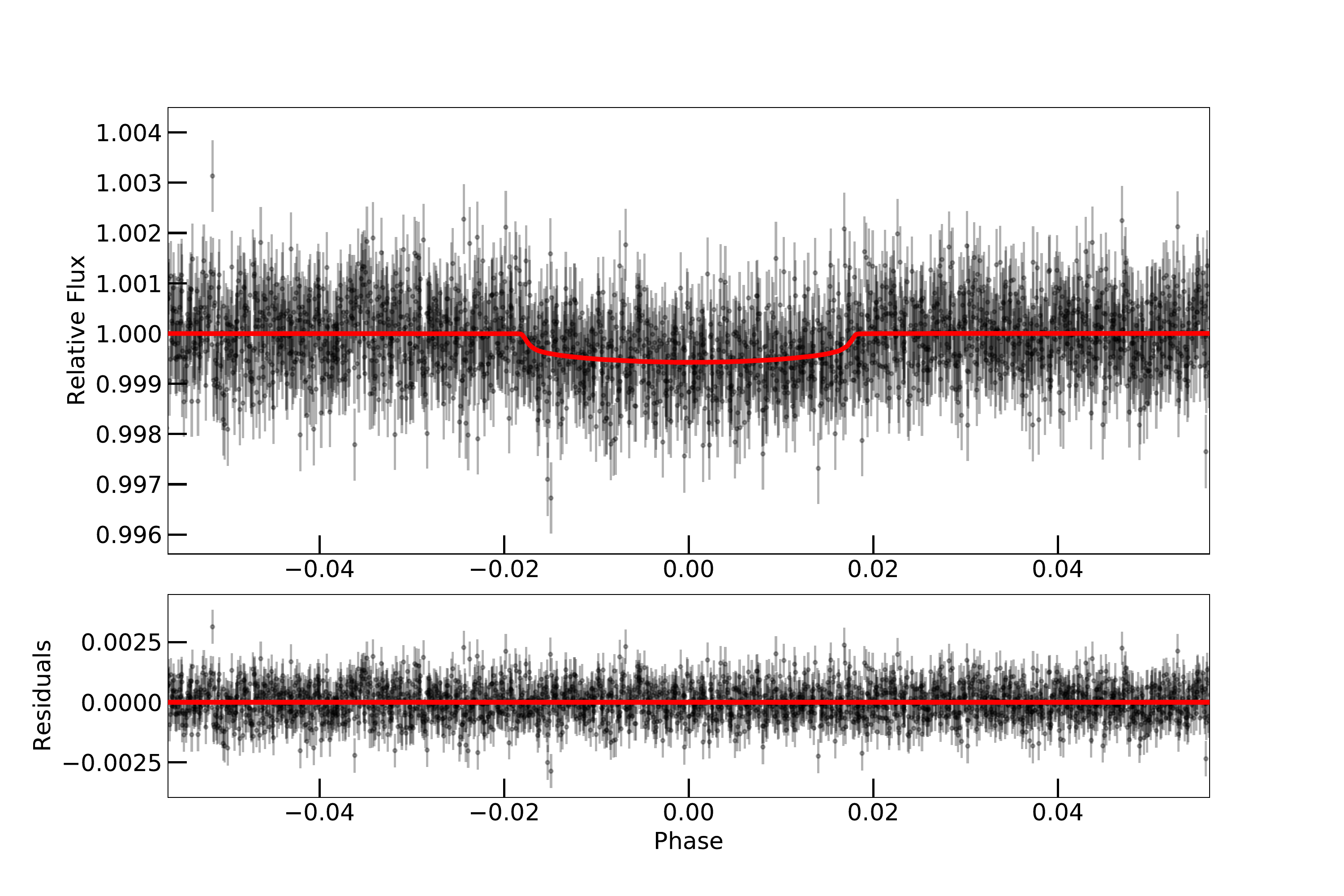}
\figsetgrpnote{The phase-folded curve of TOI-134.01. The best-fit transit model is shown as a red solid line. The residuals are plotted below.}
\figsetgrpend

\figsetgrpstart
\figsetgrpnum{12.2}
\figsetgrptitle{TOI-141.01}
\figsetplot{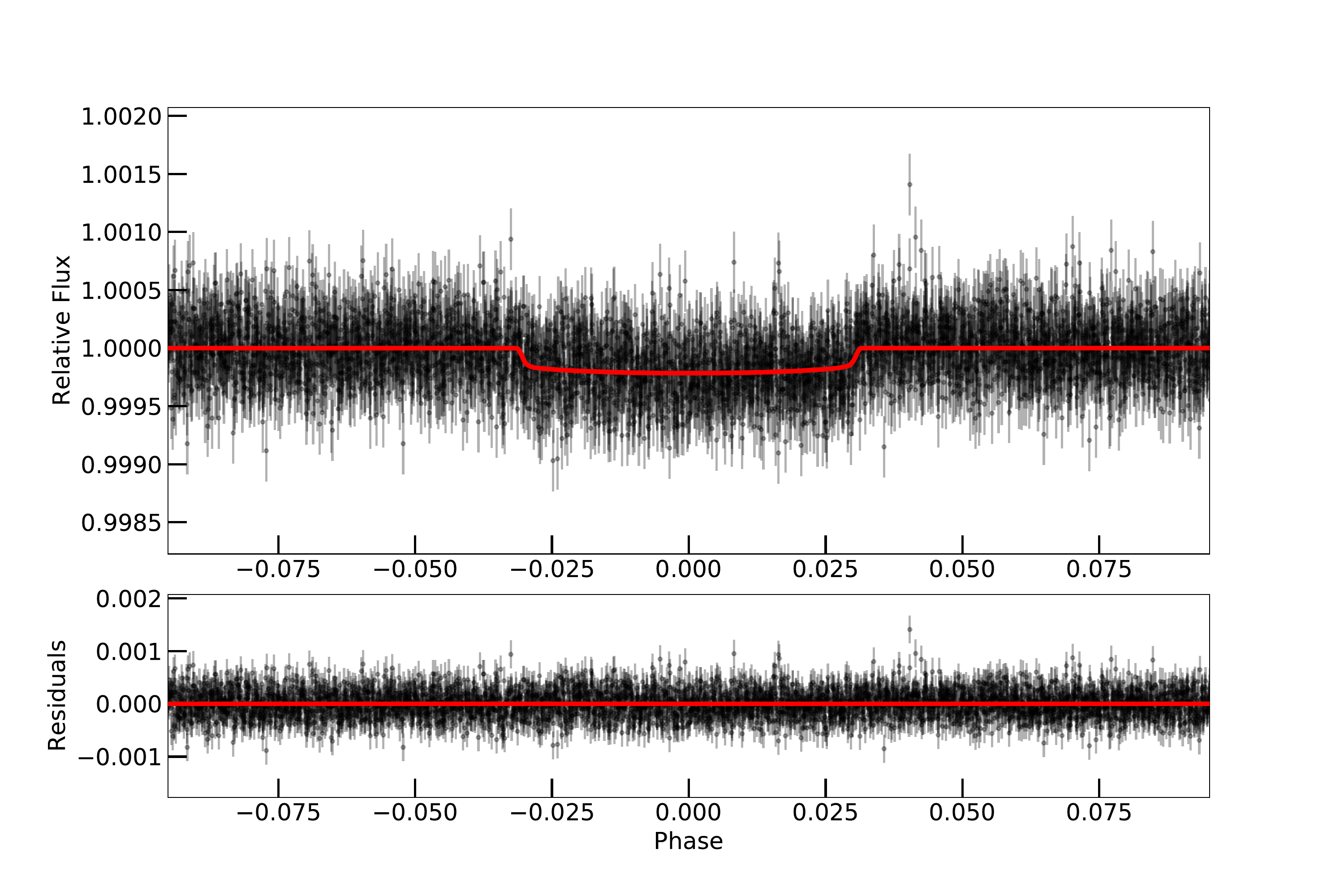}
\figsetgrpnote{The phase-folded curve of TOI 141.01. The best-fit transit model is shown as a red solid line. The residuals are plotted below.}
\figsetgrpend

\figsetgrpstart
\figsetgrpnum{12.3}
\figsetgrptitle{TOI-144.01}
\figsetplot{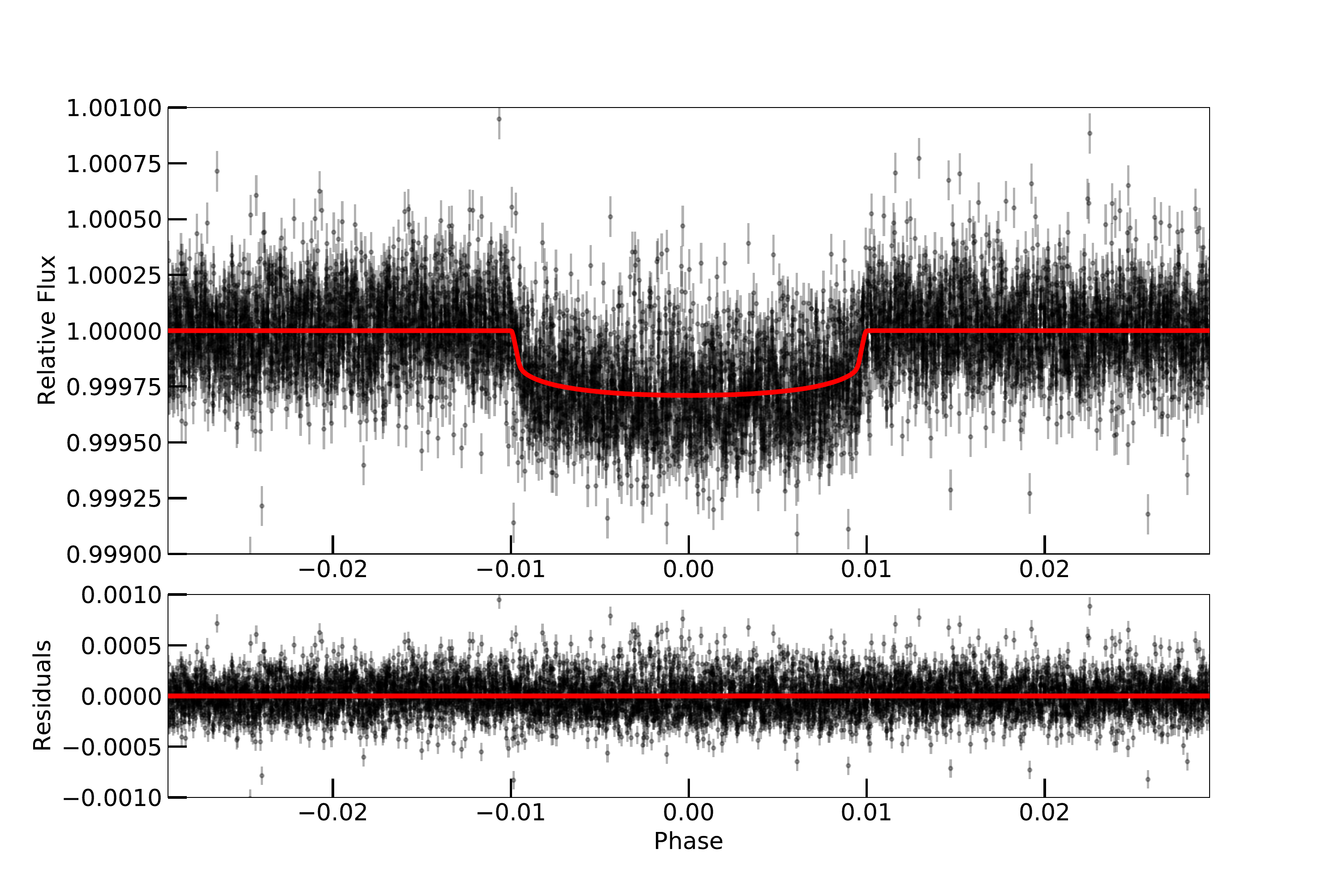}
\figsetgrpnote{The phase-folded curve of TOI 144.01. The best-fit transit model is shown as a red solid line. The residuals are plotted below.}
\figsetgrpend

\figsetgrpstart
\figsetgrpnum{12.4}
\figsetgrptitle{TOI-174.01}
\figsetplot{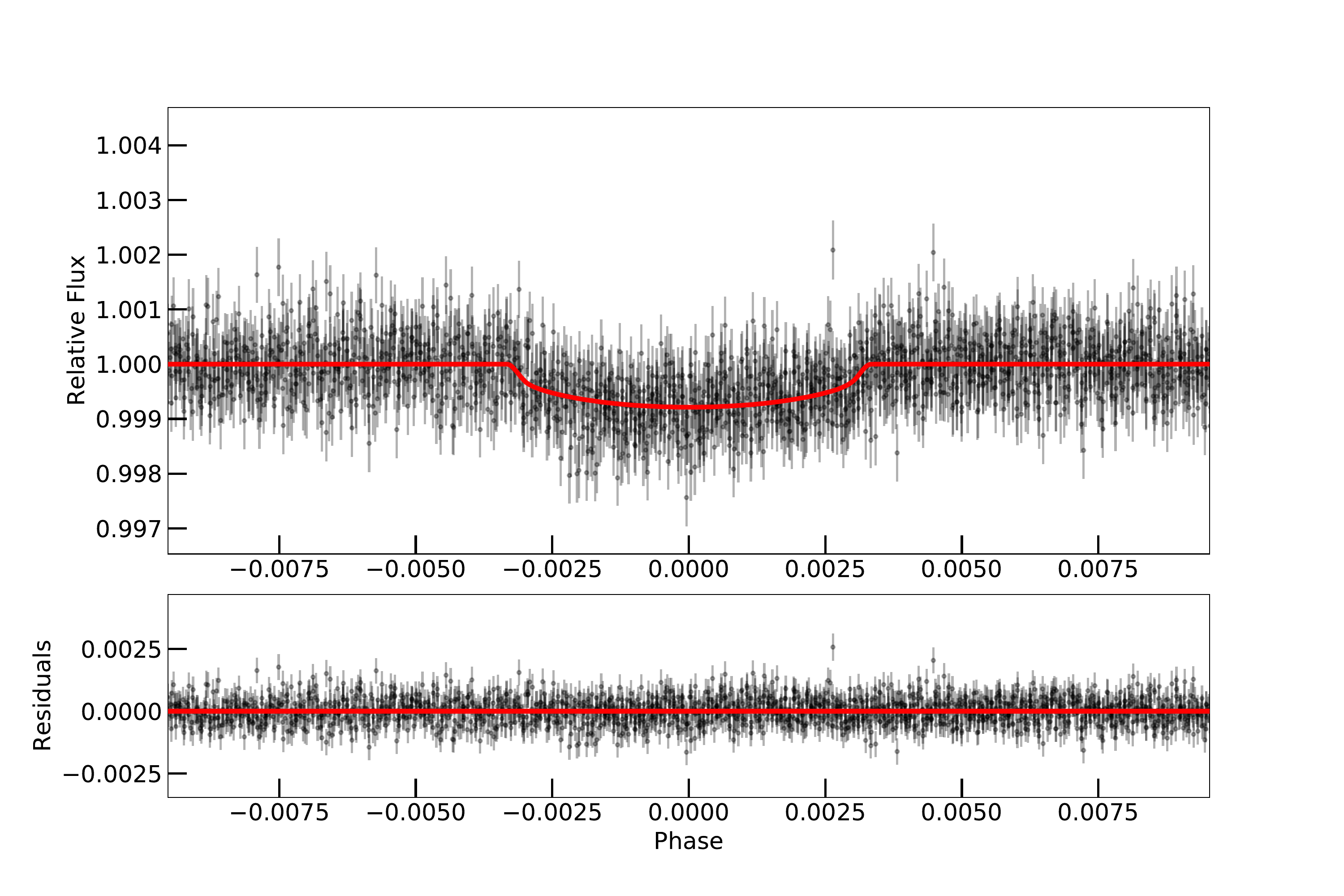}
\figsetgrpnote{The phase-folded curve of TOI 174.01. The best-fit transit model is shown as a red solid line. The residuals are plotted below.}
\figsetgrpend

\figsetgrpstart
\figsetgrpnum{12.5}
\figsetgrptitle{TOI-174.02}
\figsetplot{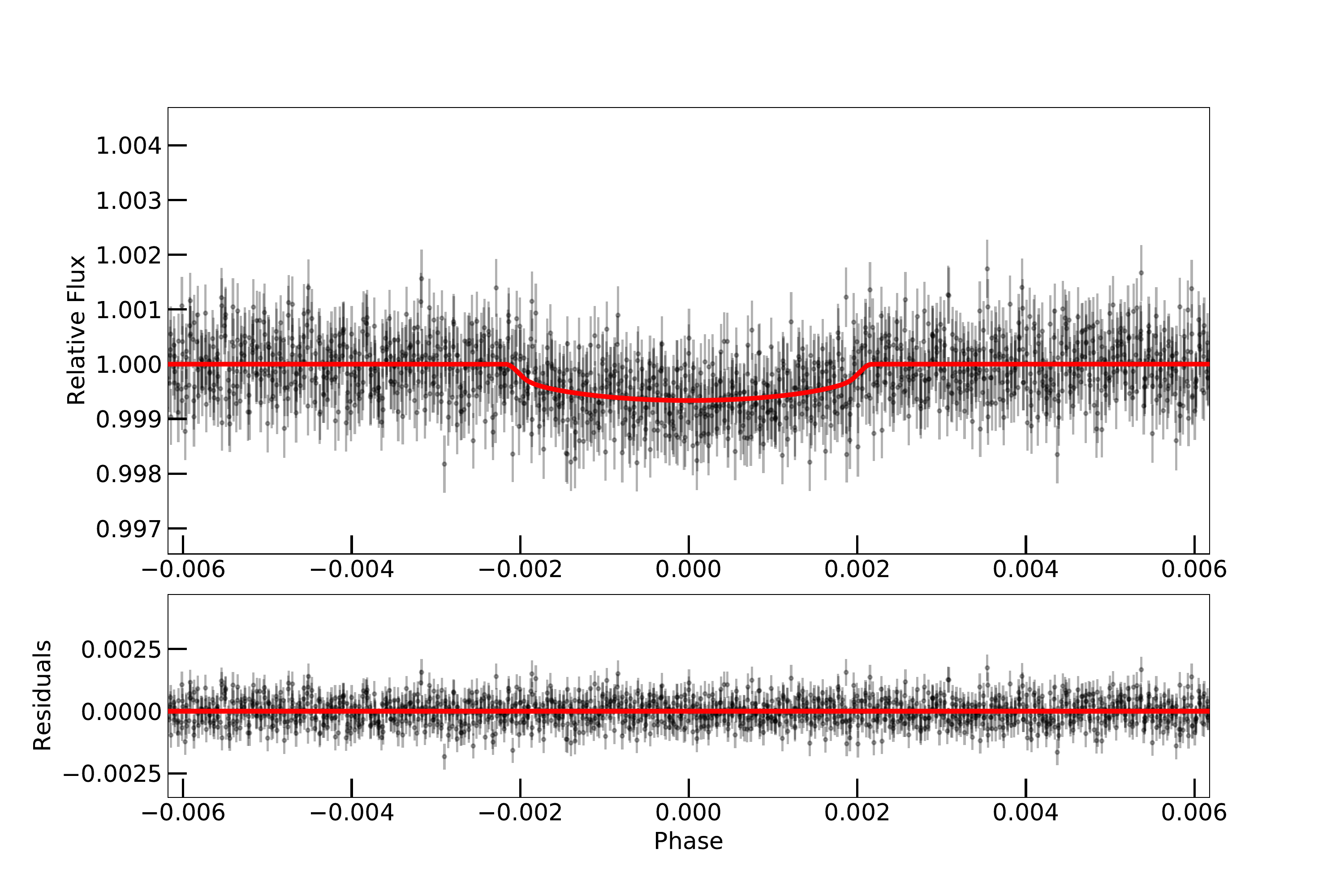}
\figsetgrpnote{The phase-folded curve of TOI 174.02. The best-fit transit model is shown as a red solid line. The residuals are plotted below.}
\figsetgrpend

\figsetgrpstart
\figsetgrpnum{12.6}
\figsetgrptitle{TOI-174.03}
\figsetplot{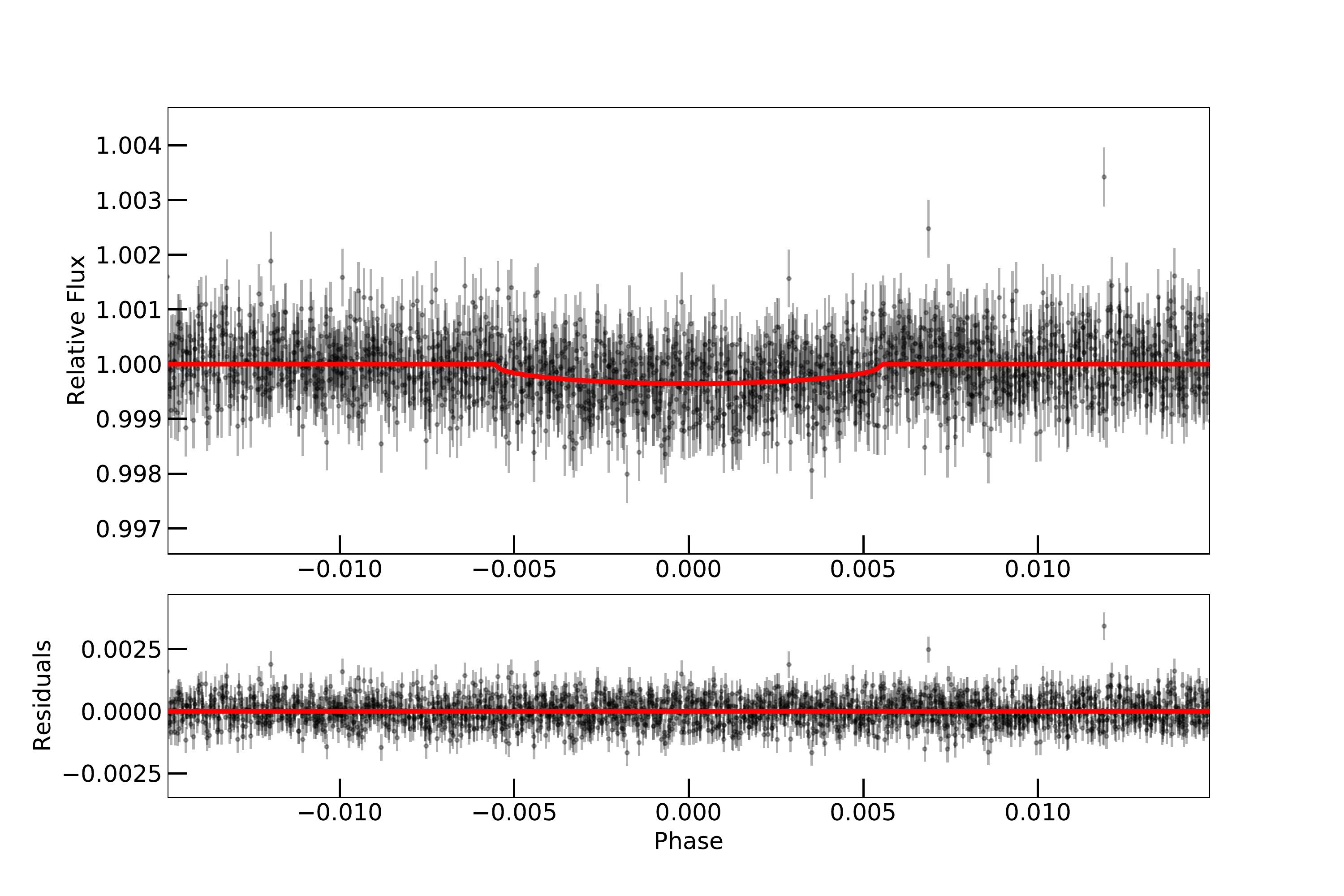}
\figsetgrpnote{The phase-folded curve of TOI 174.03. The best-fit transit model is shown as a red solid line. The residuals are plotted below.}
\figsetgrpend

\figsetgrpstart
\figsetgrpnum{12.7}
\figsetgrptitle{TOI-175.01}
\figsetplot{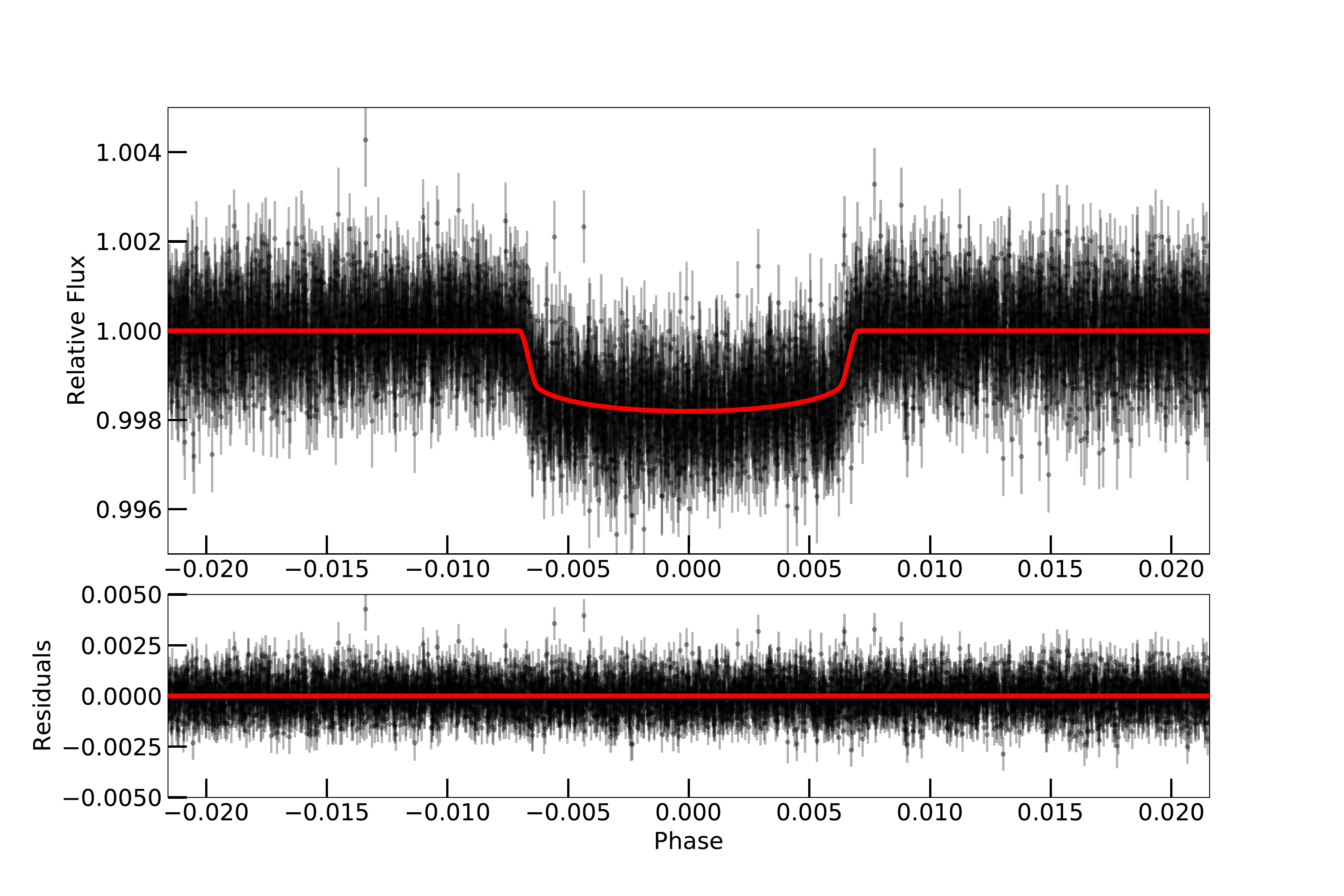}
\figsetgrpnote{The phase-folded curve of TOI 175.01. The best-fit transit model is shown as a red solid line. The residuals are plotted below.}
\figsetgrpend

\figsetgrpstart
\figsetgrpnum{12.8}
\figsetgrptitle{TOI-175.02}
\figsetplot{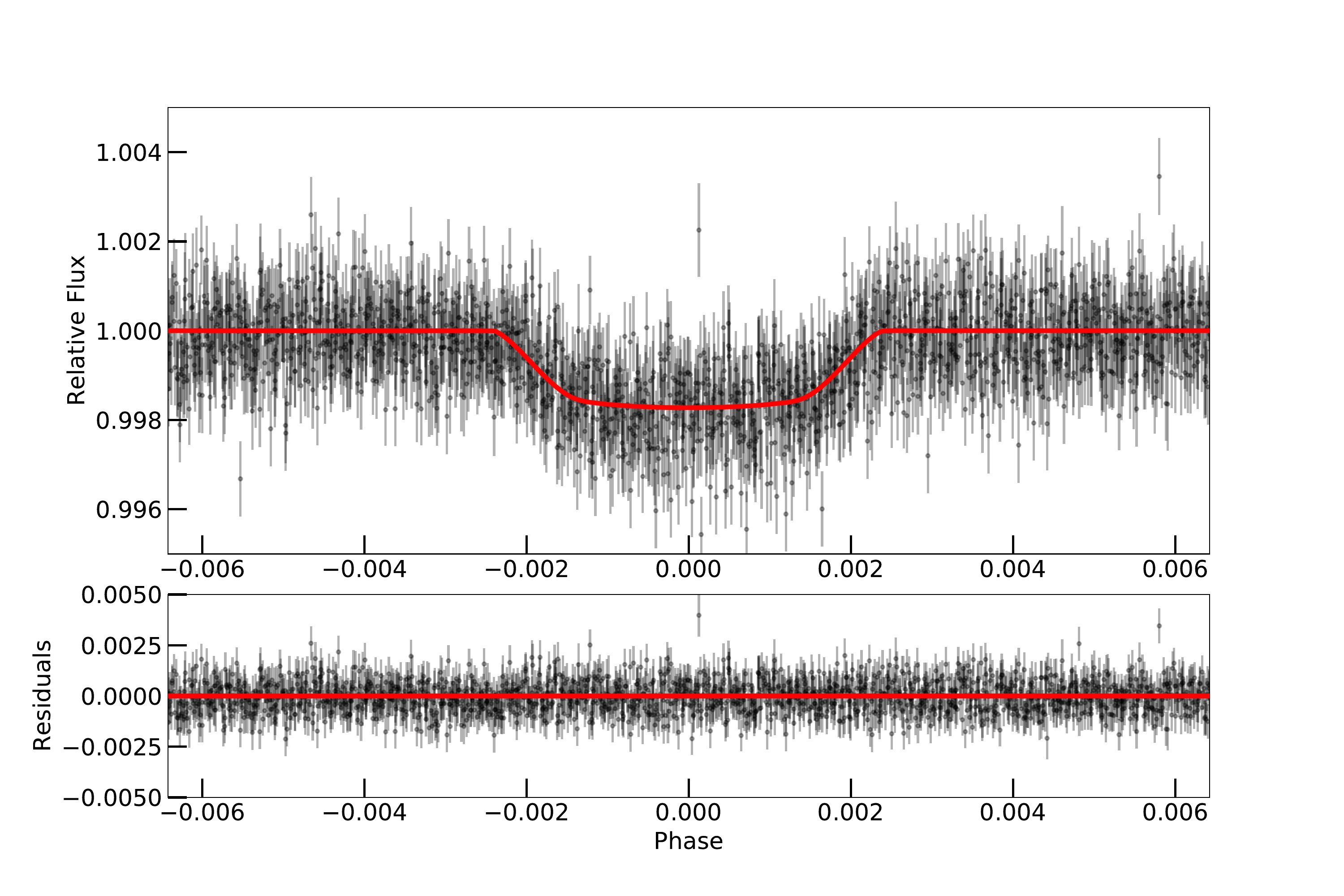}
\figsetgrpnote{The phase-folded curve of TOI 175.02. The best-fit transit model is shown as a red solid line. The residuals are plotted below.}
\figsetgrpend

\figsetgrpstart
\figsetgrpnum{12.9}
\figsetgrptitle{TOI-175.03}
\figsetplot{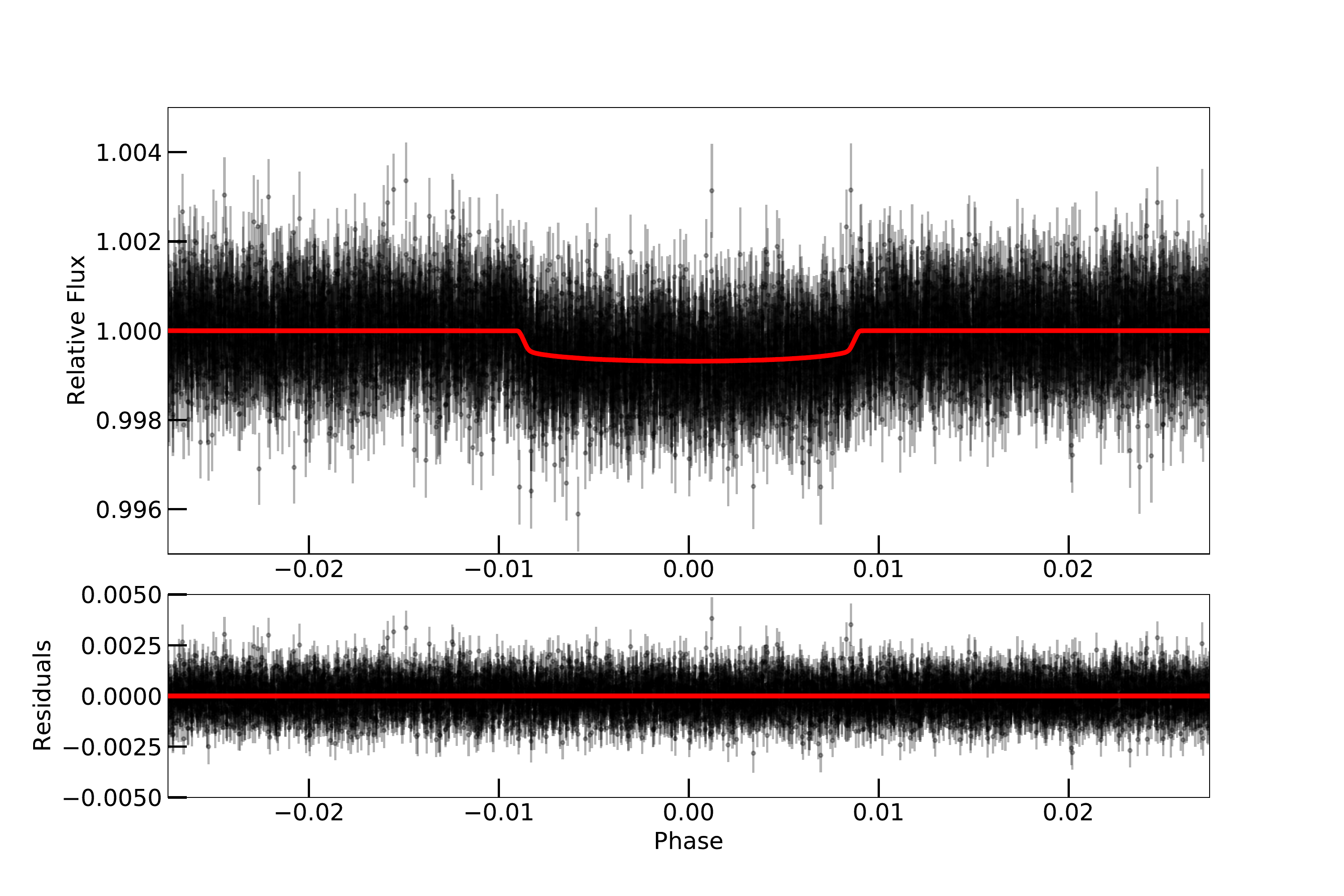}
\figsetgrpnote{The phase-folded curve of TOI 175.03. The best-fit transit model is shown as a red solid line. The residuals are plotted below.}
\figsetgrpend

\figsetgrpstart
\figsetgrpnum{12.10}
\figsetgrptitle{TOI-186.01}
\figsetplot{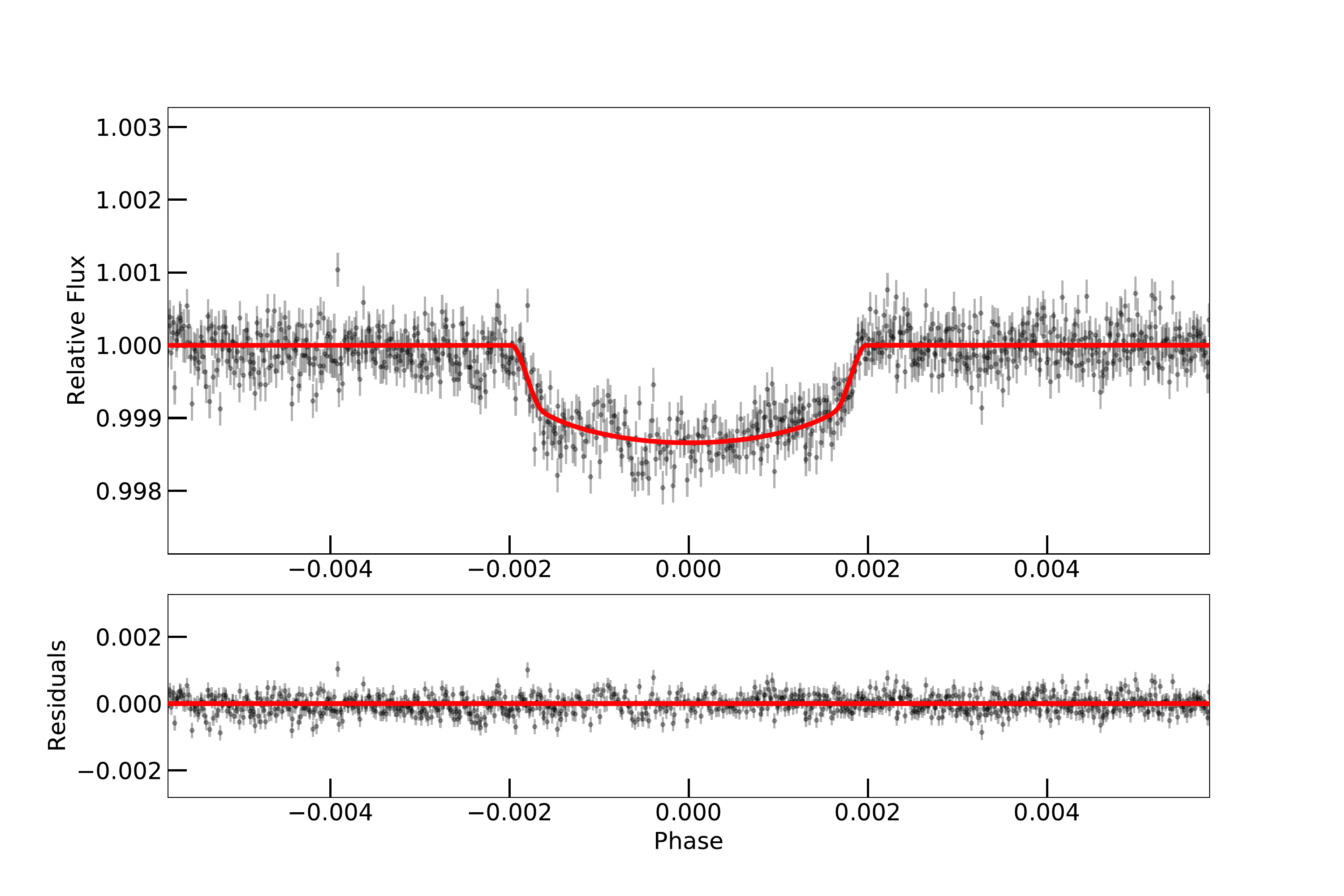}
\figsetgrpnote{The phase-folded curve of TOI 186.01. The best-fit transit model is shown as a red solid line. The residuals are plotted below.}
\figsetgrpend

\figsetgrpstart
\figsetgrpnum{12.11}
\figsetgrptitle{TOI-186.02}
\figsetplot{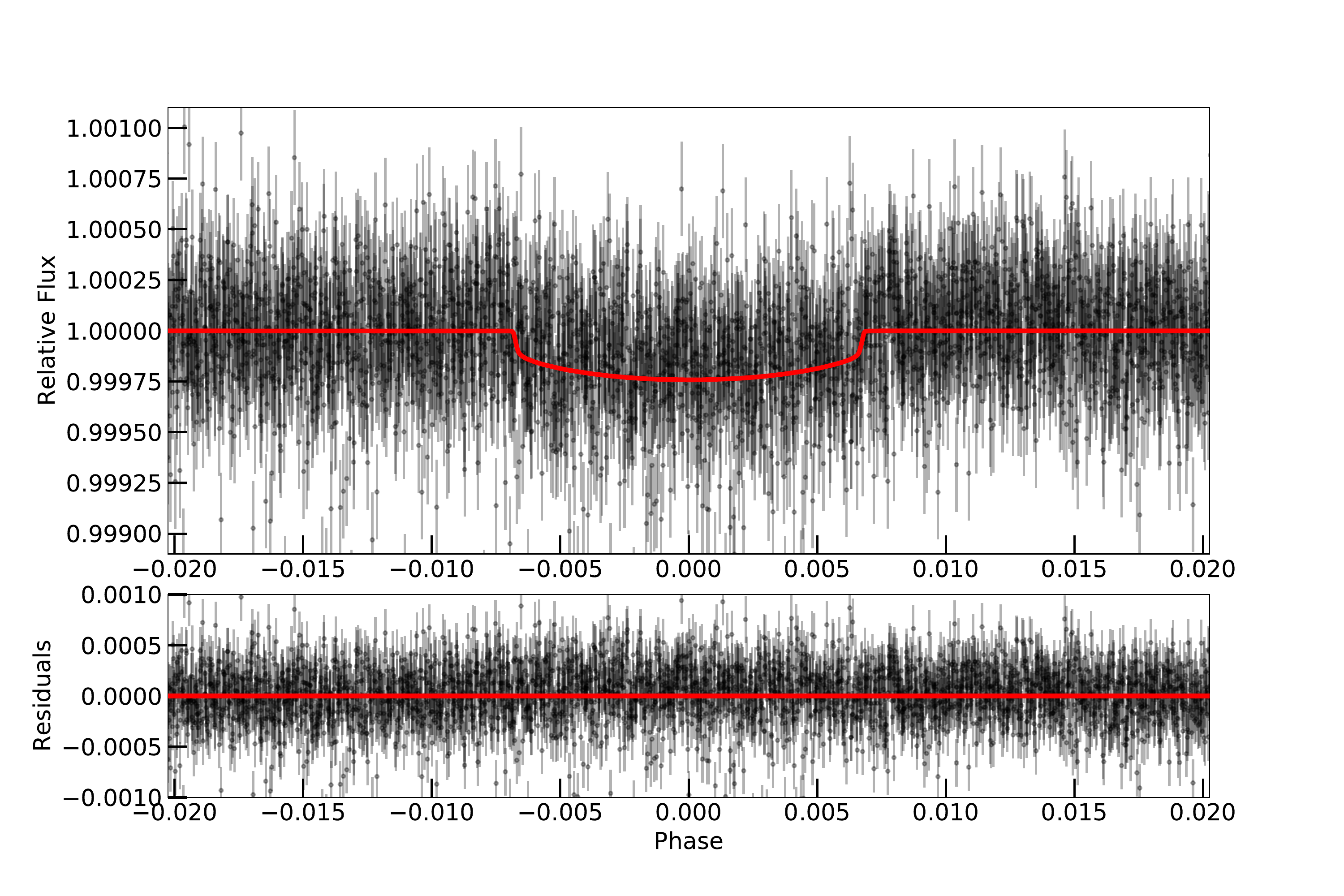}
\figsetgrpnote{The phase-folded curve of TOI 186.02. The best-fit transit model is shown as a red solid line. The residuals are plotted below.}
\figsetgrpend

\figsetgrpstart
\figsetgrpnum{12.12}
\figsetgrptitle{TOI-260.01}
\figsetplot{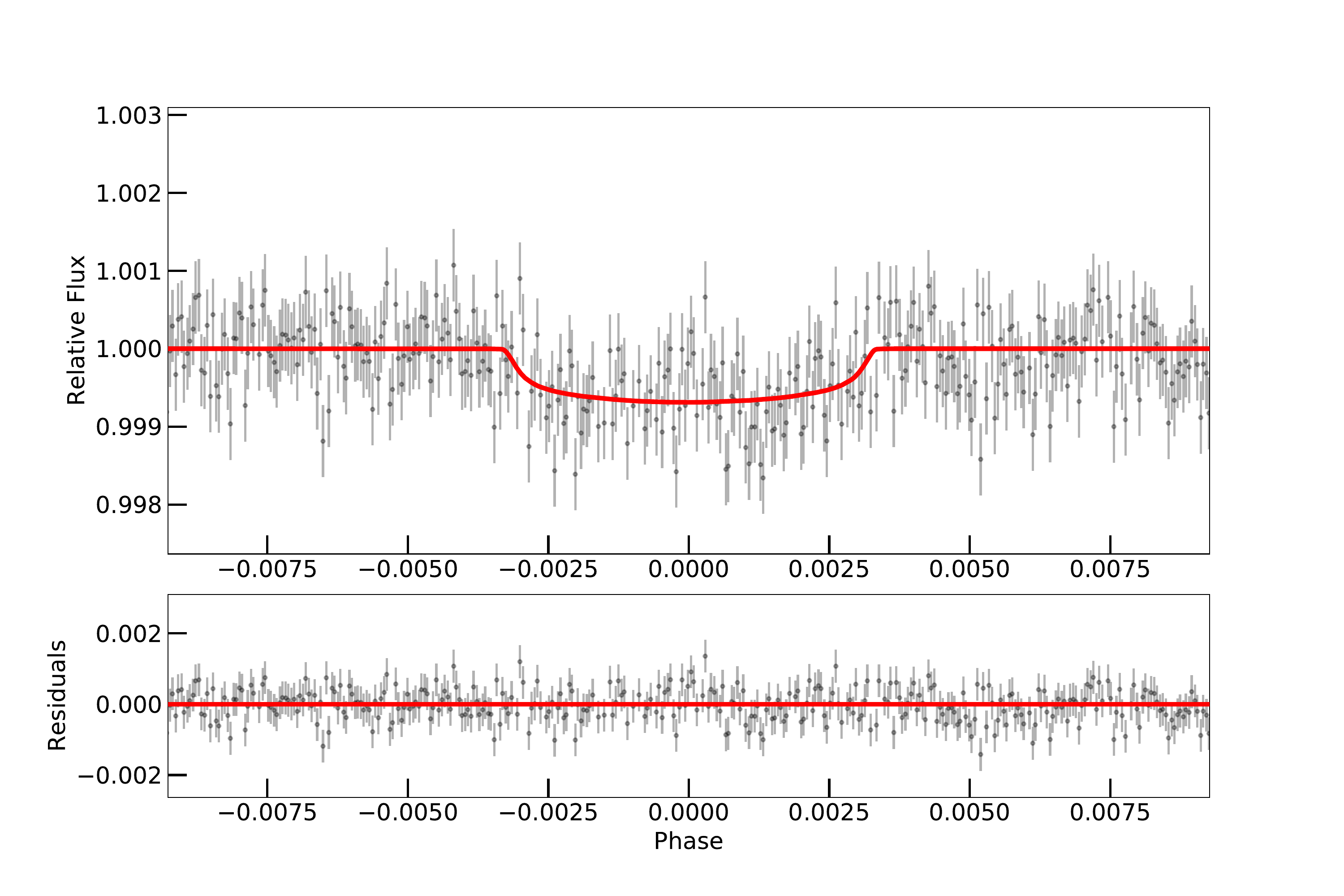}
\figsetgrpnote{The phase-folded curve of TOI 260.01. The best-fit transit model is shown as a red solid line. The residuals are plotted below.}
\figsetgrpend

\figsetgrpstart
\figsetgrpnum{12.13}
\figsetgrptitle{TOI-402.01}
\figsetplot{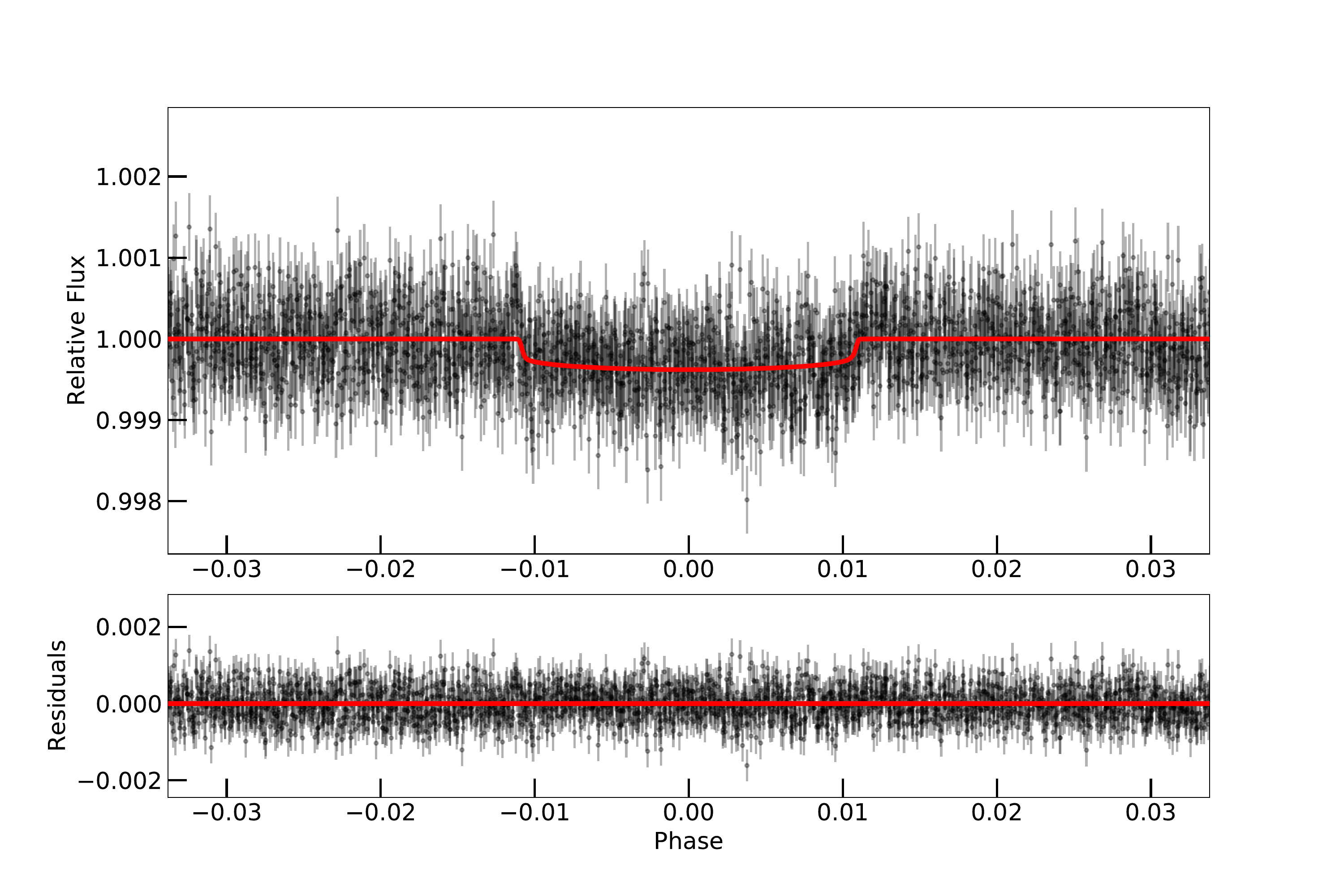}
\figsetgrpnote{The phase-folded curve of TOI 402.01. The best-fit transit model is shown as a red solid line. The residuals are plotted below.}
\figsetgrpend

\figsetgrpstart
\figsetgrpnum{12.14}
\figsetgrptitle{TOI-402.02}
\figsetplot{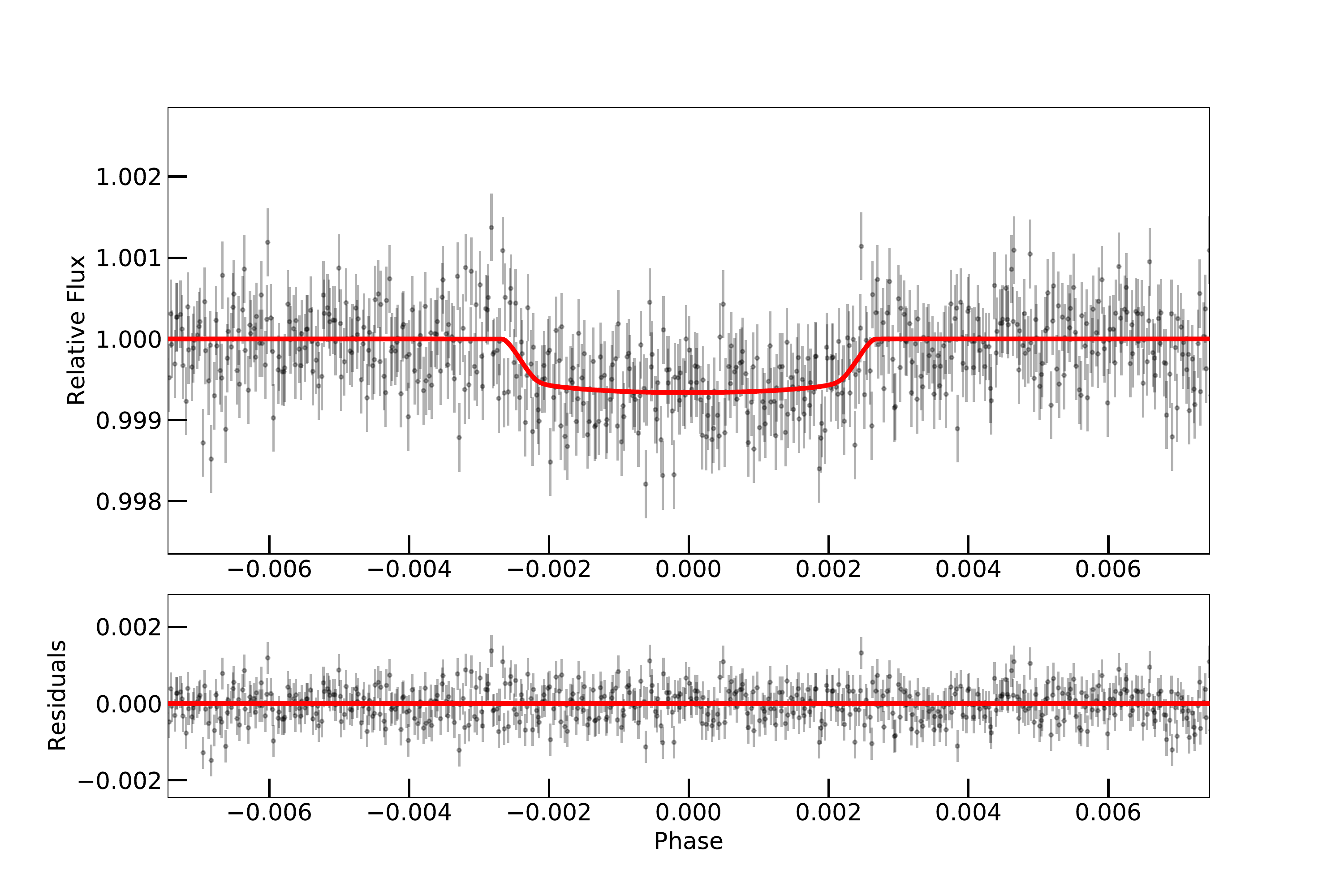}
\figsetgrpnote{The phase-folded curve of TOI 402.02. The best-fit transit model is shown as a red solid line. The residuals are plotted below.}
\figsetgrpend

\figsetgrpstart
\figsetgrpnum{12.15}
\figsetgrptitle{TOI-431.01}
\figsetplot{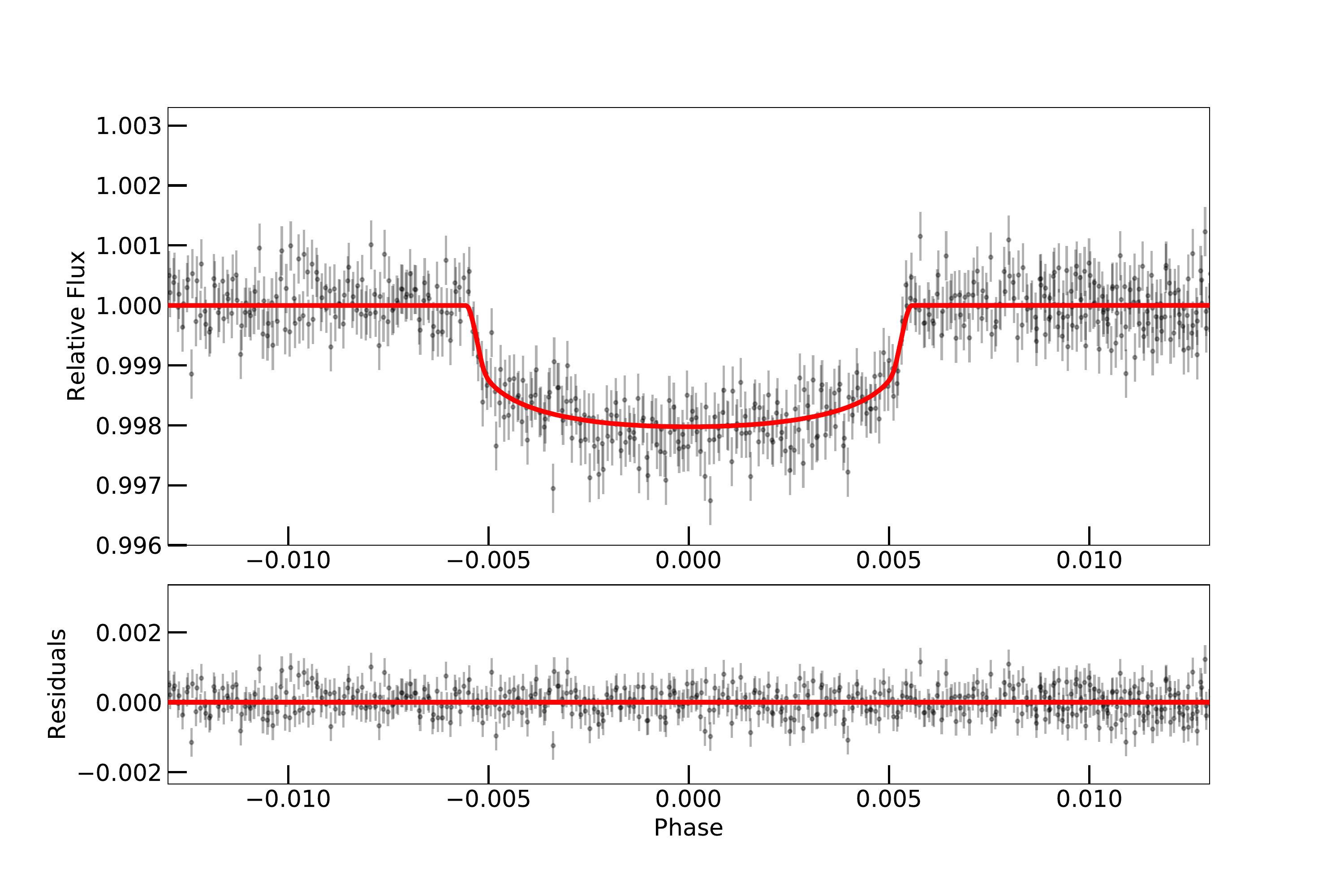}
\figsetgrpnote{The phase-folded curve of TOI 431.01. The best-fit transit model is shown as a red solid line. The residuals are plotted below.}
\figsetgrpend

\figsetgrpstart
\figsetgrpnum{12.16}
\figsetgrptitle{TOI-431.02}
\figsetplot{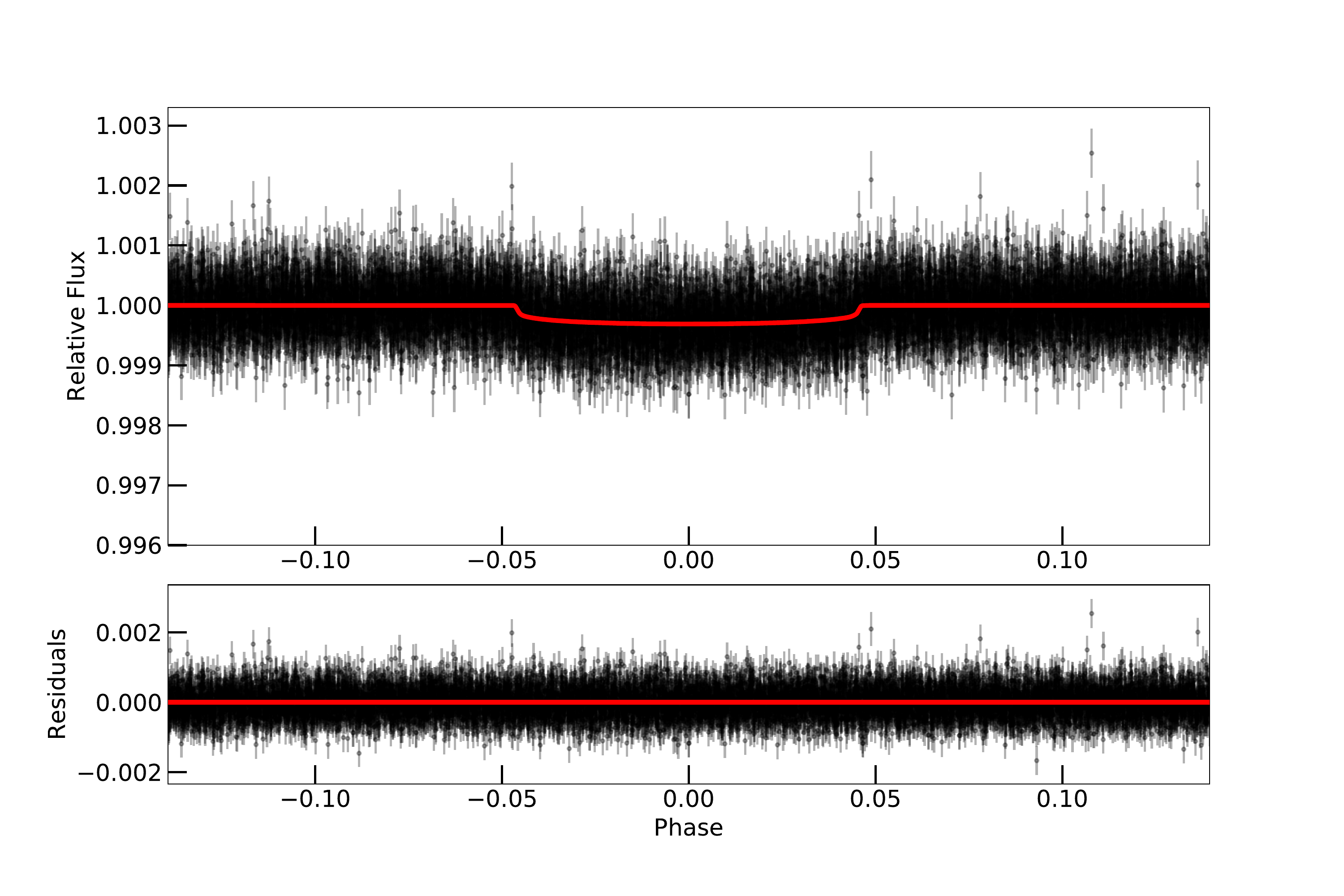}
\figsetgrpnote{The phase-folded curve of TOI 431.02. The best-fit transit model is shown as a red solid line. The residuals are plotted below.}
\figsetgrpend

\figsetgrpstart
\figsetgrpnum{12.17}
\figsetgrptitle{TOI-455.01}
\figsetplot{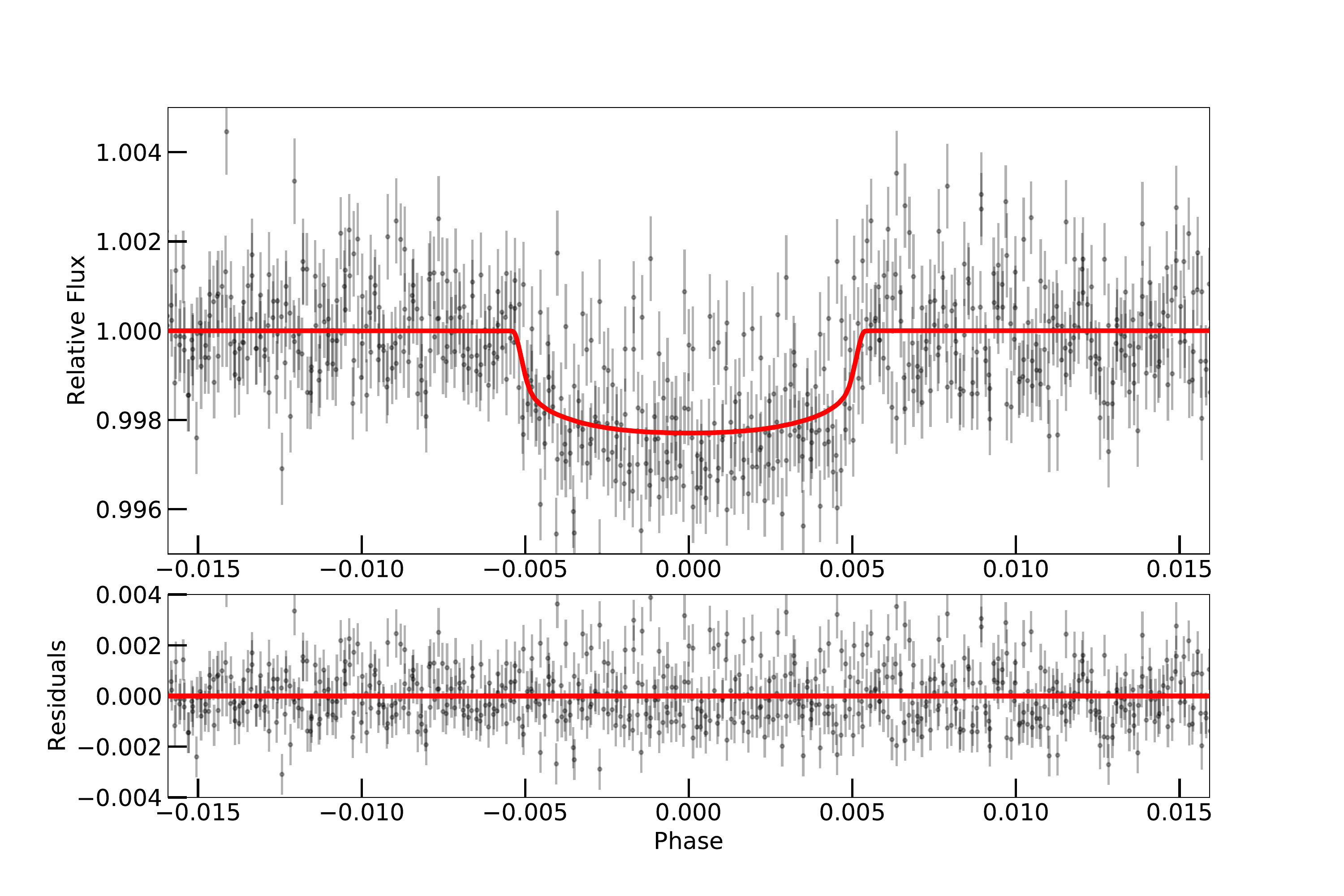}
\figsetgrpnote{The phase-folded curve of TOI 455.01. The best-fit transit model is shown as a red solid line. The residuals are plotted below.}
\figsetgrpend

\figsetgrpstart
\figsetgrpnum{12.18}
\figsetgrptitle{TOI-562.01}
\figsetplot{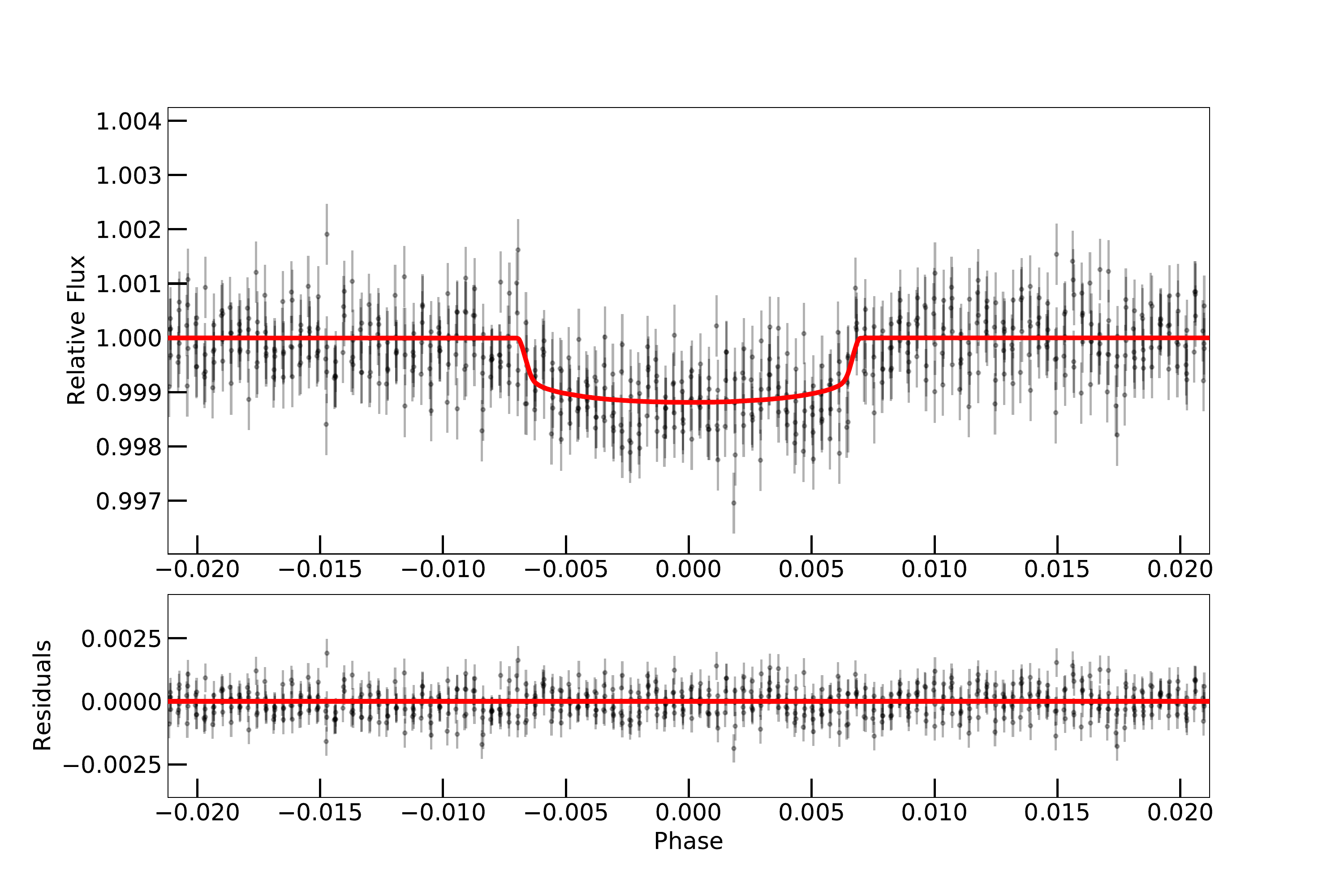}
\figsetgrpnote{The phase-folded curve of TOI 562.01. The best-fit transit model is shown as a red solid line. The residuals are plotted below.}
\figsetgrpend

\figsetgrpstart
\figsetgrpnum{12.19}
\figsetgrptitle{TOI-652.01}
\figsetplot{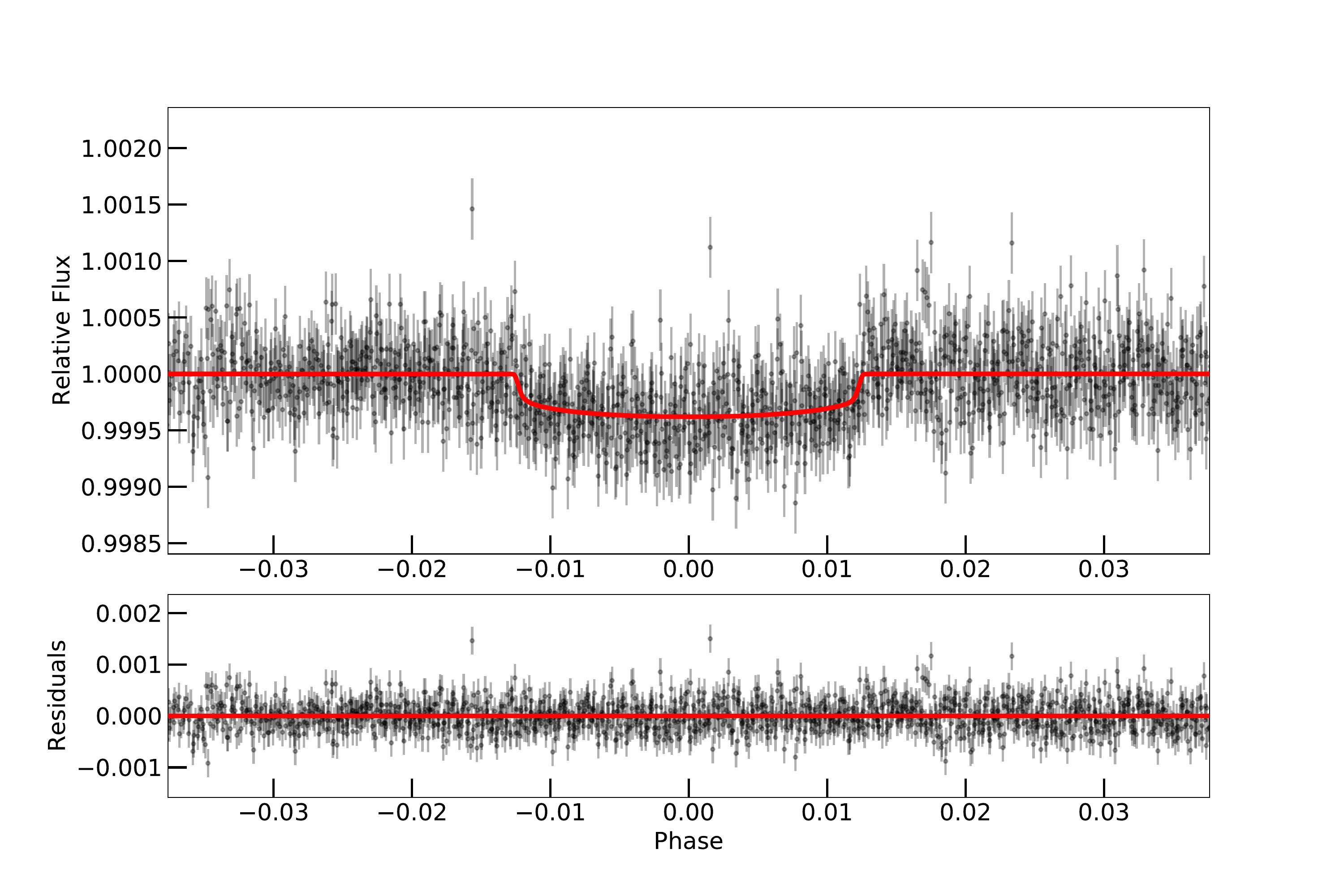}
\figsetgrpnote{The phase-folded curve of TOI 652.01. The best-fit transit model is shown as a red solid line. The residuals are plotted below.}
\figsetgrpend

\figsetgrpstart
\figsetgrpnum{12.20}
\figsetgrptitle{TOI-784.01}
\figsetplot{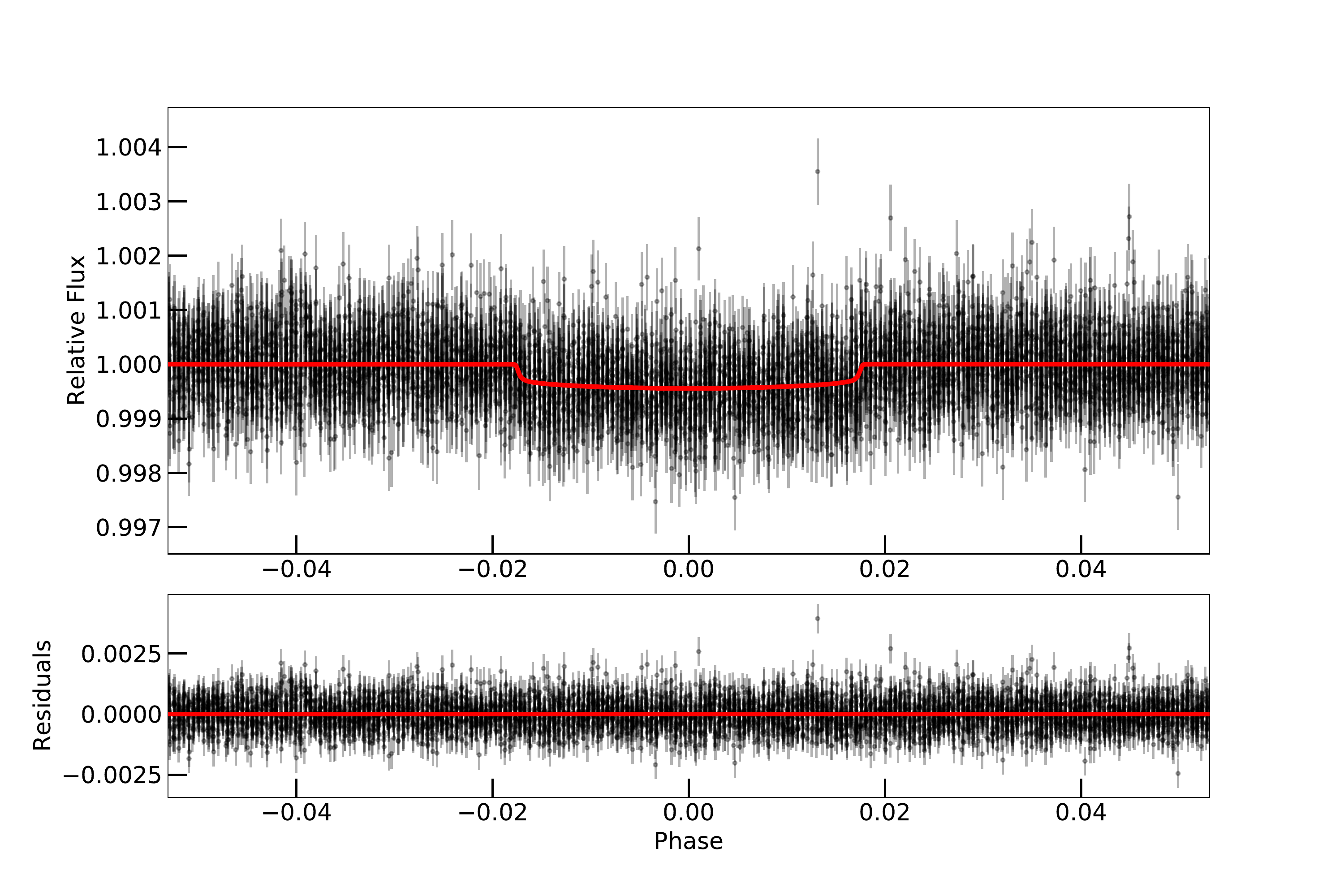}
\figsetgrpnote{The phase-folded curve of TOI 784.01. The best-fit transit model is shown as a red solid line. The residuals are plotted below.}
\figsetgrpend

\figsetgrpstart
\figsetgrpnum{12.21}
\figsetgrptitle{TOI-836.01}
\figsetplot{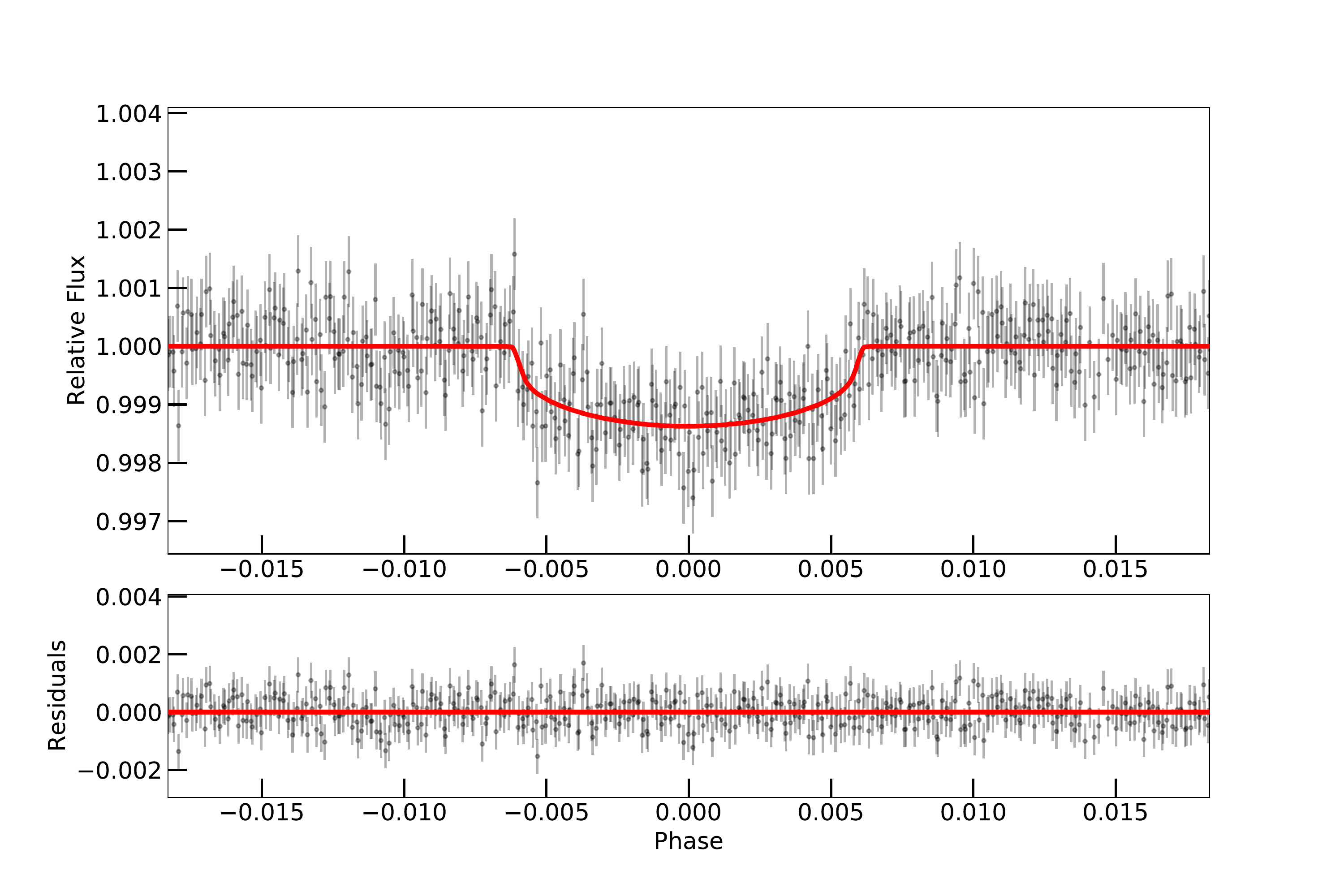}
\figsetgrpnote{The phase-folded curve of TOI 836.01. The best-fit transit model is shown as a red solid line. The residuals are plotted below.}
\figsetgrpend

\figsetgrpstart
\figsetgrpnum{12.22}
\figsetgrptitle{TOI-836.02}
\figsetplot{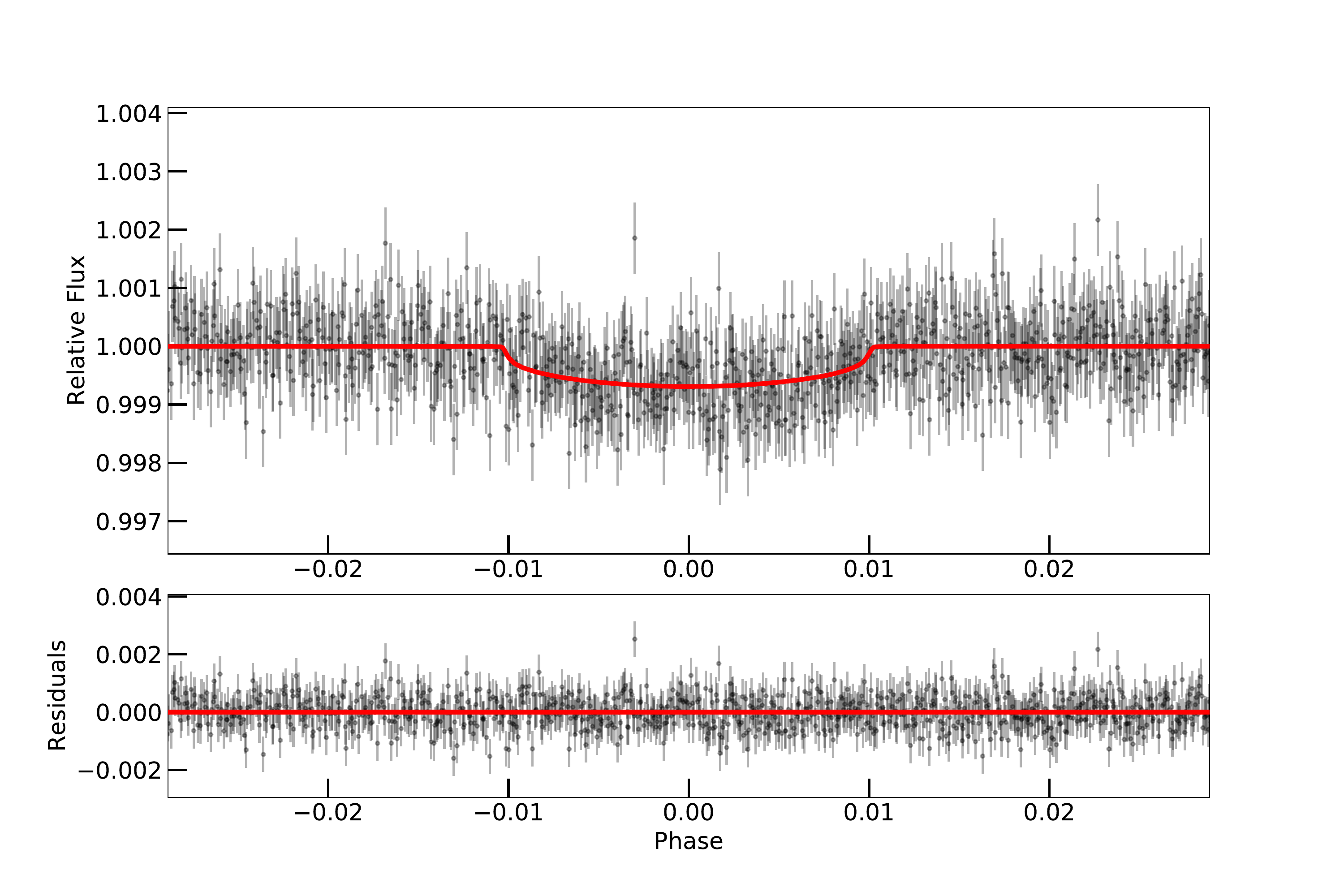}
\figsetgrpnote{The phase-folded curve of TOI 836.02. The best-fit transit model is shown as a red solid line. The residuals are plotted below.}
\figsetgrpend

\figsetgrpstart
\figsetgrpnum{12.23}
\figsetgrptitle{TOI-1233.01}
\figsetplot{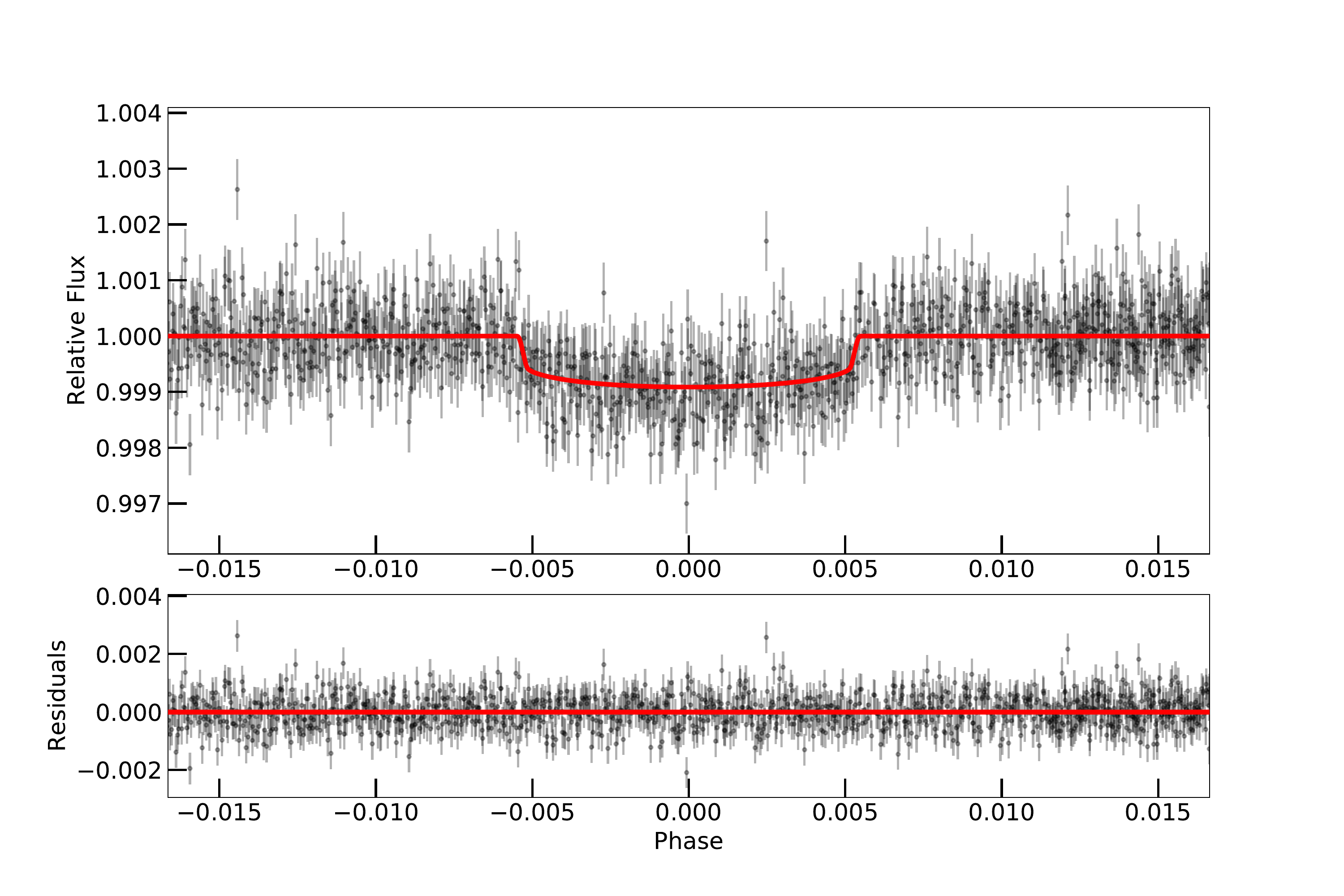}
\figsetgrpnote{The phase-folded curve of TOI 1233.01. The best-fit transit model is shown as a red solid line. The residuals are plotted below.}
\figsetgrpend

\figsetgrpstart
\figsetgrpnum{12.24}
\figsetgrptitle{TOI-1233.02}
\figsetplot{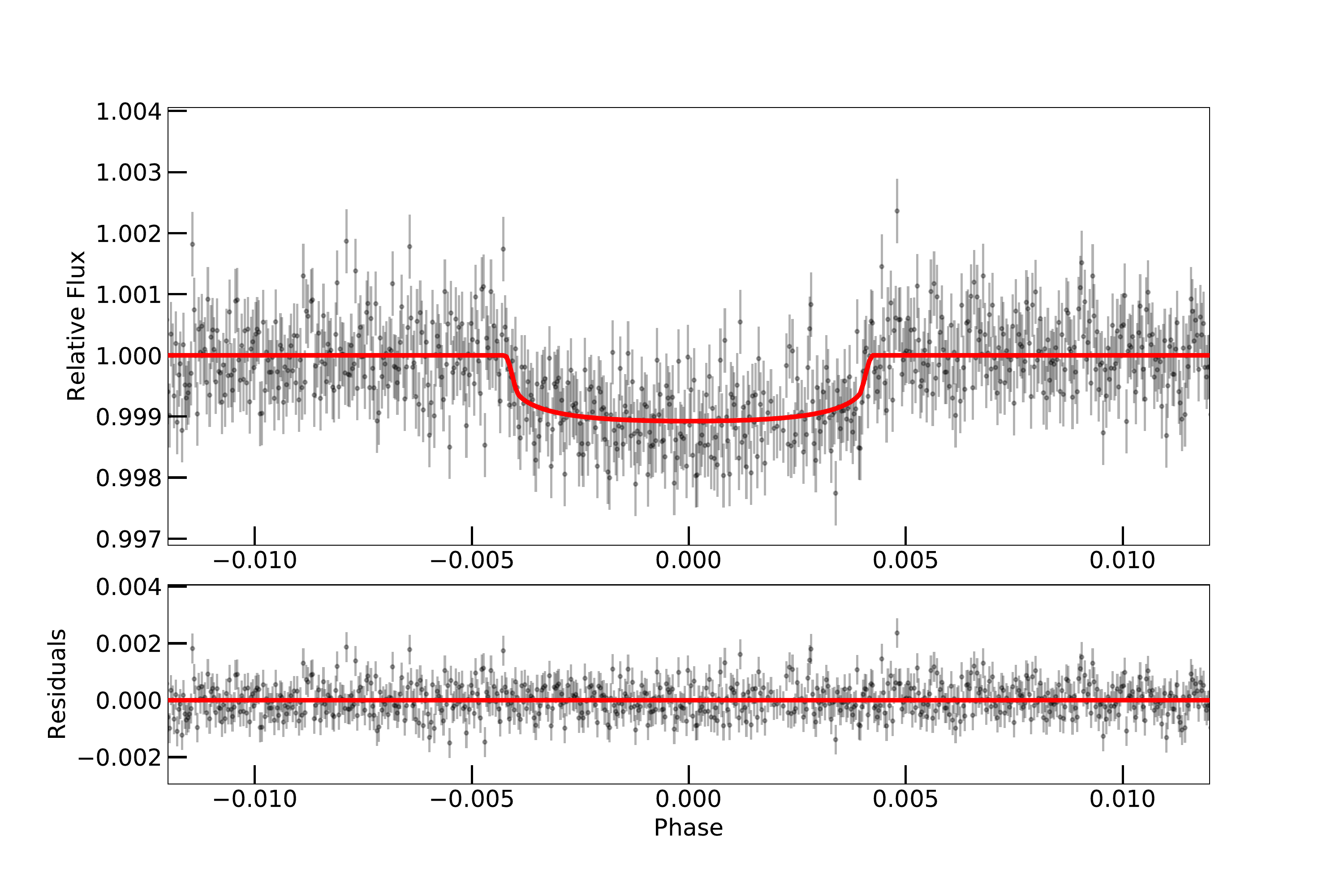}
\figsetgrpnote{The phase-folded curve of TOI 1233.02. The best-fit transit model is shown as a red solid line. The residuals are plotted below.}
\figsetgrpend

\figsetgrpstart
\figsetgrpnum{12.25}
\figsetgrptitle{TOI-1233.03}
\figsetplot{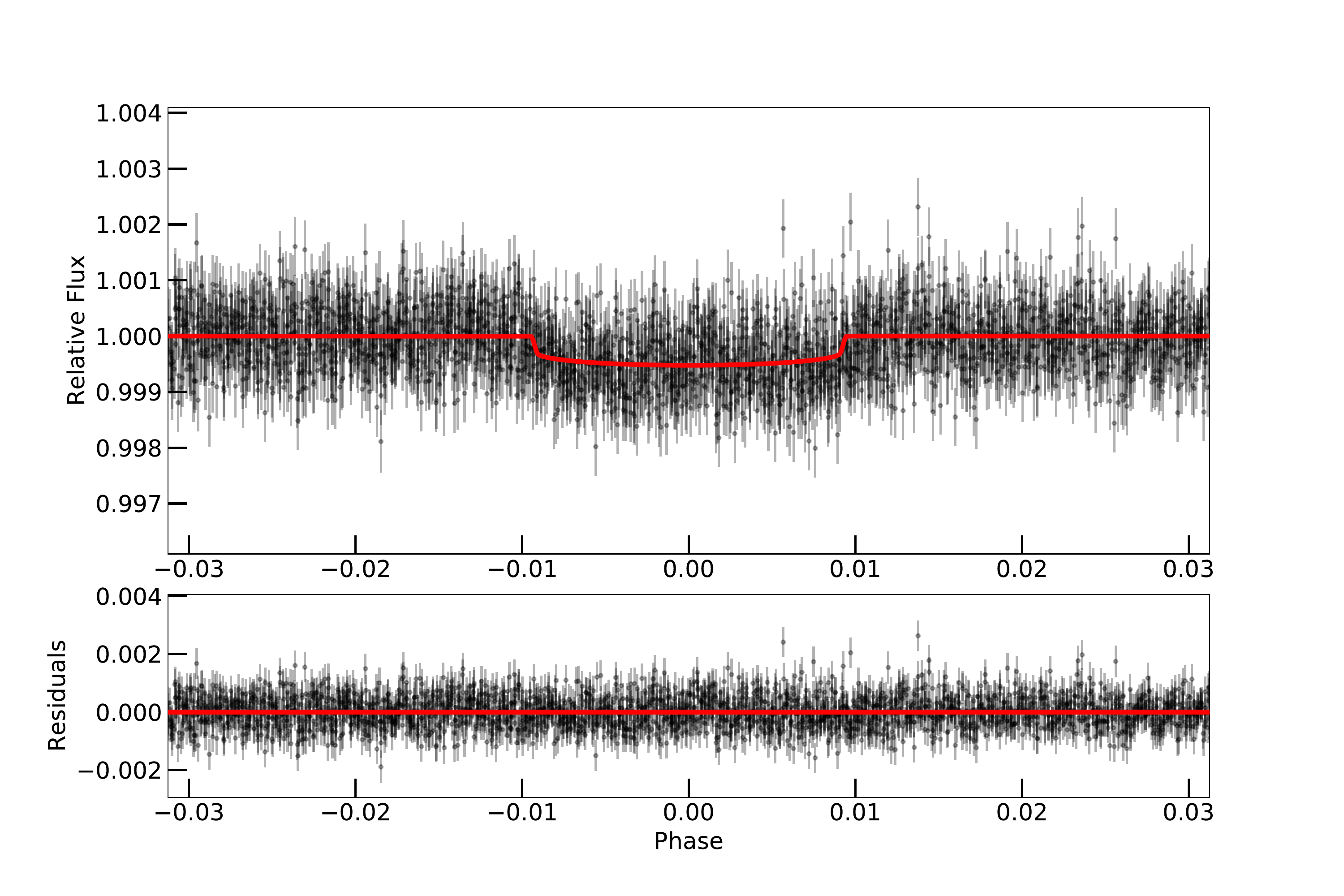}
\figsetgrpnote{The phase-folded curve of TOI 1233.03. The best-fit transit model is shown as a red solid line. The residuals are plotted below.}
\figsetgrpend

\figsetgrpstart
\figsetgrpnum{12.26}
\figsetgrptitle{TOI-1233.04}
\figsetplot{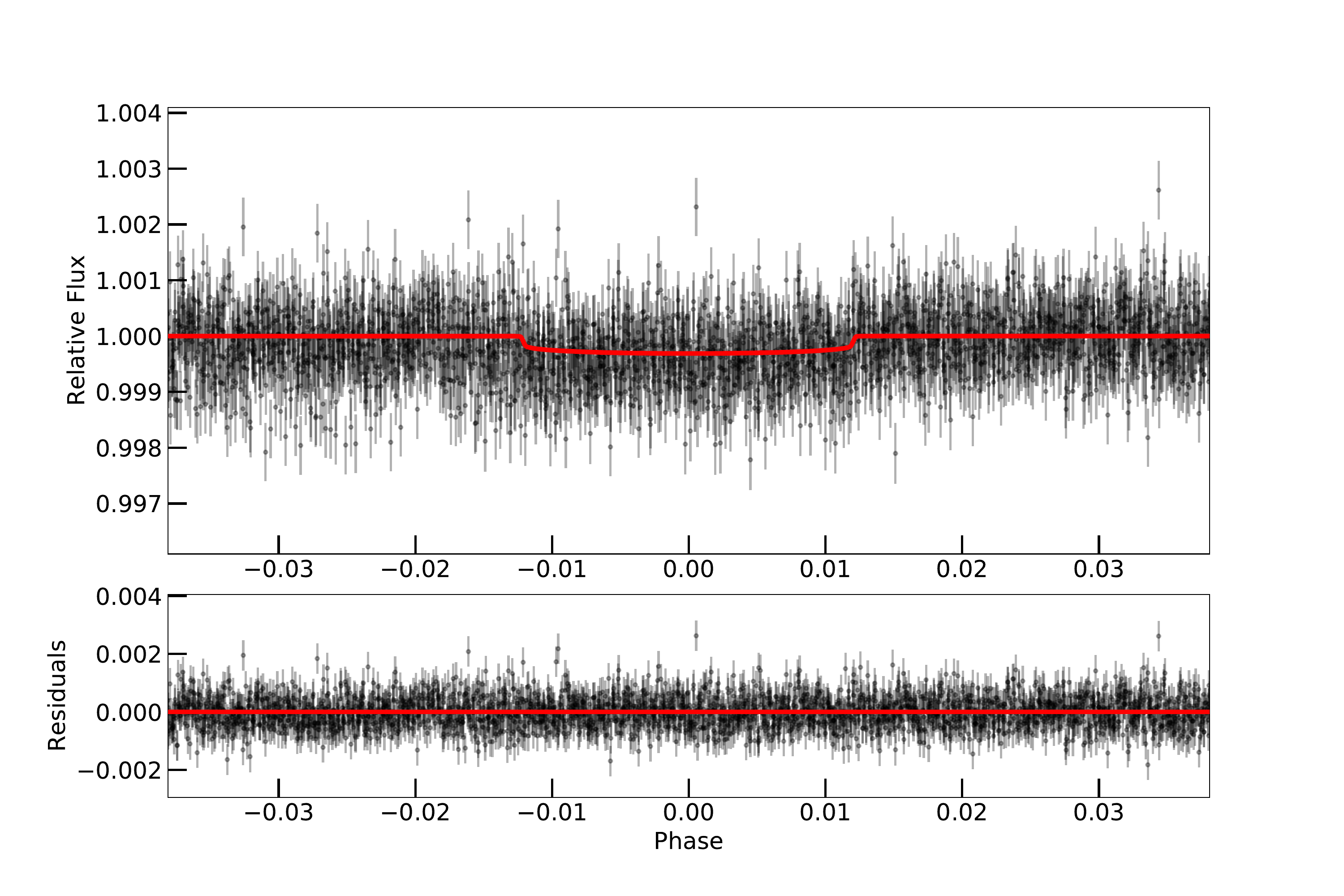}
\figsetgrpnote{The phase-folded curve of TOI 1233.04. The best-fit transit model is shown as a red solid line. The residuals are plotted below.}
\figsetgrpend

\figsetgrpstart
\figsetgrpnum{12.27}
\figsetgrptitle{TOI-1233.05}
\figsetplot{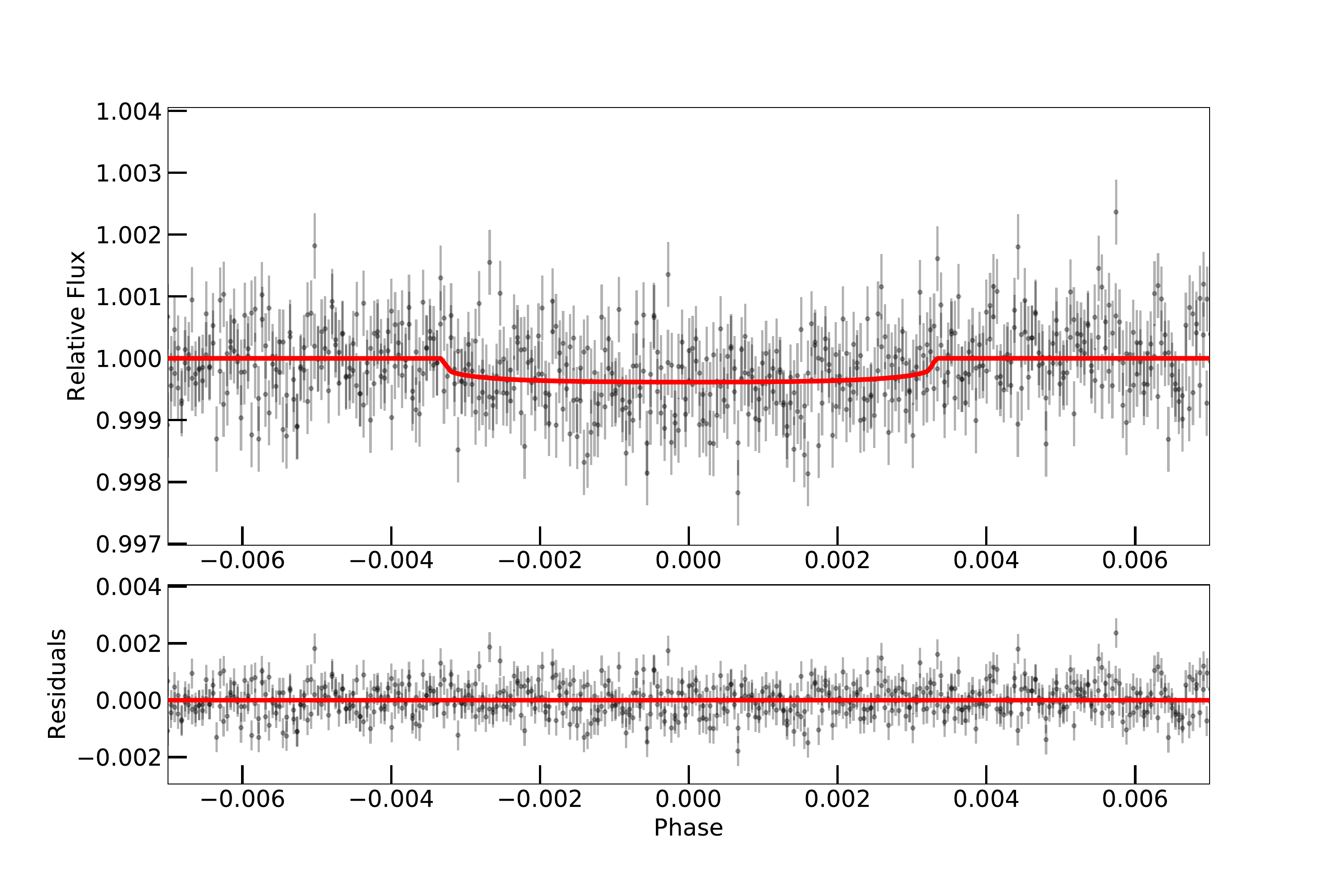}
\figsetgrpnote{The phase-folded curve of TOI 1233.05. The best-fit transit model is shown as a red solid line. The residuals are plotted below.}
\figsetgrpend

\figsetend

\begin{figure*}
\figurenum{12}
\plotone{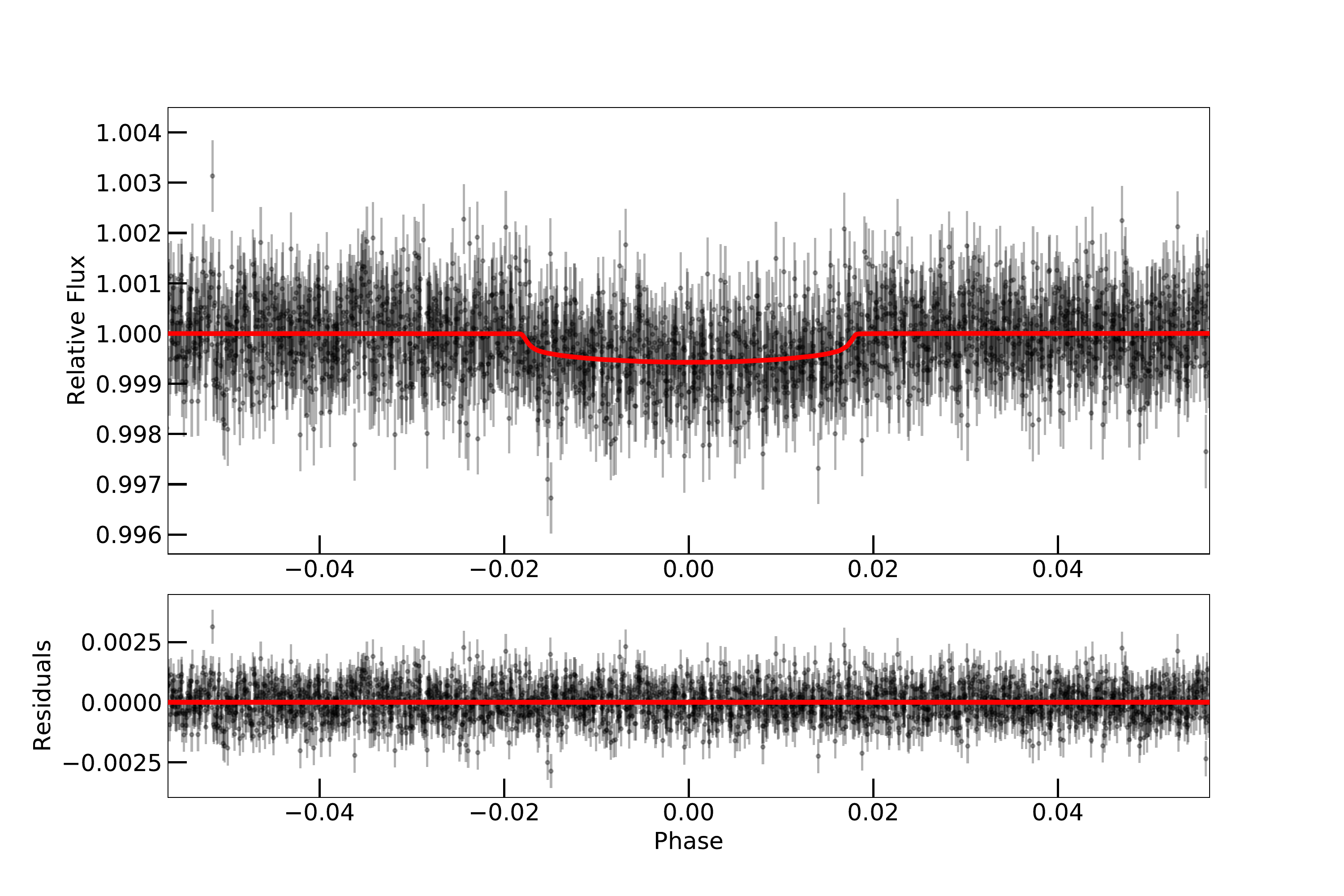}
\caption{The phase-folded curve of TOI-134.01. The best-fit transit model is shown as a red solid line. The residuals are plotted below.}
\end{figure*}

\clearpage

\newpage

\section{Radial Velocity Fits}\label{sec:appendix_rv_fits}
\setcounter{figure}{12}

\begin{figure*}[htbp]
 \begin{minipage}[t]{0.8\columnwidth}
 \vspace{20pt}
\includegraphics[width=\linewidth]{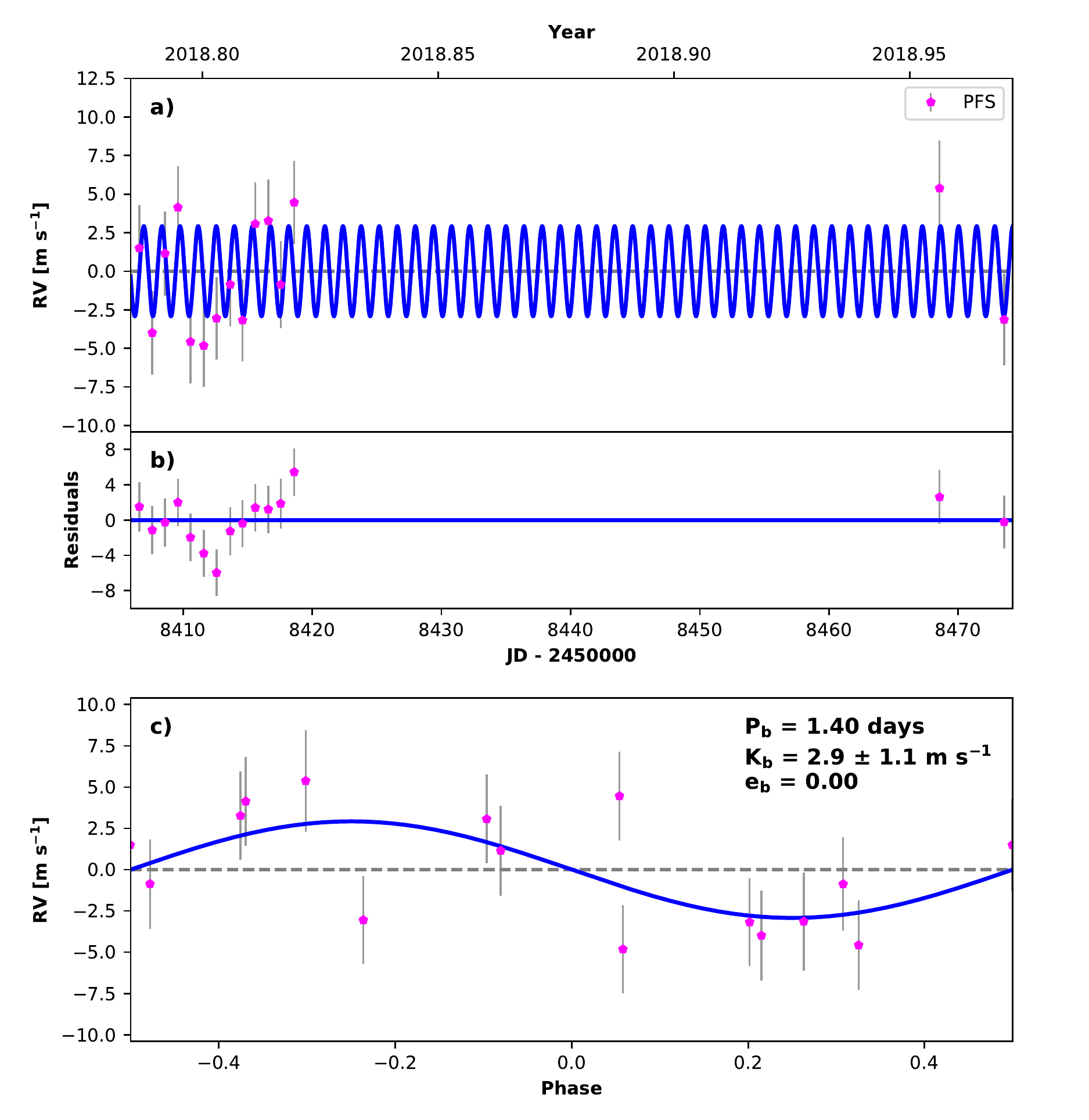}
\end{minipage}
 \begin{minipage}[t]{0.8\columnwidth}
\vspace{10pt}
\includegraphics[width=\linewidth]{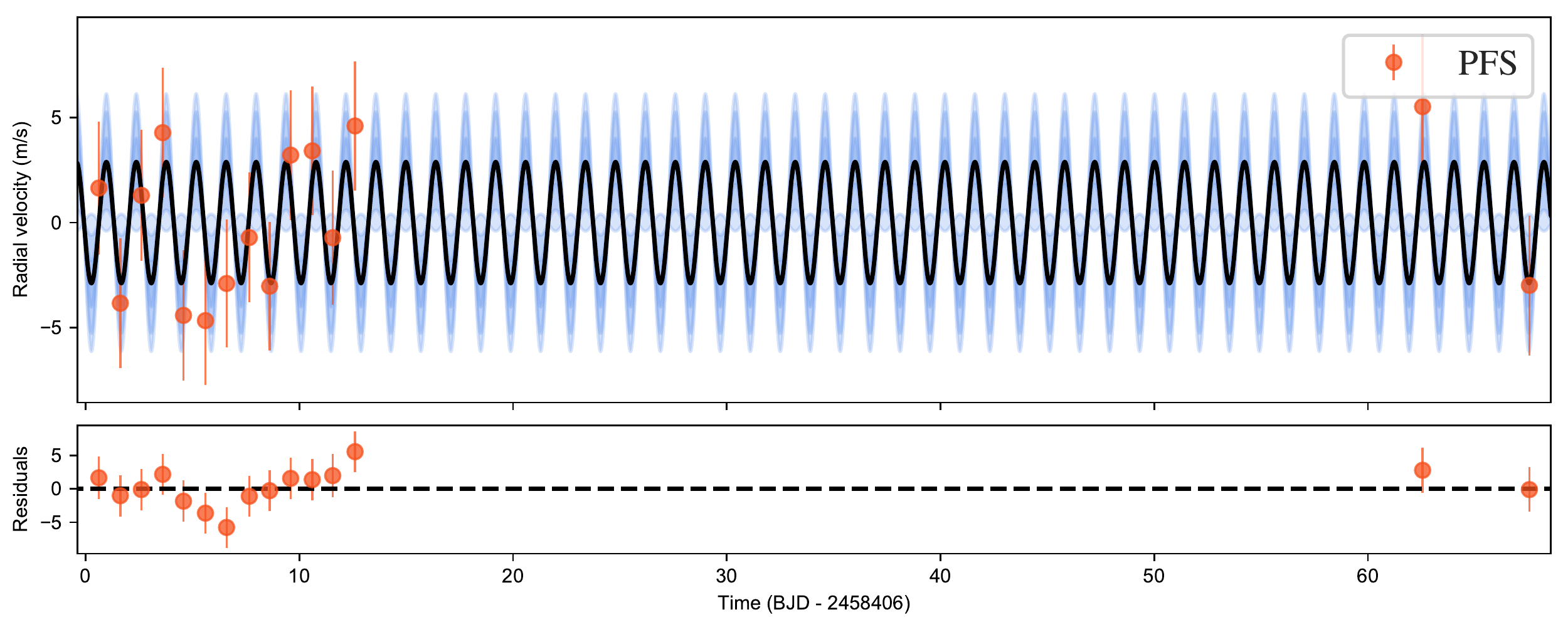} \\
\includegraphics[width=0.9\linewidth]{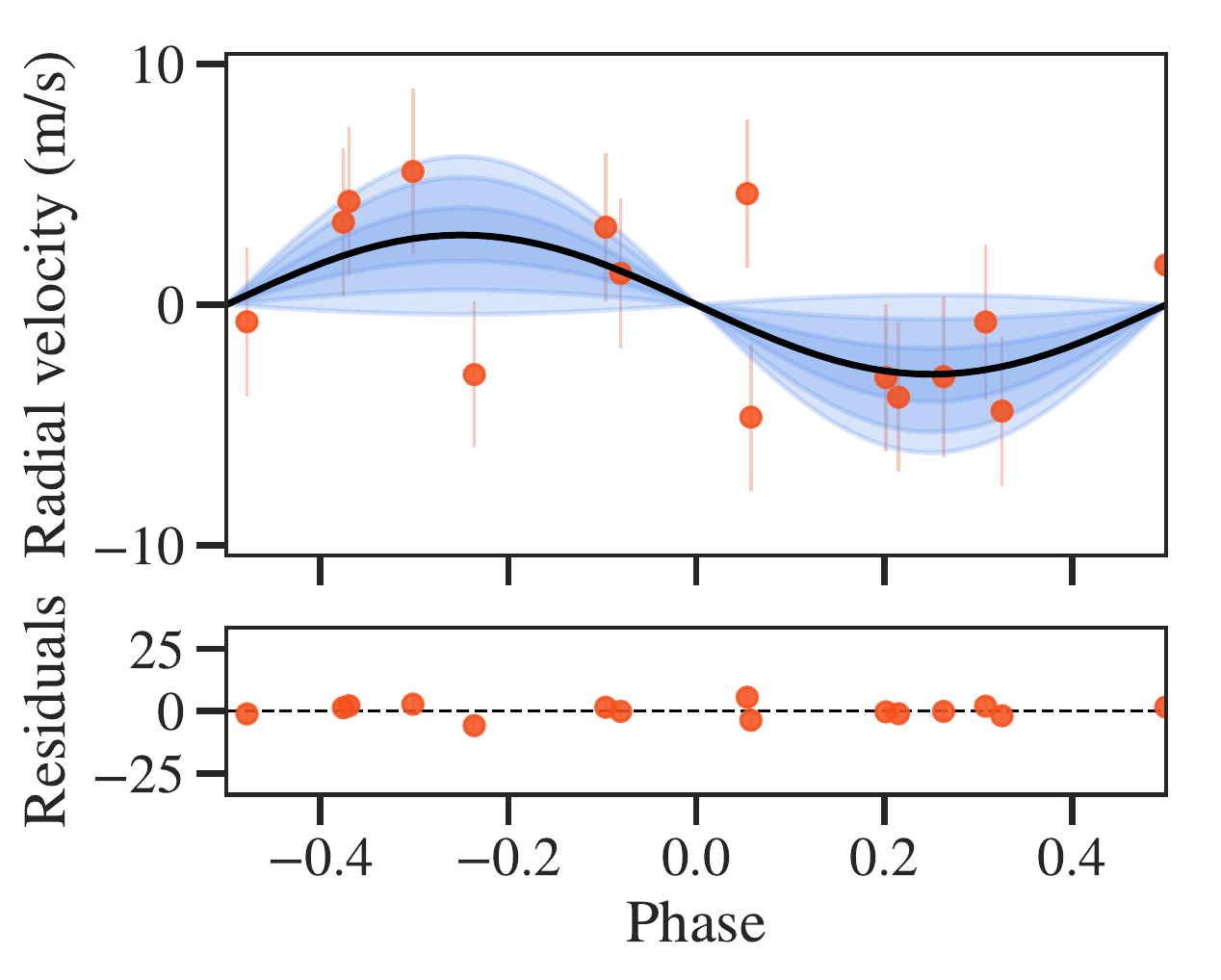} 
\end{minipage}
\caption{The best fit RV models (as listed in Table \ref{tab:pfs_K}) for TOI-134. The left/right column shows the \texttt{Radvel}/\texttt{juliet} results.}
\label{TOI134_rv_plots}
\end{figure*}

\begin{figure*}[htbp]
 \begin{minipage}[t]{0.8\columnwidth}
 \vspace{20pt}
\includegraphics[width=\linewidth]{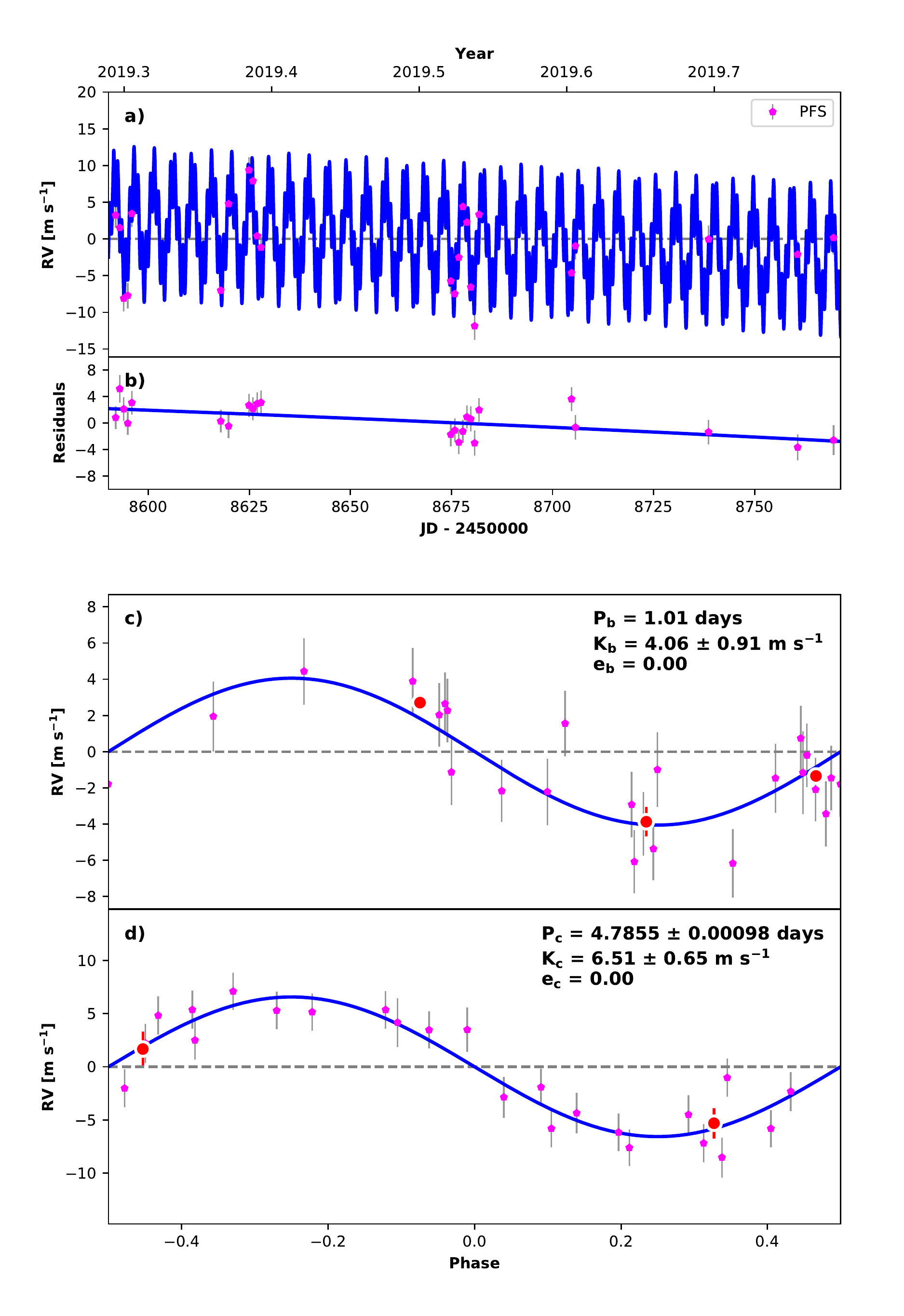}
\end{minipage}
 \begin{minipage}[t]{0.8\columnwidth}
\vspace{80pt}
\includegraphics[width=\linewidth]{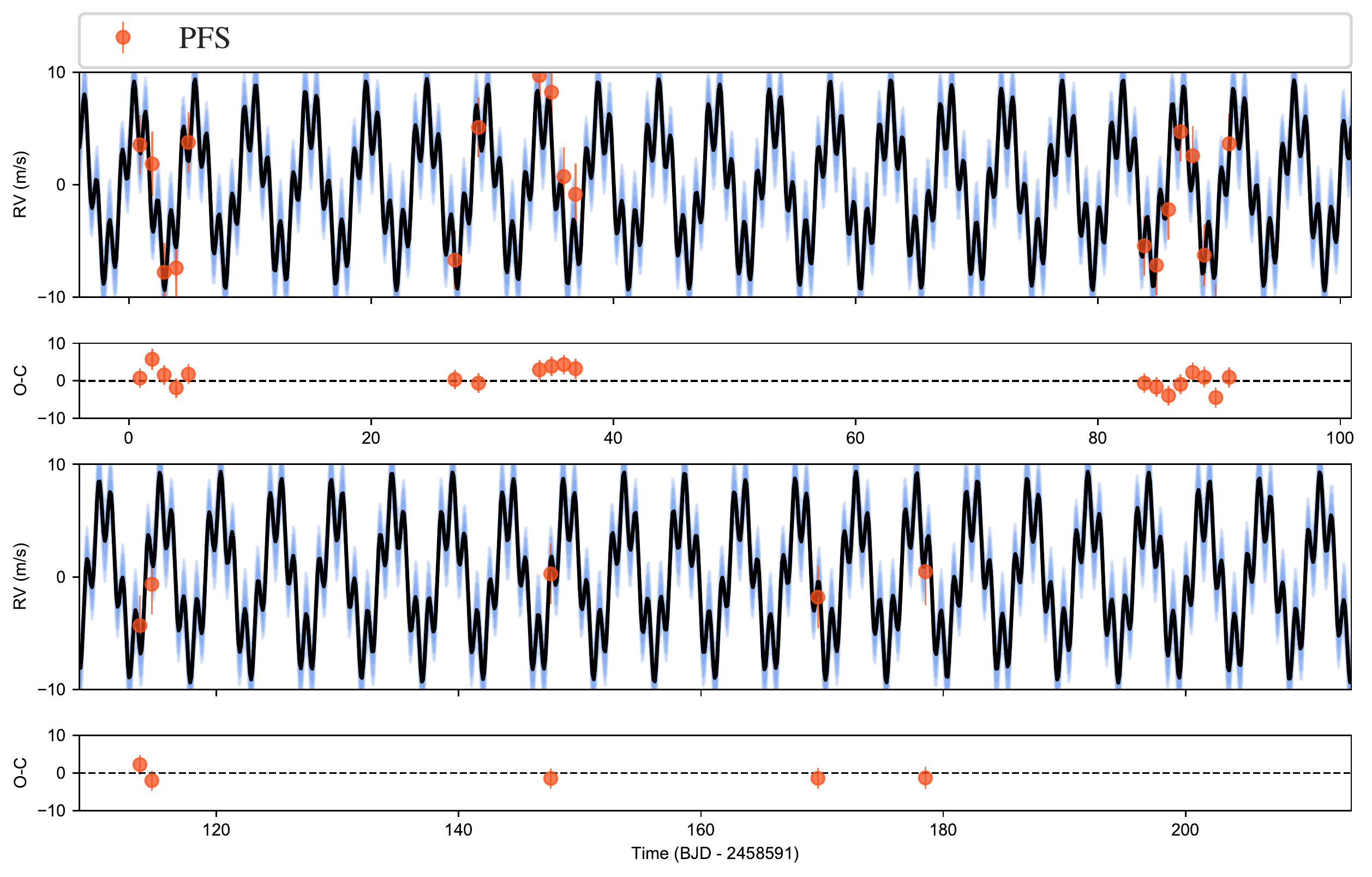} \\
\includegraphics[width=\linewidth]{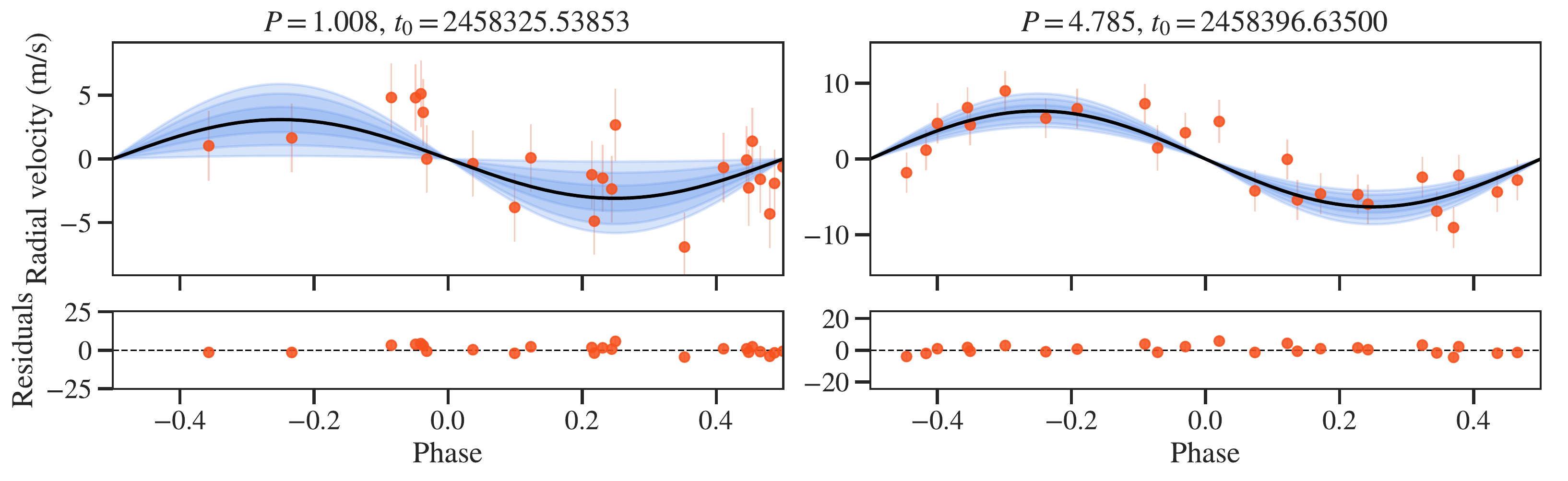} 
\end{minipage}
\caption{The best fit RV models (as listed in Table \ref{tab:pfs_K}) for TOI-141. The left/right column shows the \texttt{Radvel}/\texttt{juliet} results.}
\label{TOI141_rv_plots}
\end{figure*}

\begin{figure*}[htbp]
 \begin{minipage}[t]{0.8\columnwidth}
 \vspace{20pt}
\includegraphics[width=\linewidth]{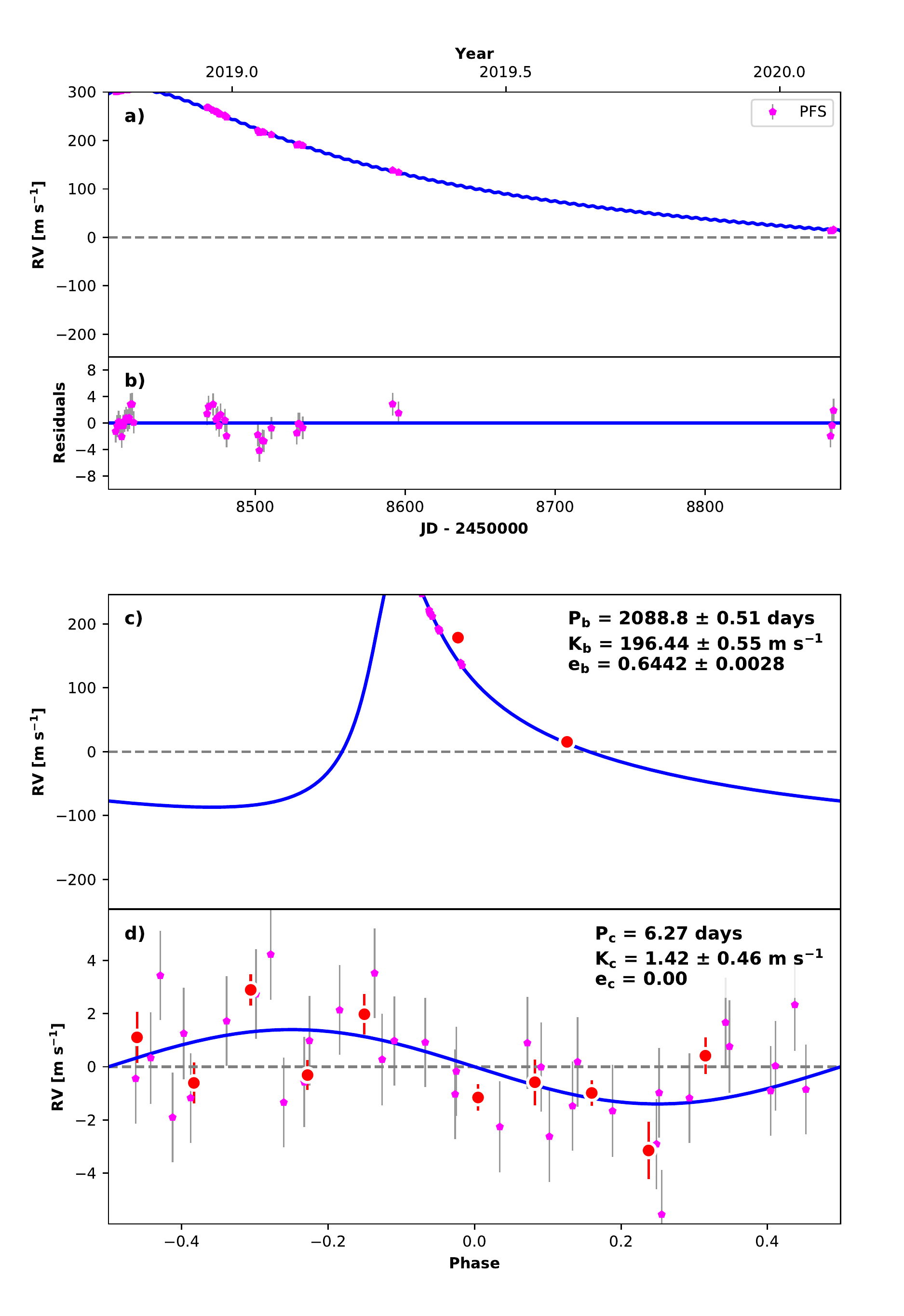}
\end{minipage}
 \begin{minipage}[t]{0.8\columnwidth}
\vspace{40pt}
\includegraphics[width=\linewidth]{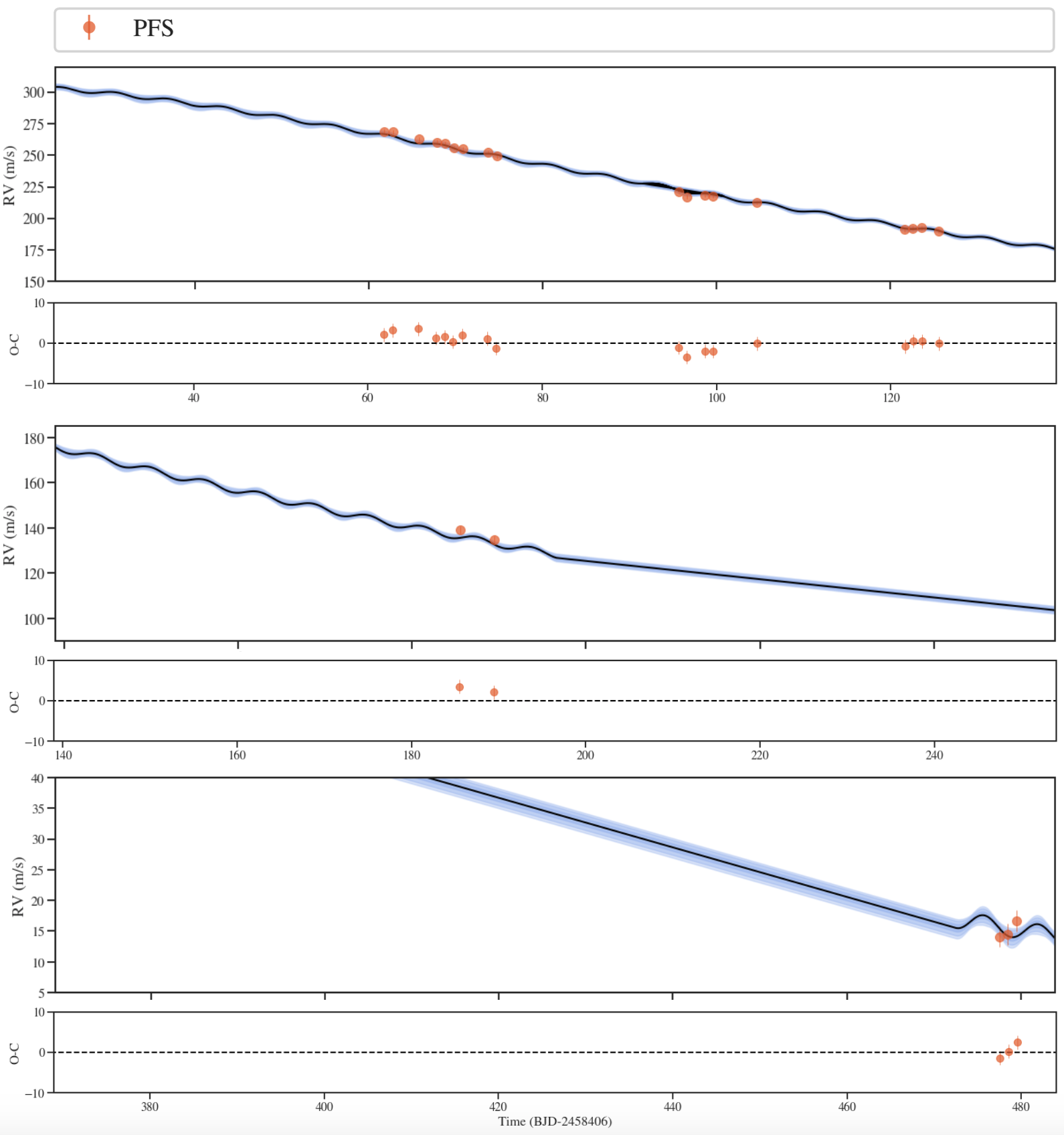} \\
\includegraphics[width=\linewidth]{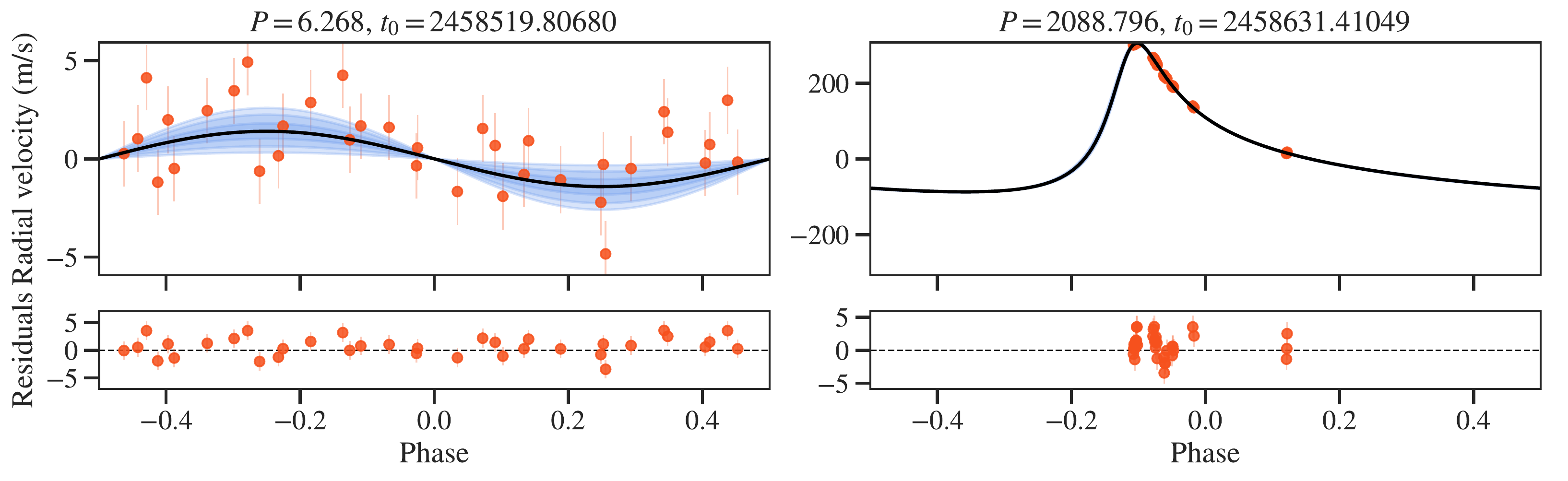} 
\end{minipage}
\caption{The best fit RV models (as listed in Table \ref{tab:pfs_K}) for TOI-144. The left/right column shows the \texttt{Radvel}/\texttt{juliet} results. \\ }
\end{figure*}


\begin{figure*}[htbp]
 \begin{minipage}[t]{0.8\columnwidth}
 \vspace{20pt}
\includegraphics[width=\linewidth]{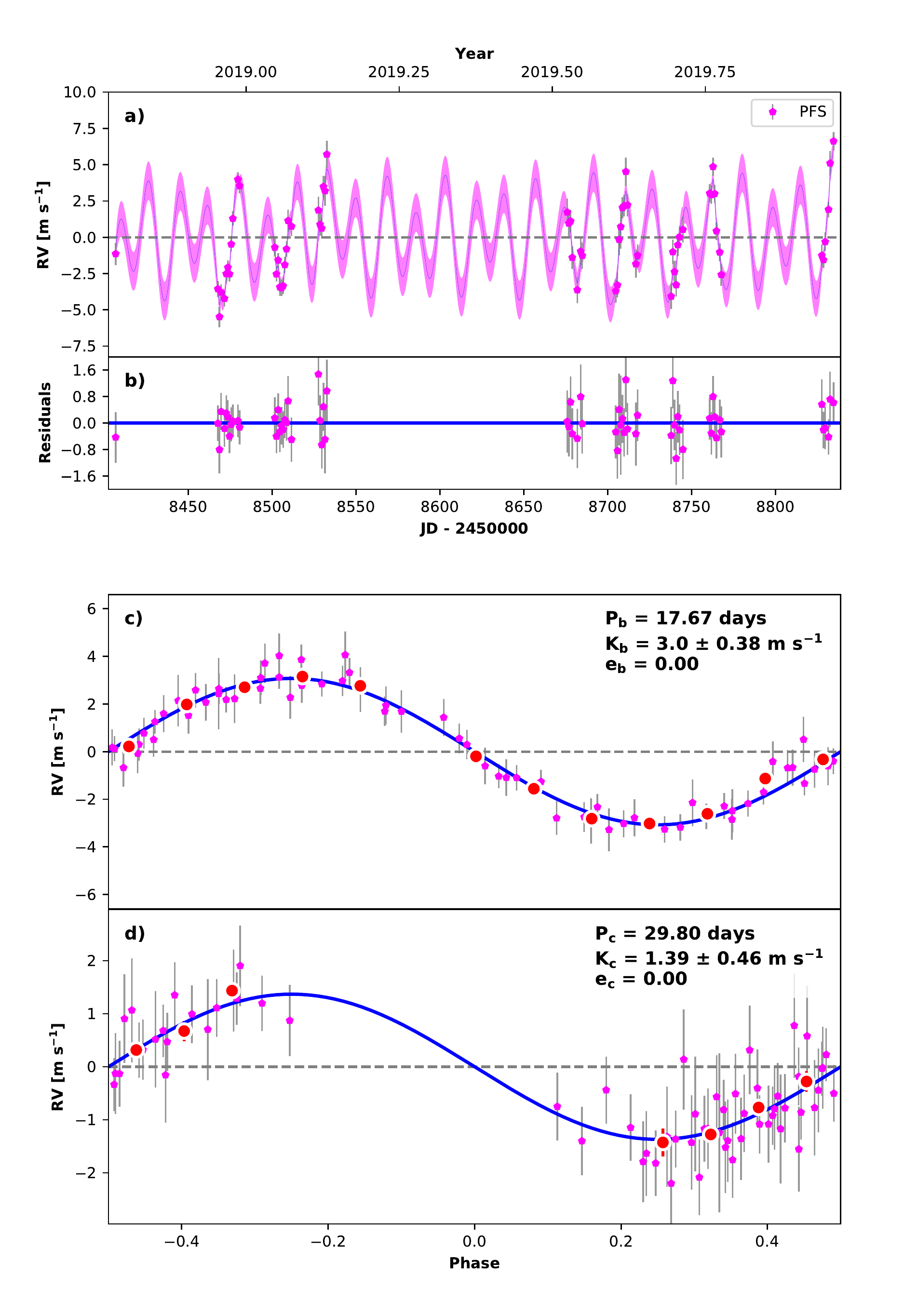}
\end{minipage}
 \begin{minipage}[t]{0.8\columnwidth}
\vspace{40pt}
\includegraphics[width=\linewidth]{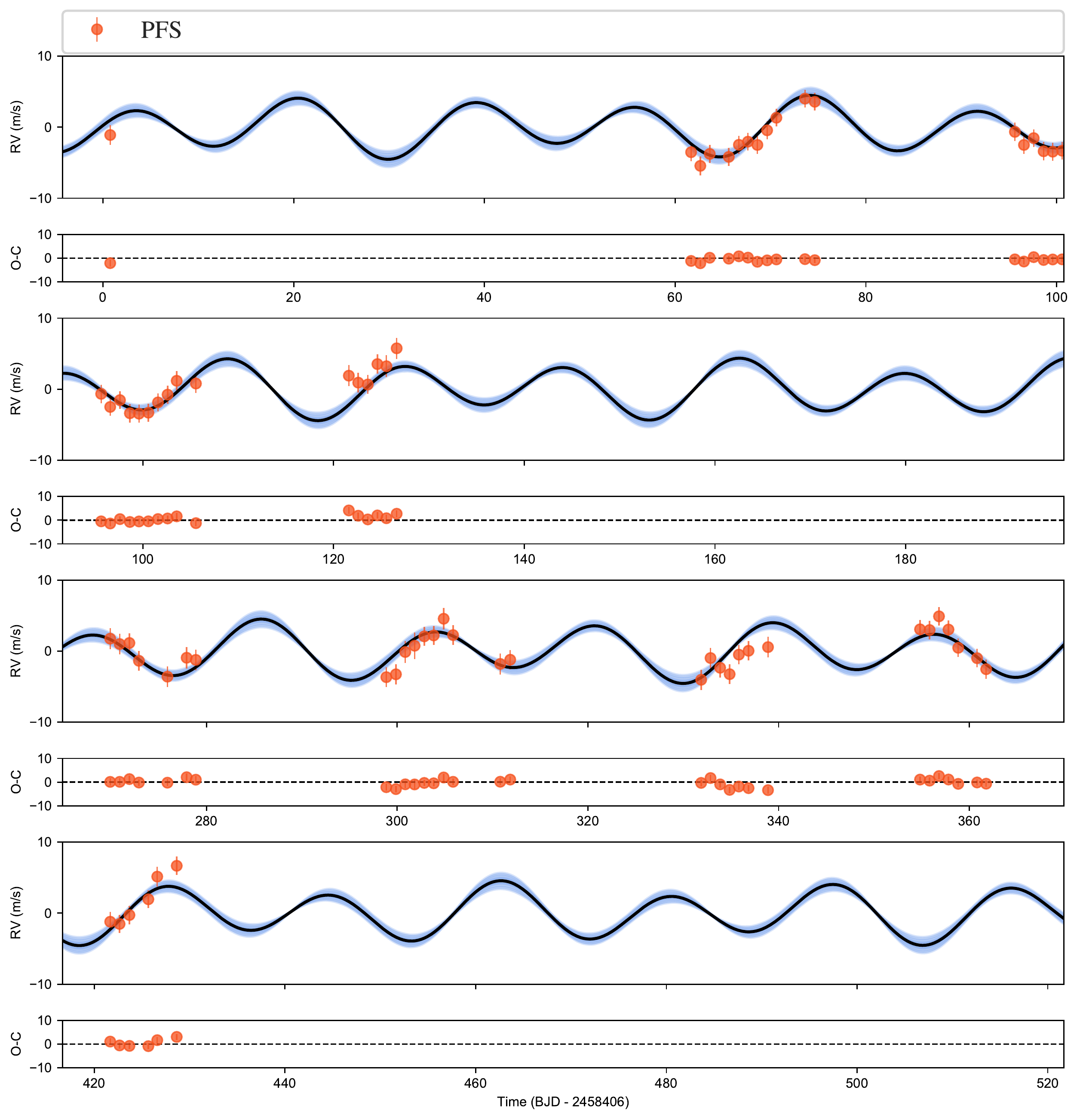} \\
\includegraphics[width=\linewidth]{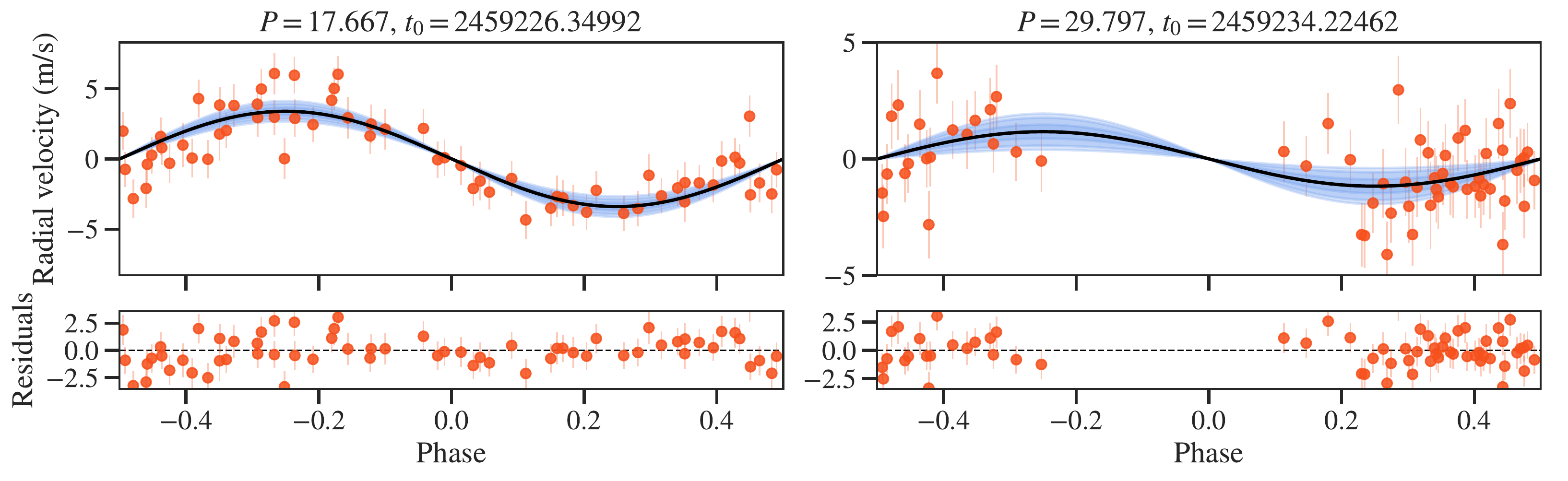} 
\end{minipage}
\caption{The best fit RV models (as listed in Table \ref{tab:pfs_K}) for TOI-174. The left/right column shows the \texttt{Radvel}/\texttt{juliet} results.}
\label{TOI174_rv_plots}
\end{figure*}

\begin{figure*}[htbp]
 \begin{minipage}[t]{0.8\columnwidth}
 \vspace{20pt}
\includegraphics[width=\linewidth]{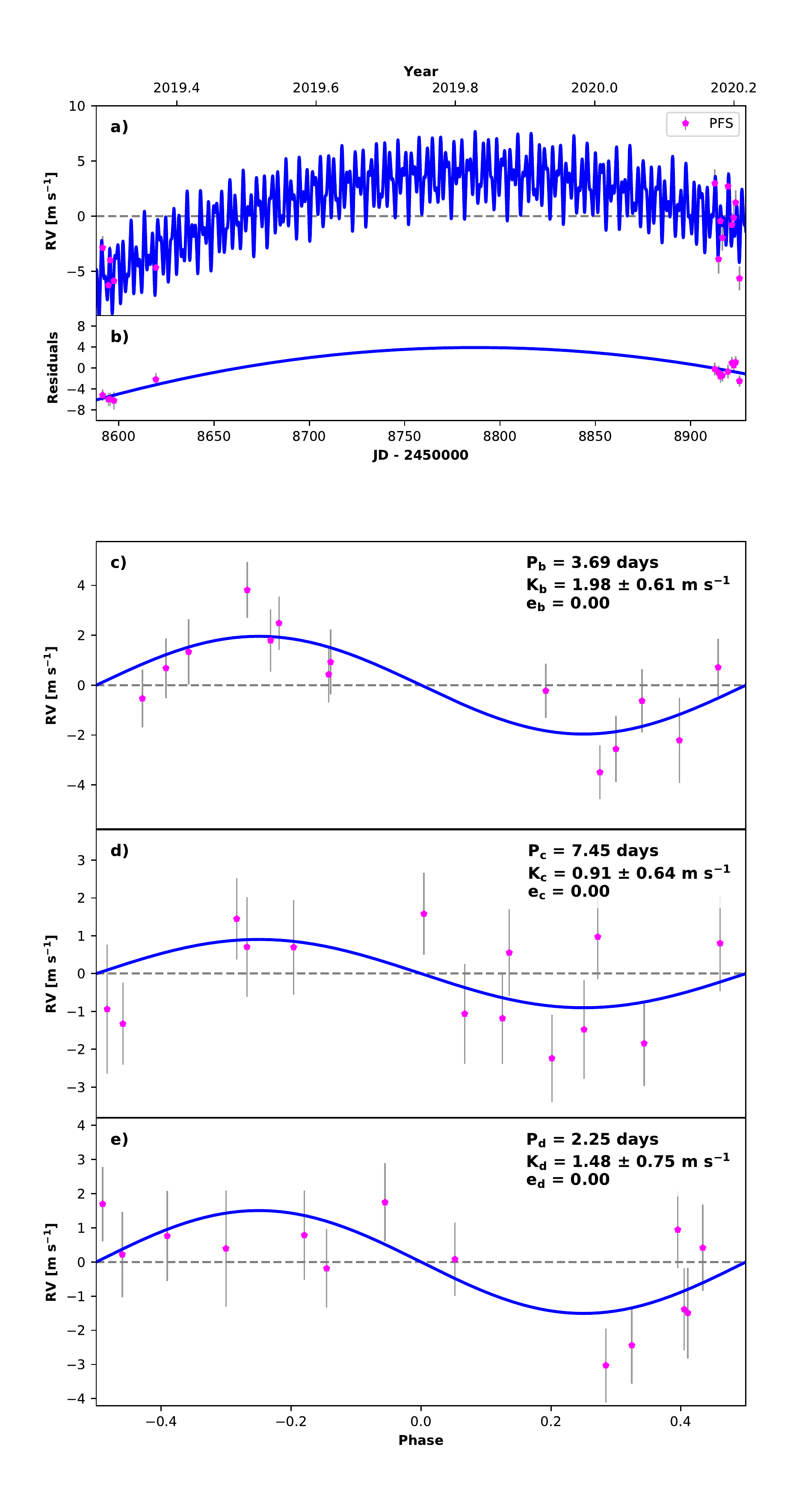}
\end{minipage}
 \begin{minipage}[t]{0.8\columnwidth}
\vspace{90pt}
\includegraphics[width=\linewidth]{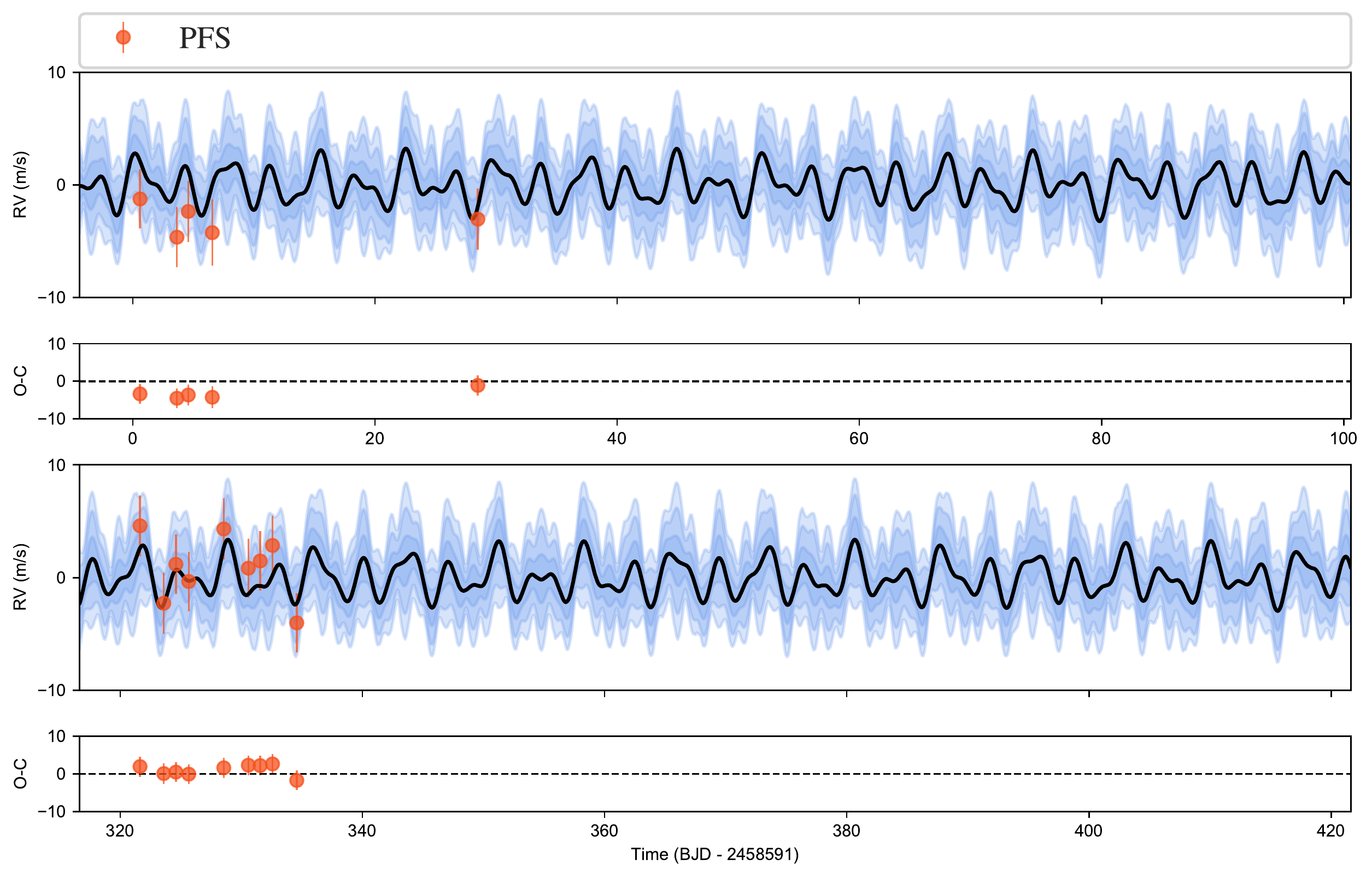} \\
\includegraphics[width=\linewidth]{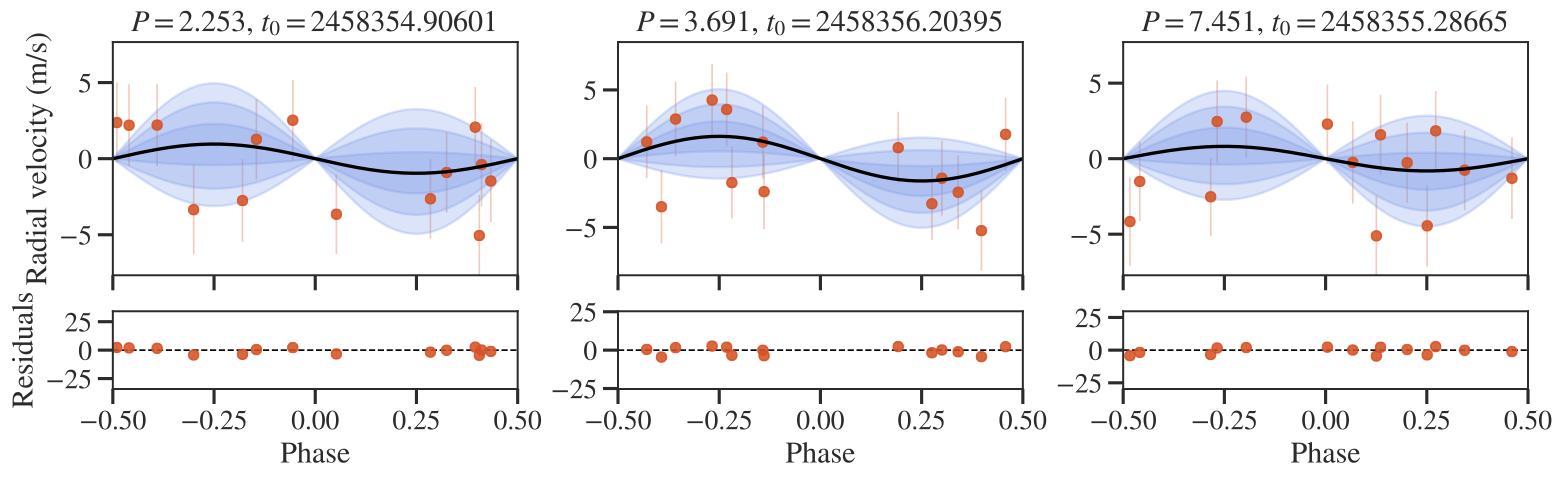} 
\end{minipage}
\caption{The best fit RV models (as listed in Table \ref{tab:pfs_K}) for TOI-175. The left/right column shows the \texttt{Radvel}/\texttt{juliet} results.}
\label{TOI175_rv_plots}
\end{figure*}

\begin{figure*}[htbp]
 \begin{minipage}[t]{0.8\columnwidth}
 \vspace{20pt}
\includegraphics[width=\linewidth]{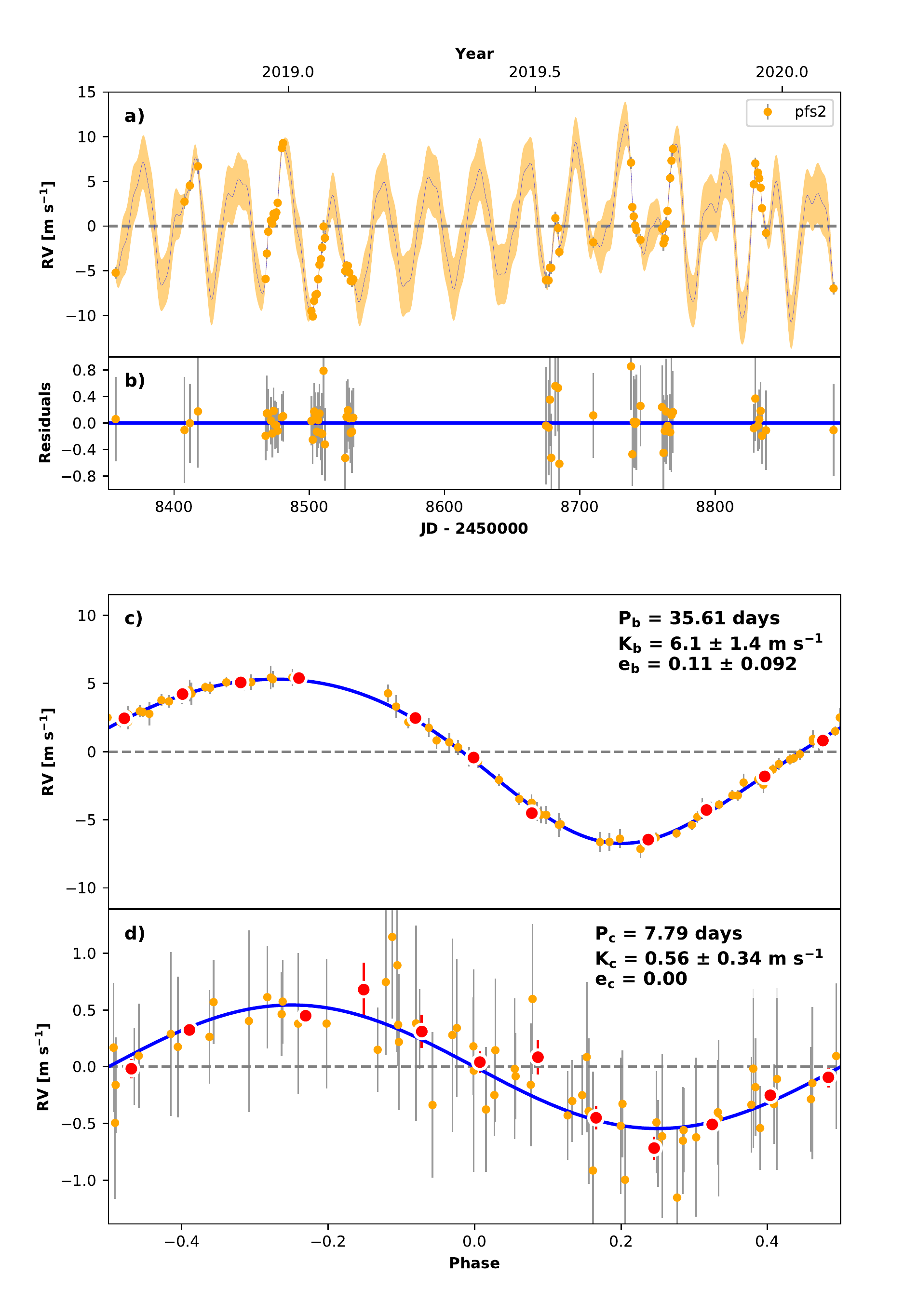}
\end{minipage}
 \begin{minipage}[t]{0.8\columnwidth}
\vspace{40pt}
\includegraphics[width=\linewidth]{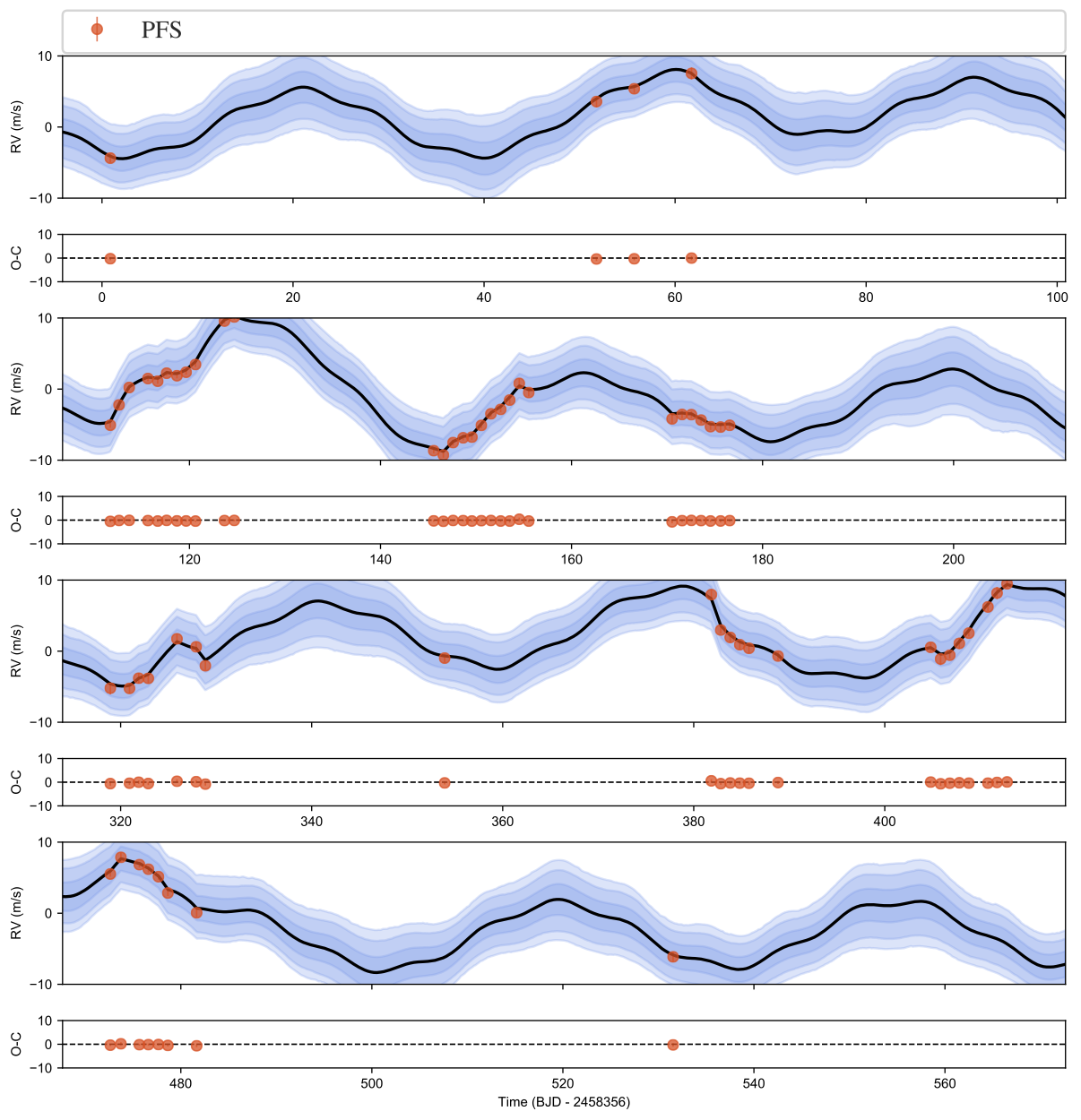} \\
\includegraphics[width=\linewidth]{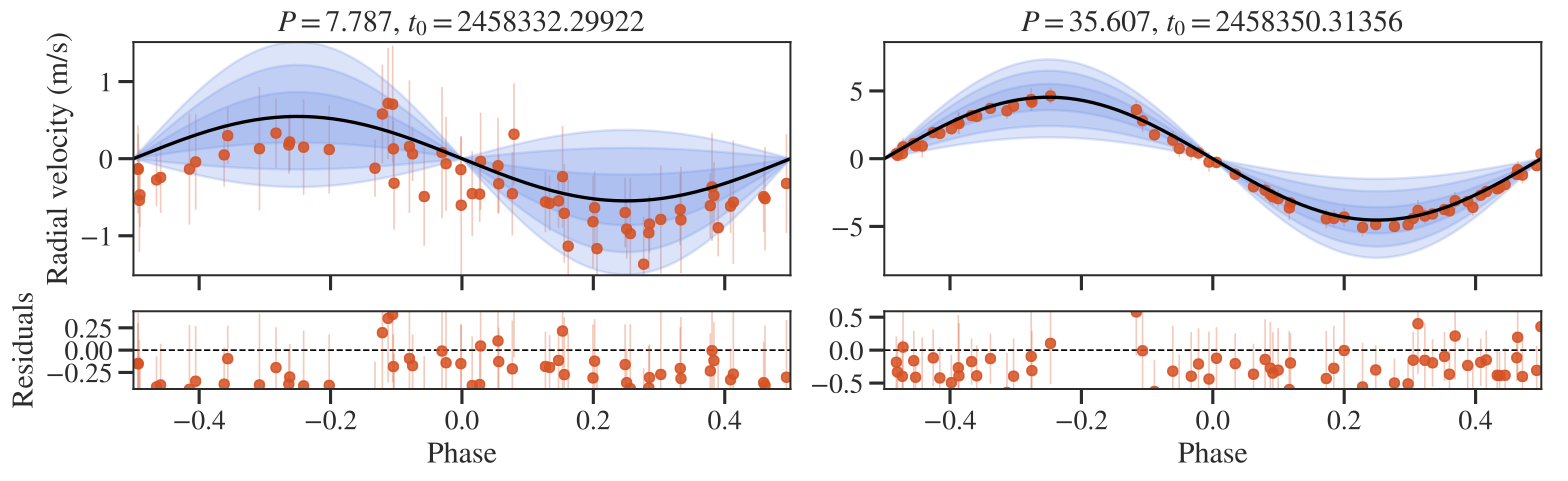} 
\end{minipage}
\caption{The best fit RV models (as listed in Table \ref{tab:pfs_K}) for TOI-186. The left/right column shows the \texttt{Radvel}/\texttt{juliet} results.}
\label{TOI186_rv_plots}
\end{figure*}

\begin{figure*}[htbp]
 \begin{minipage}[t]{0.8\columnwidth}
\vspace{30pt}
\includegraphics[width=\linewidth]{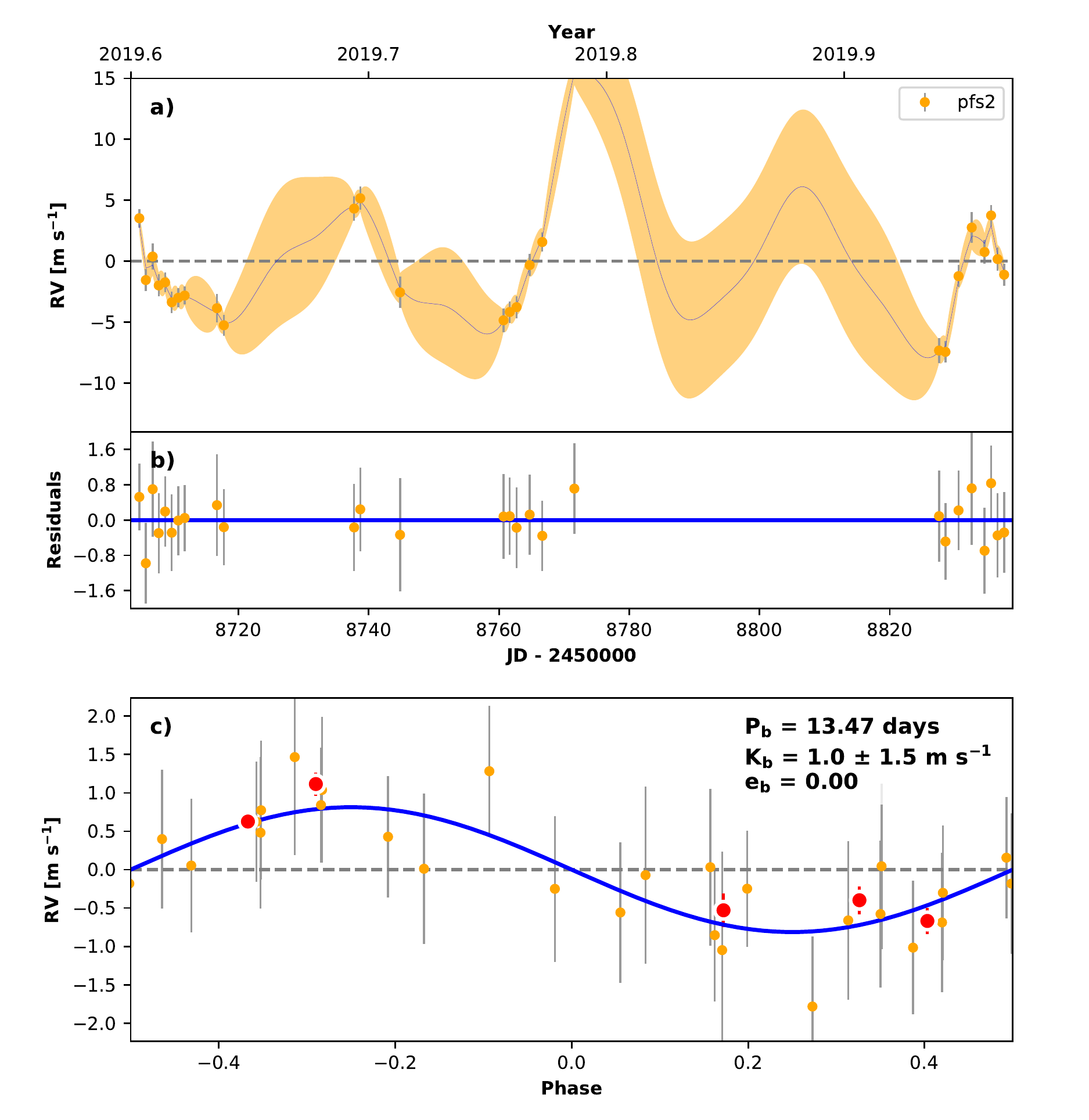}
\end{minipage}
 \begin{minipage}[t]{0.8\columnwidth}
\vspace{40pt}
\includegraphics[width=\linewidth]{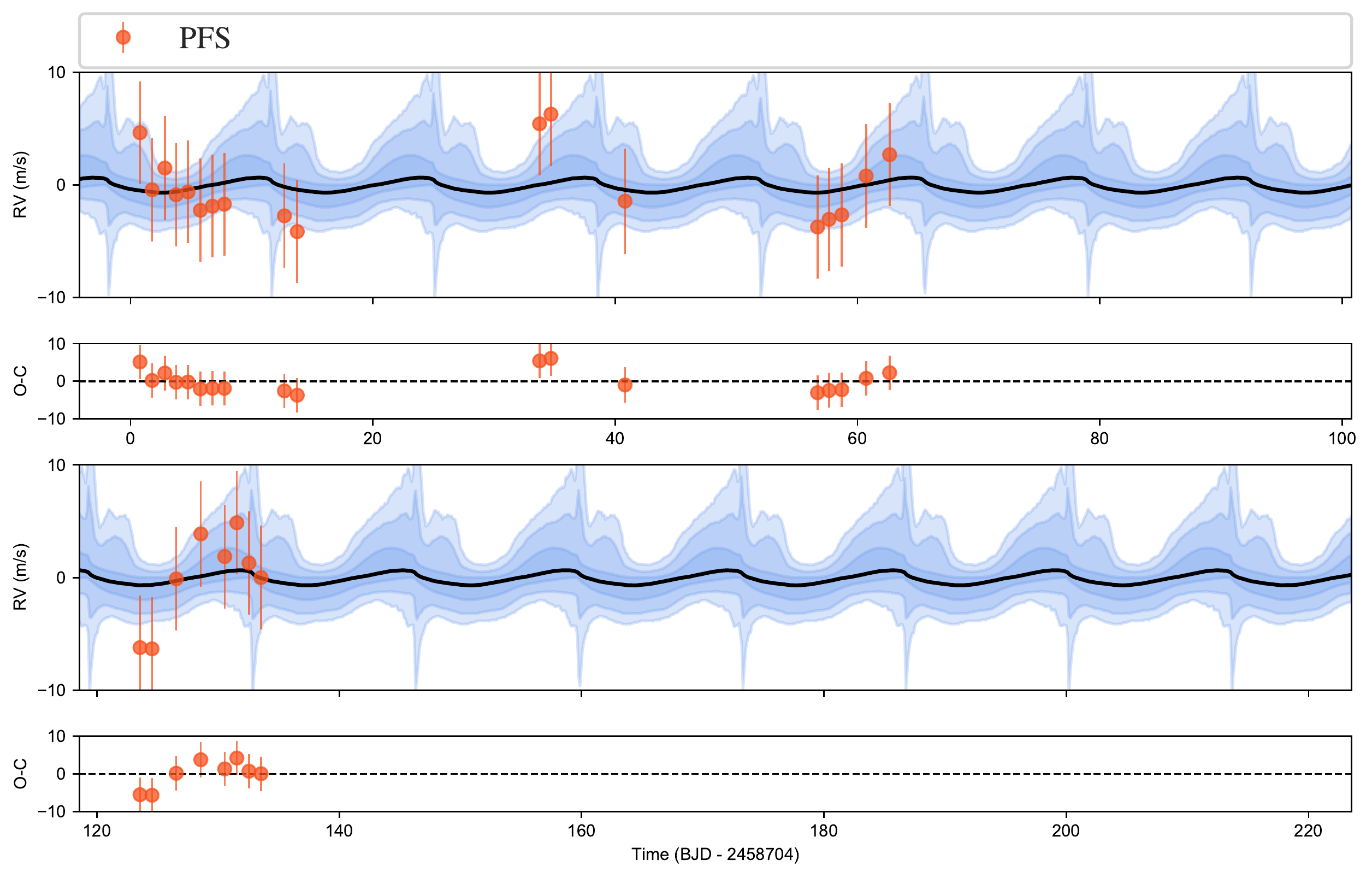} \\
\includegraphics[width=0.9\linewidth]{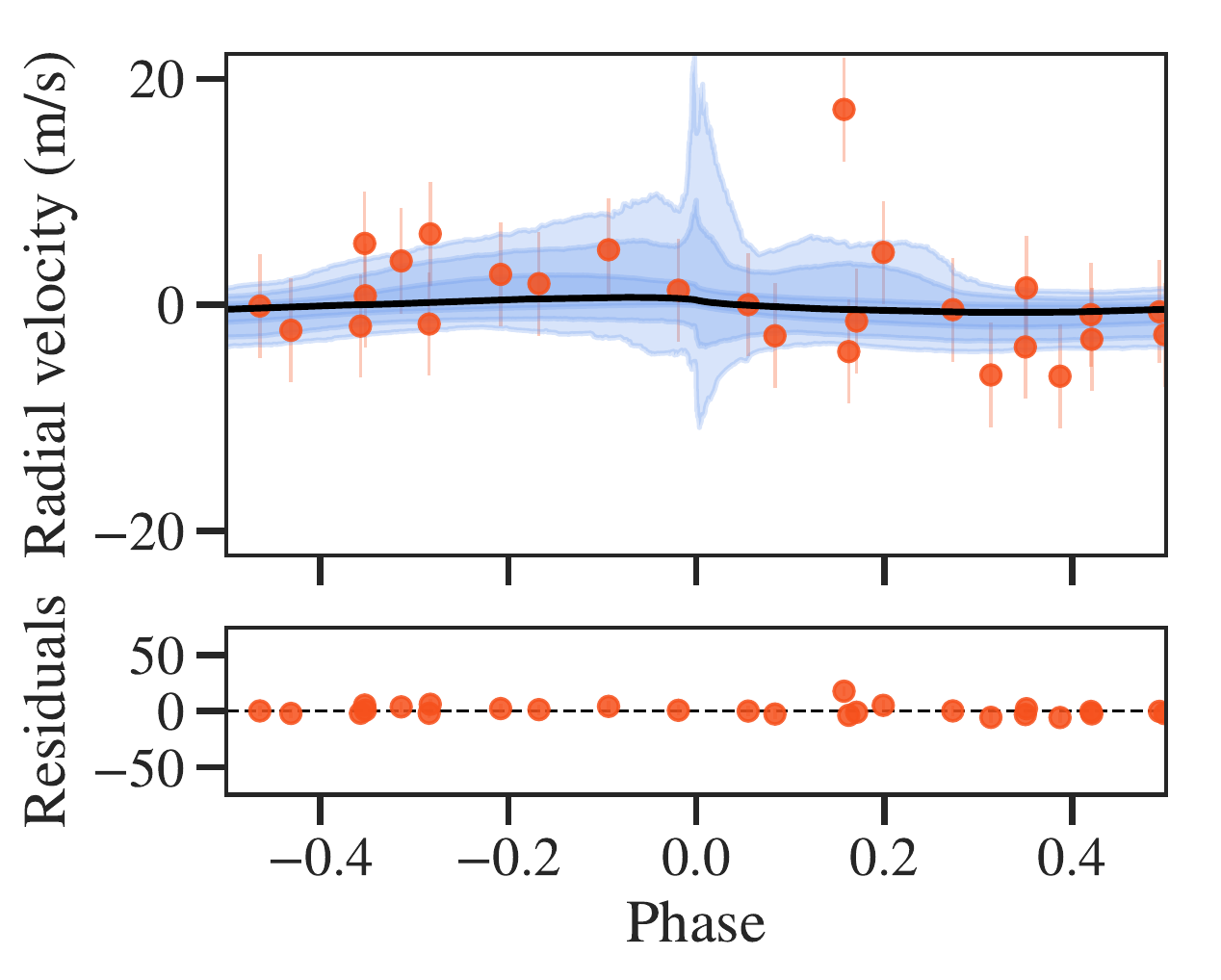} 
\end{minipage}
\caption{The best fit RV models (as listed in Table \ref{tab:pfs_K}) for TOI-260. The left/right column shows the \texttt{Radvel}/\texttt{juliet} results.}
\label{TOI260_rv_plots}
\end{figure*}

\begin{figure*}[htbp]
 \begin{minipage}[t]{0.8\columnwidth}
 \vspace{55pt}
\includegraphics[width=\linewidth]{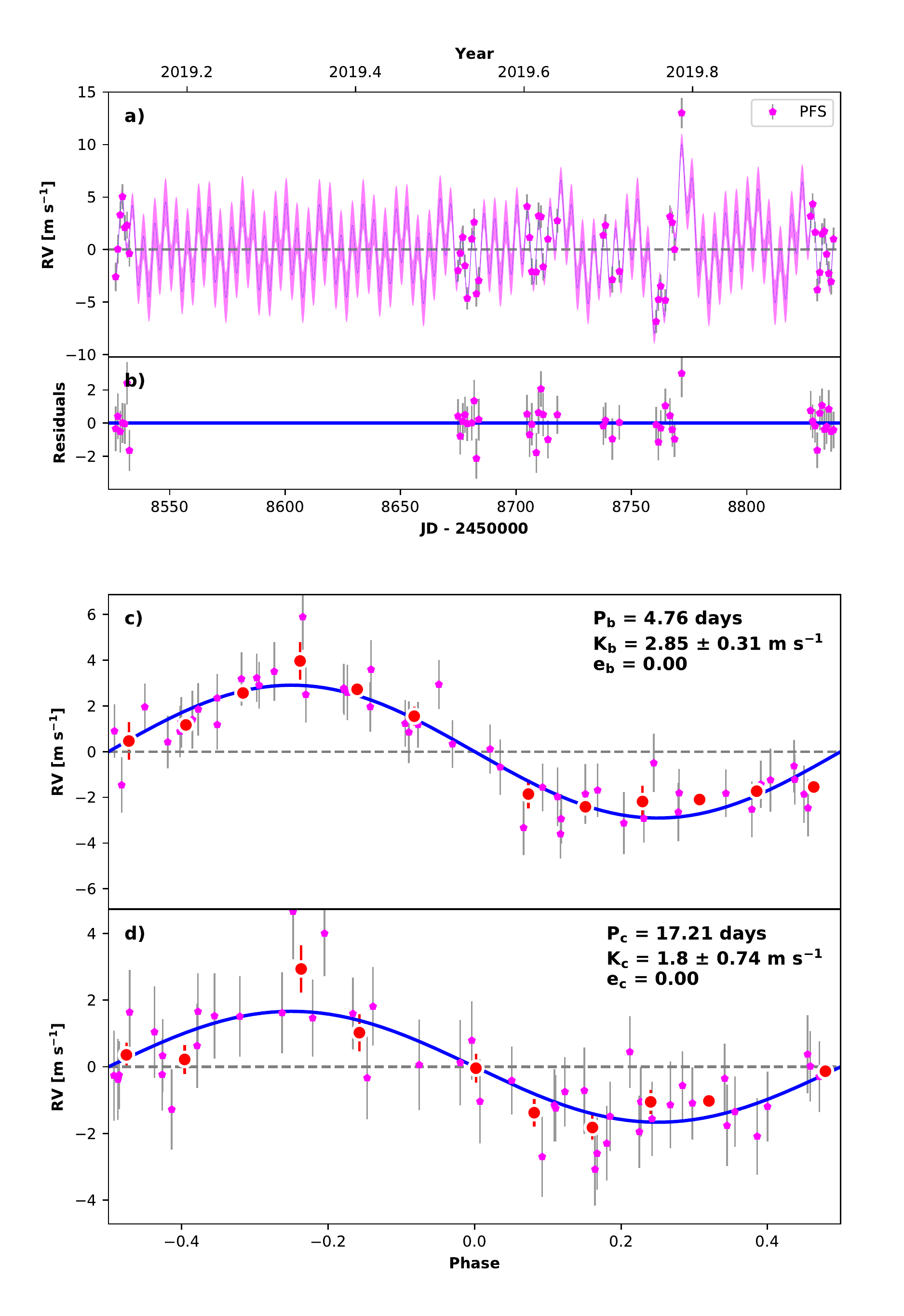}
\end{minipage}
 \begin{minipage}[t]{0.8\columnwidth}
\vspace{40pt}
\includegraphics[width=\linewidth]{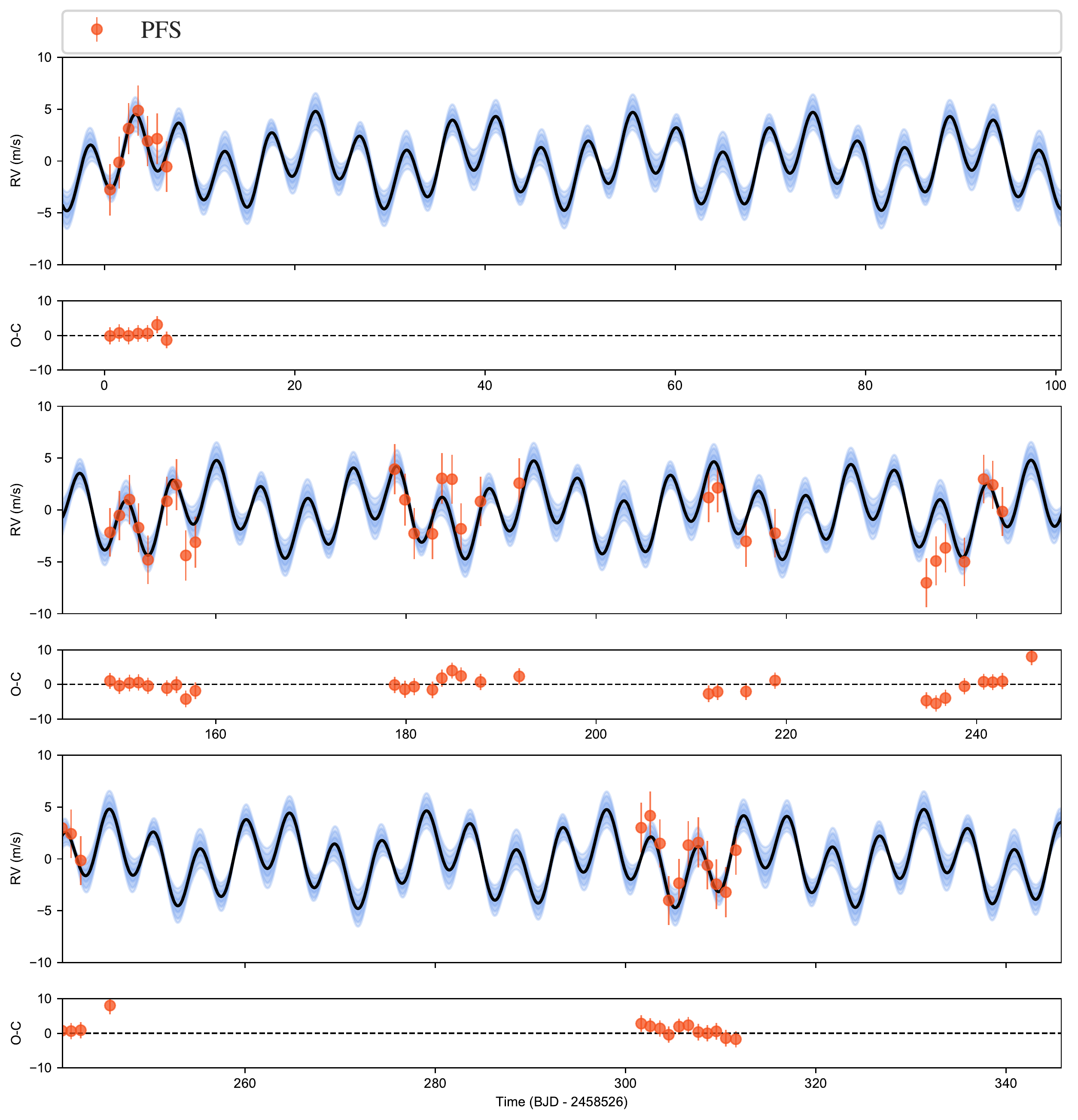} \\
\includegraphics[width=\linewidth]{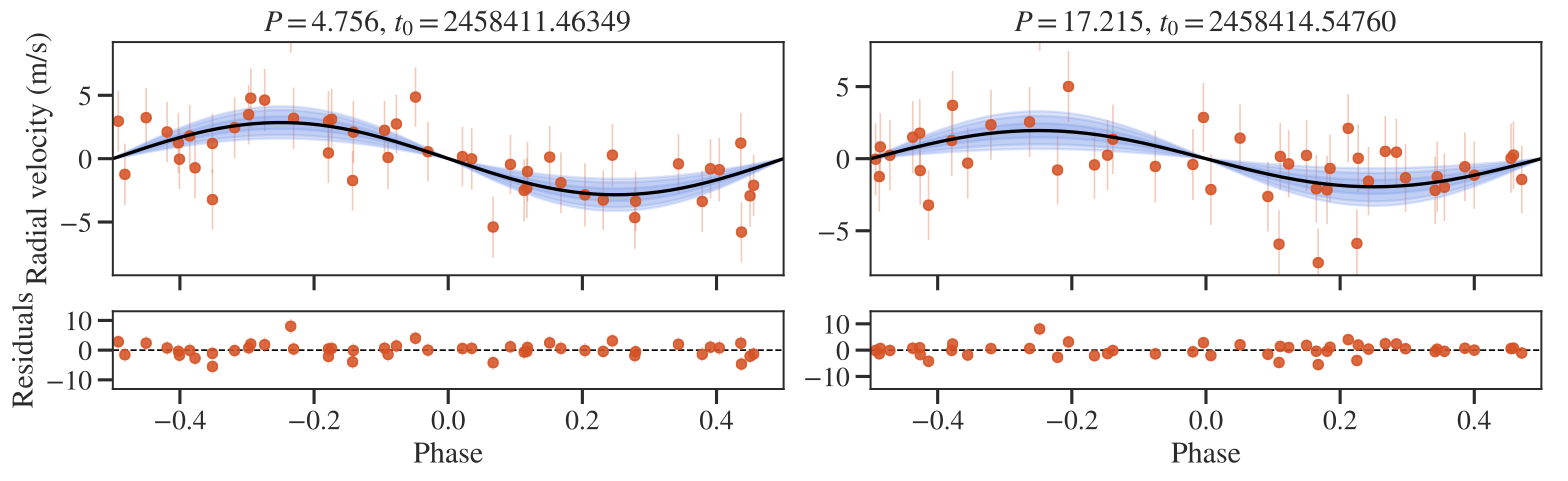} 
\end{minipage}
\caption{The best fit RV models (as listed in Table \ref{tab:pfs_K}) for TOI-402. The left/right column shows the \texttt{Radvel}/\texttt{juliet} results.}
\label{TOI420_rv_plots}
\end{figure*}

\begin{figure*}[htbp]
 \begin{minipage}[t]{0.8\columnwidth}
 \vspace{30pt}
\includegraphics[width=\linewidth]{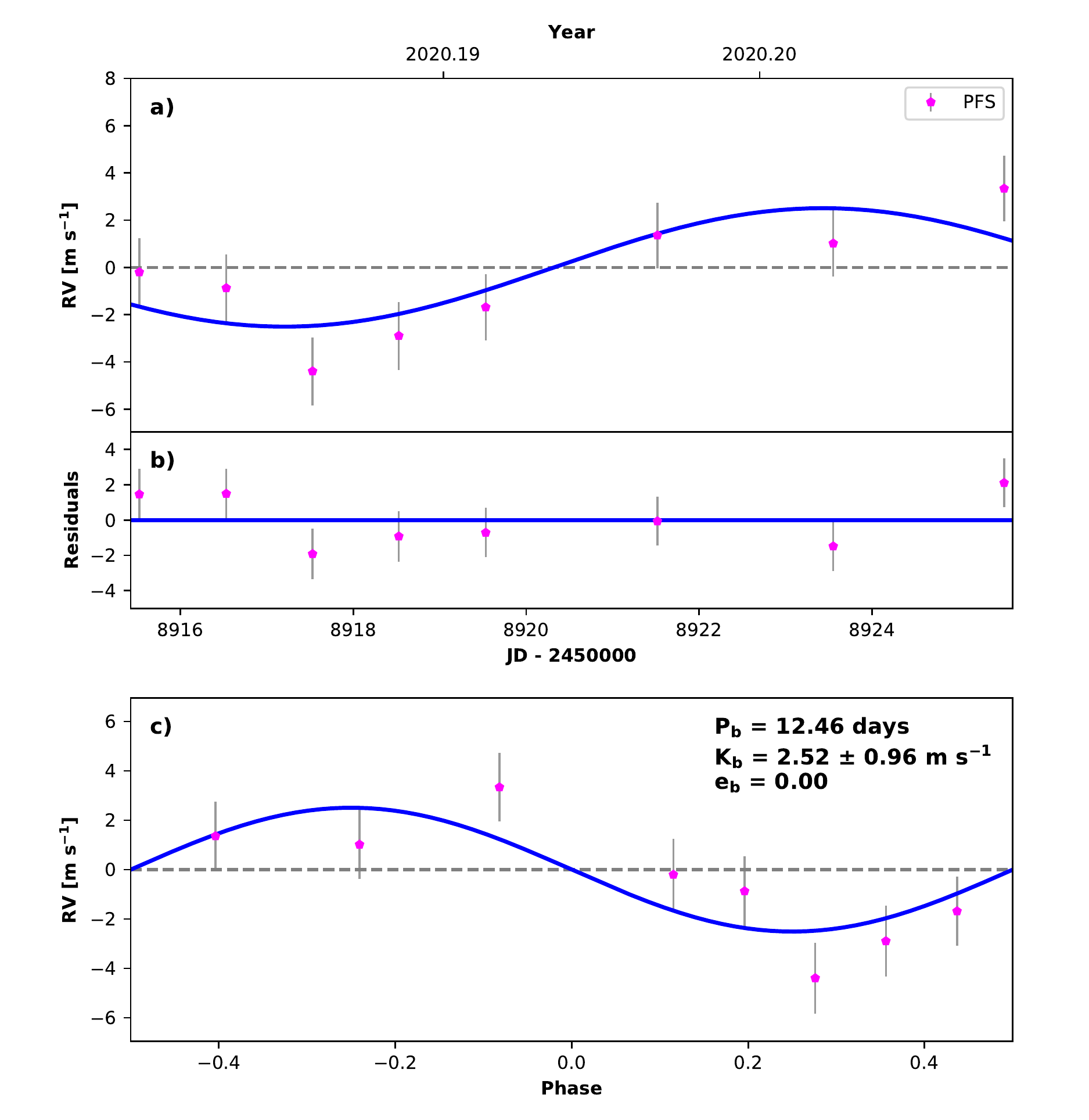}
\end{minipage}
 \begin{minipage}[t]{0.8\columnwidth}
\vspace{40pt}
\includegraphics[width=\linewidth]{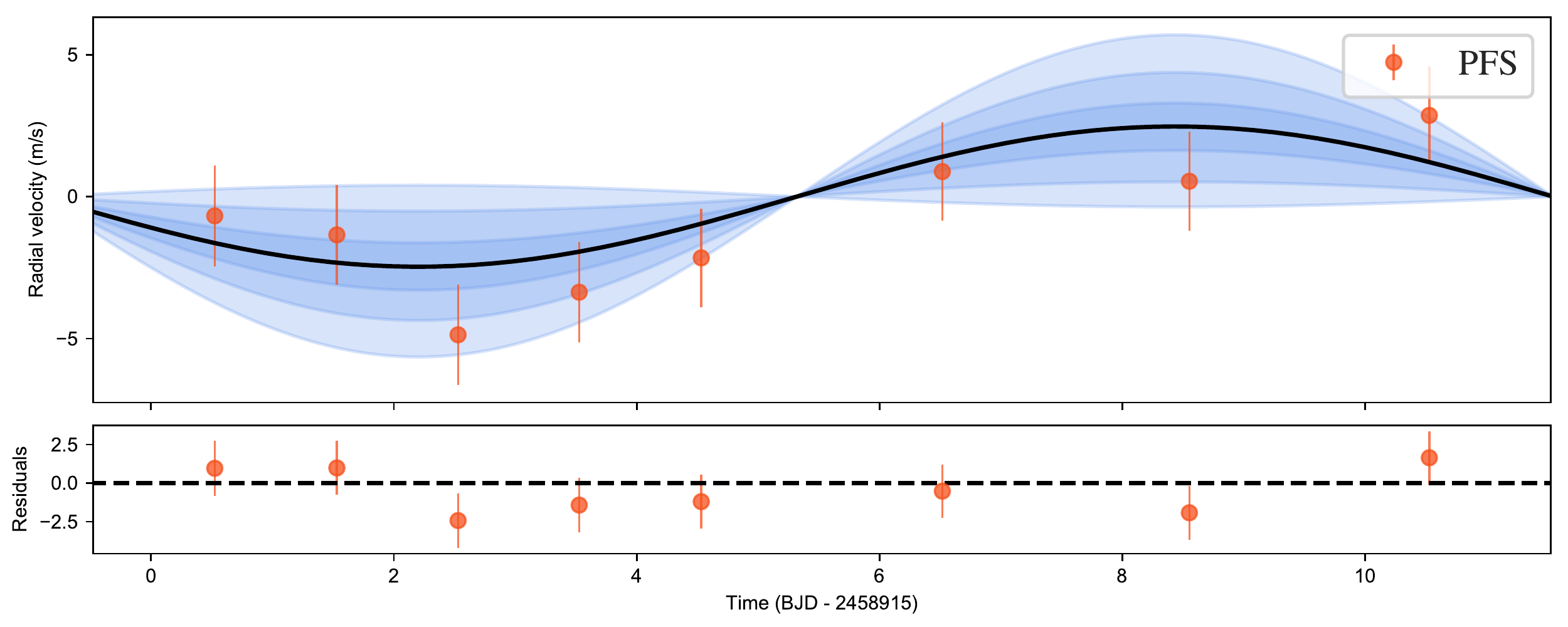} \\
\includegraphics[width=\linewidth]{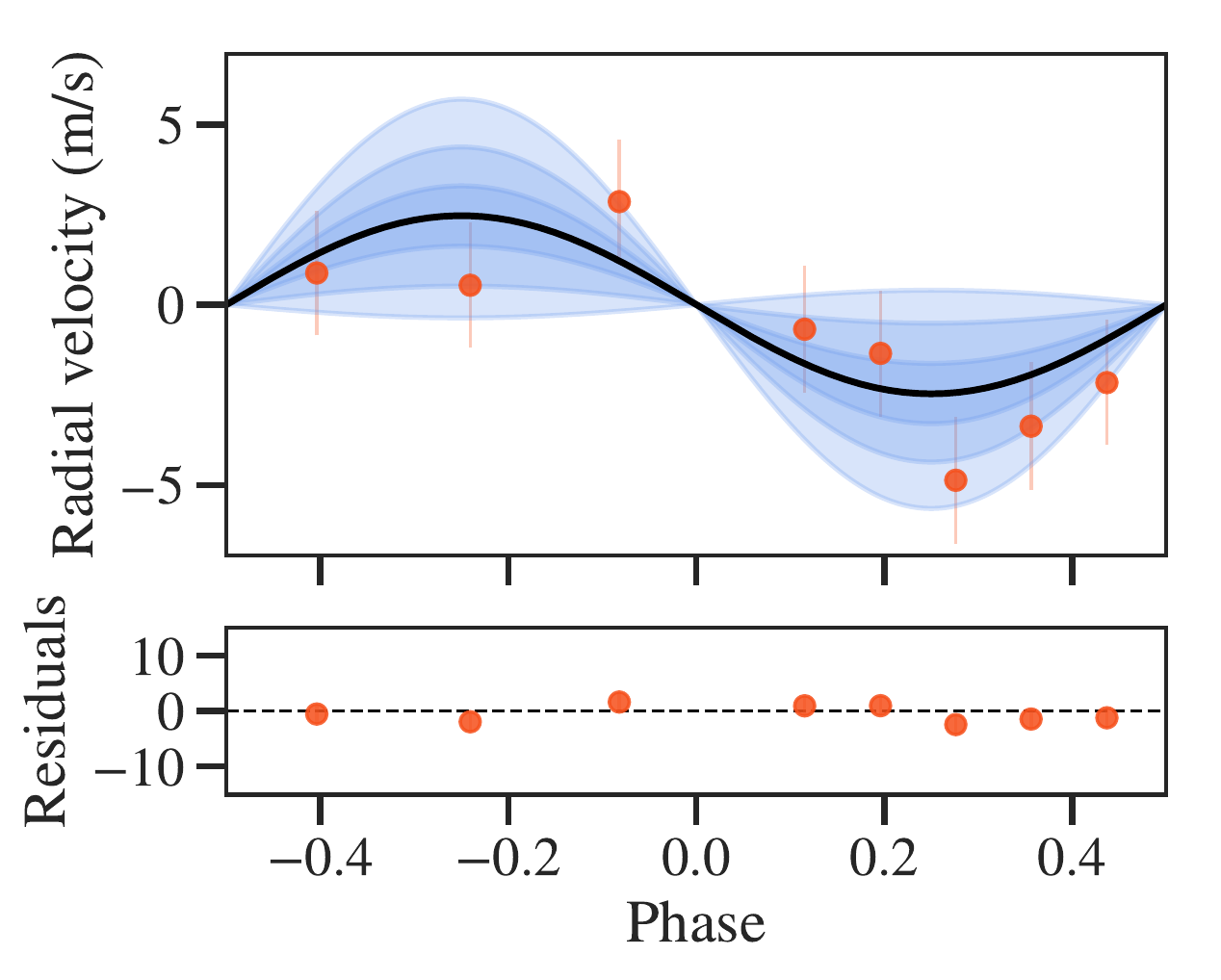} 
\end{minipage}
\caption{The best fit RV models (as listed in Table \ref{tab:pfs_K}) for TOI-431. The left/right column shows the \texttt{Radvel}/\texttt{juliet} results.}
\label{TOI431_rv_plots}
\end{figure*}

\begin{figure*}[htbp]
 \begin{minipage}[t]{0.8\columnwidth}
 \vspace{30pt}
\includegraphics[width=\linewidth]{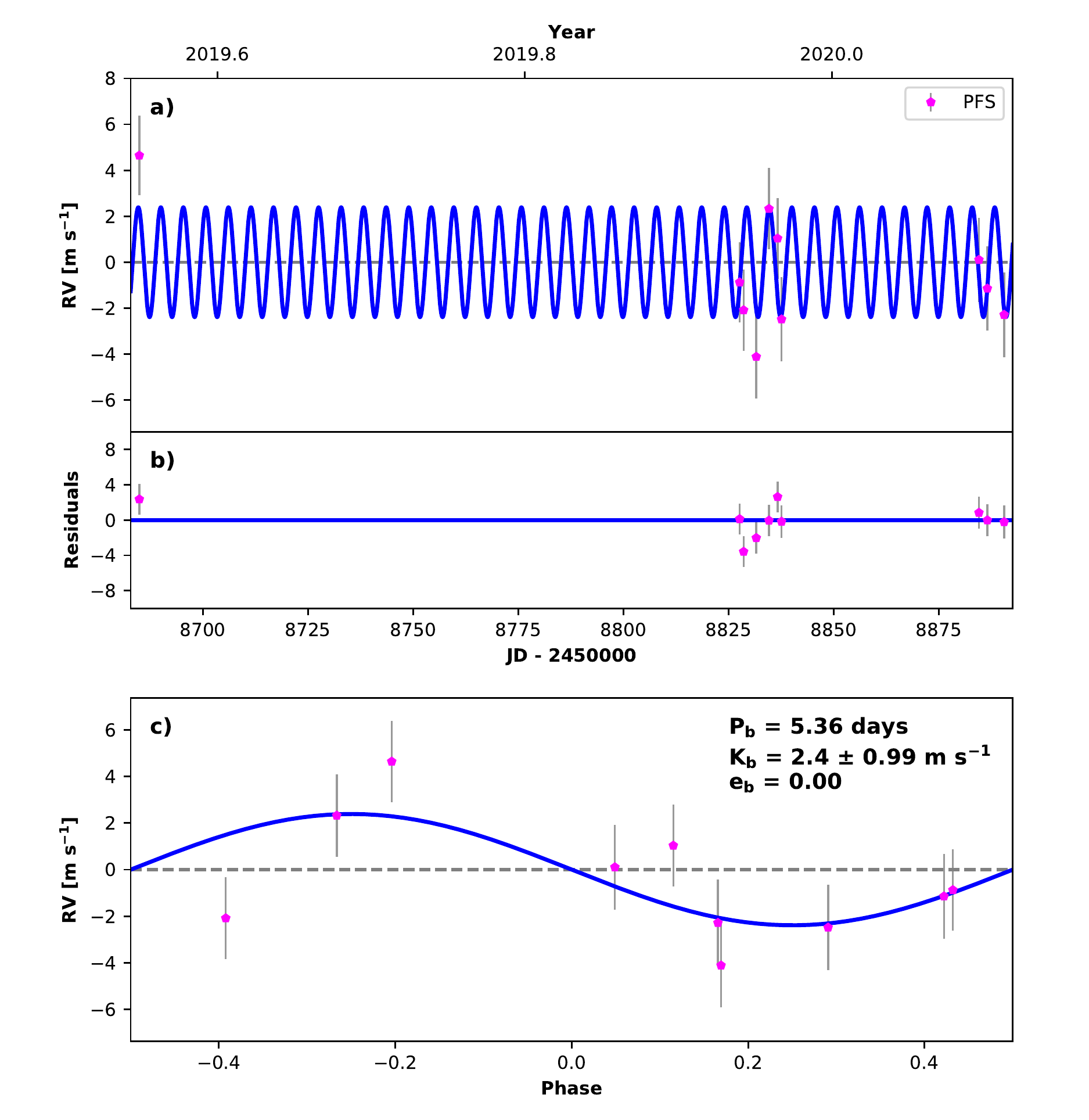}
\end{minipage}
 \begin{minipage}[t]{0.8\columnwidth}
\vspace{40pt}
\includegraphics[width=\linewidth]{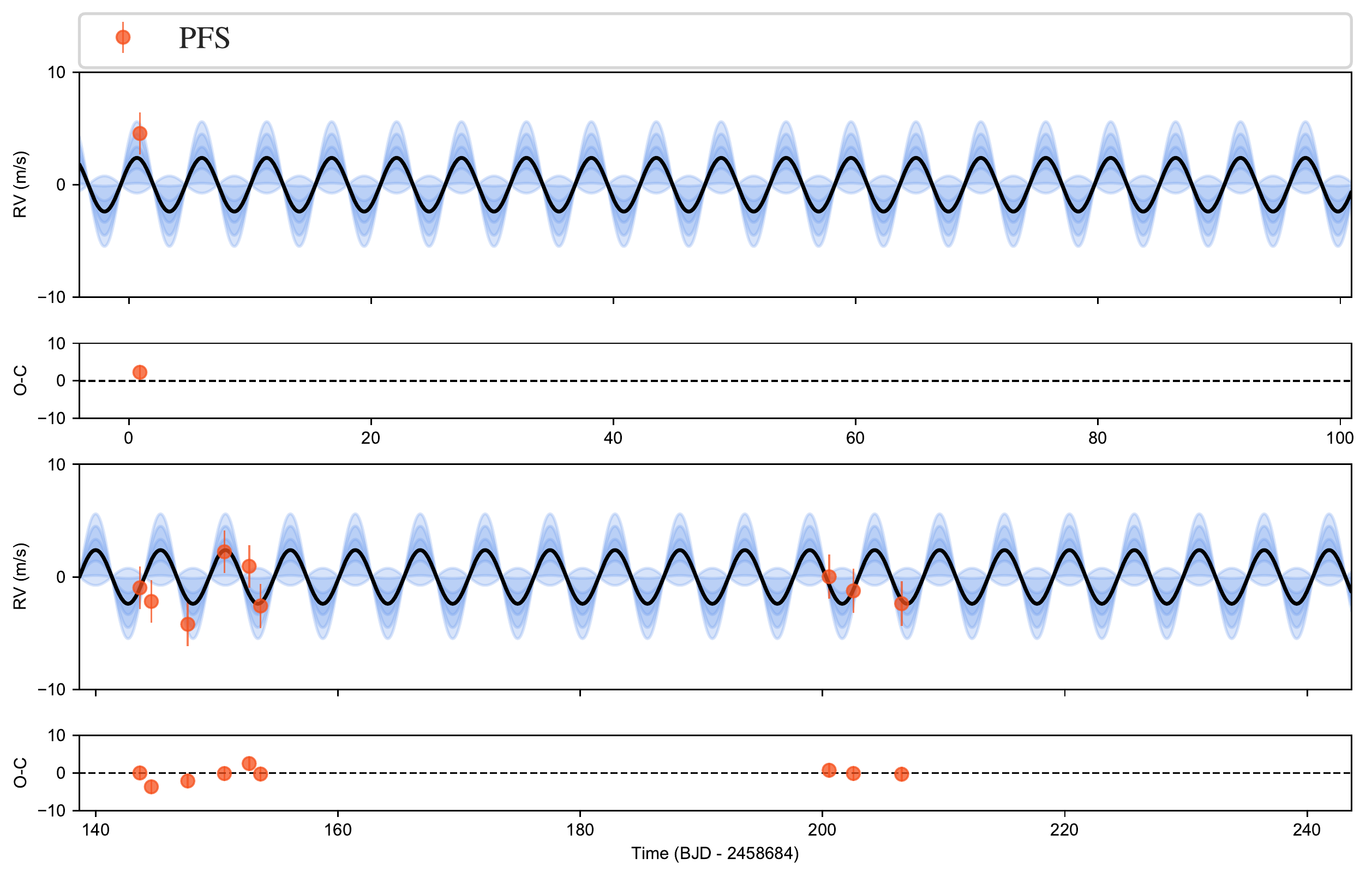} \\
\includegraphics[width=0.9\linewidth]{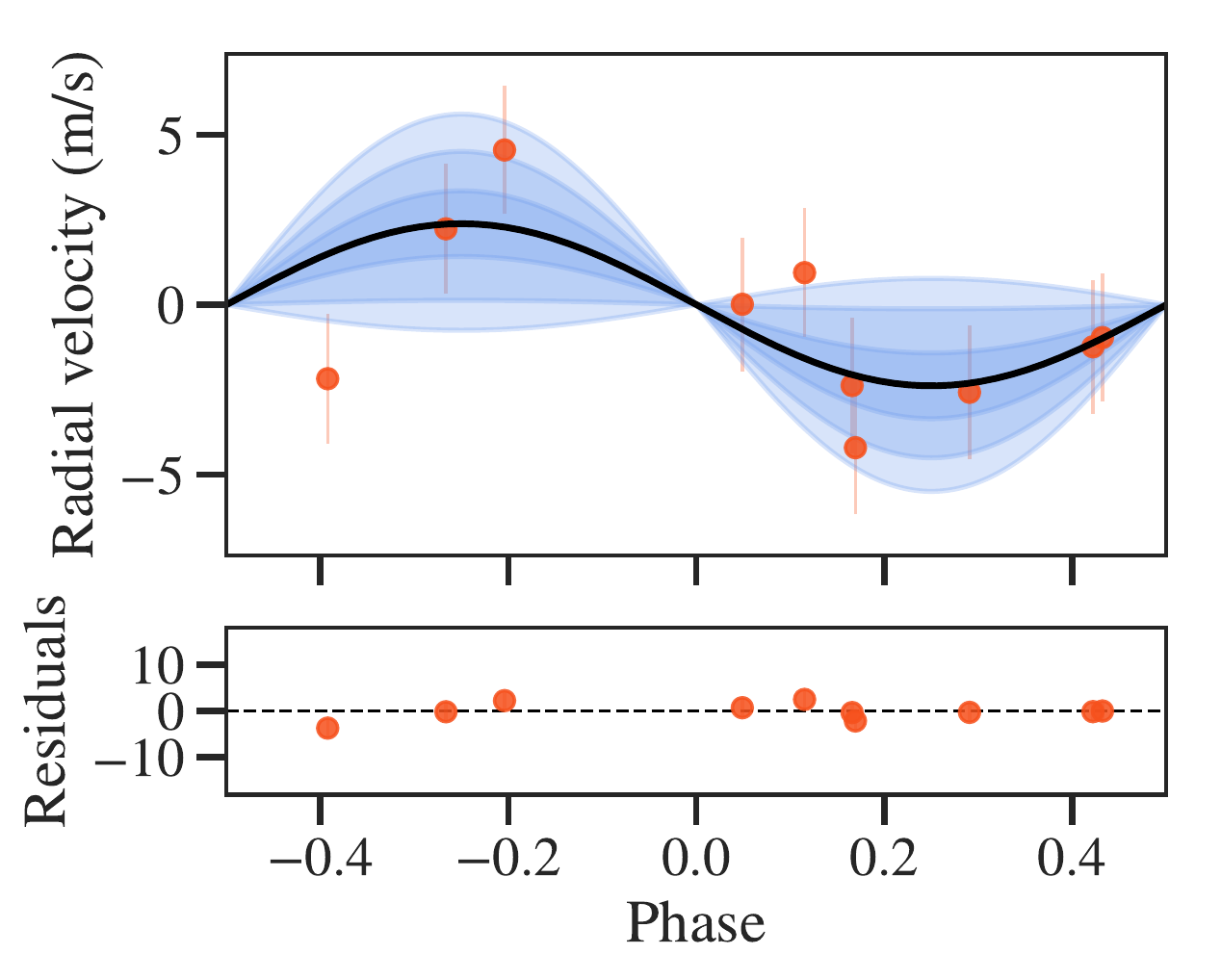} 
\end{minipage}
\caption{The best fit RV models (as listed in Table \ref{tab:pfs_K}) for TOI-455. The left/right column shows the \texttt{Radvel}/\texttt{juliet} results.}
\label{TOI455_rv_plots}
\end{figure*}

\begin{figure*}[htbp]
 \begin{minipage}[t]{0.8\columnwidth}
 \vspace{20pt}
\includegraphics[width=\linewidth]{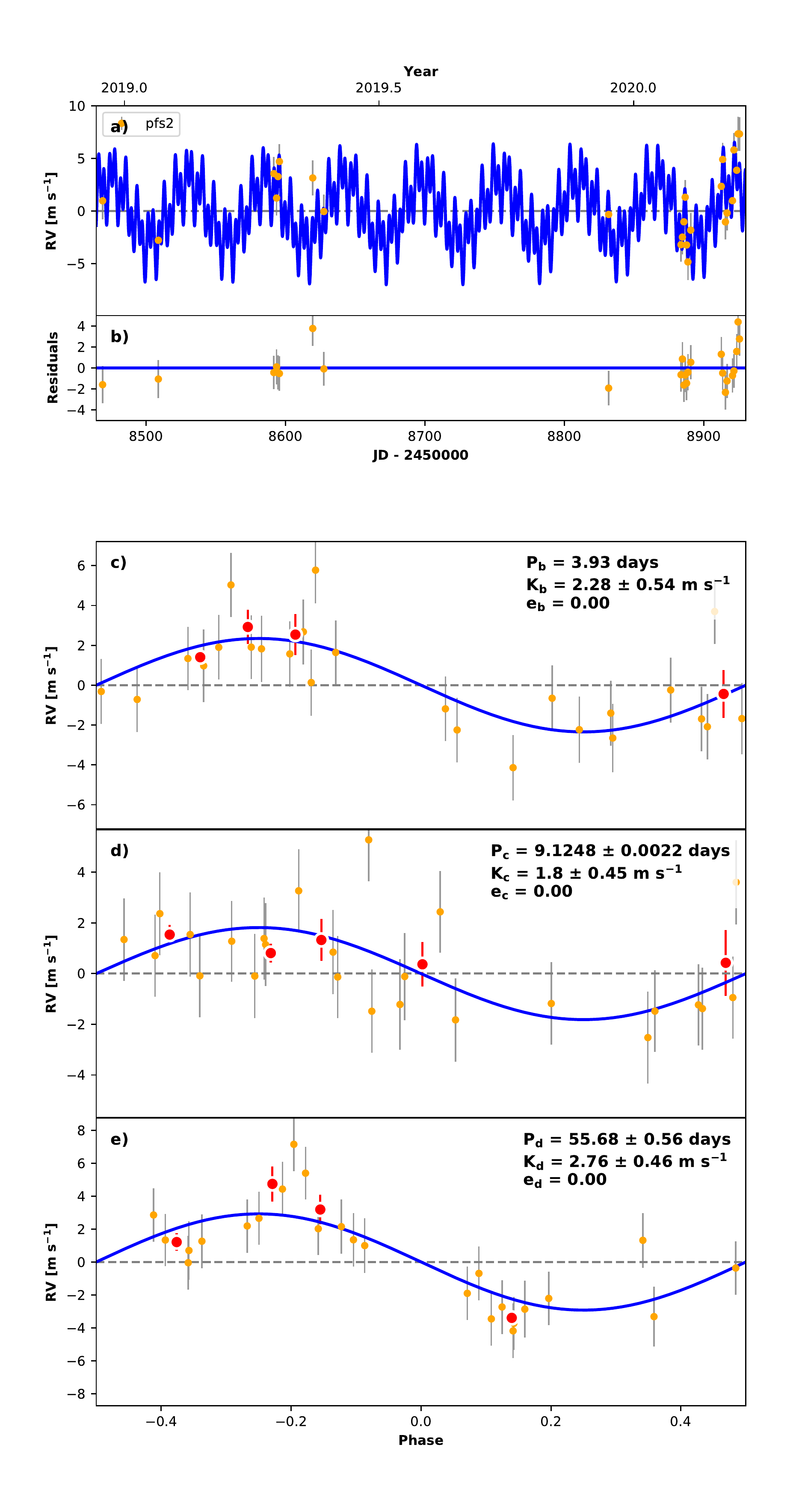}
\end{minipage}
 \begin{minipage}[t]{0.8\columnwidth}
\vspace{60pt}
\includegraphics[width=\linewidth]{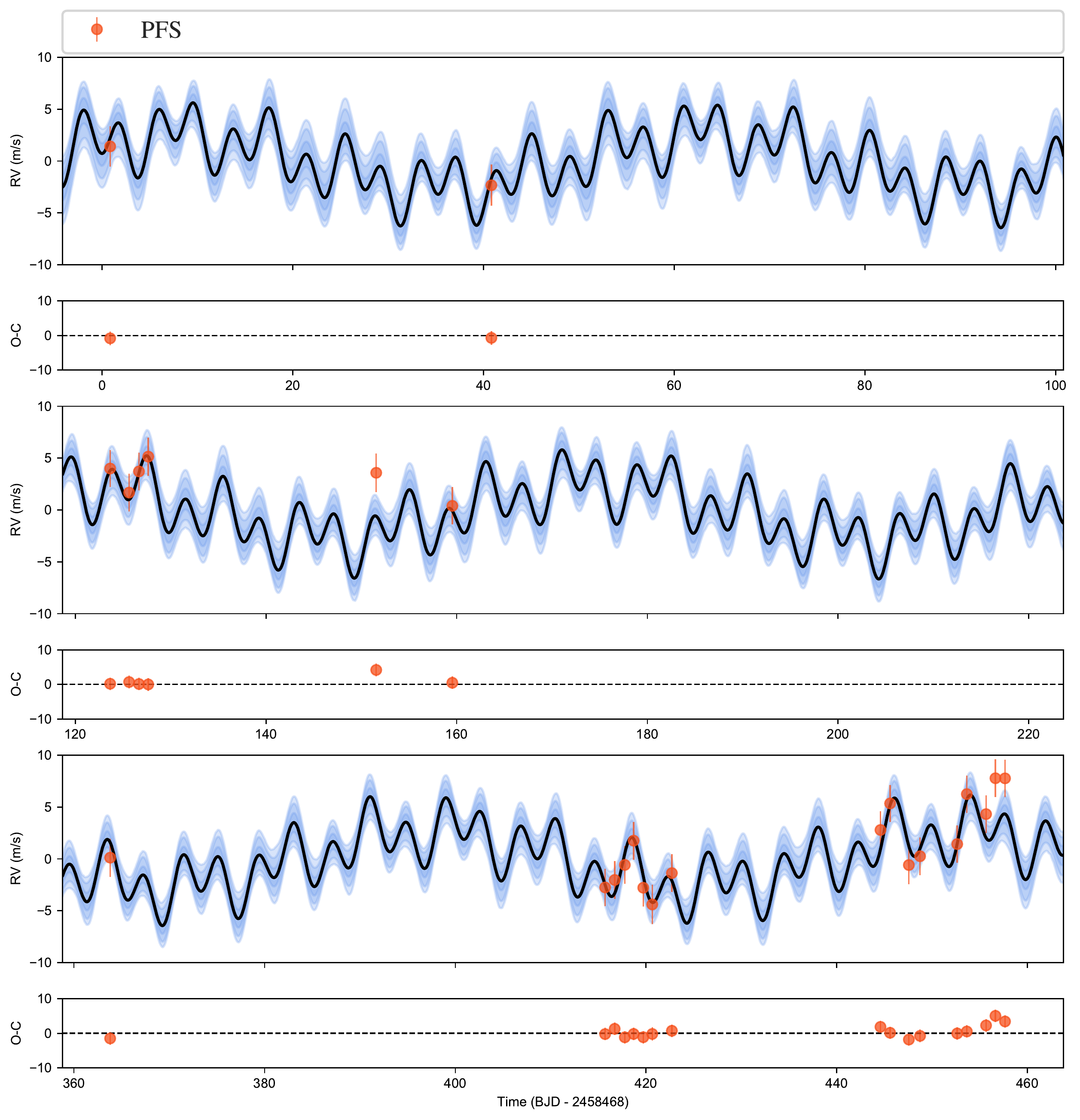} \\
\includegraphics[width=\linewidth]{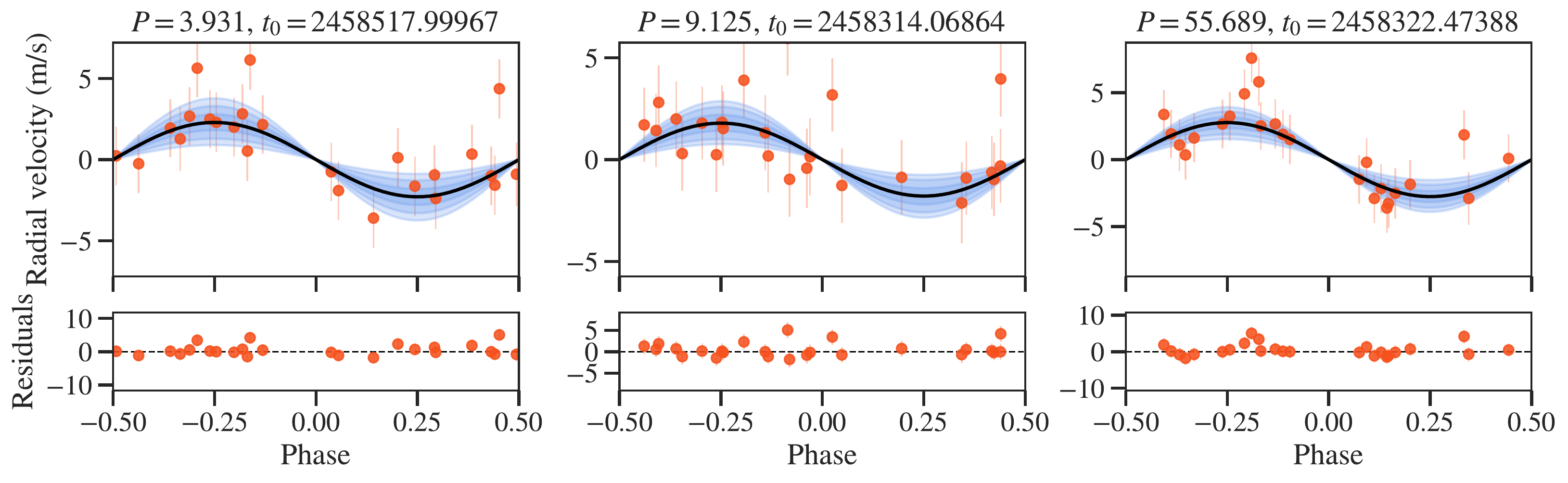} 
\end{minipage}
\caption{The best fit RV models (as listed in Table \ref{tab:pfs_K}) for TOI-562. The left/right column shows the \texttt{Radvel}/\texttt{juliet} results. }
\label{TOI562_rv_plots}
\end{figure*}

\begin{figure*}[htbp]
 \begin{minipage}[t]{0.8\columnwidth}
 \vspace{0pt}
\includegraphics[width=\linewidth]{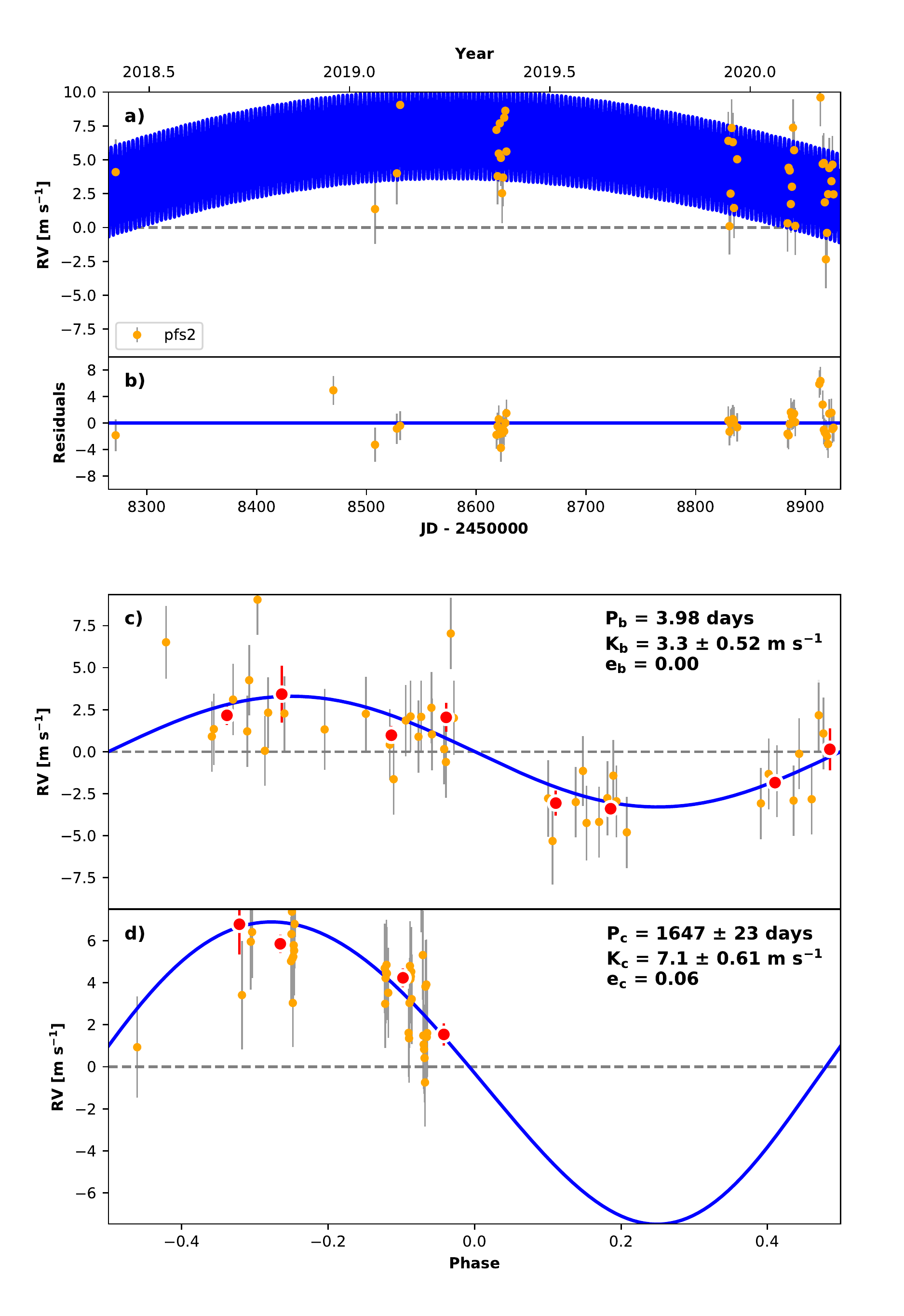}
\end{minipage}
 \begin{minipage}[t]{0.8\columnwidth}
\vspace{20pt}
\includegraphics[width=\linewidth]{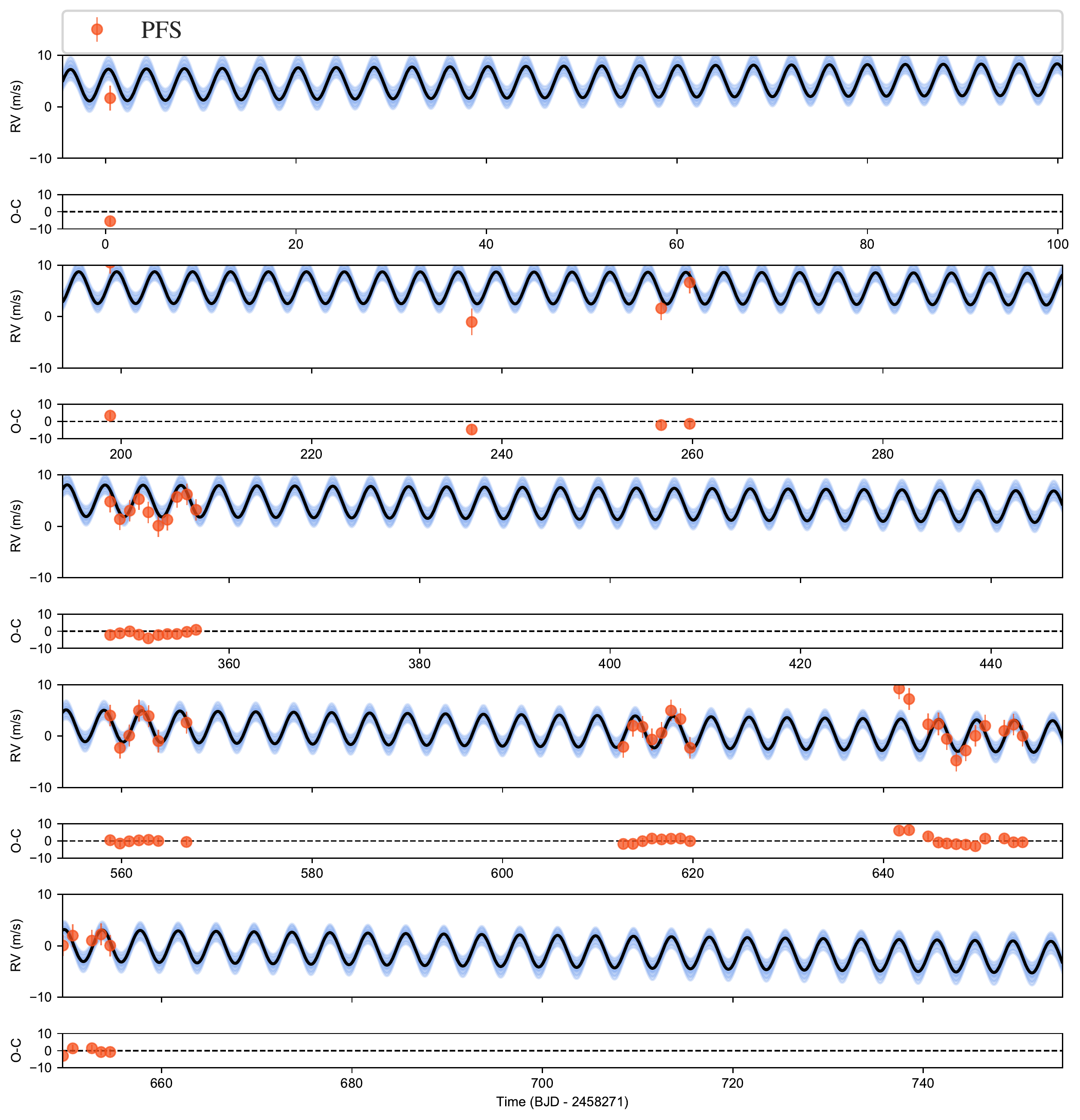} \\
\includegraphics[width=\linewidth]{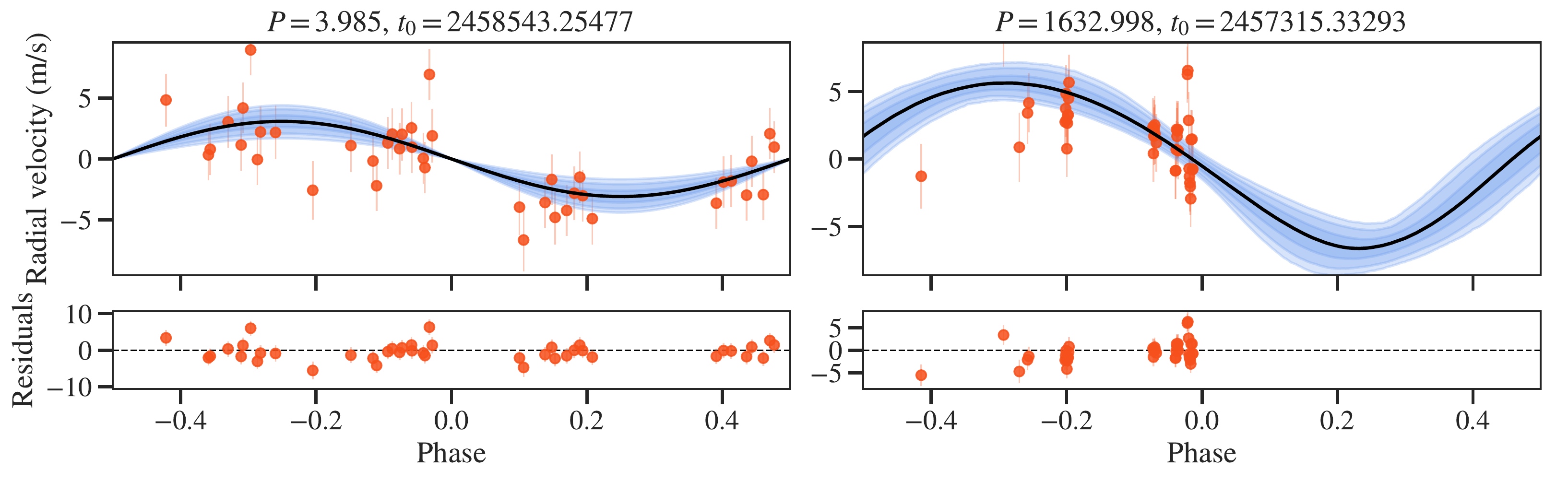} 
\end{minipage}
\caption{The best fit RV models (as listed in Table \ref{tab:pfs_K}) for TOI-652. The left/right column shows the \texttt{Radvel}/\texttt{juliet} results. }
\label{TOI652_rv_plots}
\end{figure*}

\begin{figure*}[htbp]
 \begin{minipage}[t]{0.8\columnwidth}
 \vspace{20pt}
\includegraphics[width=\linewidth]{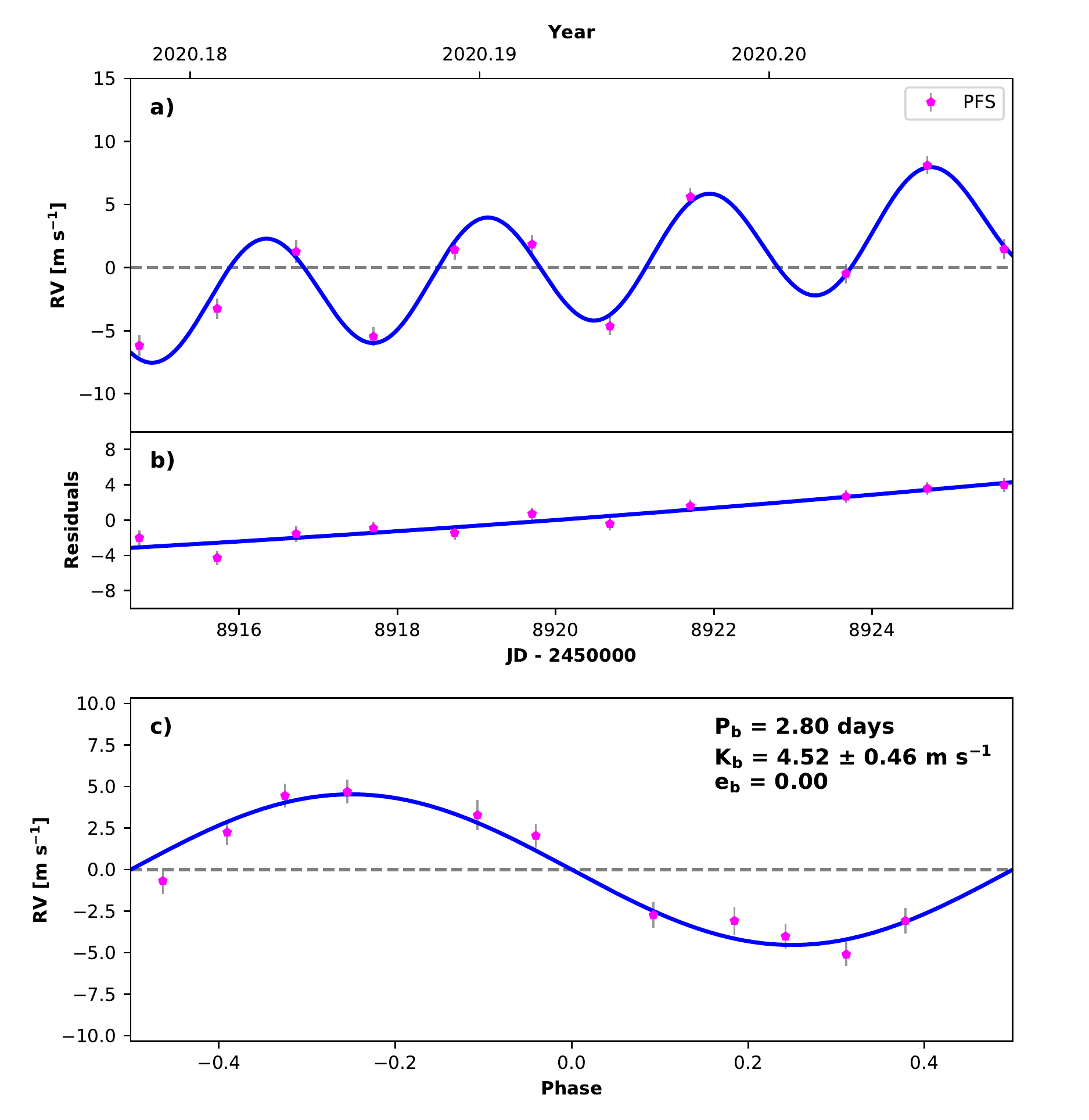}
\end{minipage}
 \begin{minipage}[t]{0.8\columnwidth}
\vspace{30pt}
\includegraphics[width=\linewidth]{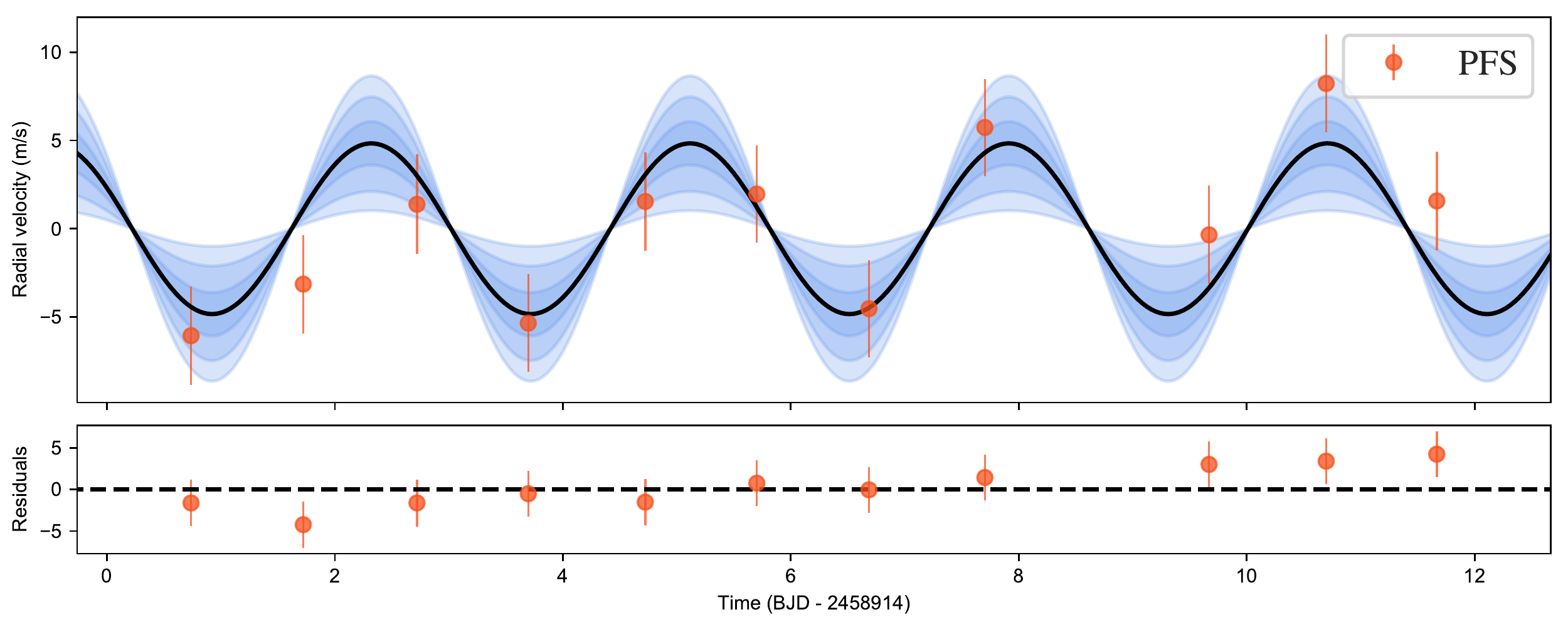} \\
\includegraphics[width=0.9\linewidth]{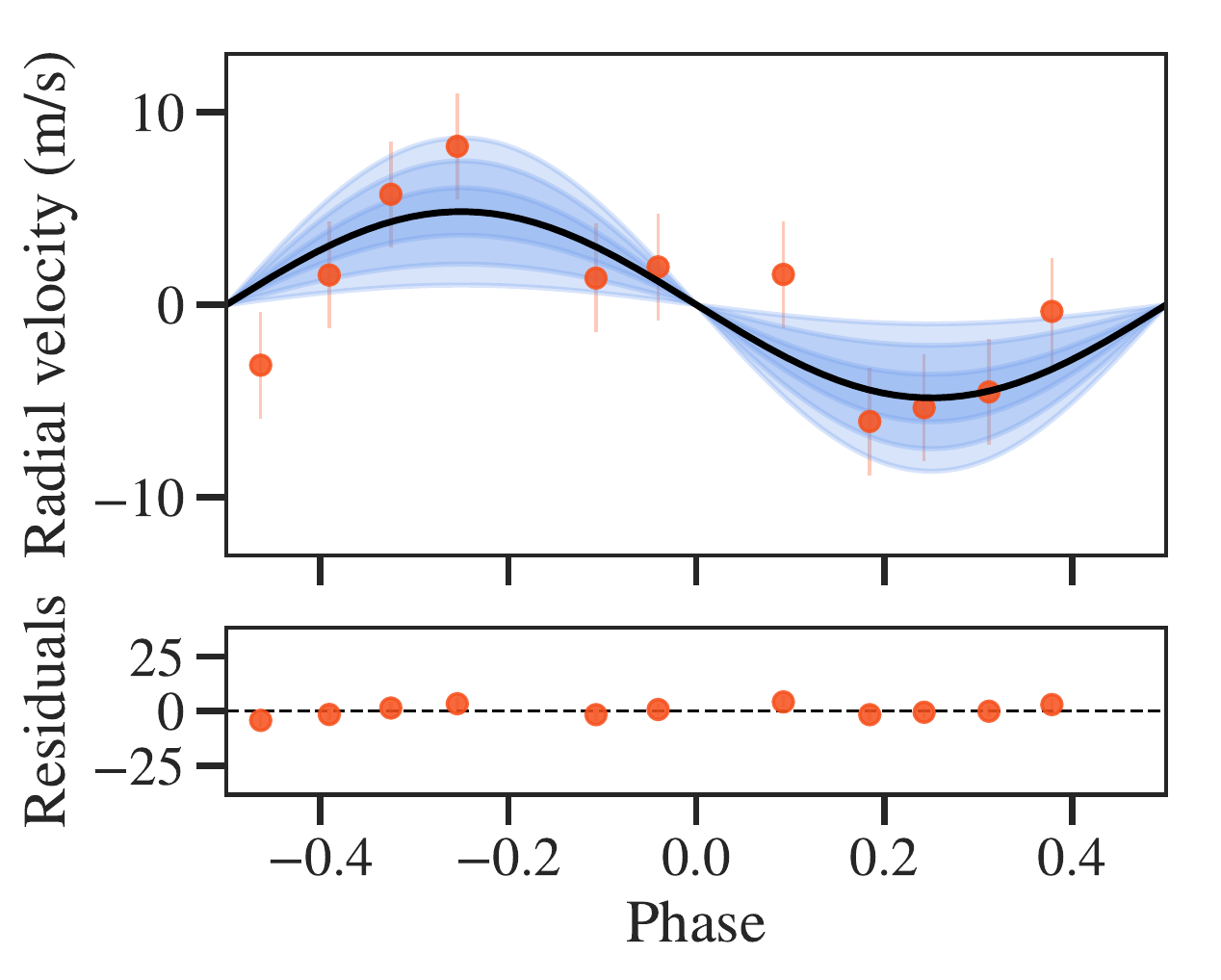} 
\end{minipage}
\caption{The best fit RV models (as listed in Table \ref{tab:pfs_K}) for TOI-784. The left/right column shows the \texttt{Radvel}/\texttt{juliet} results.}
\label{TOI784_rv_plots}
\end{figure*}

\begin{figure*}[htbp]
 \begin{minipage}[t]{0.8\columnwidth}
 \vspace{20pt}
\includegraphics[width=\linewidth]{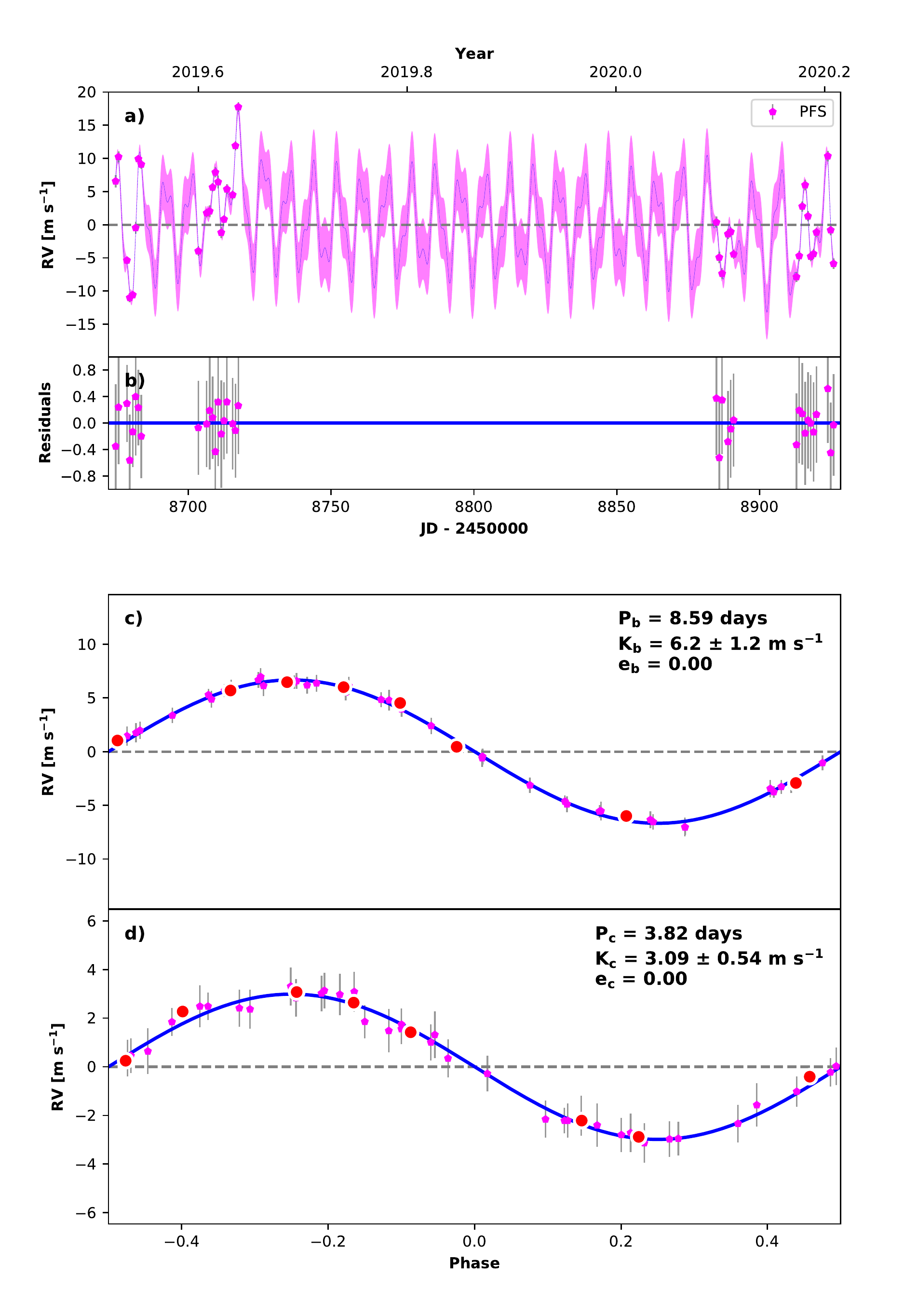}
\end{minipage}
 \begin{minipage}[t]{0.8\columnwidth}
\vspace{30pt}
\includegraphics[width=\linewidth]{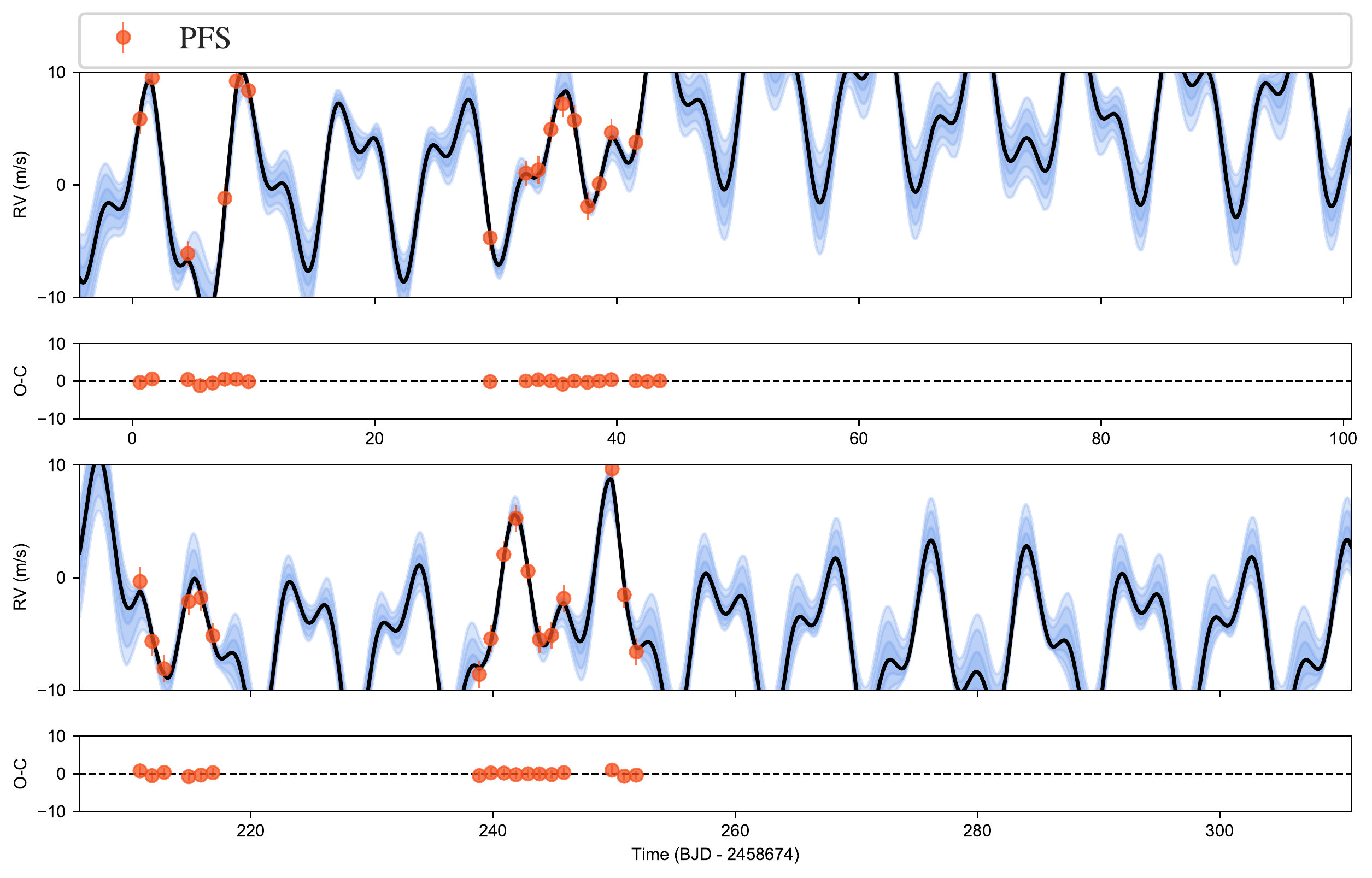} \\
\includegraphics[width=\linewidth]{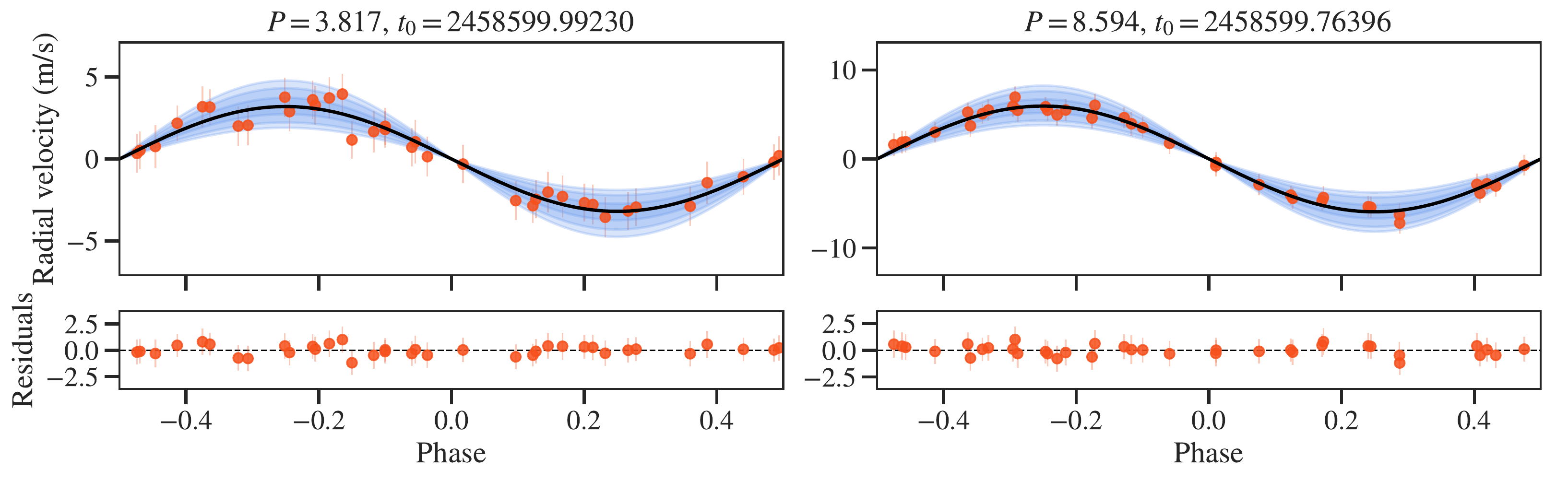} 
\end{minipage}
\caption{The best fit RV models (as listed in Table \ref{tab:pfs_K}) for TOI-836. The left/right column shows the \texttt{Radvel}/\texttt{juliet} results.}
\label{TOI836_rv_plots}
\end{figure*}

\begin{figure*}[htbp]
 \begin{minipage}[t]{0.8\columnwidth}
 \vspace{20pt}
\includegraphics[width=\linewidth]{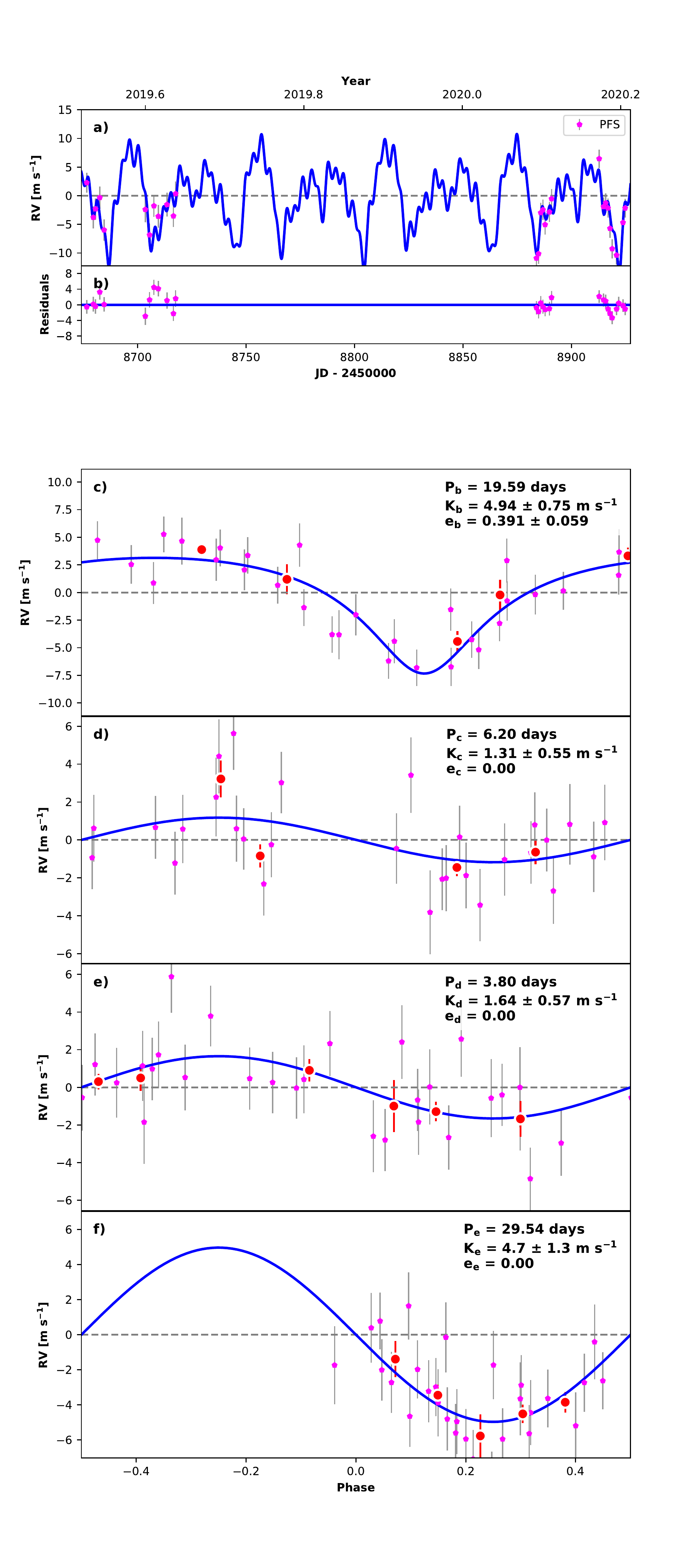}
\end{minipage}
 \begin{minipage}[t]{0.8\columnwidth}
\vspace{50pt}
\includegraphics[width=\linewidth]{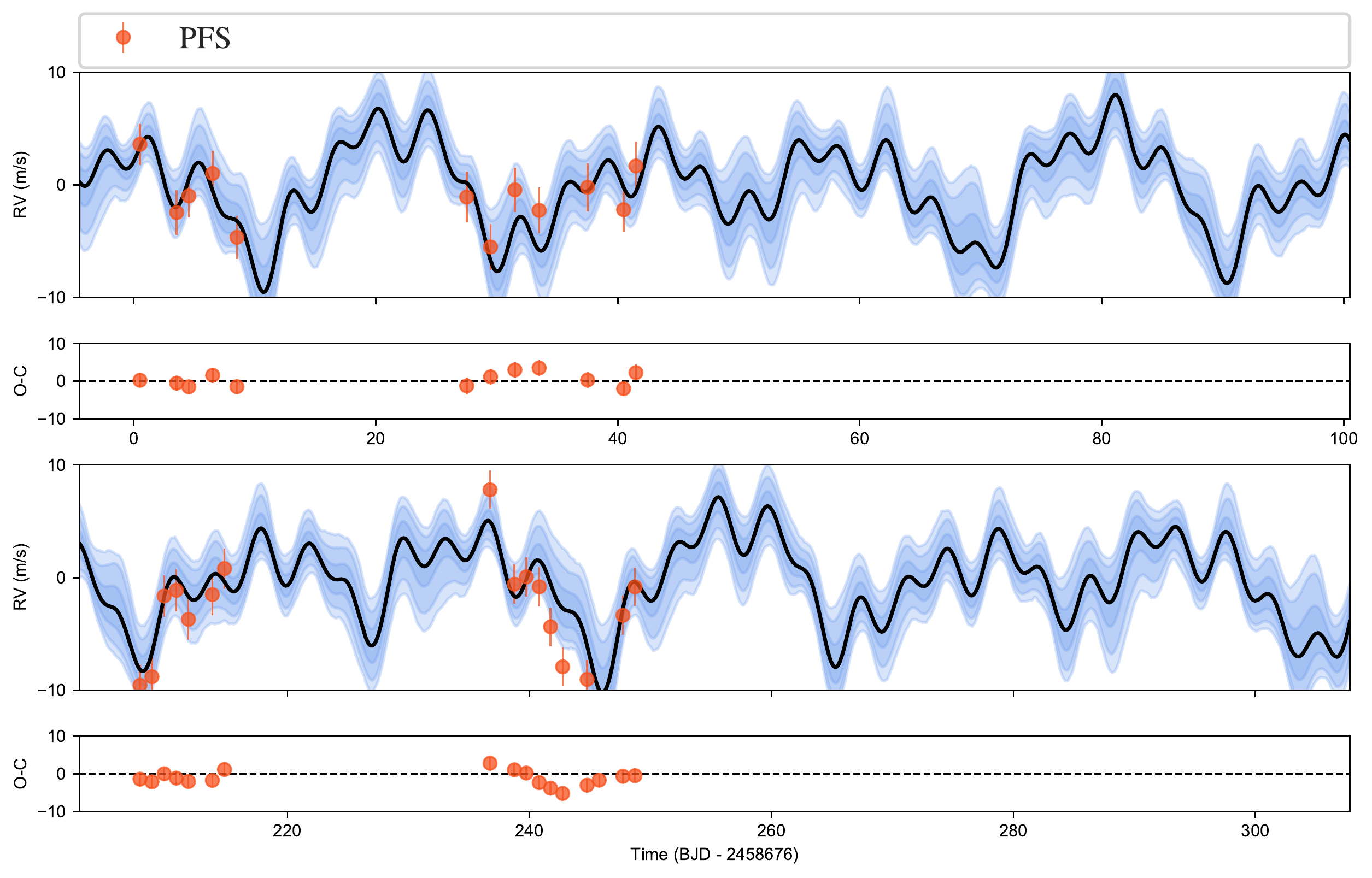} \\
\includegraphics[width=\linewidth]{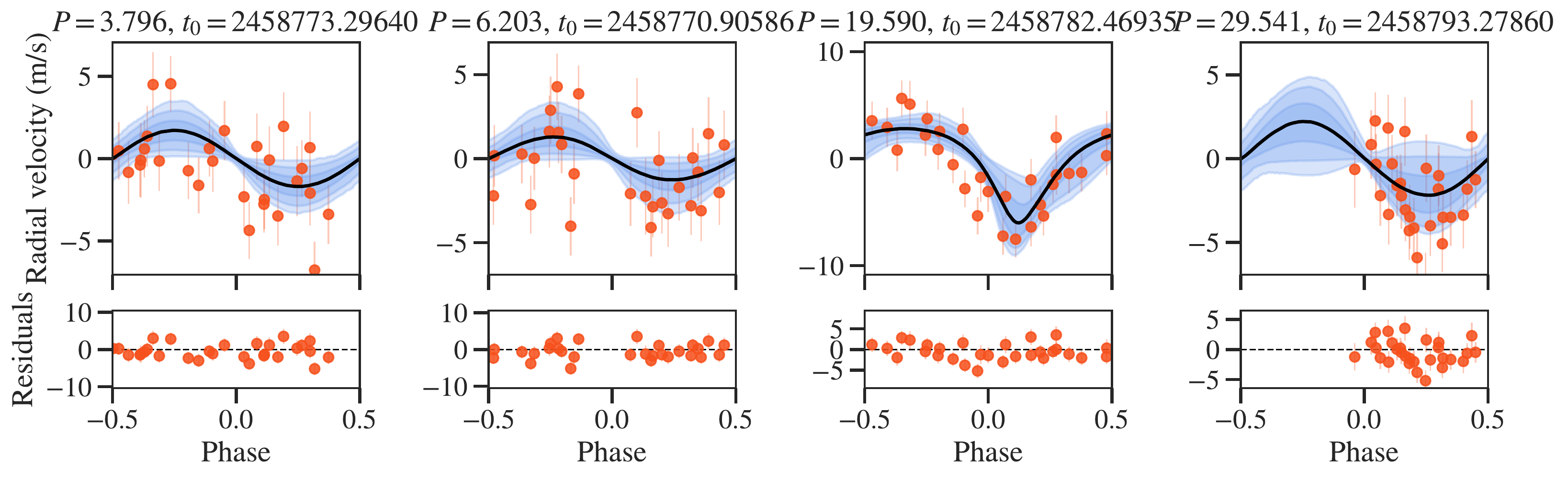} 
\end{minipage}
\caption{The best fit RV models (as listed in Table \ref{tab:pfs_K}) for TOI-1233. The left/right column shows the \texttt{Radvel}/\texttt{juliet} results. }
\label{TOI1233_rv_plots}
\end{figure*}

\begin{figure*}[htbp]
\includegraphics[width=\linewidth]{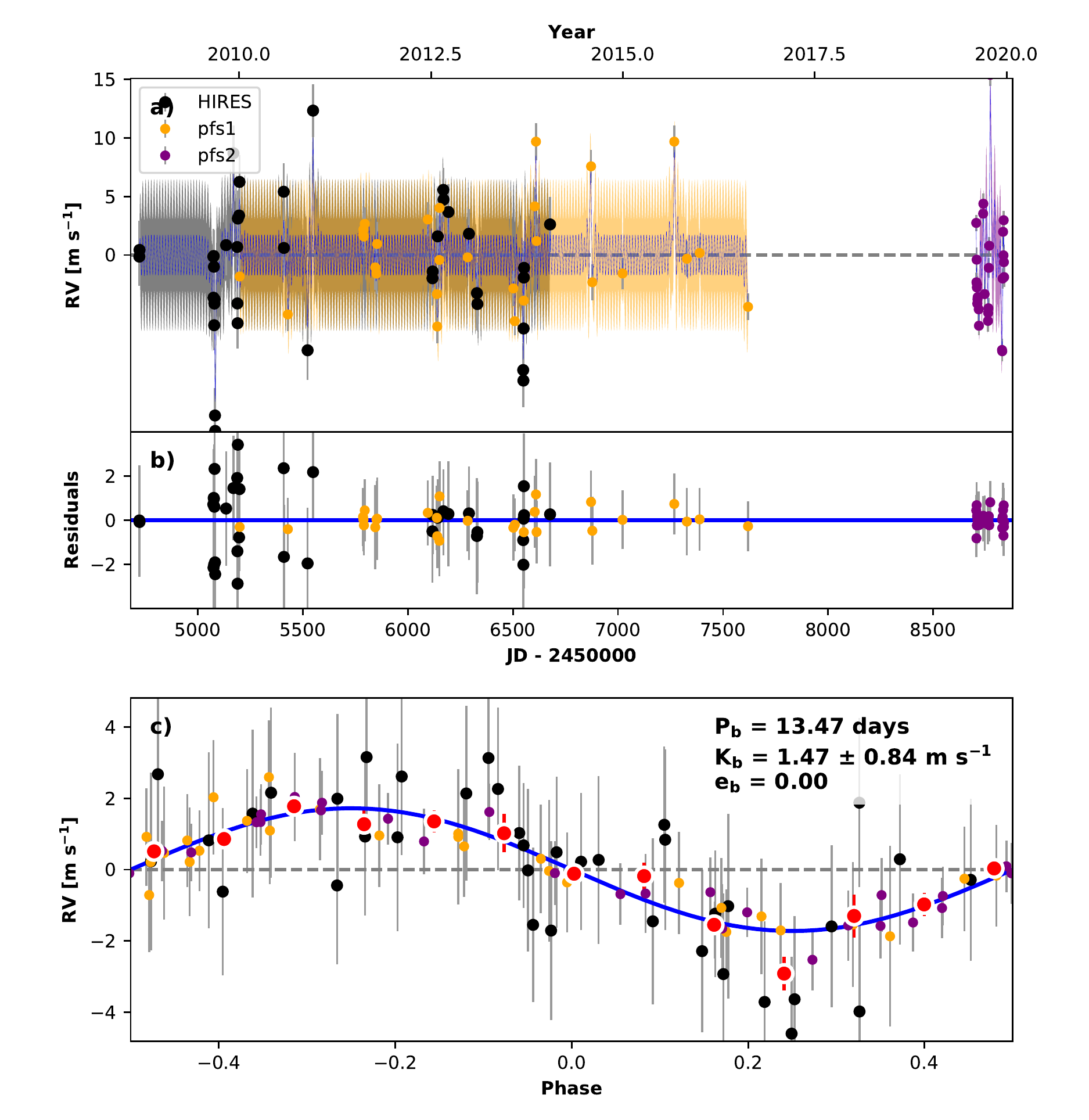}
\caption{The best fit \texttt{Radvel} RV model (as listed in Table \ref{tab:pfs_K}) for TOI-260 including PFS and HIRES data.}
\label{TOI260all_rv_plots_radvel}
\end{figure*}

\begin{figure*}[htbp]
\centering
\includegraphics[width=0.8\linewidth]{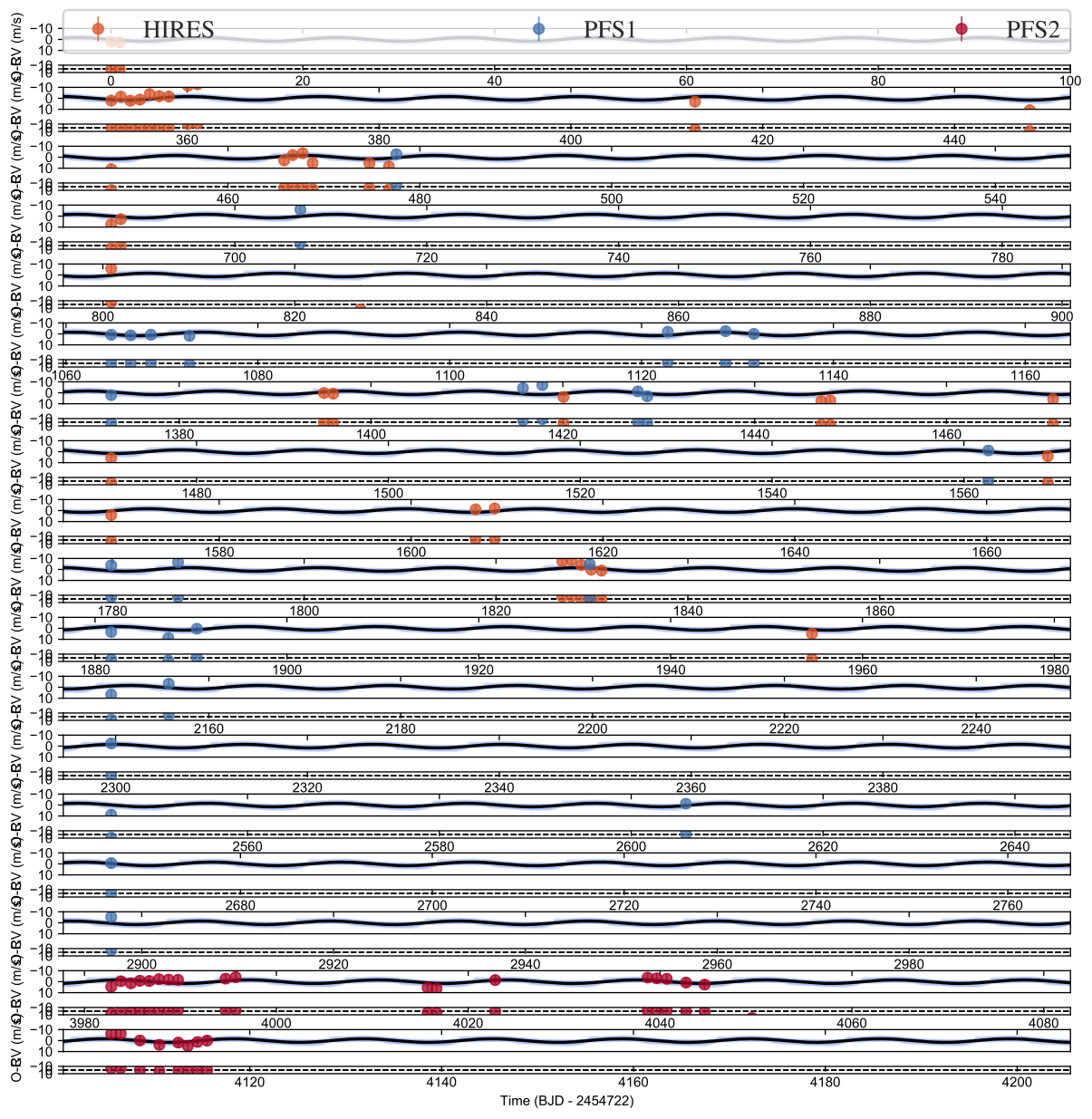}\\
\includegraphics[width=0.5\linewidth]{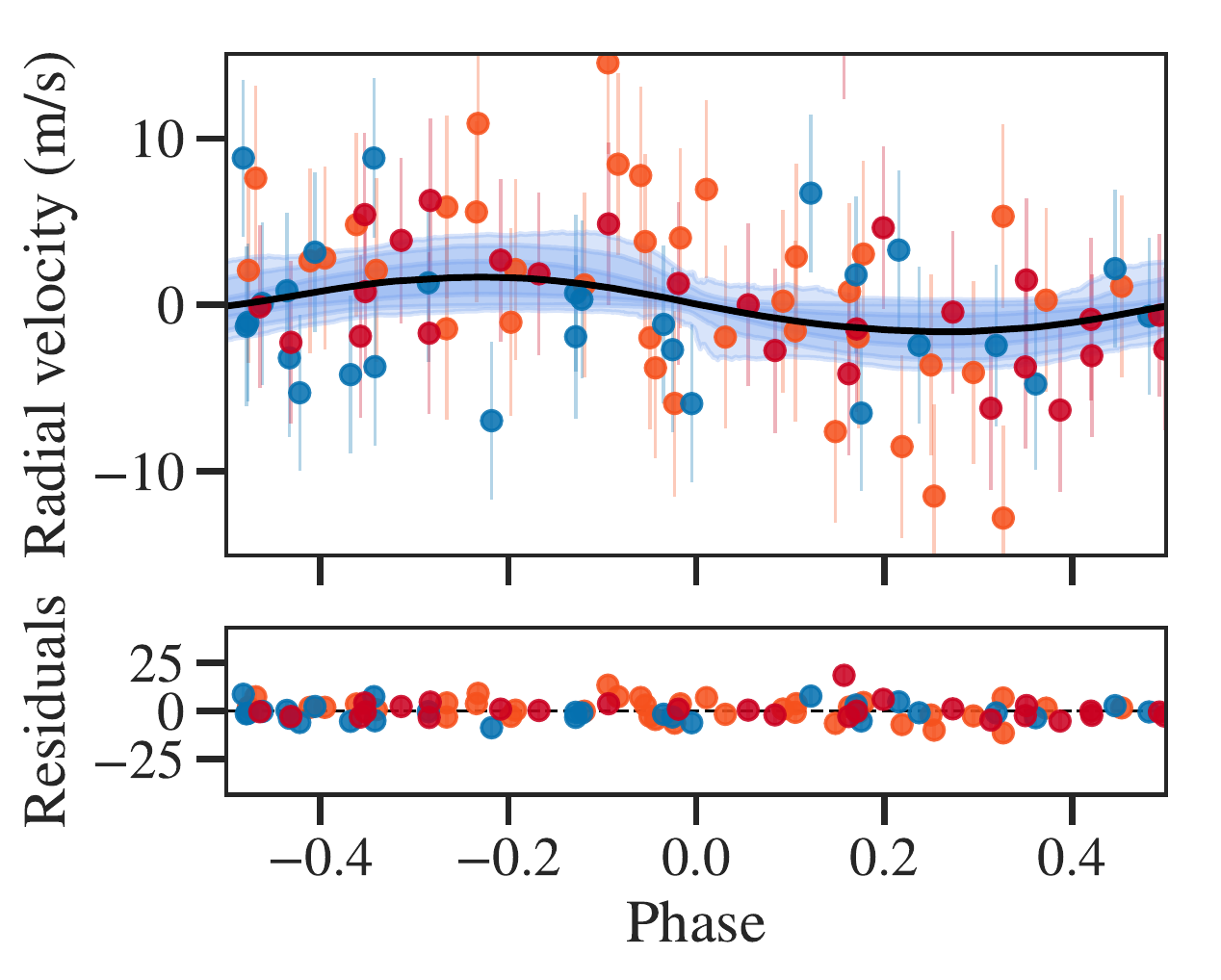} 
\caption{The best fit \texttt{juliet} RV model (as listed in Table \ref{tab:pfs_K}) for TOI-260 including PFS and HIRES data.}
\label{TOI260all_rv_plots_juliet}
\end{figure*}

\begin{figure*}[htbp]
 \begin{minipage}[t]{0.8\columnwidth}
 \vspace{20pt}
\includegraphics[width=\linewidth]{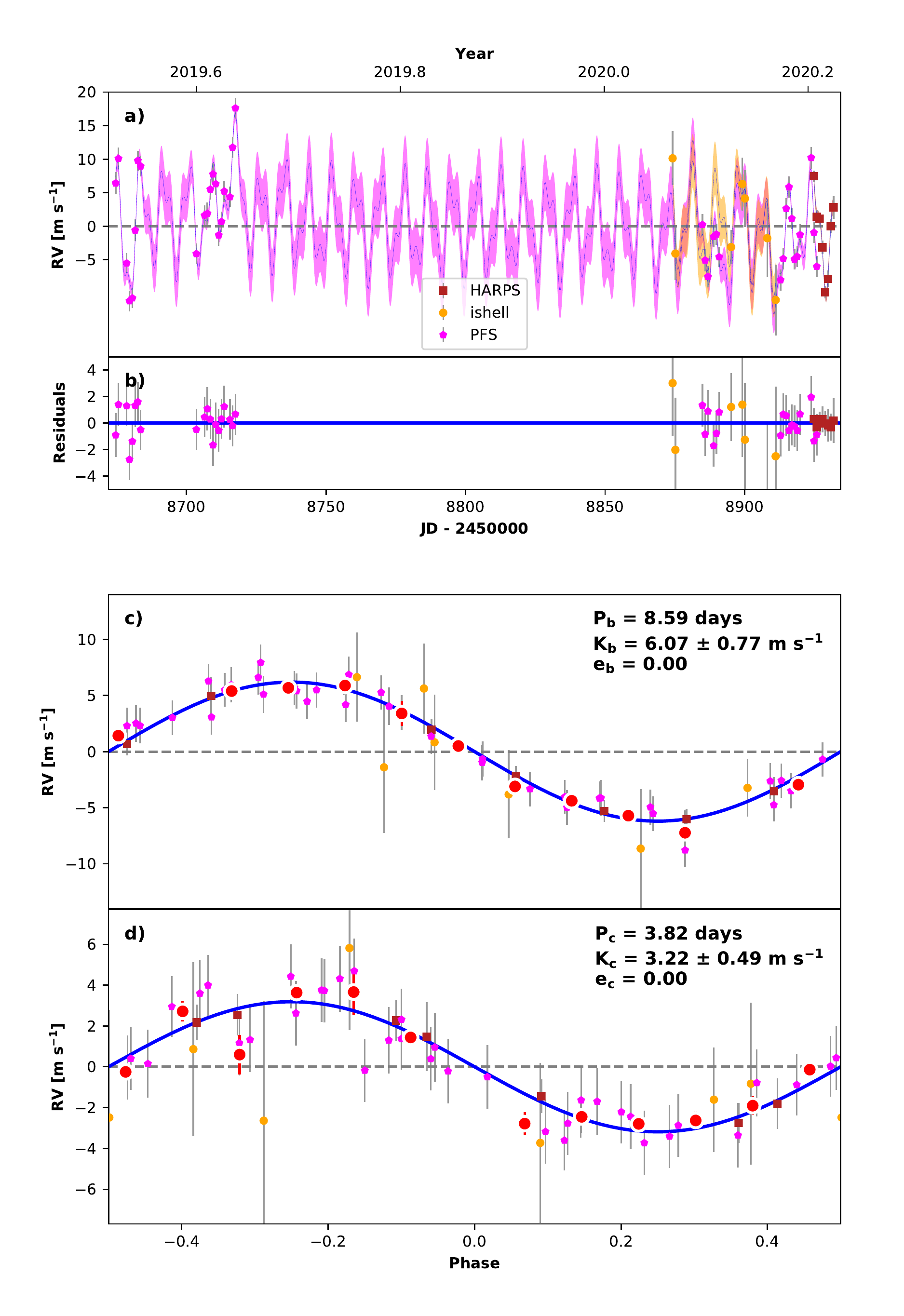}
\end{minipage}
 \begin{minipage}[t]{0.8\columnwidth}
\vspace{50pt}
\includegraphics[width=\linewidth]{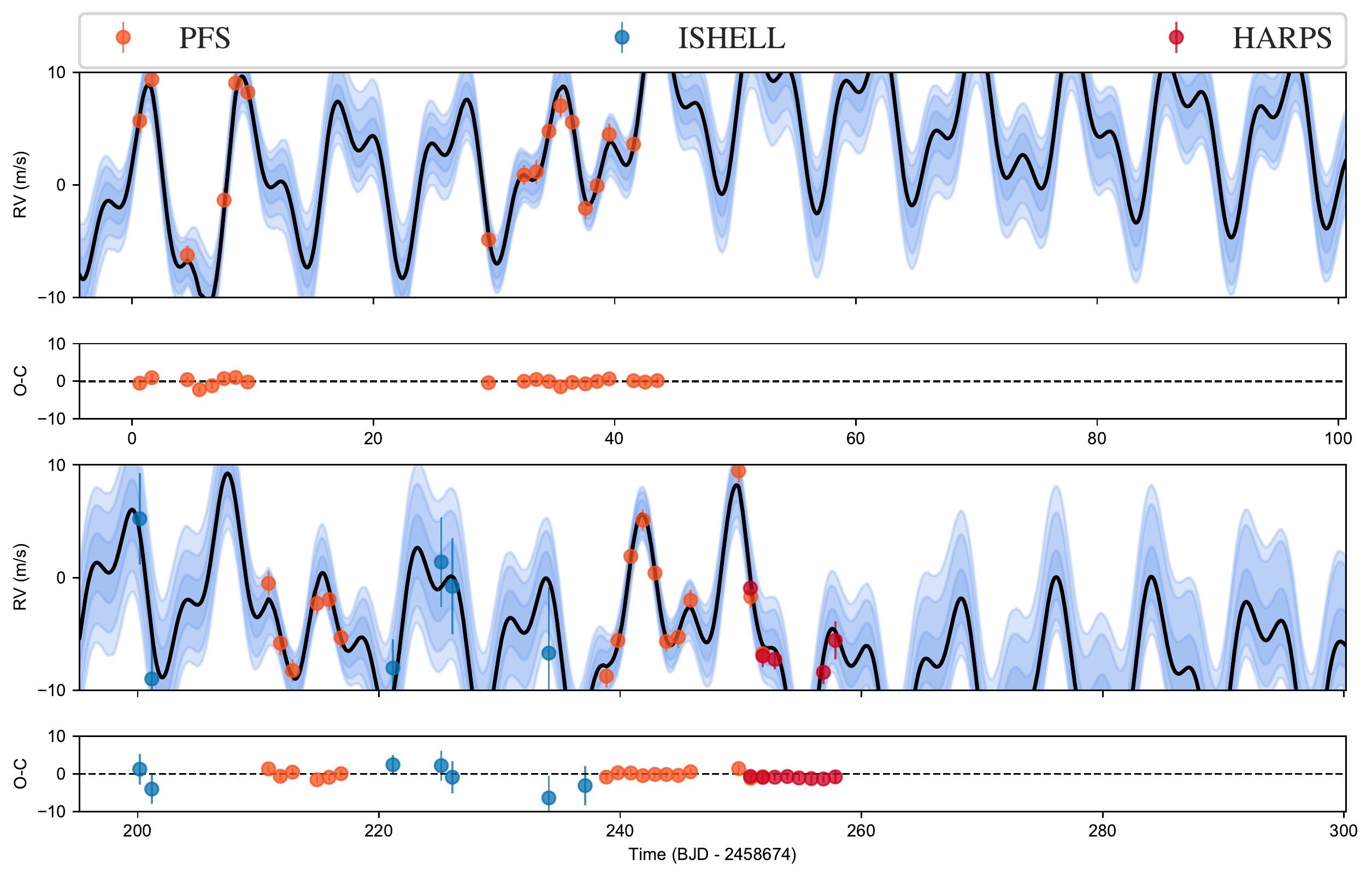} \\
\includegraphics[width=\linewidth]{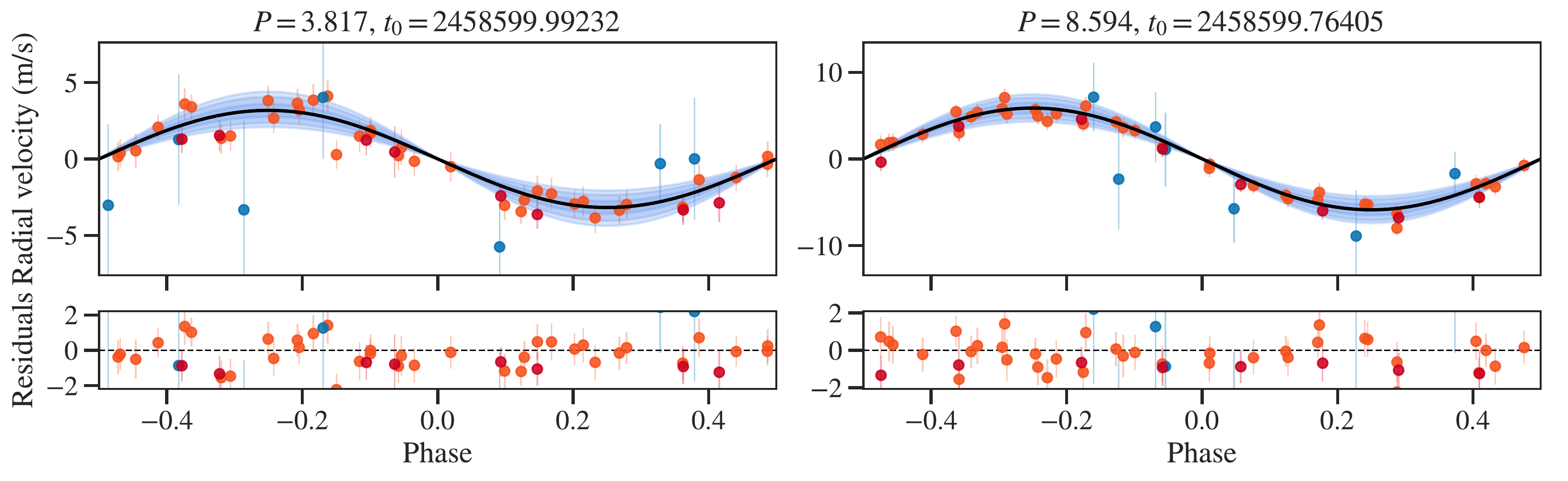} 
\end{minipage}
\caption{The best fit RV models (as listed in Table \ref{tab:pfs_K}) for TOI-836 including PFS, HARPS, and iSHELL data. The left/right column shows the \texttt{Radvel}/\texttt{juliet} results. }
\label{TOI836all_rv_plots}
\end{figure*}


\clearpage

\section{Ancillary Parameter Tables}\label{sec:ancillary_tables}

\begin{table*}[htbp]
    \centering
    \caption{Median and the 68\% credibility of the posterior distribution for additional parameters in the transit fit of each system.}
    \begin{tabular}{cccccccc}
        \hline\hline
        TOI &$r_{1}$ &$r_{2}$ &$q_{1}$ &$q_{2}$ &$\rho_{\star}$ ($\rm kg\ m^{-3}$)
        &$M_{\rm TESS}$ (ppm) &$\sigma_{\rm TESS}$ (ppm)\\\hline
        
        $134.01$ &$0.7153\pm0.0506$ &$0.0224\pm0.0010$ &$0.52\pm0.29$ &$0.46\pm0.31$ &$3894\pm194$ &$0.07\pm5.33$ &$0.19\pm0.10$\\
        
        $141.01$ &$0.7981\pm0.0438$ &$0.0144\pm0.0004$ &$0.19\pm0.21$ &$0.45\pm0.30$ &$1047\pm188$ &$-0.17\pm1.92$ &$35.16\pm13.26$\\
        
        $144.01$ &$0.7361\pm0.0336$ &$0.0163\pm0.0003$ &$0.36\pm0.14$ &$0.22\pm0.15$ &$1075\pm201$ &$-0.40\pm0.57$ &$156.33\pm0.47$\\
        
        $174.01$ &$0.7995\pm0.0196$ &$0.0273\pm0.0008$ &$0.72\pm0.18$ &$0.60\pm0.23$ &$2849\pm213$ &$-0.67\pm1.86$ &$89.63\pm8.58$\\
        
        $174.02$ &$0.8344\pm0.0177$ &$0.0260\pm0.0010$ & & & & &\\
        
        $174.03$ &$0.5021\pm0.1051$ &$0.0161\pm0.0007$ & & & & &\\
        
        $175.01$ &$0.6391\pm0.0133$ &$0.0402\pm0.0004$ &$0.23\pm0.14$ &$0.39\pm0.31$ &$13306\pm262$ &$-0.67\pm1.86$ &$89.63\pm8.58$\\
        
        $175.02$ &$0.9403\pm0.0039$ &$0.0447\pm0.0009$ & & & & &\\
        
        $175.03$ &$0.6998\pm0.0140$ &$0.0250\pm0.0004$ & & & & &\\
        
        $186.01$ &$0.7966\pm0.0281$ &$0.0382\pm0.0008$ &$0.35\pm0.12$ &$0.70\pm0.32$ &$3794\pm614$ &$0.57\pm1.29$ &$166.66\pm1.33$\\
        
        $186.02$ &$0.5332\pm0.1208$ &$0.0141\pm0.0005$ & & & & &\\
        
        $260.01$ &$0.8285\pm0.0604$ &$0.0261\pm0.0017$ &$0.54\pm0.32$ &$0.45\pm0.32$ &$4063\pm1239$ &$0.63\pm3.97$ &$0.01\pm0.01$\\
        
        $402.01$ &$0.5345\pm0.1135$ &$0.0182\pm0.0006$ &$0.19\pm0.11$ &$0.43\pm0.28$ &$2298\pm376$ &$1.00\pm2.53$ &$108.55\pm7.42$\\
        
        $402.02$ &$0.9118\pm0.0116$ &$0.0268\pm0.0011$ & & & & &\\
        
        $431.01$ &$0.4287\pm0.0809$ &$0.0410\pm0.0005$ &$0.46\pm0.24$ &$0.28\pm0.23$ &$3134\pm219$ &$-0.47\pm2.32$ &$96.46\pm6.78$\\
        
        $431.02$ &$0.5060\pm0.0783$ &$0.0161\pm0.0003$ & & & & &\\
        
        $455.01$ &$0.5514\pm0.0536$ &$0.0442\pm0.0013$ &$0.40\pm0.33$ &$0.29\pm0.19$ &$16743\pm134$ &$-0.23\pm8.52$ &$714.94\pm9.13$\\
        
        $562.01$ &$0.6800\pm0.0197$ &$0.0328\pm0.0008$ &$0.24\pm0.14$ &$0.31\pm0.21$ &$10463\pm295$ &$1.28\pm4.84$ &$0.22\pm0.22$\\
        
        $652.01$ &$0.7122\pm0.0486$ &$0.0186\pm0.0005$ &$0.32\pm0.17$ &$0.37\pm0.25$ &$1446\pm243$ &$-0.39\pm2.43$ &$126.18\pm4.12$\\
        
        $784.01$ &$0.4638\pm0.1139$ &$0.0196\pm0.0006$ &$0.19\pm0.11$ &$0.48\pm0.31$ &$1720\pm239$ &$0.22\pm3.37$ &$130.78\pm11.64$\\
        
        $836.01$ &$0.6251\pm0.1056$ &$0.0333\pm0.0013$ &$0.58\pm0.22$ &$0.56\pm0.25$ &$3553\pm727$ &$0.52\pm5.16$ &$0.01\pm1.14$\\
        
        $836.02$ &$0.6680\pm0.1269$ &$0.0240\pm0.0014$ & & & & &\\
        
        $1233.01$ &$0.4101\pm0.0619$ &$0.0280\pm0.0005$ &$0.22\pm0.04$ &$0.44\pm0.04$ &$2475\pm108$ &$-0.78\pm2.82$ &$160.04\pm7.03$\\
        
        $1233.02$ &$0.5459\pm0.0313$ &$0.0305\pm0.0006$ & & & & &\\
        
        $1233.03$ &$0.3663\pm0.0314$ &$0.0211\pm0.0005$ & & & & &\\
        
        $1233.04$ &$0.5612\pm0.0774$ &$0.0166\pm0.0006$ & & & & &\\
        
        $1233.05$ &$0.3459\pm0.0123$ &$0.0181\pm0.0008$ & & & & &\\
        
        \hline\hline 
    \end{tabular}
    \label{trasit_fit_posteriors_2}
\end{table*}

\begin{deluxetable*}{lcccccccccccc}
\setlength{\tabcolsep}{1.5pt} 
\renewcommand{\arraystretch}{0.95} 
\rotate
\tablecolumns{13}
\tabletypesize{\scriptsize}
 \tablecaption{Ancillary RV Fitting Parameters from \texttt{RadVel} \label{tab:radvel_ancilliary}}  
 \tablehead{
           \colhead{TOI System} & \colhead{P} & \colhead{$t_c$} & \colhead{$\sigma_{\rm{PFS}}$} & \colhead{$\gamma$} & \colhead{$\dot{\gamma}$} &\colhead{$\ddot{\gamma}$} & \colhead{$\sqrt{e}\cdot$sin($\omega$)} &	\colhead{$\sqrt{e} \cdot$cos($\omega$)}  & \colhead{GP$_{B}$} & \colhead{GP$_{C}$} & \colhead{GP$_{L}$} & \colhead{GP$_{\rm{Prot}}$}\\[-0.3cm]
             \colhead{}	& \colhead{d} & \colhead{d} &\colhead{m~s$^{-1}$} & \colhead{m~s$^{-1}$} & \colhead{m~s$^{-1}$~day$^{-1}$} &\colhead{m~s$^{-1}$~day$^{-2}$} & \colhead{} &	\colhead{}  & \colhead{mm$^{2}$~s$^{-2}$} & \colhead{} & \colhead{1/day} & \colhead{1/day}\\[-0.3cm]
            }
\startdata
134.01\tablenotemark{a}  &  1.401461 & 2458326.033176 &  $2.99^{+0.80}_{-0.57}$ & $0.35^{+0.82}_{-0.84}$ &   &  &  &  &  &  &  & \\
141.01 & 1.008035 & 2458325.538528 & $2.08^{+0.48}_{-0.36}$ & $0.02\pm 0.67$  & $-0.03\pm 0.01$ & &  &  &  &  &  &  \\
 non-transiting 4.78d  &  $4.7855^{+0.00097}_{-0.00098}$ & $2458396.749^{+0.073}_{-0.075}$  &  & &  &  &  &  &  &  &  &  \\
     \hspace{4pt} planet in TOI-141\tablenotemark{a}  &  &  &  &  &  &  &  &  &  &  &  &  \\
144.01 &  6.267852 & 2458519.8068 & $1.82^{+0.27}_{-0.23}$ & $-262.54\pm 0.82$  &  &   &   &   &   &   &   &   \\
    non-transiting 2088d  & $2088.80^{+0.51}_{-0.50}$ &$2458631.4^{+1.0}_{-0.98}$ &  &  &  & & $-0.40^{+0.004}_{-0.004}$ & $0.69^{+0.001}_{-0.001}$ &  &  &  &  \\
     \hspace{4pt} planet in TOI-144\tablenotemark{b}   &  &  &  &  &  &  &  &  &  &  &  &  \\
 174.01  & 17.667156 & 2459226.349919 &  $0.69^{+0.24}_{-0.29}$ & $0.40^{+2.4}_{-2.6}$ & &   &   &   & $26^{+50}_{-21}$ & $2.08^{+97000}_{-2.10} \times 10^{14}$ & $254^{+180}_{-220}$ & $30\pm 21$  \\
 174.02 & 29.797307 & 2459234.224619 &  &   &   &  &   &   &  &  &  &  \\
 174.03 (not in final fit) & 12.16099 & 2458335.751351&   &   &   &  &  &   &  &  &  &  \\
 175.01 & 3.690638 & 2458356.203945& $1.52^{+0.63}_{-0.42}$ & $2.38^{+0.77}_{-0.78}$ & $-0.06\pm0.07$ & $\sim 0$ &  &  &   &   &   &   \\
 175.02 & 7.450804 & 2458355.286650& &  &  &  &  &  &  &  &  &  \\
 175.03\tablenotemark{b}  & 2.253089 & 2458354.906006&  &  & &   &  &  &  &  &  &  \\
186.01  & 35.61253 & 2458350.313556 & $0.00^{+0.22}_{0.00}$& $0.83^{+1.59}_{-1.76}$ &  &  & $0.13^{+0.22}_{-0.27}$ & $-0.10^{+0.27}_{-0.24}$ & $23.08^{+14.31}_{-9.09}$ & $\sim 0$ & $51.95^{+29.58}_{-23.46}$ & $20.66^{+6.84}_{-6.62}$ \\
 186.02 & 7.787068 & 2458332.299222 &  &  &  &  &  &  &  &  &  &  \\
 260.01  & 13.47 & 2458392.2944 & $0.7^{+0.66}_{-0.48}$ & $0.6^{+2.9}_{-3.4}$ &  &  &  &  & $101^{+100}_{-48}$ &  $0.0081^{+340}_{-0.0081}$ & $58^{+28}_{-29}$  & $33^{+13}_{-14}$ \\
 402.01  & 4.7558 & 2458411.463484& $0.72^{+0.38}_{-0.45}$ &  $-0.9^{+1.2}_{-1.4}$  &   &  &   &   & $15.1^{+19.0}_{-7.5}$ &  $0.00027^{+2.8}_{-0.00027}$ &  $41^{+37}_{-26}$  &  $30.7^{+17.0}_{-6.5}$ \\
 402.02 & 17.214800 & 2458414.547596 &  &   &  &  &  &  &  &  &  &  \\
 431.01 (2pl)\tablenotemark{b}  & 12.459410 & 2458440.631380 & $1.87^{+1.0}_{-0.56}$& $1.1 \pm 1.3$ &  &   &   &   &   &  &   &    \\
 431.01 (1pl)\tablenotemark{b}  & 12.459410 & 2458440.631380 & $1.85^{+0.89}_{-0.51}$& $0.2^{+0.71}_{-0.73}$ &    &    &    &    &    &    & &    \\
 431.02 (not in final fit) & 0.490058 & 2458438.379781& &   &   &   &   &   &   &   &   &   \\
 455.01 & 5.358807 & 2458412.708786& $2.13^{+0.08}_{-1.60}$& $-0.11^{+0.74}_{-0.75}$ &   &   &   &   &   &   &   &   \\
562.01  & 3.930792 &  2458517.999665 & $1.8^{+0.41}_{-0.31}$ & $-0.9^{+0.45}_{-0.43}$ & & &   &   &   &   &   &   \\
    non-transiting 9.12d & $9.1248\pm 0.0022$&   $2458314.06^{+0.45}_{-0.41}$ &  &  & & &   &   &   &   &   &   \\
    \hspace{4pt} planet in TOI-562\tablenotemark{b}   &  &  &  &  &  &  &  &  &  &  &  &  \\
    non-transiting 55.96d & $55.68^{+0.54}_{-0.57}$ &  $2458322.7^{+5.4}_{-4.9}$ &  &  & & &   &   &   &   &   &   \\
    \hspace{4pt} planet in TOI-562\tablenotemark{b} &    &  &  &  &  &  &  &  &  &  &  &  \\
  652.01 &  3.984627 &   2458543.254775& $2.14^{+0.31}_{-0.26}$ & $-4.0^{+0.73}_{-0.59}$ & & &   &   &   &   &   &   \\
    non-transiting 1651d & $1647\pm 23$ &  $2457376^{+29}_{-31}$&  &  & & & $-0.17^{+0.35}_{-0.36}$ & $-0.18 \pm 0.35$ &   &   &   &   \\
    \hspace{4pt} planet in TOI-652\tablenotemark{b}  &    &  &  &  &  &  &  &  &  &  &  &  \\    
784.01 & 2.796723 & 2458570.224864&  $0.99^{+0.48}_{-0.31}$&  $-0.7\pm 0.5$  &  $0.66^{+0.097}_{-0.1}$ &  $0.014\pm 0.032$ &  &  &  &  &  &  \\
 836.01  & 8.593935 & 2458599.763960& $0.88^{+0.78}_{-0.61}$ &   $-0.5^{+2.8}_{-2.5}$ &  &  &  &  &  $54^{+70}_{-26}$  &  $6.5^{+11000}_{-6.5} \times 10^{-6}$  & $33^{+38}_{-20}$  & $21^{+24}_{-11}$  \\
 836.02& 3.817155 & 2458599.992303& &   &  &  &  &  &  &  &  &  \\
 1233.01\tablenotemark{a} (not in final fit)  & 14.175685 & 2458769.79682&  &   &  &   &  &  &  &  &  &  \\
 1233.02\tablenotemark{b}  & 19.590025 & 2458782.46935& $1.9^{+0.46}_{-0.37}$  & $3.37^{+0.91}_{-1.1}$ &  &  &0.03$^{+0.16}_{-0.17}$ & -0.60$^{+0.07}_{-0.04}$  &  &  &  &  \\
 1233.03 & 6.203449 & 2458770.90586&  &  &  &  &  &  &  &  &  &  \\
 1233.04 &                     3.795963 &   2458773.2964&  &  &  &  &   &   &   &   &   &  \\
 1233.05\tablenotemark{b}   &  29.541159 &   2458793.2786&  &  &  &  &   &   &   &   &   &   \\
      \enddata
 \begin{flushleft}
        \tablenotetext{a}{Median metric rank $>30$.}
         \vspace{-8pt}
       \tablenotetext{b}{Does not constrain the mass-radius relationship.}
        \end{flushleft}
    \end{deluxetable*}

\begin{deluxetable*}{lcccccccccccc}
\setlength{\tabcolsep}{1.5pt} 
\renewcommand{\arraystretch}{0.95} 
\rotate
\tablecolumns{13}
\tabletypesize{\scriptsize}
 \tablecaption{Ancillary RV Fitting Parameters from \texttt{juliet} \label{tab:juliet_ancilliary}}  
 \tablehead{
           \colhead{TOI System} & \colhead{P} & \colhead{$t_c$} & \colhead{$\sigma_{\rm{PFS}}$} & \colhead{$\gamma$} & \colhead{$\dot{\gamma}$} &\colhead{$\ddot{\gamma}$} & \colhead{$\sqrt{e}\cdot$sin($\omega$)} &	\colhead{$\sqrt{e} \cdot$cos($\omega$)}  & \colhead{GP$_{B}$} & \colhead{GP$_{C}$} & \colhead{GP$_{L}$} & \colhead{GP$_{\rm{Prot}}$}\\[-0.3cm]
             \colhead{}	& \colhead{d} & \colhead{d} &\colhead{m~s$^{-1}$} & \colhead{m~s$^{-1}$} & \colhead{m~s$^{-1}$~day$^{-1}$} &\colhead{m~s$^{-1}$~day$^{-2}$} & \colhead{} &	\colhead{}  & \colhead{mm$^{2}$~s$^{-2}$} & \colhead{} & \colhead{1/day} & \colhead{1/day}\\[-0.3cm]
            }
\startdata
134.01 & 1.401461  &  2458326.033176 & $3.00^{+0.78}_{-0.57}$ & $0.35^{+0.82}_{-0.81}$  & & & &  & &  &  & \\
141.01 &  1.007908 &   2458325.538528 & 3.04$^{+0.57}_{-0.46}$ & -0.07$^{+0.67}_{-0.68}$  & & & &  & &  &  & \\
 non-transiting 4.78d  &  4.78503 & 2458396.635& & & & &  & &  &  & & \\
  \hspace{4pt} planet in TOI-141\tablenotemark{a}   &  &  &  &  &  &  &  &  &  &  &  &  \\
  144.01 & 6.267852 &  2458519.8068& 1.78$^{+0.25}_{-0.20}$ & -0.110$^{+8.72}_{-8.63}$ &  -1.06$^{+0.005}_{-0.004}$ &  &   &   & 4.94$^{+3.12}_{-3.16} \times 10^{6}$ & 5.16$^{+3.03}_{-3.19} \times 10^{6}$ & $5015^{+3049}_{-3124}$ & $30\pm 18$ \\
  non-transiting 2088d  &  $2088^{+0.45}_{-0.42}$ & $2458631^{+0.85}_{-0.81}$&  &  &   &  & $-0.405\pm0.003$ & $0.694\pm0.003$ &  &  &  &  \\
     \hspace{4pt} planet in TOI-144\tablenotemark{a}   &  &  &  &  &  &  &  &  &  &  &  &  \\
     174.01  &  17.667156 &  2459226.349919& 1.34$^{+0.16}_{-0.14}$ & 0.33$^{+0.19}_{-0.20}$ &    &    &    &    &    &    &    &    \\
    174.02 & 29.797307& 2459234.224619&   &  &    &    &    &    &    &    &    &    \\
    174.03 (not in final fit) & 12.16099 & 2458335.751351&  &  &   &    &    &    &    &    &    &    \\
    175.01   &  3.690638& 2458356.203945& 3.07$^{+0.95}_{-0.66}$ & 0.27$^{+0.85}_{-0.88}$ &  & &   & &   &  &   &   \\
    175.02  & 7.450804 & 2458355.286650& &  &   &  &   &   &   &   &   &   \\
    175.03\tablenotemark{a}  &  2.253089 & 2458354.906006& &  &   &  &   &   &   &   &   &   \\
    186.01 &  35.607 & 2458350.313556& 0.23$^{+0.23}_{-0.15}$& -0.09$^{+1.72}_{-1.69}$ & &  &  &  & 5447$^{+2875}_{-2737}$ & 4.41$^{+3.43}_{-2.81}\times$10$^{6}$ & 5669$^{+2443}_{-2937}$ & 32$^{+17}_{-16}$ \\
    186.02 & 7.78706 & 2458332.299222&  &  &  &  &  &  &  &  &  &  \\
    260.01 & 13.470018 & 2458392.294409& $4.72^{+0.78}_{-0.61}$ & $0.85^{+0.78}_{-0.80}$ &  &  & $-0.11^{+0.58}_{-0.52}$ & $0.21^{+0.45}_{-0.78}$ &  &  &  &  \\
    402.01 &4.755833 &  2458411.463494& $2.26^{+0.30}_{-0.26}$ & $-0.31\pm0.36$ & & &  &  &  &  &  &  \\
    402.02  &  17.2148& 2458414.547596&  &  & & &  &  &  &  &  &  \\
     431.01 (2pl)\tablenotemark{a}  & 12.45941  & 2458440.63138 & 1.81$^{+0.94}_{-0.52}$& 0.31$^{+0.48}_{0.69}$ & & &  &  &  &  &  &  \\
    431.01 (1pl)\tablenotemark{a} &12.45941 & 2458440.63138& $1.74^{+0.76}_{-0.45}$& $0.15^{+0.49}_{-0.54}$ & & &  &  &  &  &  &  \\
    431.02 (not in final fit) &  0.490058 & 2458438.379781&  & & & &  &  &  &  &  &  \\
  455.01  & 5.358807 & 2458412.708786& 2.14$^{+0.84}_{-0.53}$ & $-0.09^{+0.72}_{-0.77}$ & & &   &   &   &   &   &   \\
    562.01  & 3.930792 &  2458517.999665 & 1.80$^{+0.39}_{-0.29}$ & $-0.88\pm0.41$ & & &   &   &   &   &   &   \\
    non-transiting 9.12d & $9.125\pm0.002$&   $2458314.07 \pm 0.42$ &  &  & & &   &   &   &   &   &   \\
    \hspace{4pt} planet in TOI-562\tablenotemark{a}  &  &  &  &  &  &  &  &  &  &  &  &  \\
    non-transiting 55.96d & $55.759^{+0.494}_{0.539}$ &   $2458322^{+5}_{-4}$ &  &  & & &   &   &   &   &   &   \\
    \hspace{4pt} planet in TOI-562\tablenotemark{a}&    &  &  &  &  &  &  &  &  &  &  &  \\
  652.01 &  3.984627 &   2458543.254774& $2.32^{+0.33}_{-0.27}$ & $-1.79^{+0.28}_{-0.15}$ & & &   &   &   &   &   &   \\
    non-transiting 1651d & $1633 \pm 22$ &   $2457315^{+27}_{-28}$&  &  & & & $-0.16^{+0.23}_{-0.21}$ & $-0.30^{-0.14}_{+0.10}$ &   &   &   &   \\
    \hspace{4pt} planet in TOI-652\tablenotemark{a} &    &  &  &  &  &  &  &  &  &  &  &  \\
    784.01&  2.796723 &  2458570.224864& $2.94^{+1.0}_{-0.62}$ & $-0.45^{+0.91}_{-0.91}$ & & &   &   &   &   &   &   \\
    836.01 & 8.593935 &  2458599.763960& $0.85^{+0.77}_{-0.53}$ & held at 0 & & &   &   &$ 1.65^{+1.11}_{-0.92} \times 10^{4}$ & $4.75^{+3.18}_{-3.02} \times 10^{6}$ & 5879$^{+2402}_{-2892}$ & 32$^{+16}_{-15}$ \\
    836.02& 3.817155 &  2458599.992303& &  & & &   &   &  &  &  &  \\
    1233.01\tablenotemark{a} (not in final fit) & 14.1761 &  24585571.3375 &  &  &  &  &   &   &   &   &   &  \\
    1233.02\tablenotemark{a} & 19.5917 & 24585586.5679& $2.15^{+0.54}_{-0.40}$&  $1.32^{+0.75}_{-0.79}$& & &  0.06$\pm0.16$ & -0.60$^{+0.13}_{-0.06}$  &  &  &  &  \\
    1233.03  &  6.20374&  24585572.3950&   &  & & &  0.02$^{+0.23}_{-0.22}$  & 0.02$\pm 0.21$ &   &   &   &   \\
    1233.04  &  3.79519 &   2458572.1138&  &  & & &  -0.04$^{+0.22}_{-0.23}$ & 0.01$^{+0.18}_{-0.19}$  &   &   &   &   \\
    1233.05\tablenotemark{a}  &  29.541159 &   2458793.2786&  &  & & & -0.05$\pm 0.21$  & 0.03$^{+0.20}_{-0.19}$  &   &   &   &   \\
      \enddata
 \begin{flushleft}
       \tablenotetext{a}{Not included in our mass-radius modeling.}
        \end{flushleft}

        \end{deluxetable*}

\section{TOIs 262, 271, and 560\label{sec:eliminated_tois}}

Three TOIs -- 262, 271 and 560 -- were observed under MTS as they were originally in our target list but later excluded. TOI-560 was found out to be a fast rotator with a period shorter than 15 days later in our survey as more ancillary data become available. A 12-day modulation is detected consistently across four seasons by WASP (see ExoFOP-TESS). The SAP light curve also shows a clear rotation signal. The RVs of TOI-560 are in Table \ref{tab:eliminate_TOI}, while the RV data of TOI-271 and TOI-262 will be published elsewhere. 

TOI-271 was initially ranked higher than 30 but dropped below our threshold as more TESS targets were released. In order to start the MTS observations as soon as possible, we selected 2--3 targets from each sector in the early days of TESS before Year 1 of the survey was finished. When new TOIs were released in later sectors, there were targets with higher merit scores according to our target selection function and thus TOI-271 fell below 30 in our ranking. 

TOI-262 was initially selected as an MTS target using the reported period when calculating its ranking merit. However, since it was only detected with two transit events in TESS Year 1 data, the reported period was ambiguous: the signals could be two transits of the same planet or one transit each of two similarly sized but distinct planets. With the TESS Year 3 data on this target, we performed a transit fit and determined its period to indeed be 11.15 days, the same as reported in the Year 1 data release. However, we only select targets based on the Year 1 data for the purpose of finalizing our target list as early as possible and also simplifying the completeness assessment in the future. TOI-262 is not a validated target based on the Year 1 data only, and thus we do not include TOI-262 in our sample. Moreover, in addition to TOI-262, ignoring the TESS Year 3 data will eliminate only 2--3 targets that are not fully vetted yet from our list of top 30 targets. 
 
    \begin{table}
        \centering
        \caption{RV data of TOI-560}
        \label{tab:eliminate_TOI}
        \begin{tabular}{lcccc} 
        TOI & JD-2450000 	&	RV & $\sigma_{\rm RV}$  &Note\\
        \hline
               & (- 2450000)& (\mps) & (\mps)&\\
        \hline 
        560	&8591.58957	&-3.92	&0.63	&PFS2 \\ 
        560	&8591.60409	&-4.24	&0.6	&PFS2 \\
        560	&8592.54942	&0.19	&0.72	&PFS2 \\
        560	&8592.56387	&0.39	&0.79	&PFS2 \\
        560	&8593.60634	&8.27	&0.69	&PFS2 \\
        560	&8593.62062	&3.57	&0.75	&PFS2 \\	 	
        ...&...&...&...&...\\
        \hline
        \end{tabular}
    \tablecomments{This table is published in its entirety in the machine-readable format. A portion is shown here for guidance regarding its form and content.}
    \end{table}


\end{document}